\documentclass[10pt,twocolumn,journal]{IEEEtran}
\usepackage{setspace}
\usepackage{amsmath}
\usepackage{algorithmic} 
\usepackage{graphicx}
\usepackage{epstopdf}
\usepackage{algorithm}
\usepackage{amssymb}
\usepackage{epsfig}
\usepackage{array}
\usepackage{subfigure}
\usepackage{bm}
\usepackage{color}
\usepackage{latexsym}
\usepackage{multirow}
\usepackage{diagbox}
\usepackage{float}
\usepackage{cite}
\usepackage{makecell}

\newcommand\norm[1]{\left\lVert#1\right\rVert}
\newcommand{\RNum}[1]{\uppercase\expandafter{\romannumeral #1\relax}}

\def\swone{0.99\linewidth}
\def\swtwo{0.49\linewidth}

\def\swfour{0.25\linewidth}
\def\emptsp{-0.06in}
\def\mS{\mathcal{S}}
\def\mI{\mathcal{I}}
\def\mP{\mathcal{P}}
\def\mM{\mathcal{M}}
\def\mO{\mathcal{O}}
\def\360{$360^\circ$}

\begin{document}
% can use linebreaks  \\ within to get better formatting as desired
\title{Landmarking for Navigational Streaming of Stored High-Dimensional Media}

\author{Yuan Yuan~\IEEEmembership{Member,~IEEE},
Gene Cheung~\IEEEmembership{Fellow,~IEEE},
Pascal Frossard~\IEEEmembership{Fellow,~IEEE},\\
H. Vicky Zhao~\IEEEmembership{Senior Member,~IEEE},
Jiwu Huang~\IEEEmembership{Fellow,~IEEE}

\begin{small}

\thanks{This work was supported in part by the Guangdong Natural Science Foundation (2020A1515110781), and in part by National Science Foundations of China (U19B2022, U1636202, 61701310). 
}

\thanks{Y. Yuan is with School of Computer Science, Guangdong Polytechnic Normal University (e-mail:yuanyustc@hotmail.com)}

\thanks{G. Cheung is with the Department of EECS, York University, Toronto, Canada, M3J 1P3 (e-mail: genec@yorku.ca).}

\thanks{P. Frossard is with Signal Processing Laboratory (LTS4), Ecole Polytechnique F\'{e}d\'{e}rale de Lausanne (EPFL), CH-1015 Lausanne, Switzerland (e-mail: pascal.frossard@epfl.ch).}

\thanks{V. Zhao is with Department of Automation, Tsinghua University, State Key Lab of Intelligent Technologies and Systems, Tsinghua National Laboratory for Information and Science and Technology (TNList), Beijing, P.R.China (e-mail: vzhao@tsinghua.edu.cn).}

\thanks{Jiwu Huang is with the Guangdong Key Laboratory of Intelligent Information Processing and Shenzhen Key Laboratory of Media Security, Shenzhen University, Shenzhen 518060, China, and also with the Shenzhen Institute of Artificial Intelligence and Robotics for Society, Shenzhen 518172, China (e-mail: jwhuang@szu.edu.cn, \textit{corresponding author}).}

\thanks{Copyright©20xx IEEE. Personal use of this material is permitted. However, permission to use this material for any other purposes must be obtained from the IEEE by sending an email to pubs-permissions@ieee.org.}

\end{small}
}

\maketitle

\begin{abstract}
Modern media data such as $360^\circ$ videos and light field (LF) images are typically captured in much higher dimensions than the observers' visual displays.
To efficiently browse high-dimensional media, a navigational streaming model is considered: 
a client navigates the media space by dictating a navigation path to a server, who in response transmits the corresponding pre-encoded media data units (MDU) to the client one-by-one in sequence. 
Assuming that the MDU quality is pre-chosen and fixed, the problem resides in selecting and storing redundant representations of MDUs at the server in order to best trade off storage and transmission costs, while enabling adequate user's random access.
We address this problem with a landmark-based MDU optimization framework.
The media space is divided into neighborhoods, each containing one landmark (a chosen MDU). 
MDUs in a neighborhood use the associated landmark as a predictor for inter-coding.
Thus, for any MDU transition within the same neighborhood, only one inter-coded MDU transmission is required when the landmark is already in the decoder buffer.
It results in lower transmission cost and enables navigational random access.
To optimize an MDU structure, we employ tree-structured vector quantizer (TSVQ) to first optimize landmark locations, then iteratively add P-MDUs as refinements using a fast branch-and-bound technique.   
Taking interactive LF images and viewport adaptive $360^\circ$ images as illustrative applications, 
%and I-, P- and previously proposed merge frames to intra- and inter-code MDUs,
and I-, P- and previously proposed merge frames to intra- and inter-code MDUs,
we show experimentally that landmarked MDU structures can noticeably reduce the expected transmission cost compared with MDU structures without landmarks.

%
%To address this fundamental compression problem in navigational streaming, we propose landmarking---the selection of key sparse samples from the high-dimensional media, from which nearby MDUs in local neighborhoods are differentially coded, resulting in a predictive structure that facilitates MDU navigation.
%Landmarking means that any requested MDU can be decoded by transmitting only a landmark and an inter-coded MDU in the worst case.
%In the typical case when the landmark is in the decoder buffer, only one inter-coded MDU is required.
%To build a landmarked MDU structure, we employ tree-structured vector quantizer (TSVQ) to first optimize landmark locations, then greedily add/subtract inter-coded MDUs using a fast branch-and-bound technique.   
%Using interactive LF images and viewport adaptive $360^\circ$ images as example applications and I-, P- and previously proposed merge frames to intra- and inter-code MDUs, we show experimentally that landmarked MDU structures can enable noticeable lower expected transmission cost compared with MDU structures without landmarks.

%, given the same storage cost at the server.
%save up to $26.92\%$ and $71.85\%$ storage bits for the same expected transmission cost.
% have noticeably lower expected transmission cost for the same storage than structures generated by a previous greedy algorithm.

\end{abstract}


\IEEEpeerreviewmaketitle

\vspace{\emptsp}
\section{Introduction}
\label{sec:intro}
\begin{figure}[htb]
\centering
\centerline{\includegraphics[width=\swone]{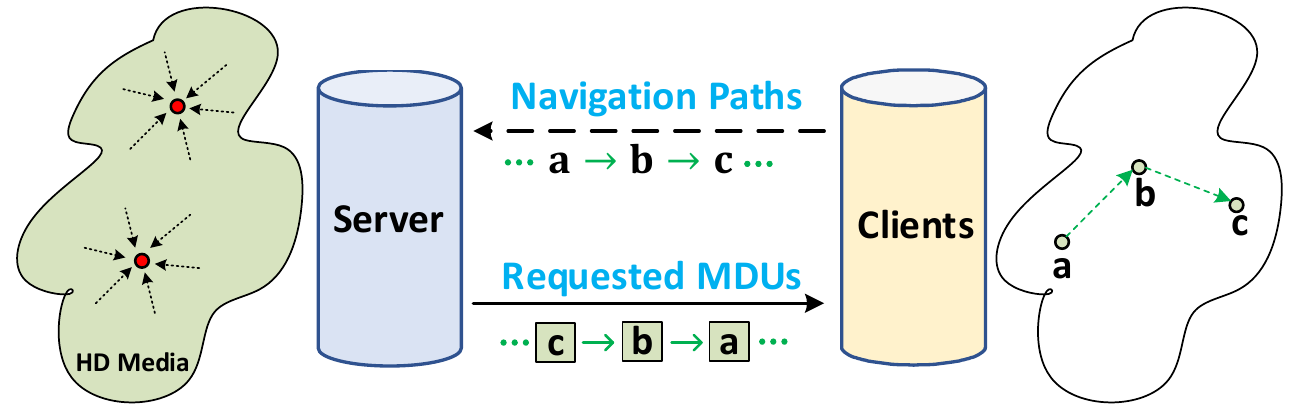}}
%\vspace{-0.1in}
\caption{Navigational streaming system for stored HD media. 
A server pre-encodes and stores MDUs locally. 
When a client dictates a navigation path to the server, the server then in response transmits the corresponding requested MDUs to the client one-by-one in sequence.
The red dots on the server-side are two example landmark MDUs.
}
\label{fig:sysModel}
\end{figure}
The rapid advance in camera technologies means that high-dimensional (HD) media can now be captured cheaply and are widely available for mass consumption.
In contrast, physical visual displays usually exist in much lower dimensions, and as a result, users typically consume HD media on these displays via a \textit{media navigation} model.
For example, \textit{light field} (LF) cameras such as Lytro\footnote{Lytro illum: https://illum.lytro.com/} can capture multiple viewpoint LF images of a 3D scene~\cite{rerabek2016new}, while a user browses these viewpoint images on a conventional 2D display monitor one-by-one in successive order.
Similarly, an omnidirectional camera rig can produce an immersive \360 video~\cite{tan2018360}, but an observer commonly wears a head-mounted display to observe one \textit{field-of-view} (FoV) of the $360^\circ$ sphere (also called a \textit{viewport}) at a given time as she/he rotates his head.

A na\"{i}ve approach is to download the entire HD media content from a server before user navigation.
However, it would incur a large start-up delay and a waste of network resources. 
An alternative strategy is \textit{navigational streaming}:  download the desired HD media portion-by-portion successively per request.
As shown in Fig.\;\ref{fig:sysModel}, a client navigates the media space by dictating a chosen \textit{navigational path}---a sequence of desired \textit{media data units} (MDU), \textit{e.g.}, viewpoint images in LF and viewports in $360^\circ$ videos---to the server.
In response, the server transmits the corresponding requested MDUs to the client one-by-one in sequence. 

Navigational streaming has been studied extensively for specific media types.
For multiview videos~\cite{cheung09pv,cheung09icip,cheung11tip},
a suitable set of camera views are interactively transmitted for playback or view synthesis of a 3D scene.
Similarly, works in \cite{aaron04,cai11,cai12,motz16icip} propose an interactive LF streaming framework, where a user periodically requests a desired view, and in response a server transmits a pre-synthesized and pre-encoded viewpoint image to the user for observation. 
Orthogonally, works in \cite{zhang2018cooperative,shafi2020360,hu2020tvg,maniotis2021tile,
pyramid2017facebook,van2017AHG8,van2020viewport}
investigate viewport adaptive \360 video streaming, where the content of the requested viewport is transmitted to a user as FoV changes.

To facilitate navigational streaming, 
we consider a server that pre-encodes and stores MDUs locally to satisfy subsequent client streaming requests~\cite{zhang2018cooperative}\cite{pyramid2017facebook}.
In general, \textit{intra-coding} an MDU (I-MDU) incurs a large coding cost\footnote{By coding cost, we mean the coding bitrate.}, but I-MDU can be randomly accessed.
\textit{Inter-coding} an MDU (P-MDU) using another MDU as a predictor incurs a small coding cost, but imposes an order where the predictor must be first transmitted and decoded.
The coding costs of MDUs directly affect the transmission costs during streaming.
Given that at encoding time clients' navigation paths are unknown, the technical challenge is:
how to reduce transmission cost via inter-coding, while enabling enough random access for satisfactory user navigation. 
One brute-force approach is to inter-code all MDU pairs to satisfy all possible MDU switch requests.
However, for an HD media containing $N$ MDUs, this means $\mathcal{O}(N^2)$ inter-coded MDUs pre-encoded in the server, resulting in a large storage cost.

In this paper, to control the storage cost of inter-coded MDUs while facilitating random access, we propose a landmark-based MDU optimization framework with \textit{redundant representation}.  
By redundant, we mean that each MDU can be coded \textit{both} as I-MDU \textit{and} one or more P-MDUs.
The media space is divided into neighborhoods, each containing one landmark (a chosen MDU in a neighborhood).
MDUs in a neighborhood use the associated landmark as a predictor/target for inter-coding.
Clearly, smaller neighborhoods mean smaller P-MDU sizes.

A landmark operates like an airline hub in commercial aviation: 
by creating flights to/from a designated hub for all cities---$\mathcal{O}(2N)$ flights for $N$ cities---a passenger can travel from any city to any other via only two flights (one connecting flight at the hub).
Similarly, any MDU can transition to another in the same neighborhood by only two/one P-MDU transmissions, depending on whether the landmark is available in the decoder buffer. 
The number of stored inter-coded MDUs is then reduced to $\mathcal{O}(N)$ by using landmarking. 
However, when an MDU transitions to another in a different neighborhood, a new landmark or I-MDU must be transmitted, resulting in a larger transmission cost. 
Thus, the optimization of landmarks and neighborhoods for a given media space is essential.

We solve the design problem for landmarks and neighborhoods as follows.
We first derive recursive equations to compute the expected transmission cost given a stored MDU structure. 
We then use a \textit{tree-structured vector quantizer} (TSVQ) method~\cite{franti1997} to recursively select a set of landmarks and their associated neighborhoods. 
We initialize an MDU structure by inter-coding each MDU using its closest landmark as a predictor, and inter-coding to/from any two landmarks. 
We next iteratively add inter-coded MDUs using a fast branch-and-bound (B$\&$B) method \cite{narendra1977branch} to refine the MDU structure.
Taking interactive LF images and viewport adaptive $360^\circ$ images as illustrative applications, 
and using I-, P- and previously proposed merge (M-) frames~\cite{dai2016merge} to encode MDUs, 
we show experimentally that landmark-based MDU structures can noticeably reduce the expected transmission cost compared to MDU structures without landmarks.

In summary, the main contributions of our paper include:
\begin{enumerate}
\item We propose landmarking to build an MDU structure with redundant representation, which can achieve compression gain via inter-coding while enabling adequate random access. 
\item We derive recursive equations to compute the expected transmission rate for a given MDU structure, and formulate the MDU structure design problem that best trades off storage cost and the expected transmission rate.
\item We combine TSVQ and Lloyd's~\cite{vq92} algorithms to find locally optimal landmark locations and their associated local neighborhoods. 
\item We employ a fast B$\&$B algorithm to refine the initialized MDU structure.
\item We demonstrate the efficiency of our abstract MDU framework for two concrete applications: interactive LF images and viewport adaptive \360 images.
\end{enumerate}

The outline of the paper is as follows. 
We first overview related works in Section \ref{sec:related}. 
We then discuss the navigational streaming system in Section \ref{sec:overview}.
We compute the expected transmission cost for a given MDU structure and formulate our MDU structure design problem in Section \ref{sec:trans}. 
We optimally insert landmarks in Section~\ref{sec:landmark} and discuss a structure design algorithm in Section~\ref{sec:algorithm}. 
Implementation details and results are discussed in Section \ref{sec:details} and \ref{sec:results}, respectively.
We finally conclude the paper in Section \ref{sec:conclude}.

\vspace{\emptsp}
\section{Related Work}
\label{sec:related}
%\vspace{-0.1in}
\subsection{Interactive Light Field Streaming (ILFS)}

ILFS was first studied in \cite{aaron04}\cite{chang2004rate}\cite{ramanathan07}.
In \cite{aaron04}, new switching mechanisms to adjacent views based on Wyner-Ziv coding were proposed. However, the navigation model only permitted switching to horizontal and vertical adjacent views.
We discuss a more general view navigation model in Section \ref{sec:details}. 
\cite{chang2004rate}\cite{ramanathan07} proposed rate-distortion optimized ILFS, where bitstream packetization and packet scheduling for LF streaming were studied.
Several works \cite{tong2003interactive, mehajabin2019efficient, verhack2019steered} also focused on improving compression efficiency while facilitating random access to LF images.
These works are orthogonal to our work, where we optimize redundant MDU representation to trade off expected transmission rate with storage cost.

Two recent studies~\cite{cai11}\cite{motz16icip} focused on the use of~\textit{distributed source coding} (DSC) frames~\cite{mcheung09pcs} and merge frames~\cite{dai2016merge} for view-switching without coding drift. 
\cite{cai12} discussed the general notion of landmark to be used in ILFS but did not provide a formal optimization framework. 
In our previous work~\cite{yuan2017icip}, we dealt with this problem by greedily adding landmarks one at a time. 
To avoid undesirable local minima, in this paper we adopt a TSVQ method to recursively find optimal landmarks. 
Further, instead of modeling the lifetime of an ILFS session as a fixed constant, we assume more generally that the lifetime follows a Poisson distribution.   

\vspace{-0.1in}
\subsection{Interactive Multi-view Video Streaming}

The MDU structures generated by our design can also be
used for interactive multi-view video streaming (IMVS) \cite{cheung09pv,cheung09icip,cheung11tip}.
Given multiview video captured by multiple closely spaced cameras, users can send requests periodically to a server to switch to different views while streaming video is played back uninterrupted in time. 
The work in \cite{cheung09pv} optimized IMVS frame structures by appropriately selecting I- and redundant P-frames. 
Later, works in \cite{cheung09icip}\cite{cheung11tip} investigated how DSC frames can improve compression performance. 
Though the notion of interactive view navigation is similar, we study the more general media navigation problem in the abstract. 
Unlike IMVS, an LF image or $360^\circ$ viewport can be revisited, creating possible loops in a navigation path and making the MDU design problem more challenging.

\vspace{-0.09in}
\subsection{$360^\circ$ Video Streaming}
To address the high bandwidth requirements of $360^\circ$ video streaming, viewport adaptive streaming was studied recently. 
We categorize previous works into two groups: i) \textit{tile-based} streaming, and ii) \textit{viewport-based} streaming.

\begin{figure}[htb]
\centering
\centerline{\includegraphics[width=7.2cm]{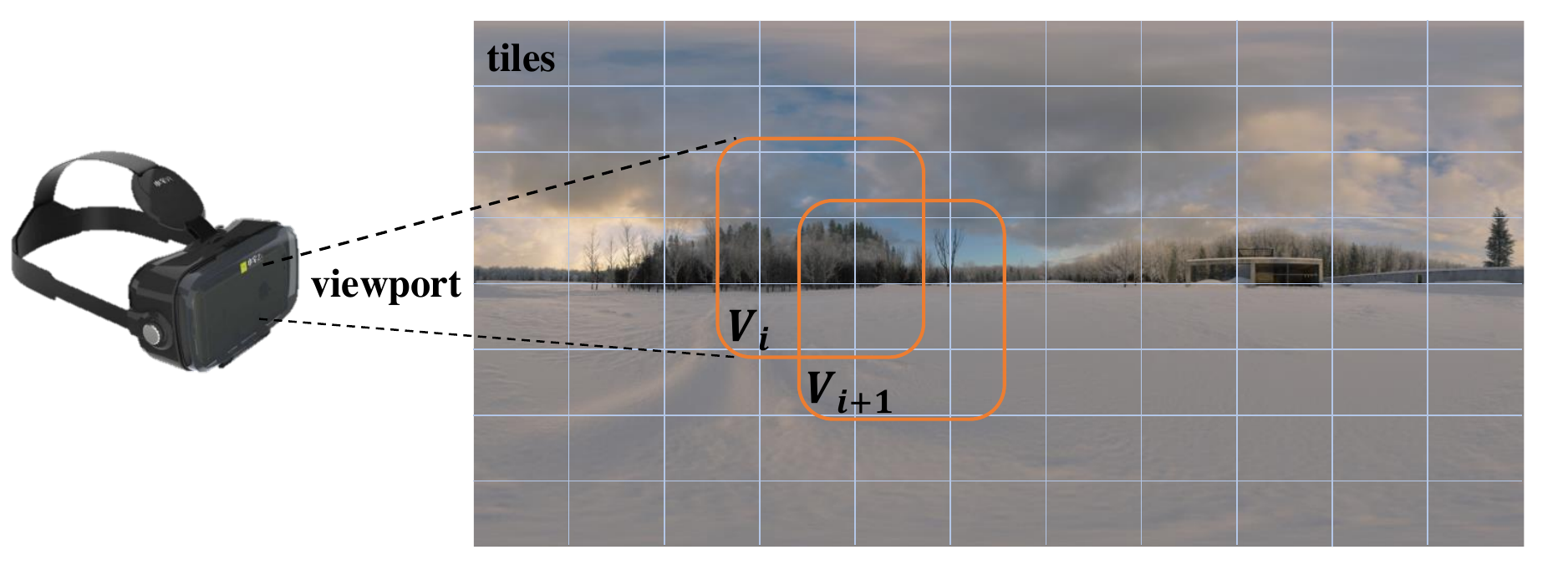}}
\vspace{-0.15in}
\caption{Visualization of tiles and viewports. Each viewport spans multiple tiles.}
\label{fig:visual}
\end{figure}

In tile-based streaming \cite{zhang2018cooperative,shafi2020360,hu2020tvg,maniotis2021tile}, 
the \360 sphere is spatially partitioned into non-overlapping tiles, each of which is pre-encoded independently of other tiles.
When a viewport $V_i$ is requested by a user, the server transmits a subset of tiles that compose the requested viewport. 
If an adjacent viewport $V_{i+1}$ is requested next (typically overlapping with the original $V_i$), only tiles that compose the newly exposed spatial region $V_{i+1} \setminus V_i$ are additionally transmitted. 
See Fig.\;\ref{fig:visual} for an illustration.
While the view-switching cost is low, the encoding of a viewport worth of visual data is inefficient, due to the lack of inter-tile coding to exploit spatial correlation among tiles.
%\vspace{-0.1in}

For viewport-based streaming \cite{pyramid2017facebook,van2017AHG8,van2020viewport}, a \360 sphere is projected to different \textit{overlapping} viewports. 
Each viewport is encoded independently from other viewports to facilitate switching from one viewport to another.
While the coding of a single viewport-worth of spatial information is efficient (thanks to intra-prediction within one viewport), the view-switching cost is high---the overlapped region between the two neighboring viewports, $V_i \cap V_{i+1}$ needs to be transmitted again due to independent coding of viewports.

In contrast, our proposed MDU-based navigational streaming framework combines the advantages of both tile-based streaming and viewport-based streaming.
We encode one viewport-worth of data as one MDU, thus achieving good coding efficiency due to intra-prediction.
At the same time, we achieve low view-switching cost thanks to the inter-coding of neighboring MDUs.
The price is the redundant representation---a given spatial region is encoded more than once into multiple MDUs. 
We formally optimize the tradeoff between expected transmission rate and storage cost in our framework. 

\vspace{\emptsp}
\section{System Overview}
\label{sec:overview}
We first overview our navigational streaming system of stored HD media and briefly introduce landmarking.
We then present an MDU navigation model that captures a typical user's navigation behaviors. 
Finally, we describe different MDU representation types in our coding structure.

\vspace{-0.1in}
\subsection{System Model}

The navigational system of stored media is shown in Fig.\;\ref{fig:sysModel}. 
A media is abstractly represented by a collection of atomic units called MDUs, \textit{e.g.}, viewpoint images in LF, and viewports in $360^\circ$ videos.
A server pre-encodes these MDUs offline into an MDU structure $\Theta$ with redundant representation, where each MDU can be coded into intra-coded I-MDU, one or more inter-coded P-MDUs and merge-coded M-MDU (to be discussed in Section~\ref{sec:merge}).
The server stores these redundantly coded representations locally, using which it serves the streaming clients per request subsequently. 
This encoding results in a storage cost at the server.

A client navigates the media space by interactively requesting MDUs one-by-one from the server. 
In response, the server transmits each requested MDU to the client in one of three transmission types: 0-hop, 1-hop or 2-hop (to be discussed in Section~\ref{sec:trans}), resulting in different transmission costs. 
In general, more P-MDUs stored at the server means lower expected transmission rate for user navigation.
Assuming that MDUs are required by applications to be encoded in high visual quality (thus a tradeoff between rate and distortion is not possible), the problem is to find an optimal MDU structure $\Theta^*$ pre-encoded at the server---optimally trading off the total storage cost with the expected transmission rate.

\vspace{-0.1in}
\subsection{Landmarking}
\label{sec:lmkr}
\vspace{-0.05in}
\begin{figure}[htb]
\centering
\centerline{\includegraphics[width=3.7cm]{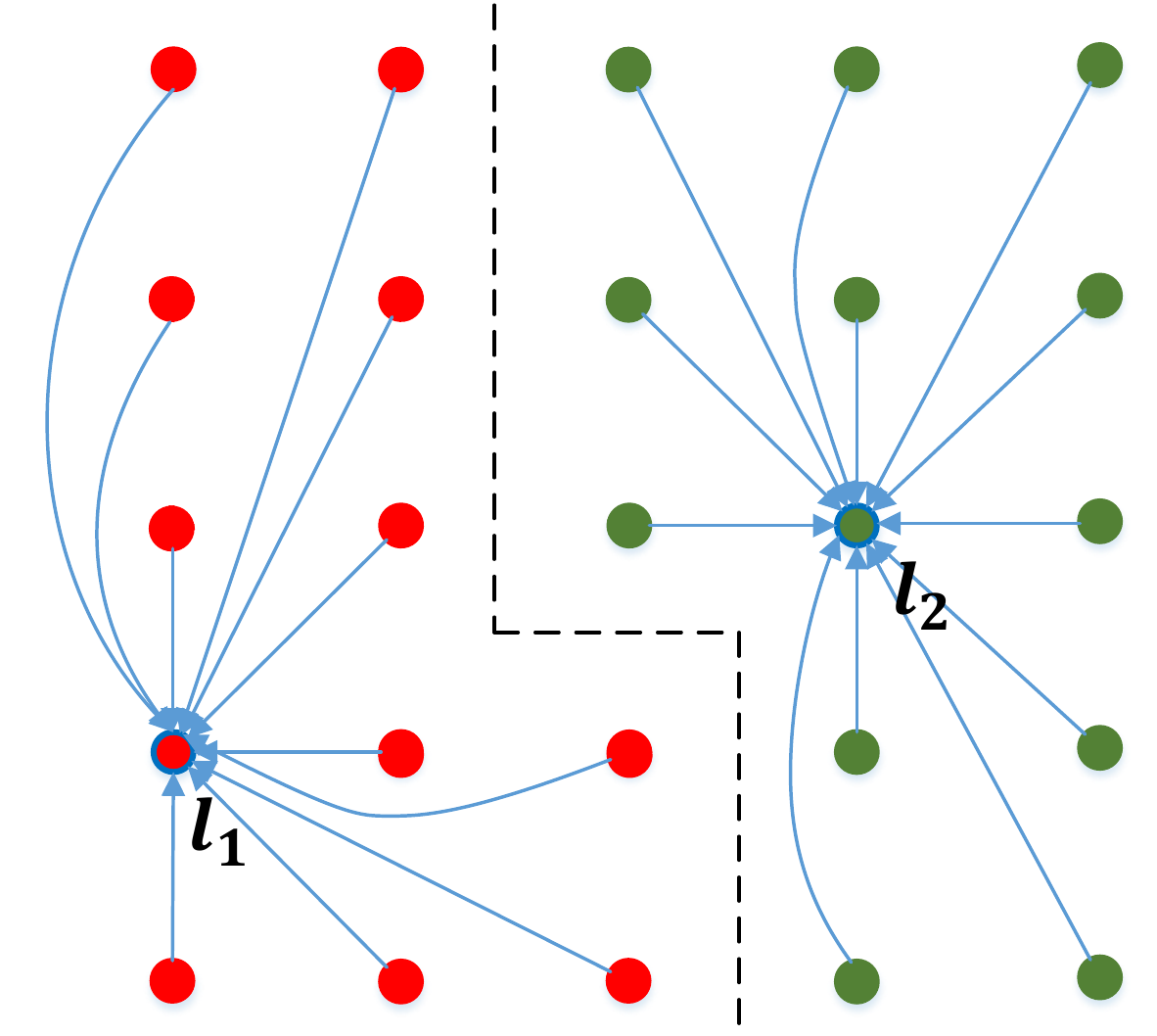}}
\vspace{-0.1in}
\caption{Two key images $l_1$ and $l_2$ are selected as landmarks among a $5 \times 5$ 2D grid of sub-aperture LF images, which are divided into two neighborhoods. 
Arrows mean the neighborhood images are inter-coded using the landmark inside as a predictor.}
\label{fig:landmark}
\end{figure}
A landmark is a chosen MDU with connections to MDUs in a local neighborhood, and is used as an intermediary to a destination MDU. 
As an illustration, Fig.\;\ref{fig:landmark} shows an example of landmarking for a 2D grid of sub-aperture images (MDUs) in LF imaging. 
In this case, sub-aperture images are divided into two neighborhoods.
In each neighborhood, a key image is selected as a landmark, which is used as a predictor to inter-code MDUs in the neighborhood.
%In this case, sub-aperture images are divided into two neighborhoods, where landmarks are used as predictors to inter-code MDUs in each neighborhood.
Assuming the landmark is in the decoder buffer, a sub-aperture image $i$ can switch to any other image $j$ in the same neighborhood by decoding an inter-coded MDU $P_j$ using the landmark as a predictor.

\vspace{-0.1in}
\subsection{MDU Navigation Model}
\label{sec:viewModel}
%\vspace{-0.1in}
Denote by $N$ the number of MDUs in a media space. 
We assume that a client, starting a streaming session at time $t = 0$ at a given initial MDU $s$, switches MDUs at each discrete time instant until a lifetime of $T$ MDU-switches are performed, upon which he exits the session. 
As often done in lifetime modeling, we assume random variable $T$ follows a Poisson distribution; \textit{i.e.}, the probability of $T = m$ MDU-switches is
\begin{eqnarray}
p(T = m) = \frac{\mu^m e^{-\mu}}{m!}
\label{eq:Poisson}
\end{eqnarray}
where $\mu$ is the expected lifetime of a navigational streaming session and $m!$ is $m$ factorial. 
The Poisson probability will be very small when $m$ becomes large.
To avoid a heavy computation load, we define a large $T_{\max}$ as the maximum lifetime, beyond which the lifetime probability will be negligible.
 
The probability $g(t)$ that there are at least $t$ MDU-switches in a navigational streaming session can be defined as
\begin{eqnarray}
\label{eq:ltProb}
g(t) & = & \mathbf{1}(t \leq T_{\max})\sum_{m = t}^{T_{\max}} p(T = m) \\ \nonumber
&=&  \mathbf{1}(t \leq T_{\max}) e^{-\mu} \sum \limits_{m=t}^{T_{\max}} \frac{\mu^m}{m!}
\end{eqnarray}
where $\mathbf{1}(x)$ is an indicator function that equals to 1 if clause $x$ is true and 0 otherwise. 
In other words, $g(t)$ is the probability that the lifetime $T$ is no smaller than $t$.

To capture a user's tendency to select the same navigation direction in consecutive time instants, we assume a \textit{1-step memory} user behavior model to assign a probability for each feasible MDU-switch. 
Depending on previous MDU $k$, denote by $p_{k,i,j}$ the probability of switching from current MDU $i$ to MDU $j$.
%(shown as green arrows in Fig.\;\ref{fig:2Dgrid}).
Alternatively, in a memoryless behavior model, denote by $p_{s,j}$ the probability of switching from starting MDU $s$ to MDU $j$.
Probabilities $p_{s,j}$ and $p_{k,i,j}$ are used to compute the expected transmission rate in Section\;\ref{sec:trans}.
Their specific definitions are application-dependent; we present our definitions in Section\;\ref{sec:details}.

\vspace{-0.1in}
\subsection{Representation Types in Coding Structure} 
\label{sec:merge}

Let first consider the case where an MDU $j$ has an independently intra-coded representation (I-MDU), denoted by $I_j$.
An I-MDU does not require any predictor MDU for decoding.
To more efficiently facilitate an MDU-switch from $i$ to $j$, MDU $j$ may contain a predictively inter-coded representation (P-MDU), denoted by $P_j(i)$, which uses intra-coded $I_i$ of MDU $i$ as a predictor. 
Thus, $P_j(i)$ is transmitted only if the client has $I_i$ in his decoder buffer. 
Note that a ``further-away" MDU $j$ from MDU $i$ (\textit{e.g.}, an image viewpoint $j$ that is further from $i$ in LF, or a spatial viewport $j$ that is further from $i$ in $360^\circ$ videos) entails a bigger prediction residual, and hence a larger coding bit-rate for $P_j(i)$.

MDU $j$ may subsequently be used as a predictor for future MDU-switches. 
In general, MDU $j$ may be reconstructed from different inter-coded $P_j(i)$'s, stemming from MDU-switches from different MDUs $i$. 
The reconstructed P-MDU $P_j(i)$'s for different $i$ differ slightly due to transform domain quantization of different prediction residuals. 
Hence, to avoid \textit{coding drift} in the following prediction MDU sequence, we require an \textit{identical} MDU reconstruction from these different $P_j(i)$'s.

\begin{figure}[htb]
\centering
\centerline{\includegraphics[width=5.5cm]{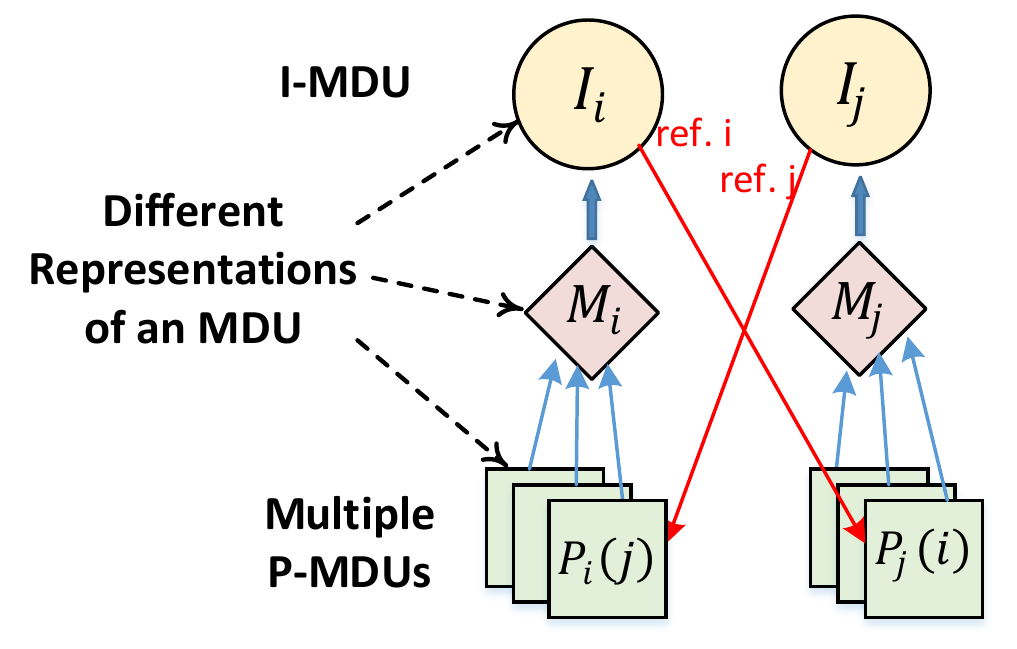}}
\vspace{-0.1in}
\caption{Example for MDU representation types of MDU $i$ and $j$, with I-(circle), P-(square) and M-MDUs (diamonds). Multiple P-MDUs $P_j(i)$ are merged identically to the I-MDU $I_j$ via M-MDU $M_j$.}
\label{fig:frameStr}
\end{figure}

To accomplish this, a \textit{merge}-coded MDU (M-MDU) $M_j$ is employed to ``merge" different reconstructions of $P_j(i)$ identically to I-MDU $I_j$.
Many previous works have studied the merge operator, including SP-frame in H.264~\cite{karczewicz03}, distributed source coding (DSC) in~\cite{mcheung09pcs}, and the merge frame in~\cite{dai2016merge}.
We discuss implementation of our M-MDU based on \cite{dai2016merge} in Section\;\ref{subsec:merge}.
The size of a merge frame by \cite{dai2016merge} is roughly 3-4 times of an inter-coded frame, depending on the media content and quantization parameters.
Thus, an M-MDU $M_j$ \textit{plus} any decoded P-MDU $P_j(i)$ will result in an identically reconstructed $I_j$.
See Fig.\;\ref{fig:frameStr} for an illustration.
Because of its usefulness, we assume that $M_j$ is pre-encoded for every $j$ by default in our structure.

\vspace{-0.03in}

Thus, when a client requests switching from MDUs $i$ to $j$, the server can either transmit an I-MDU $I_j$, or a P-MDU $P_j(i)$ \textit{plus} an M-MDU $M_j$ if I-MDU $I_i$ is available in the decoder buffer. 
\textit{The challenge is how to select an appropriate set of I- and P-MDUs for pre-encoding in the MDU structure, given the server's storage limitation.} 
We focus on this problem in the sequel. 
We first describe how to compute the expected transmission cost given an MDU structure $\Theta$. 
We then discuss how to insert ``landmarks" into an MDU structure $\Theta^*$, facilitating MDU-switches.

\vspace{\emptsp}
\section{Expected Transmission Cost}
\label{sec:trans}
We derive recursive equations to compute the expected transmission cost of a media navigational streaming session for a given MDU structure $\Theta$, assuming a fixed or flexible decoder buffer in order.
We then formulate the MDU structure design problem.

\vspace{-0.1in}
\subsection{Transmission Cost for Fixed One-MDU Buffer}
\label{sec:fixed}
To keep the computation of the expected transmission cost tractable, we make different simplifying assumptions about the size and function of the client's decoder buffer.
For intuition, we first assume a \textit{fixed} one-MDU buffer: the size of the reference buffer is a single MDU. 
In practice, a user's reference buffer can be larger, and thus we are computing an upper bound of the expected transmission cost with this assumption.
Our optimization is thus to minimize a mathematically tractable upper bound of the actual transmission cost. 

A fixed one-MDU reference buffer means that the currently displayed MDU is always stored in the buffer as a reference for the next MDU-switch. 
When a user observing MDU $i$ switches to MDU $j$, we consider two different transmission types: \textit{0-hop} and \textit{1-hop} transmissions. 
0-hop transmission means that the server transmits an independent reconstruction of MDU $j$, regardless of what MDU is stored at a user's buffer.
For example, it could be an I-MDU $I_j$ or an I-MDU $I_\gamma$ 
\textit{plus} a P-MDU $P_j(\gamma)$ \textit{plus} an M-MDU $M_j$. 
The overhead is marked as $r_j^I$.
1-hop transmission means that considering the buffered MDU as a predictor, a P-MDU $P_j(i)$ is transmitted along with an M-MDU $M_j$, resulting in an overhead $r_j^P(i)$.

\subsubsection{Expected Transmission Cost}

Given an MDU structure $\Theta$, with a fixed one-MDU buffer, the expected transmission cost $c_{i|k}^{(t)}$ for a client who is currently at MDU $i$ at time instant $t$ and previously at MDU $k$ can be written as:
\begin{eqnarray}
c_{i|k}^{(t)} = \sum \limits_{j\in \mathcal{N}(i)} p_{k,i,j} \; \min \left[h_i^{(t)}(j), \dot{h}_i^{(t)}(j) \right].
\label{eq:objFiexed}
\end{eqnarray}
where $j\in \mathcal{N}(i)$ is the set of MDUs in $i$'s neighborhood to which a user can switch.
$p_{k,i,j} $ is the MDU-switch probability from MDUs $i$ to $j$ given previous MDU $k$. 
$h_i^{(t)}(j)$ and $\dot{h}_i^{(t)}(j)$ are the costs of 0-hop and 1-hop transmissions, respectively. 

The 0-hop transmission cost $h_i^{(t)}(j)$ is the sum of coding cost $r_j^I$ \textit{plus} the recursive cost $c_{j|i}^{(t+1)}$ at MDU $j$ at instant $t+1$, if the maximum lifetime $T_{\max}$ has not been reached. 
We write $h_i^{(t)}(j)$ as
\begin{eqnarray}
h_i^{(t)}(j) = r_j^I + g(t+1) \; c_{j|i}^{(t+1)}.
\label{eq:0hopFixed}
\end{eqnarray}

The 1-hop transmission cost can be written recursively as the sum of differential coding cost $r_j^P(i)$ plus the future cost $c_{j|i}^{(t+1)}$:
\begin{eqnarray}
\dot{h}_i^{(t)}(j) =  r_j^P(i) + g(t+1)\;  c_{j|i}^{(t+1)}.
\label{eq:1hopFixed}
\end{eqnarray}
The exact definitions of $r_j^I$ and $r_j^P(i)$ depending on particular applications; we describe ours in Section~\ref{sec:details}.

\vspace{-0.1in}
\begin{figure}[htb]
\centering
\centerline{\includegraphics[width=7.0cm]{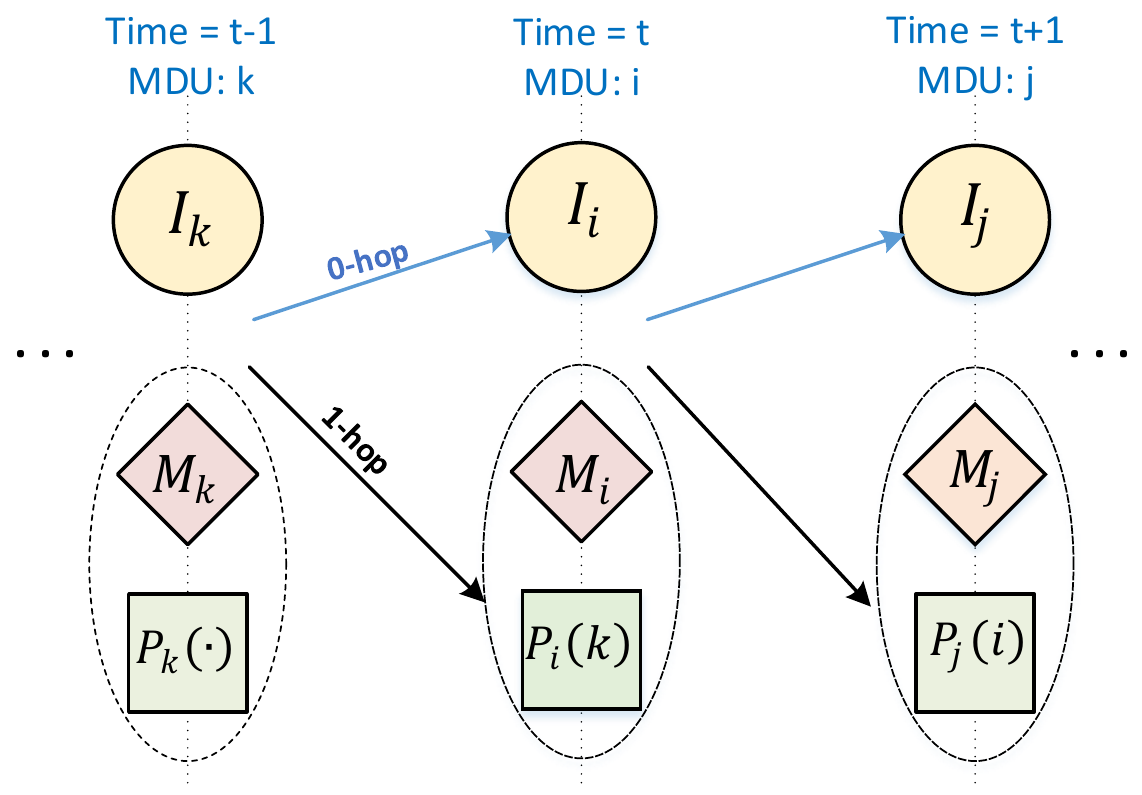}}
\vspace{-0.15in}
\caption{An example MDU transmission for MDU-switching with fixed one-MDU reference buffer, with I-, P- and M-MDUs are shown with solid circles, squares and diamonds, respectively. The 0-hop transmission (we take an I-MDU as an example) and 1-hop transmission (a combo of M- and P-MDUs) are shown with blue and black arrows, respectively.
}
\label{fig:fixedBuffer}
\end{figure}

Fig.\;\ref{fig:fixedBuffer} shows possible sequences of transmitted MDUs assuming a fixed one-MDU reference buffer. 
Here we assume that the I-MDU is transmitted for 0-hop transmission.
For three time instants, a user switches from MDU $k$ to MDU $i$, then to MDU $j$. 
Each MDU-switch can be implemented by either transmitting an I-MDU (0-hop) \textit{or} a combo of P- and M-MDUs (1-hop), assuming that I- and P-MDUs are pre-encoded into the structure prior to streaming.
Both the I-MDU and the combo can reconstruct exactly to the requested I-MDU, thanks to the aforementioned merging operation in M-MDU.

For a given MDU structure $\Theta$ and with a fixed one-MDU reference buffer, the expected transmission cost $c_s^{(0)}$ starting at a given initial MDU $s$ then can be computed as  
\begin{eqnarray}
c_s^{(0)} = r_s^I + \sum \limits_{j\in \mathcal{N}(s)} p_{s,j} \; \min \left[h_s^{(0)}(j), \dot{h}_s^{(0)}(j) \right],
\label{eq:objFixed2}
\end{eqnarray}
where $r_s^I$ is the 0-hop transmission cost of the initial MDU $s$.

\subsubsection{Complexity Analysis}

The recursive computation of $c_s^{(0)}$ can be efficiently solved using \textit{dynamic programming} (DP). Specifically, each time (\ref{eq:objFiexed}) is solved for $c_{i|k}^{(t)}$, the result is stored in entry $[t][i][k]$ of a DP table. 
When the same sub-problem $c_{i|k}^{(t)}$ is called again next time, one can simply look up the DP table for a solution. 
The complexity of a DP algorithm is upper-bounded by the size of the DP table multiplied by the complexity to compute each table entry. 
In this case, the DP table is bounded as $\mathcal{O}(KNT_{\max})$, where $K$ is a constant representing the maximum size of $\mathcal{N}(i)$ for all MDU $i$. 
Each table entry can be computed in $\mathcal{O}(1)$. 
Hence, the order of complexity is $\mathcal{O}(NT_{\max})$.

\vspace{-0.06in}
\subsection{Transmission Cost for Flexible One-MDU Buffer}
We next generalize our previous assumption to a \textit{flexible} one-MDU reference buffer.
A flexible one-MDU reference buffer means that given an MDU $\gamma$ currently in the one-MDU reference buffer and an MDU $i$ currently being displayed, the user can decide freely which MDU between $\gamma$ and $i$ should reside in the one-MDU reference buffer next for future decoding. 
Thus, the flexible one-MDU reference buffer allows a user to store the more valuable of two available MDUs as the reference---one that helps reduce the expected transmission cost as the user continues to navigate the media space. 

\vspace{0.05in}
\noindent
\textbf{Remark}: 
\textit{One can use the flexible one-MDU buffer model to persistently store a landmark MDU to facilitate MDU-switching.}
For instance, if there is a landmark MDU $\gamma$ currently in the user's reference buffer, then using the flexible one-MDU buffer a user can switch from MDU $i$ to $j$ by directly decoding $P_j(\gamma)$, resulting in a 1-hop transmission cost.
\vspace{0.05in}

When computing the expected transmission cost using a flexible one-MDU buffer, previously discussed 0-hop and 1-hop transmission types during an MDU-switch can also be applied. 
For 0-hop transmission, the server can send the same independent reconstruction of MDU $j$ with overhead $r_j^I$. 
For 1-hop transmission, one can choose between the buffered MDU $\gamma$ and $i$ as a reference to decode an inter-coded representation for target MDU $j$, \textit{i.e.}, either $P_j(\gamma)$ or $P_j(i)$ along with an M-MDU $M_j$, resulting in an overhead $r_j^P(\gamma)$ or $r_j^P(i)$, respectively. 

Further, in addition to these two transmission types, we consider also a \textit{2-hop} transmission:
an inter-coded $P_\eta(\gamma)$ or $P_\eta(i)$ is transmitted along with a merge-coded $M_\eta$ to transition to an \textit{intermediate} MDU $\eta$ first, then the inter-coded $P_j(\eta)$ and $M_j$ are transmitted to arrive at the target MDU $j$. 
The overhead is thus $r^P_\eta(\gamma)$ or $r^P_\eta(i)$, \textit{plus} $r^P_j(\eta)$.

\vspace{0.05in}
\noindent
\textbf{Remark}: 
\textit{One can use the 2-hop transmission to proactively insert a useful landmark MDU (\textit{e.g.} MDU $\eta$) into the buffer, facilitating future MDU-switching.}
\vspace{0.05in}

\subsubsection{Expected Transmission Cost}
Given an MDU structure $\Theta$ and with a flexible one-MDU reference buffer, we derive equations to compute the expected transmission cost.
Denote by $c_{i|k}^{(t)}(\gamma)$ the expected transmission cost from current instant $t$ to the end of a media navigational streaming session, given a user is at MDU $i$ and was previously at MDU $k$, and with MDU $\gamma$ in the reference buffer. 
Generalizing \eqref{eq:objFiexed}, we write $c_{i|k}^{(t)}(\gamma)$ recursively as

%\vspace{-0.05in}
\begin{small}
\begin{align}
c_{i|k}^{(t)}(\gamma) = \sum \limits_{j\in \mathcal{N}(i)} p_{k,i,j} \min \left [h_i^{(t)}(\gamma,j), \dot{h}_i^{(t)}(\gamma,j),\ddot{h}_i^{(t)}(\gamma,j) \right]
\label{eq:cost}
\end{align}
\end{small}\noindent
where $h_i^{(t)}(\gamma,j)$, $\dot{h}_i^{(t)}(\gamma,j)$ and $\ddot{h}_i^{(t)}(\gamma,j)$ are the costs of 0-hop, 1-hop and 2-hop transmissions, respectively.

The 0-hop transmission cost $h_i^{(t)}(\gamma,j)$ is the sum of $r_j^I$ \textit{plus} the recursive cost $c_{j|i}^{(t+1)}(\gamma')$ if the maximum lifetime $T_{\max}$ has not been reached. 
MDU $\gamma'$ stored in the reference buffer for the $(t+1)$-th instant is selected between MDUs $\gamma$ and $i$ to minimize subsequent transmission cost:
\begin{eqnarray}
h_i^{(t)}(\gamma,j) = r_j^I + g(t+1) \min \limits_{\gamma' \in \{\gamma,i\}} c_{j|i}^{(t+1)}(\gamma').
\label{eq:0hop}
\end{eqnarray}

The 1-hop transmission cost is the sum of either $r_j^P(\gamma)$ or $r_j^P(i)$ \textit{plus} the recursive cost $c_{j|i}^{(t+1)}(\gamma')$. 
MDU $\gamma$ or $i$ that is used as a predictor to MDU $j$ will become the new reference in the recursive term. 
Thus we write $ \dot{h}_i^{(t)}(\gamma,j)$ as
%
%\begin{small}
\begin{align}
\dot{h}_i^{(t)}(\gamma,j) = \min \limits_{\gamma' \in \{\gamma,i\}} \left [ r_j^P(\gamma') + g(t+1) c_{j|i}^{(t+1)}(\gamma') \right ].
\label{eq:1hop}
\end{align}
%\end{small}

The 2-hop transmission cost is, for an intermediate MDU $\eta$, the sum of either differential coding cost $r_\eta^P(\gamma)$ or $r_\eta^P(i)$, \textit{plus} differential coding cost $r_j^P(\eta)$, \textit{plus} recursive cost $c_{j|i}^{(t+1)}(\eta)$. 

%\vspace{-0.05in}
\begin{small}
\begin{align}
\ddot{h}^{(t)}_i(\gamma,j) \!= \!\min \limits_{\eta} \! \left[ r_j^P(\eta)\! +\! g(t+1) c_{j|i}^{(t+1)}(\eta)\! + \! \min \limits_{\tau \in \{\gamma,i\}} r_\eta^P(\tau)\right ].
\label{eq:2hop}
\end{align}
\end{small}\noindent
The reference MDU for $(t+1)$-th MDU-switch is $\eta$.

Having defined the above, $c_s^{(0)}(\emptyset)$ computes the expected transmission cost starting from initial MDU $s$ with an empty reference buffer $\emptyset$. 
The definition of $c_s^{(0)}(\emptyset)$ is similar to (\ref{eq:objFixed2}):
%\begin{small}
\begin{align}
\label{eq:objFlex}
c_s^{(0)}(\emptyset) = \! r_s^I + \! \! \sum \limits_{j \in \mathcal{N}(s)}\!\! p_{s,j} \min \left [h_s^{(0)}(\emptyset,j), \dot{h}_s^{(0)}(\emptyset,j),\ddot{h}_s^{(0)}(\emptyset,j) \right]
\end{align}
%\end{small}

\subsubsection{Complexity Analysis}

$c_s^{(0)}(\emptyset)$ can also be computed by a DP algorithm.
In this case, the size of DP table is bounded by $\mathcal{O}(KN^2 T_{\max})$. 
The steps required to compute (\ref{eq:cost}), (\ref{eq:0hop}), (\ref{eq:1hop}) and (\ref{eq:2hop}) are bounded by $\mathcal{O}(N)$ (to find the intermediate MDU for the 2-hop transmission). 
Thus the overall complexity of computing the expected transmission cost is $\mathcal{O}(N^3 T_{\max})$.

\vspace{-0.1in}
\subsection{MDU Structure Design Problem Formulation}
We now focus on the problem to determine which I- and P-MDUs to pre-encode to minimize the expected transmission cost given a storage constraint.
For a given MDU structure $\Theta$, we define the storage cost $b(\Theta)$ as the total size of all the pre-encoded I-MDUs and differential P-MDUs:
\begin{eqnarray}
 b(\Theta) = \sum \limits_{I_j \in \Theta} |I_j| + \sum \limits_{P_j(i) \in \Theta} |P_j(i)| ,
\label{eq:storage}
\end{eqnarray} 
where $|\cdot|$ stands for the coding bitrate of an MDU.
M-MDUs are not considered since they are pre-encoded by default into the structure $\Theta$.

Having defined the expected transmission cost and the storage cost for a given structure $\Theta$,
we can next define the optimal MDU structure design problem: find a structure $\Theta^*$ that optimally trades off the expected transmission cost and storage cost, \textit{i.e.}
\begin{eqnarray}
\Theta^* = \arg \min \limits_\Theta c(\Theta) + \lambda b(\Theta),
\label{eq:object}
\end{eqnarray}
where $\lambda$ is a tradeoff parameter.
In the case of the fixed one-MDU reference buffer, expected transmission cost $c(\Theta) = c_s^{(0)}$ is computed using (\ref{eq:objFixed2}) given $\Theta$. 
In the case of the flexible one-MDU reference buffer, it is computed using (\ref{eq:objFlex}).

\vspace{\emptsp}
\section{Optimizing Landmark Insertions}
\label{sec:landmark}
Because of the combinatorial nature of the problem, optimizing the objective (\ref{eq:object}) is difficult. 
We propose to design an MDU structure using a carefully selected set of landmarks as the initial structure $\Theta$, then subsequently add P-MDUs incrementally as refinements. 
Assuming the flexible one-MDU reference buffer model, 
we first investigate how to optimally insert landmarks into an MDU structure.

Each landmark is associated with a local neighborhood of MDUs; all MDUs in a neighborhood are initially coded as P-MDUs using the associated landmark as a predictor.  
Thus, an MDU can transition to another in the same neighborhood by just one P-MDU transmission, assuming the landmark is available in the buffer. 
Having multiple landmarks means the neighborhood associated with each landmark is smaller.
Clearly, a smaller neighborhood implies smaller distances between target MDUs and the predictor landmark, resulting in smaller inter-coded P-MDUs.
However, an MDU transitioning to another in a different neighborhood would require the transmission of another landmark or intra-coded target MDU, both of which are costly.
The challenge then is to identify the appropriate number and locations of landmarks and their respective neighborhoods, in order to minimize the total transmission and storage cost \eqref{eq:object}.

To select landmarks, we adopt a \textit{tree-structured vector quantizer} (TSVQ) approach \cite{franti1997}, which we define recursively as follows.
At a recursive instant, a landmark $l$ is associated with a neighborhood of MDUs or \textit{partition} $\Psi$; 
the landmark has an I-MDU $I_l$ and all MDUs inside the partition have P-MDUs from the landmark.
We define two cost functions $\phi(\cdot)$ and $\delta(\cdot)$ to decide whether to split partition $\Psi$ into two non-overlapping sub-partitions $\Psi_1$ and $\Psi_2$ (with corresponding landmarks $l_1 \in \Psi_1$ and $l_2 \in \Psi_2$), where $\Psi = \Psi_1 \cup \Psi_2$. 
Specifically, $\phi(\cdot)$ defines the average cost of an MDU transition within a partition, and $\delta(\cdot)$ defines the average cost of an MDU transition across two different partitions. 
If 
\begin{eqnarray}
\label{eq:splitCond}
\phi(\Psi_1, l_1) + \phi(\Psi_2, l_2) + \delta(\Psi_1,\Psi_2,l_1,l_2) < \phi(\Psi, l),
\end{eqnarray}
then splitting would be more beneficial, resulting in lower average cost.
$\Psi$ will be replaced by $\Psi_1$ and $\Psi_2$, each of which may further be divided recursively following the same splitting strategy. 
The definitions of $\phi(\cdot)$ and $\delta(\cdot)$ are discussed next.
  
\vspace{-0.08in}  
\subsection{Definitions of $\phi(\cdot)$ and $\delta(\cdot)$}

We write the cost function $\phi(\Psi,l)$ as a variant of (\ref{eq:object}) computed for MDUs in a partition $\Psi$, combining expected transmission cost with a weighted storage cost.
We begin first with the definition of $q_{i,j}$, the aggregate probability of the MDU-switching event from MDUs $i$ to $j$ after an expected lifetime of $\mu$ MDU-switches. $q_{i,j}$ is computed as follows. 

We can interpret the MDU-switching process for $N$ MDUs as a $N^2$-state discrete Markov chain, where state $(k,i)$ denotes an MDU-switching event from MDU $k$ to $i$. 
MDU-switching probability $p_{k,i,j}$ is then the state transition probability from state $(k,i)$ to state $(i,j)$.
Suppose we represent state $(i,j)$ as the $(i-1)N+j$-th entry in a length-$N^2$ vector.  
Given $p_{k,i,j}$, we can correspondingly write a state transition probability matrix $\mathbf{P}$ of size $N^2 \times N^2$. 

Given initial MDU $s$ and initial MDU-switch probabilities $p_{s,j}$, we define the initial state probability as a $1 \times N^2$ canonical unit vector $\mathbf{v_s}$ with all zeros except for $K$ entries corresponding to states $(s,j)$, $j \in \mathcal{N}(s)$, where each has probability $p_{s,j}$. 
We can now compute the aggregate MDU-switch event probability $\mathbf{q}$ after $\mu$ MDU-switches as:
\begin{eqnarray}
\mathbf{q} = \sum_{t=1}^\mu g(t) \; \mathbf{v_s} \mathbf{P}^t,
\label{eq:Q}
\end{eqnarray}
where $\mathbf{P}^t$ is the $t$-step transition matrix and $q_{i,j}$ is the $(i-1)N + j$-th entry of $\mathbf{q}$.

Given $q_{i,j}$ and assuming landmark $l$ is available in the reference buffer, the cost function $\phi(\Psi,l)$ can be written as:
%\begin{small}
\begin{equation}
\small
\label{eq:costWithin}
\phi(\Psi, l) =  \sum_{i \in \Psi} 
\sum_{j \in \mathcal{N}(i) | j \in \Psi} q_{i,j} \cdot r^P_j(l) +  w \left(|I_l| + \sum_{i \in \Psi} |P_i(l)| \right) 
\end{equation}
%\end{small}
\noindent 
where the first term is the normalized expected transmission cost for MDU-switches within $\Psi$, and
the rest is the weighted storage cost of I-MDU $I_l$ and P-MDUs from landmark $l$ to all the MDUs in $\Psi$. 
$w = \lambda/ \mu $ is a normalized weight between the storage cost and the expected transmission cost.

\begin{figure*}[ht]
\begin{small}
\begin{align}
\label{eq:boundary}
\delta(\Psi_1,\Psi_2,l_1,l_2) = 
\underbrace{\sum_{i\in \Psi_1} \; \sum_{j \in \mathcal{N}(i) | j \in \Psi_2} q_{i,j} \cdot \left[ r^P_{l_2}(l_1) + r^P_{j}(l_2)  \right] }_{\text{term 1}} 
  + \underbrace{\sum_{j\in \Psi_2}\; \sum_{i \in \mathcal{N}(j) | i \in \Psi_1} q_{j,i}  \cdot \left[  r^P_{l_1}(l_2) + r^P_i(l_1)  \right]}_{\text{term 2}} 
+ \underbrace{w \left( \, |P_{l_1}(l_2)| + |P_{l_2}(l_1)| \, \right)}_{\text{term 3}}
\end{align}
\end{small}
\hrulefill
\end{figure*}
We next define the cost function $\delta(\Psi_1,\Psi_2,l_1,l_2)$ of MDU-switches across neighboring partitions. 
We assume that P-MDUs connecting landmarks of adjacent partitions, \textit{i.e.}, $P_{l_2}(l_1)$ and $P_{l_1}(l_2)$, are pre-encoded in the MDU structure $\Theta$, such that I-MDUs are not required when transitioning across partitions.
Consider now the case of switching from MDU $i \in \Psi_1$ to MDU $j \in \Psi_2$, and $l_1$ is in the reference buffer. 
The resulting 2-hop transmission cost is $r^P_{l_2}(l_1) + r^P_{j}(l_2)$, \textit{i.e}, the cost of switching to landmark $l_2$ then to target $j$.
Conversely, MDU-switching from MDU $j$ to $i$ has transmission cost $r^P_{l_1}(l_2) + r^P_{i}(l_1)$. 
Generalizing this case to any MDU-switch from one partition to another, we can write \eqref{eq:boundary}, which also includes the storage cost (term 3) for P-MDUs that connect landmarks across partitions.

%\vspace{-0.12in}
\subsection{Optimal Partition Splitting}

To split a partition $\Psi$ with landmark $l$, we first initialize two new landmarks $l_1$ and $l_2$. 
As done in~\cite{kaukoranta1996reallocation}, we retain the original landmark (\textit{i.e.}, $l_1 = l$) and find a new one $l_2$ with the ``furthest'' distance to the original. 
The rationale is as follows.
Examining the cost function $\phi(\Psi,l)$ in (\ref{eq:costWithin}), we see that each MDU $i \in \Psi$ contributes a storage cost of P-DMU from $l$ to $i$ and an expected transmission cost switching from MDU $i$ to MDUs in its neighborhood. 
Alternatively, if an MDU $i$ is chosen as a new landmark, it induces a new storage cost $|I_i|$.
We thus define the ``furthest'' MDU as one that contributes the largest cost difference, \textit{i.e.},
\begin{align}
\label{eq:initL}
l_2 = \arg \max_{i\in \Psi}  \Bigg\{ \sum_{j \in \mathcal{N}(i) | j\in \Psi} q_{i,j} \cdot r^P_j(l)  + w |P_i(l)| - w |I_i|  \Bigg\} .
\end{align}
\noindent
We then split partition $\Psi$ with two initial sub-partitions: i) $\Psi_2 = \{ l_2\}$ with landmark $l_2$, and ii) $\Psi_1 = \Psi \backslash \Psi_2$ with landmark $l_1$. 
%By doing so, $\phi(\Psi_2,l_2) = 0$ and we remove the largest cost from original $\phi(\Psi,l)$. 
By doing so, $\phi(\Psi_2,l_2) = w|I_{l_2}|$ and we maximize the difference between $\phi(\Psi,l)$ and $\phi(\Psi_1,l_1)$ \textit{plus} $\phi(\Psi_2,l_2)$.
Considering the splitting constraint \eqref{eq:splitCond}, with this sub-partitions initialization, we will likely have a splitting that results in lower cost. 

To refine and select the local optimal landmarks $l_1$ and $l_2$ and sub-partitions $\Psi_1$ and $\Psi_2$ during splitting, we use the Lloyd's algorithm \cite{vq92} that iterates between two alternating steps until convergence.
First, given landmarks $l_1$ and $l_2$, we assign each MDU $j$ in partition $\Psi$ to the closer of the two landmarks: 
\begin{eqnarray}
z = \arg \min_{\tau \in \{1, 2\}} |P_j(l_\tau)|
\label{eq:optPartition}
\end{eqnarray}
where $z$ is the partition to which MDU $j$ is assigned.

Second, given sub-partitions $\Psi_\tau$, $\tau \in \{1,2\}$, we find each locally optimal landmark $l_\tau$ by minimizing the following:
\begin{eqnarray}
l_\tau = \arg \min_{l \in \Psi_\tau} \phi(\Psi_\tau, l)
\label{eq:optLandmark}
\end{eqnarray}
where $\phi(\Psi_\tau, l)$ is defined in \eqref{eq:costWithin}.
In words, an optimal MDU $l_\tau \in \Psi_\tau $ that minimizes the total cost in a partition $\Psi_\tau$ is selected as a new landmark.

With the initial $l_1$ and $l_2$, we iteratively solve (\ref{eq:optPartition}) and (\ref{eq:optLandmark}) until convergence to find the optimal sub-partitions and landmarks. Empirical data show that the Lloyd's algorithm can converge quickly. 

Our proposed TSVQ algorithm to optimally select landmarks is summarized in Algorithm 1.
Given an initialized partition $\Psi_o$ with all the MDUs, we first find the landmark $l_o$ by solving (\ref{eq:optLandmark}) and put $(\Psi_o, l_o)$ into a candidate partition pool $\mathcal{X}$.
By iterating between (\ref{eq:optPartition}) and (\ref{eq:optLandmark}), we optimally split one candidate partition $(\Psi,l)\in \mathcal{X}$ into two sub-partitions $\Psi_1$ and $\Psi_2$ and find the two corresponding landmarks $l_1$ and $l_2$.
We compute (\ref{eq:costWithin}) and (\ref{eq:boundary}) given these partitions, and use (\ref{eq:splitCond}) to check if partition splitting is beneficial. 
After a beneficial partition split, we recursively apply the same splitting strategy on partitions $\Psi_1$ and $\Psi_2$ until splitting is no longer able to reduce the average cost.

\begin{table}[ht!]
\label{alg:TSVQ}
\begin{tabular}{l}
\hline
\hline
\textbf{Algorithm 1} Optimal Landmark Insertions \\
\hline
1:  Initialize $\Psi_o=\{1,\cdots,N\}$ MDUs, find $l_o$ with (\ref{eq:optLandmark}); \\
\;\;\;  Initialize $\mathcal{X} = \{(\Psi_o, l_o)\}$ and $\mathcal{Y} = \emptyset$. \\
2:  \textbf{for} $(\Psi,l) \in \mathcal{X}$ \textbf{do} \\
3:	 \quad Initialize $(\Psi_1, l_1)$ and $(\Psi_2, l_2)$ with (\ref{eq:initL}); \\
4:	 \quad Find optimal sub-partitions by alternating between \\ \;\;\; \quad (\ref{eq:optPartition}) and (\ref{eq:optLandmark}); \\
5:	 \quad Compute (\ref{eq:costWithin}) and (\ref{eq:boundary}) given $(\Psi_1, l_1)$ and $(\Psi_2, l_2)$. \\
6:	 \quad \textbf{if} (\ref{eq:splitCond}) holds \textbf{then} \\
7:	 \quad \quad Add $(\Psi_1, l_1)$ and $(\Psi_2, l_2)$ into $\mathcal{X}$. \\
8:	 \quad \textbf{else} \\
9:	 \quad \quad Add $(\Psi,l)$ into $\mathcal{Y}$. \\
10: \textbf{return} $\mathcal{Y}$.	 \\
\hline
\end{tabular}
\end{table}

\vspace{\emptsp}
\section{MDU Structure Design Algorithm}
\label{sec:algorithm}
Having identified landmarks via TSVQ, we first initialize an MDU structure $\Theta$ with I-MDUs of all the landmarks, P-MDUs connecting from landmarks to their corresponding partitions, and P-MDUs connecting any pair of landmarks.
With this initialization, a combo of an I-MDU of a landmark $I_l$ and a P-MDU $P_j(l)$ (along with an M-MDU $M_j$) can always fulfil a user's request of any MDU $j$, which is the 0-hop transmission.
Note that this initialized MDU structure $\Theta$ is consistent with the assumption to optimally insert landmarks.
But it can still be refined to trade off the storage and the expected transmission cost.
Next, we employ a greedy algorithm to further improve structure $\Theta^*$:
iteratively add P-MDUs one at a time until the objective \eqref{eq:object} cannot be further improved. 
Specifically, at each iterative step, we identify the most ``beneficial" P-MDU not in $\Theta$ that will maximally reduce objective \eqref{eq:object} upon its addition. 
We stop when no P-MDU addition will induce a decrease in the objective. 

Although our proposed DP algorithm (\ref{eq:objFlex}) to compute the expected transmission cost for a given structure $\Theta$ can be executed in polynomial time, 
to recompute the cost for each P-MDU addition to the structure is still time-consuming (the number of P-MDUs to be checked for adding is $O(N^2)$. 
We thus employ a \textit{branch-and-bound (B$\&$B) algorithm}~\cite{narendra1977branch} to reduce the computation complexity
when greedily add P-MDUs to a given MDU structure.
 
Specifically, starting from a given MDU structure $\Theta$, we first compute the expected transmission cost $c(\Theta)$ with (\ref{eq:objFlex}) and store the computed DP table as $\text{DP}_{\Theta}$. 
Denote by $J(\Theta) = c(\Theta) + \lambda b(\Theta)$ the objective (\ref{eq:object}) for $\Theta$. 
At each iteration, for each candidate MDU structure $\Theta'$ (adding one P-MDU to $\Theta$), we need to compute $J(\Theta')$ and compare it against $J(\Theta)$. 
Instead of computing $J(\Theta')$ exactly, which requires executing \eqref{eq:objFlex} to compute $c(\Theta')$, we compute a fast lower bound $c_L(\Theta')$ of $c(\Theta')$ as a first step.
We discuss how to compute $c_L(\Theta')$ of adding P-MDUs as follows.
 
Denote by $J_{\min}$ the current minimum cost, which is initialized as $J_{\min} = J(\Theta)$. 
When P-MDU $P_j(i)$ is added, the storage cost increases to $b(\Theta') = b(\Theta) + |P_j(i)|$. 
To compute lower bound $c_L(\Theta')$, we invoke recursion \eqref{eq:objFlex}.
But when we encounter a switch from MDUs $i$ to $j$, we lower-bound the recursive cost by assuming that the smallest P-MDU $|P|_{\text{min}} = \min_{P \in \Theta'} |P|$ in the current structure $\Theta'$ is used for the remaining lifetime.  
It means that $\min \left[ h_i^{(t)}(\gamma,j), \dot{h}_i^{(t)}(\gamma,j), \ddot{h}_i^{(t)}(\gamma,j) \right] \geq \sum_{\tau = t}^{T_{\max}} g(\tau) |P|_{\text{min}}$ for (\ref{eq:cost}) to avoid recursive calls when MDU transition $i$ to $j$ takes place. 
Clearly, $c_L(\Theta')$ is a lower bound of $c(\Theta')$. 
Thus, we can compute the lower bound of $J(\Theta')$ as $J_L(\Theta') = c_L(\Theta') + \lambda b(\Theta') $ and compare it with $J_{\min}$. 
We execute the original algorithm \eqref{eq:objFlex} to compute $c(\Theta')$ only if $J_L(\Theta') \leq J_{\min}$. 
$J_{\min}$ is updated by $J(\Theta')$ if $J(\Theta') < J_{\min}$.

The B$\&$B algorithm to greedily add P-MDUs is outlined in Algorithm 2. 
The resulting $\Theta^*$ is our finally computed MDU structure.
Using our proposed B$\&$B algorithm, we can save computation of the full DP algorithm if the lower bound of the objective \eqref{eq:object} for a candidate solution is larger than the current best solution. 
Our empirical results show that our B$\&$B algorithm can prevent roughly $50\%$ of candidate MDU structures from computing the full DP recursion (\ref{eq:objFlex}).

\begin{table}[ht!]\small
\label{alg:Pframe_removing}
\begin{tabular}{l}
\hline
\hline
\textbf{Algorithm 2} Greedily Add P-MDU(s) \\
\hline
1:  Given initial MDU structure $\Theta$ and $\lambda$. \\
2:  \textbf{while} \text{TRUE} \textbf{do} \\
3:	 \quad Run (\ref{eq:objFlex}) to compute $c(\Theta)$, store DP table $\text{DP}_{\Theta}$; \\
	 \;\;\; \quad Initialize $J_{\min} = c(\Theta) + \lambda b(\Theta)$; Label $\Theta^* = \Theta$. \\
4:	 \quad \textbf{for} each candidate $\Theta'$ \textbf{do} \\
5:	 \quad \quad Compute $c_L(\Theta')$ and $J_L(\Theta')$. \\
6:	 \quad \quad \textbf{if} $J_L(\Theta') \leq J_{\min}$ \textbf{then} \\
7:	 \quad \quad \quad Run (\ref{eq:objFlex}) to compute $c(\Theta')$ and $J(\Theta')$. \\
8:	  \quad \quad \quad \textbf{if} $J(\Theta') < J_{\min}$ \textbf{then} \\
9:	  \quad \quad \quad \quad $J_{\min} = J(\Theta')$ ;  \\
10:  \quad \quad \quad \quad $\Theta^*= \Theta'$. \\
11:	 \!\!\!\! \quad \textbf{if} $\Theta == \Theta^*$ \textbf{then} \\
12:	 \!\!\!\! \quad \quad Break \\
13:	 \!\!\!\! \quad \textbf{else} \\
14:	 \!\!\!\! \quad \quad $\Theta =\Theta^*$. \\
15: \textbf{return} $\Theta^*$	 \\
\hline
\end{tabular}
\end{table}

\vspace{\emptsp}
\section{Illustrative Implementation}
\label{sec:details}
To test the performance of our designed MDU structures using landmarks for navigational HD media streaming, we take LF and $360^\circ$ images as illustrative examples for experiments.
We first briefly introduce the merge frame (M-frame)~\cite{dai2016merge} that is used in this paper for identical reconstruction.
We then describe MDU definitions and the user behavior models for LF and $360^\circ$ image streaming, respectively. 
Note however that our method is generic and can be applied to other user behavior models.

\vspace{-0.1in}
\subsection{Merge Frame for Video Coding}
\label{subsec:merge}

\begin{figure}[htb]
\centering
\centerline{\includegraphics[width=6.5cm]{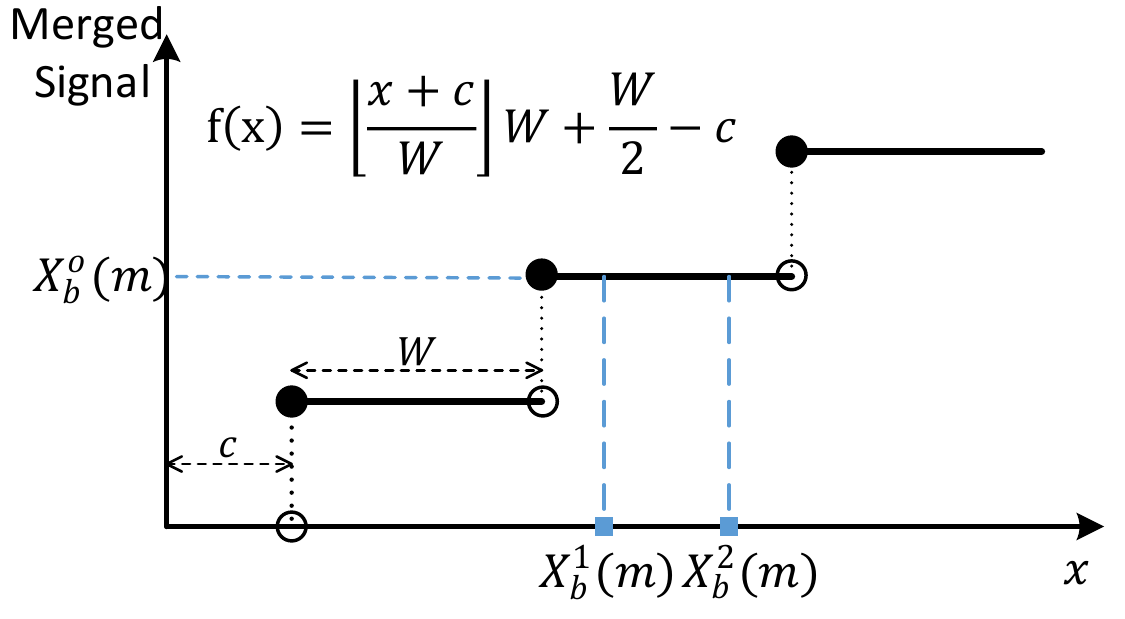}}
\vspace{-0.15in}
\caption{An example of the PWC function $f(x)$ with step size $W$ and horizontal shift $c$. $f(x)$ merges two quantized coefficients $X_b^1(m)$ and $X_b^2(m)$ into an identical value $X_b^o(m)$.}
\label{fig:stepFunc}
\end{figure}

M-frame \cite{dai2016merge} is designed to switch among pre-encoded video streams efficiently without coding drifts. 
Specifically, M-frame is a new DSC design that uses shift and rounding operations to merge multiple differentially coded P-frames into an identical reconstructed frame for video coding.
The design employs a \textit{piece-wise constant} (PWC) function $f(x)$ as the merge operator to merge quantized transform coefficients $x$ from different reconstructions to the same value, where
\begin{eqnarray}
f(x) = \left \lfloor \frac{x+c}{W} \right \rfloor W + \frac{W}{2} - c,
\label{eq:mframe}
\end{eqnarray}
where $W$ is the step size and $c$ is the horizontal shift. 
An example of $f(x)$ is shown in Fig.\;\ref{fig:stepFunc}.

Specifically, consider merging different reconstructions of frame $j$ from different P-frames $P_j(i)$ to a target I-frame $I_j$.
For each quantized transform coefficient $X_b^i(m)$ of the $m$-th frequency in a block $b$ of the reconstructed P-frame $P_j(i)$, an appropriate selection of $W$ and $c$ in $f(x)$ can ensure that coefficients $X_b^i(m)$ of different $i$'s are all mapped to a desired target value $X_b^o(m)$, \textit{i.e.}, 
%\begin{small}
\begin{eqnarray}
X_b^o(m) = f(X_b^i(m)), ~~~ \forall \; i \text{ s.t. } P_j(i) \text{ exists}
\label{eq:merge}
\end{eqnarray}
%\end{small}
where $X_b^o(m)$ is the quantized $m$-th frequency in block $b$ of I-frame $I_j$.
It means that the DCT coefficients of each frequency in each block for all the different P-frames $P_j(i)$'s can be identically merged to the corresponding coefficient of I-frame $I_j$.
The selected value of $W$ and $c$ for each coefficient of each block \cite{dai2016merge} are then entropy encoded as an M-frame, which results in much smaller coding overhead compared to previous SP-frames~\cite{karczewicz03} and DSC frames~\cite{mcheung09pcs}.
For our dataset, the size of an M-frame is roughly 3 to 4 times of a P-frame, while an I-frame is 10 to 12 times the size of a P-frame. 
By definition, an M-frame $M_j$ \textit{plus} any decoded $P_j(i)$ will result in an identically reconstructed $I_j$. 

\vspace{-0.1in}
\subsection{Interactive Light Field Streaming}

%\vspace{-0.1in}
\begin{figure}[htb]
\centering
\centerline{\includegraphics[width=2.8cm]{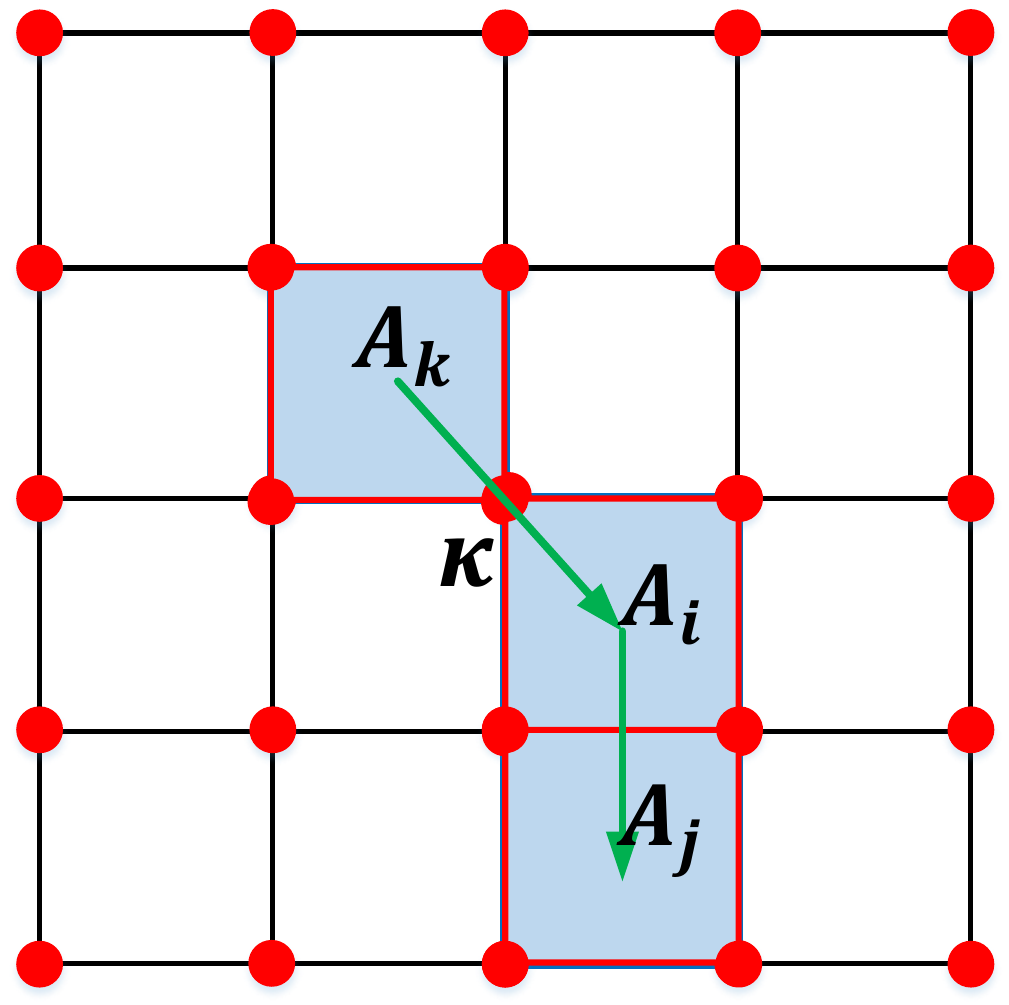}}
\vspace{-0.1in}
\caption{An example of navigational streaming for a 2D grid of $5 \times 5$ LF images (red dots). Considering four nearest neighboring LF images as an MDU, each MDU bounds a view area (blue square). 
Green arrows indicate possible MDU navigation from view area $A_i$ to $A_j$, and previously $A_k$. 
}
\label{fig:2Dgrid}
\end{figure} 

An LF camera employs a 2D array of microlenses in front of the photo sensor to capture multiple light ray intensities and directions per pixel, so that a user can navigate and observe a static 3D scene from different viewpoints post-capture.
For the sake of simplicity, we assume that the captured $N$ viewpoints (anchor views) are arranged into a $\sqrt{N} \times \sqrt{N}$ 2D grid.
Each anchor view $\kappa \in \{1, \cdots, N\}$ has a 2D coordinate $(X_\kappa, Y_\kappa)$, where $X_{\kappa}, Y_{\kappa} \in \mathbb{Z}^+$.
An example 2D grid of $5 \times 5$ LF images is shown in Fig.\;\ref{fig:2Dgrid}.

\begin{figure}[htb]
	\begin{center}
		\begin{tabular}{cc}
		
		\includegraphics[width=\swtwo]{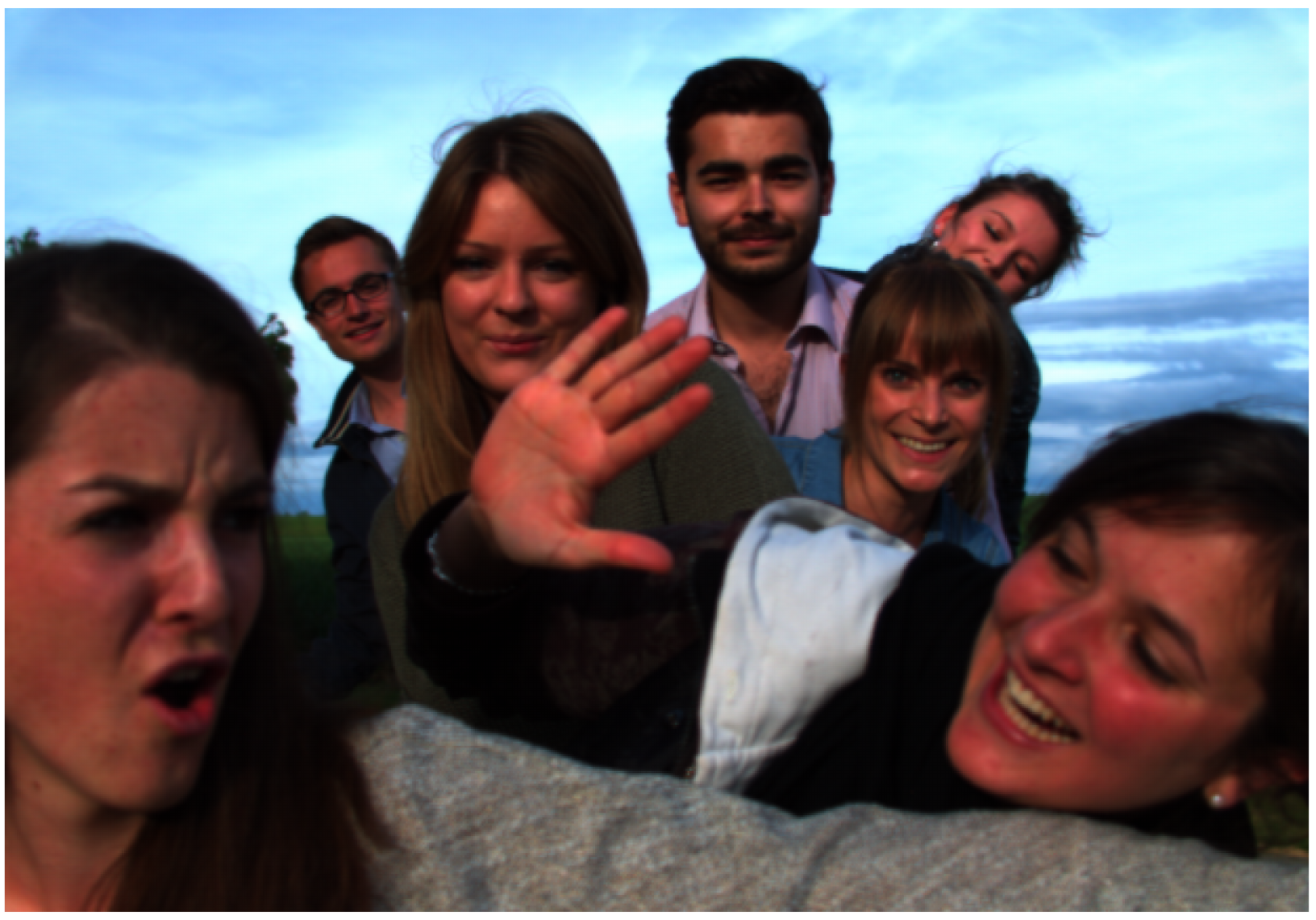} &
		\includegraphics[width=\swtwo]{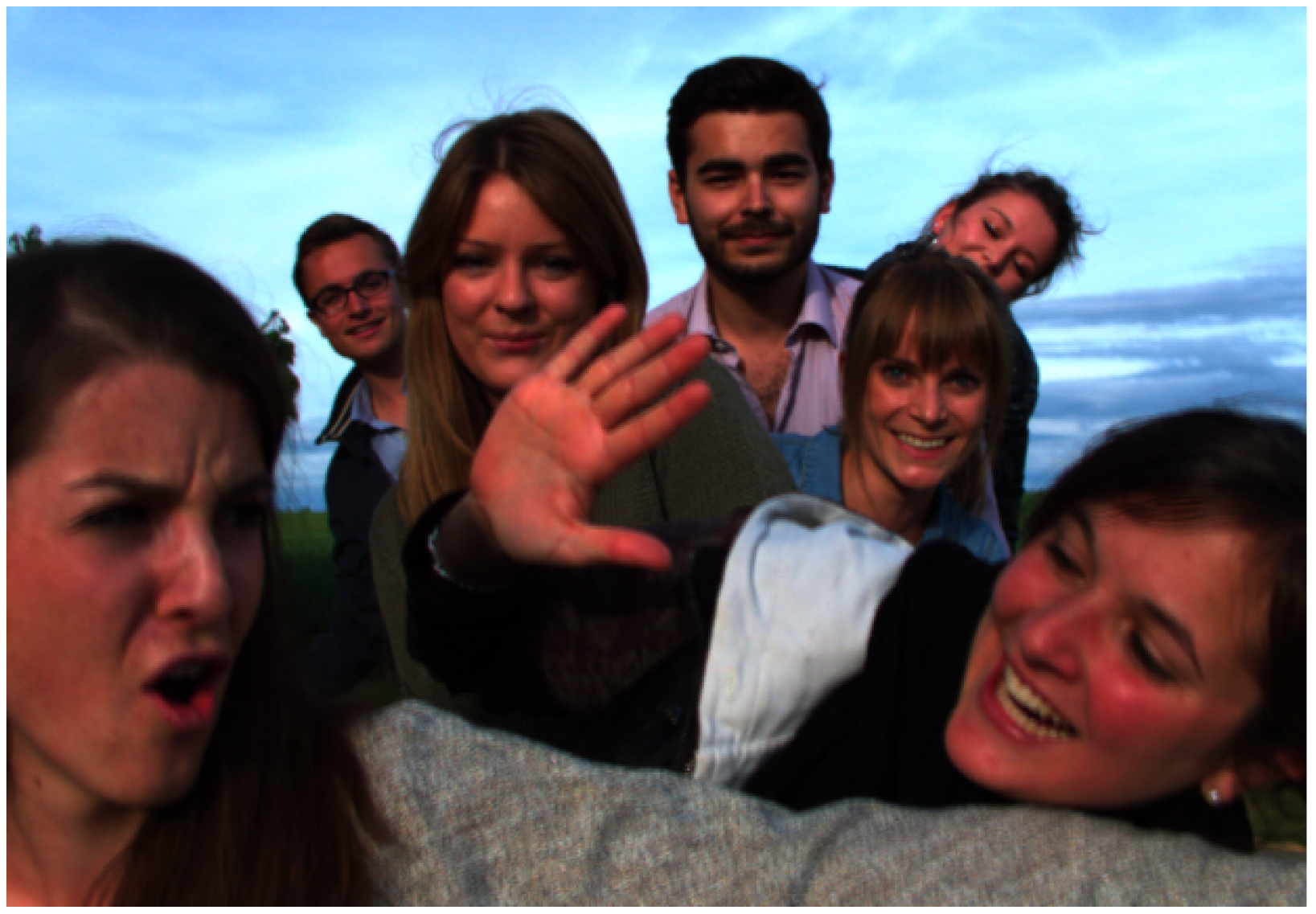} \\
		\footnotesize{(a) } & \footnotesize{(b)}  \\		

		\end{tabular}
	\end{center}
	\vspace{-0.15in}
	\caption{A virtual viewpoint of one LF image dataset is synthesized by all the anchor views (a), versus synthesized only by four nearest anchor views (b). The PSNR value between (a) and (b) is 37.06dB.}
\label{fig:LFsyn}
\end{figure}

A client observes one \textit{virtual view} $u$ with 2D coordinate $(x_u,y_u)$ at a time, where $x_u,y_u \in \mathbb{R}^+$, which in general is an arbitrary intermediate view between anchor views.
We assume that a virtual viewpoint is synthesized using only its four nearest anchor views that encircle it~\cite{ng2005light}.
While in theory using all available anchor views to synthesize a virtual view results in the best quality, it has been shown~\cite{peixoto2017progressive} that using a small set of neighboring anchor views that encircle the target virtual viewpoint is sufficient in practice.  
Fig.\,\ref{fig:LFsyn} shows two virtual view images synthesized using (a) all anchor views and (b) four nearest neighboring anchor views. 
Using (a) as ground truth, the PSNR value of (b) is as high as 37.06dB.
%\vspace{-0.1in}
\renewcommand{\tabcolsep}{.1pt}

Thus, for each requested virtual view $u$, we consider the four nearest neighboring anchor views as one MDU required to synthesize $u$.
Specifically, the $j$-th MDU contains a set $\mS_j$ of four anchor views, bounding a view-area $A_j$. 
Virtual viewpoints within $A_j$ can be synthesized using $\mS_j$.
As shown in Fig.\;\ref{fig:2Dgrid}, the $5\times5$ anchor view grid can be partitioned into $4\times4$ view-areas and corresponding MDUs, where the blue-rectangle regions are example view-areas $A_k$, $A_i$ and $A_j$ synthesized using MDUs $\mS_k$, $\mS_i$ and $\mS_j$, respectively. 
Note that two neighboring MDUs have overlapped anchor views; MDU $i$ has one anchor view overlapped with MDU $k$ and two anchor views overlapped with MDU $j$.

When navigating the LF data, if a client requests a virtual view within the view-area bounded by his current buffered MDU, the requested view can be synthesized directly without further MDU transmission from the server. 
However, if the requested viewpoint is outside the current bounded view area, a new MDU needs to be sent.
Depending on the view-area distance between the required new MDU and the buffered MDU, the server sends the non-overlapping anchor views to the client.
We next describe the MDU transmission overhead $r_j^I$ and $r_j^P(i)$, and the MDU-switch probability $p_{k,i,j}$ and $p_{s,j}$ used to compute the expected transmission cost.

\subsubsection{Transmission Overhead}

We first introduce the 1-hop transmission overhead $r_j^P(i)$, where a P-MDU $P_j(i)$ is transmitted along with an M-MDU $M_j$.
%$r_j^P(i)$ is the overhead of 1-hop transmission, where a P-MDU $P_j(i)$ is transmitted along with an M-MDU $M_j$. 
Recall that the P-MDU $P_j(i)$ means differentially encoding MDU $j$ using MDU $i$ as a predictor.
Conventionally, using multiple reference frames to differentially code a P-frame in general results in better rate-distortion performance than using a single reference frame \cite{bartelmess2016compression}.
Hence, we use the entire set of anchor views $\mS_i$ in MDU $i$ to predict anchor views $\kappa$ in MDU $j$ but not in $i$, denoted by $\mP_\kappa(\mS_i)$.
M-MDU $M_j$ is the collection of merge frames $\mM_\kappa, \kappa \in \mS_j$.  
We thus define $r_j^P(i)$ as follows:
\begin{eqnarray}
r_j^P(i) = \sum_{\kappa \in \mS_j \backslash \mS_i} \min \left \{ \; |\mI_\kappa|, \; |\mP_\kappa(\mS_i)| + |\mM_\kappa| \; \right  \},
\end{eqnarray}
where $\mI_\kappa$ is the intra-coded version of anchor view $\kappa$. $|\cdot|$ means the coding bitrate.
We define $|\mP_\kappa(\mS_i)| = \infty$ to signal a violation if P-frame $\mP_\kappa(\mS_i)$ is not pre-encoded and stored.

Given the generated MDU structure with landmarks, the overhead $r_j^I$ is the 0-hop transmission cost when an I-MDU $I_l$, \textit{plus} a P-MDU $P_j(l)$, \textit{plus} an M-MDU $M_j$ are transmitted for a requested MDU $j$, where MDU $l$ is the landmark of the neighborhood $j$ belongs to.
The anchor views within one MDU can internally be either intra- or inter-coded.
To intra-code MDU $l$, we first intra-code one anchor view and then use it as a predictor to differentially encode the remaining three anchor views, where M-frames of these three anchor views are also considered for identical reconstruction to their I-frames.
The total coding rate for all four anchor views is denoted by $R(\mS_l)$.
We thus define $r_j^I = R(\mS_l) + r_j^P(l)$.

\subsubsection{MDU-Switch Probability}
Before computing the one-step memory MDU-switch probability $p_{k,i,j}$, we first describe the MDU-switch from current MDU $i$ to MDU $j$, given previous MDU was $k$. 
Previously, a user observes a virtual view $u$ with 2D coordinates $\mathbf{u}=(x_u,y_u) \in A_k$, then switches to view $v$ in $A_i$.
Then, after possible switches to views within $A_i$, he switches from a view $v'$ inside $A_i$ to view $w$ in $A_j$.
In general, views $v$ and $v'$ can be different. 
The green arrows in Fig.\;\ref{fig:2Dgrid} show an example of MDU-switch. 

Generally, users have the tendency to select the same navigation in consecutive time instants~\cite{xie2017360probdash}: the higher the similarity between two directions from MDU $i$ to $j$ and from MDU $k$ to $i$, the larger the probability $p_{k,i,j}$.
Before computing $p_{k,i,j}$, we first define an one-step memory view-switch probability $f_{u,v,w}$ as
\begin{eqnarray}
f_{u,v,w} = \alpha \exp \left\{-\frac{\norm{ (\mathbf{v} - \mathbf{u}) - (\mathbf{w} - \mathbf{v}) }^2}{2\sigma^2} \right\},
\label{eq:fuvw}
\end{eqnarray}\noindent
where $\alpha$ is a normalization constant and $\sigma$ is a parameter.
$f_{u,v,w}$ is the probability of switching from view $v$ to view $w$ with previously at view $u$.
It captures a user's the same navigation tendency, both in direction and distance.
Neglecting the view-switch from $v$ to $v'$ (within one MDU, no transmission overhead required), we thus define the MDU-switch probability $p_{k,i,j}$ as:
\begin{align}
p_{k,i,j} = \int_{\mathbf{u} \in A_k} \int_{\mathbf{v} \in A_i} \int_{\mathbf{w} \in A_j} f_{u,v,w} \;\; d\mathbf{u}\;  d\mathbf{v} \; d\mathbf{w}.
\label{eq:pkij_revised}
\end{align}
The starting MDU switching transition probability $p_{s,j}$ can be defined similarly. We omit it here for space reason.

\vspace{-0.1in}
\subsection{Viewport Adaptive $360^\circ$ Image Streaming}
$360^\circ$ image provides an immersive $360^\circ$ viewing experience to a client wearing a head-mounted display (HMD): as the client rotating her/his head, a server sends corresponding spatial fractions (viewports) of the $360^\circ$ image content to the client for observation. 
Clients can freely navigate a $360^\circ$ image by switching the viewports.

%\vspace{-0.1in}
\renewcommand{\tabcolsep}{.1pt}
\begin{figure}[htb]
	\begin{center}
		\begin{tabular}{cc}
		
		\includegraphics[width=0.545\linewidth]{figures/0013.pdf} \;&
		\includegraphics[width=0.41\linewidth]{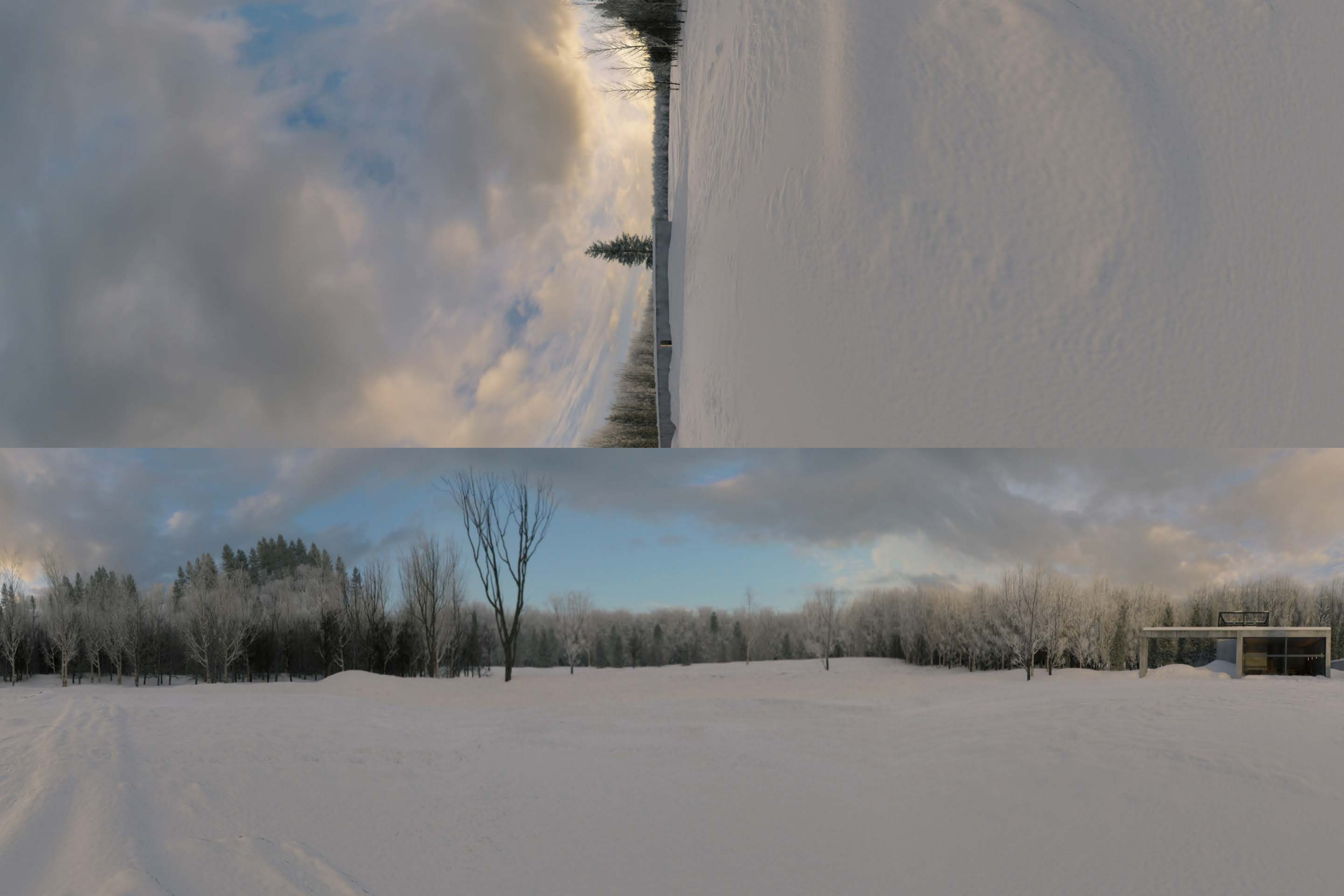} \\
		\footnotesize{(a) } \;& \footnotesize{(b)}  \\		

		\end{tabular}
	\end{center}
	\vspace{-0.15in}
	\caption{An example of an equirectangular projected $360^\circ$ image (a) and one viewport with offset cubemap projection (b).}
\label{fig:360viewport}
\end{figure}
Although tiles-based streaming improves transmitting efficiency, a client has to reconstruct the viewport from independently transmitted tiles and thus the latency may be increased~\cite{corbillon2017viewport}.
In our work, we consider the viewport-based streaming scheme for \360 image streaming.
As done by a navigational streaming setup in Facebook Inc.~\cite{pyramid2017facebook} and Qualcomm~\cite{van2017AHG8}\cite{van2020viewport}, a $360^\circ$ sphere is projected into multiple overlapping viewports.
Each viewport is then projected to a 2D image and encoded separately.
Fig.\;\ref{fig:360viewport} shows an example of (a) a $360^\circ$ sphere projected with equirectangular (ERP) format and (b) a viewport with offset cubemap projection.
Each viewport refers to an independent viewpoint image, which can be either intra-coded or inter-coded.
Different from LF data, $360^\circ$ data already captures $360^\circ$ views of a 3D scene.
Hence, there is no virtual viewpoint synthesis. 
We thus consider each single viewport as one MDU for encoding and streaming.
Transmission overhead and MDU-switch probability for navigational $360^\circ$ image streaming are as follows:

\subsubsection{Transmitting Overhead}

A P-MDU $P_j(i)$ means a viewport $j$ is inter-coded using viewport $i$ as a predictor.
Hence the 1-hop transmission overhead $r^P_j(i)$ is
\begin{equation}
r^P_j(i) = |\mP_j(i)| + |\mM_j|,
\label{eq:cost_p_360}
\end{equation}
where $\mP_j(i)$ and $\mM_j$ are the P- and M-frames of viewport $j$, respectively.
We define $|\mP_j(i)| = \infty$ if P-frame $\mP_j(i)$ is not pre-encoded and stored.
The 0-hop transmission overhead for an MDU $j$ is $r_j^I = |\mI_l| + r_j^P(l)$,
where $l$ is the landmark viewport of the neighborhood $j$ belongs to, and $\mI_l$ stands for the intra-coding I-frame of the viewport $l$.

\subsubsection{MDU Switch Probability}
MDU-switches for navigational $360^\circ$ image streaming are affected by the content and locations of viewports.
Users' head movement prediction in $360^\circ$ images is well studied in~\cite{sitzmann2018saliency}\cite{qiao2020viewport}.
The work in \cite{sitzmann2018saliency} provides a dataset recording roughly 2000 head and gaze trajectories from 169 users in 22 static stereoscopic \360 images.
The trajectories data is recorded using a head-mounted display in three observing conditions.
By projecting the \360 images into viewports and using the recorded trajectories, we estimate the MDU switch probabilities $p_{k,i,j}$ and $p_{s,j}$ of a user's head movement from one viewport to another.
By doing so, the model parameters are consistent with observed data obtained in practice.

\vspace{\emptsp}
\section{Experimentation}
\label{sec:results}
%\vspace{-0.08in}
\subsection{Experiment Setup}
\subsubsection{Light Field Images}
we downloaded four LF image sets \textsl{Bracelet}, \textsl{Cards}, \textsl{Chess} and \textsl{Lego} from Stanford archive with $17 \times 17$ anchor views, where the size of each image was about $1024 \times 1024$.
We also downloaded two larger LF image sets \textsl{Poznan} ($31 \times 31$ anchor views with size $1288 \times 1936$) and \textsl{Set2} ($33 \times 11$ anchor views with size $ 1080 \times 1920$)
for experiments. 
We used the common test condition (CTC) of HEVC HM 16.0~\cite{sullivan2012overview} with low delay (LD) configuration to encode I- and P-frames of each viewpoint image, and used \cite{dai2016merge} to encode M-frames. 
Assuming high quality was an absolute requirement for HD media navigation (to be adaptive to ideal network conditions), 
we set the quantization parameters (QP) to 28 so that the PSNR of the encoded frames was around 38dB.

For the Poisson distribution of lifetime, we set the maximum lifetime $T_{\max}$ of a navigational streaming session to $1/3$ of the number of anchor views (as done in~\cite{motz16icip}). 
The expected lifetime $\mu$ was set to $0.5 T_{\max}$.  
For the view-switching probability of the user behavior model, we set $\sigma = 0.5$ such that a user had a larger probability to switch to the 8 nearest neighboring view-areas from the current observing view-area.
Hence $K=8$ was the size of $\mathcal{N}(i)$ for each MDU. 
The starting viewpoint of each ILFS session was initialized as the center view of each 2D grid.
We varied $\lambda$ in (\ref{eq:object}) to induce different trade-offs between the expected transmission cost and storage cost.

\subsubsection{\360 Images}
Among the 22 \360 images provided by \cite{sitzmann2018saliency}, we randomly selected twelve $360^\circ$ images for demonstration.
The size of the projected ERP format was $8192 \times 4096$.
With the help of engineers from Kandao Technology~\footnote{https://www.kandaovr.com/}---a \360 VR camera company, we projected each sphere into 30 overlapping viewports (which was the same number of viewports used in a navigational streaming setup in Facebook Inc.~\cite{pyramid2017facebook}).
These viewports were then projected into 2D images using offset cubemap projection, with size $2880 \times 1920$.

To encode $360^\circ$ viewports, we adopted the latest Versatile Video Coding (VVC)~\cite{bross2018vvc} software\footnote{https://vcgit.hhi.fraunhofer.de/jvet/VVCSoftware\_VTM} (version 9.2) to encode the I- and P-frame of viewports, with CTC setup (LD configuration) and QP 19.
We set $T_{\max} = 8$ and $\mu = 3$.
We used the trajectories from standing VR conditions in \cite{sitzmann2018saliency} to compute the viewport switching probabilities.

\subsubsection{Performance Evaluation}
To test the performance of the designed MDU structures generated using our proposed landmark insertion method (labeled as \texttt{Flex-LM}), we selected two other MDU structure designation methods without using landmarks for comparison.
We first compared with a greedy algorithm proposed in~\cite{motz16icip} (labeled as \texttt{Flex-GA}), which also considered a flexible reference buffer to compute the expected transmission cost.
Specifically, starting with no P-MDU in the structure, \texttt{Flex-GA} iteratively added one or one pair of locally optimal P-MDU(s) to the structure at a time to reduce the objective function \eqref{eq:object}, until the objective could no longer be decreased. 
We also compared our proposed frame structure with the \texttt{Fixed-GA} method.
\texttt{Fixed-GA} used the same greedy algorithm as in~\cite{motz16icip} to solve (\ref{eq:object}) but with a fixed MDU reference buffer.
Given the MDU structure generated by the proposed landmark insertion method, we also evaluated the lowest transmission cost by assuming \textit{infinte} buffer size at the user side, labeled as \texttt{Inf-LM}.

Due to the absence of landmarks in \texttt{Flex-GA} and \texttt{Fixed-GA}, we initialized their structures with pre-encoded I-MDUs for every view at a server, then greedily added P-MDUs into the structure to reduce the total cost \eqref{eq:object}.
This initialization was necessary since it ensured that the server could always fulfil a user's request of MDU $j$ by transmitting $I_j$.
%, albeit at a large transmission cost.
For these two methods, a 0-hop transmission overhead was defined as $r_j^I = R(\mS_j)$ and $r_j^I = |\mI_j|$ for LF and \360 images, respectively.
For a fair comparison with \texttt{Flex-GA}, we also considered inserting landmarks when all I-MDUs were pre-stored and the 0-hop transmission was the I-MDU of the requested MDU.
We labeled it as \texttt{Flex-LM-I}.
By keeping other setups identical, the only difference between \texttt{Flex-LM-I} and \texttt{Flex-GA} was the use (or the lack) of landmarks.

\vspace{-0.08in}
\subsection{Landmarks Insertion Results}
%\vspace{-0.1in}
\renewcommand{\tabcolsep}{.1pt}
\begin{figure}[htb]
	\begin{center}
		\begin{tabular}{cc}
		
		\includegraphics[width=0.475\linewidth]{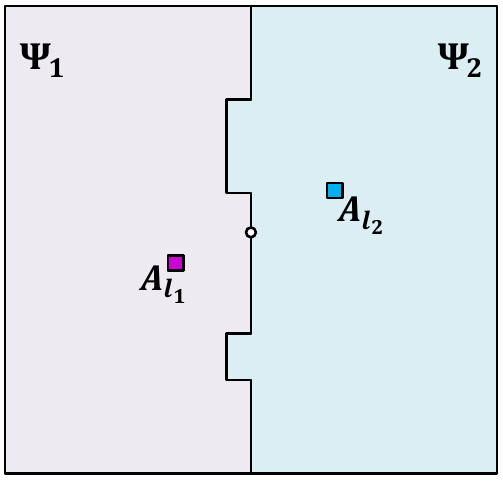} &
		\includegraphics[width=\swtwo]{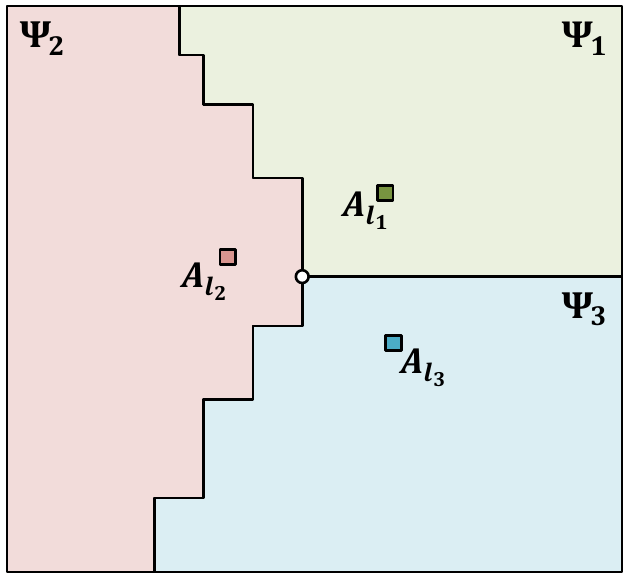} \\
		\footnotesize{(a) $\lambda = 4.5$, two landmarks} & \footnotesize{(b) $\lambda = 8.0$, three landmarks}  \\		

		\end{tabular}
	\end{center}
	\vspace{-0.15in}
	\caption{Illustrations of the view-area partitions and the corresponding landmarks for LF data \textsl{Poznan}. The circle point stands for the starting viewpoint, while the square represents the landmark view-area in each partition.}
\label{fig:lmkr}
\end{figure}

\renewcommand{\tabcolsep}{.1pt}
\begin{figure*}[htb]
	\begin{center}
		\begin{tabular}{cccc}
		
		\includegraphics[width=0.236\linewidth]{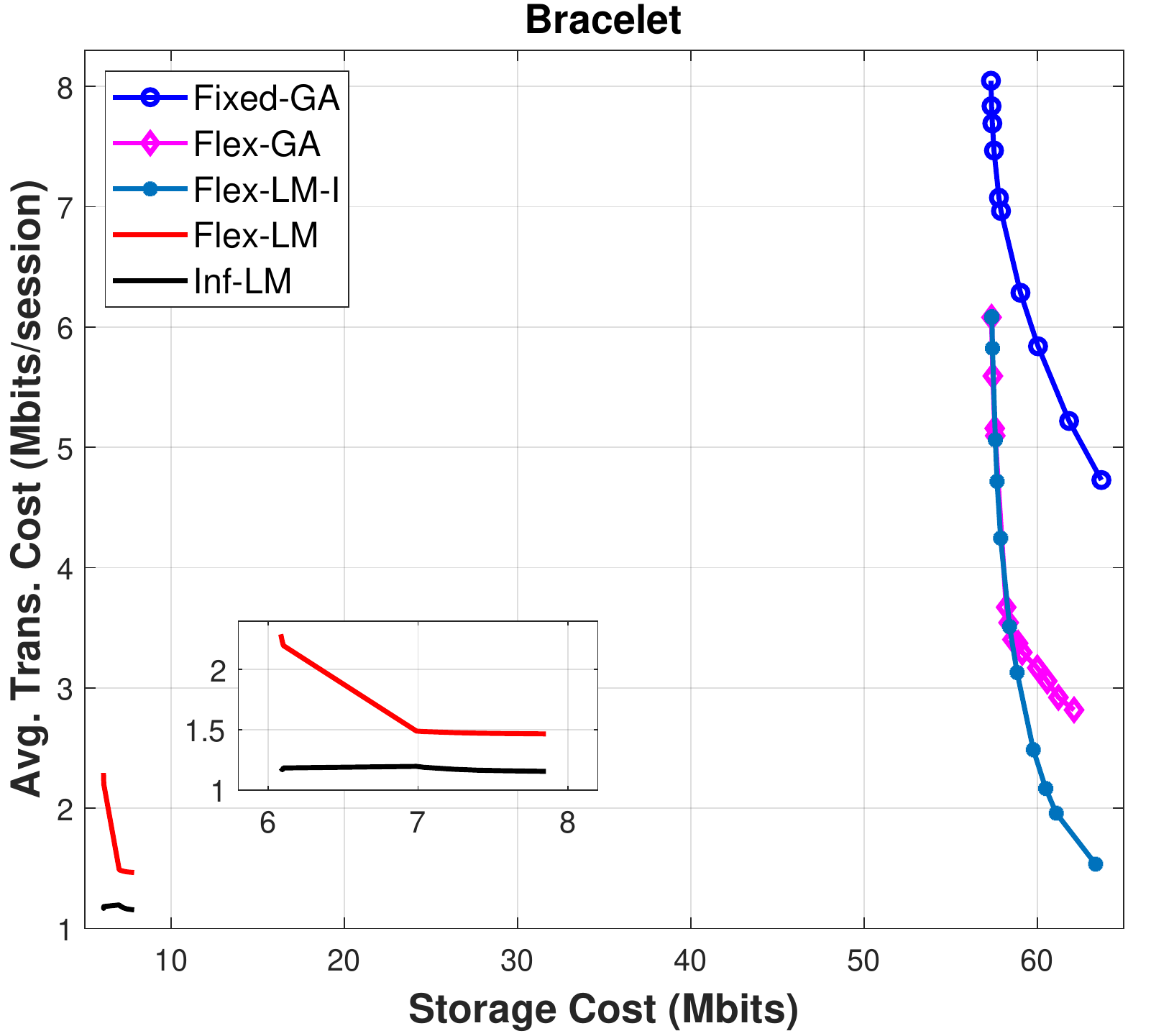} &
		\includegraphics[width=0.240\linewidth]{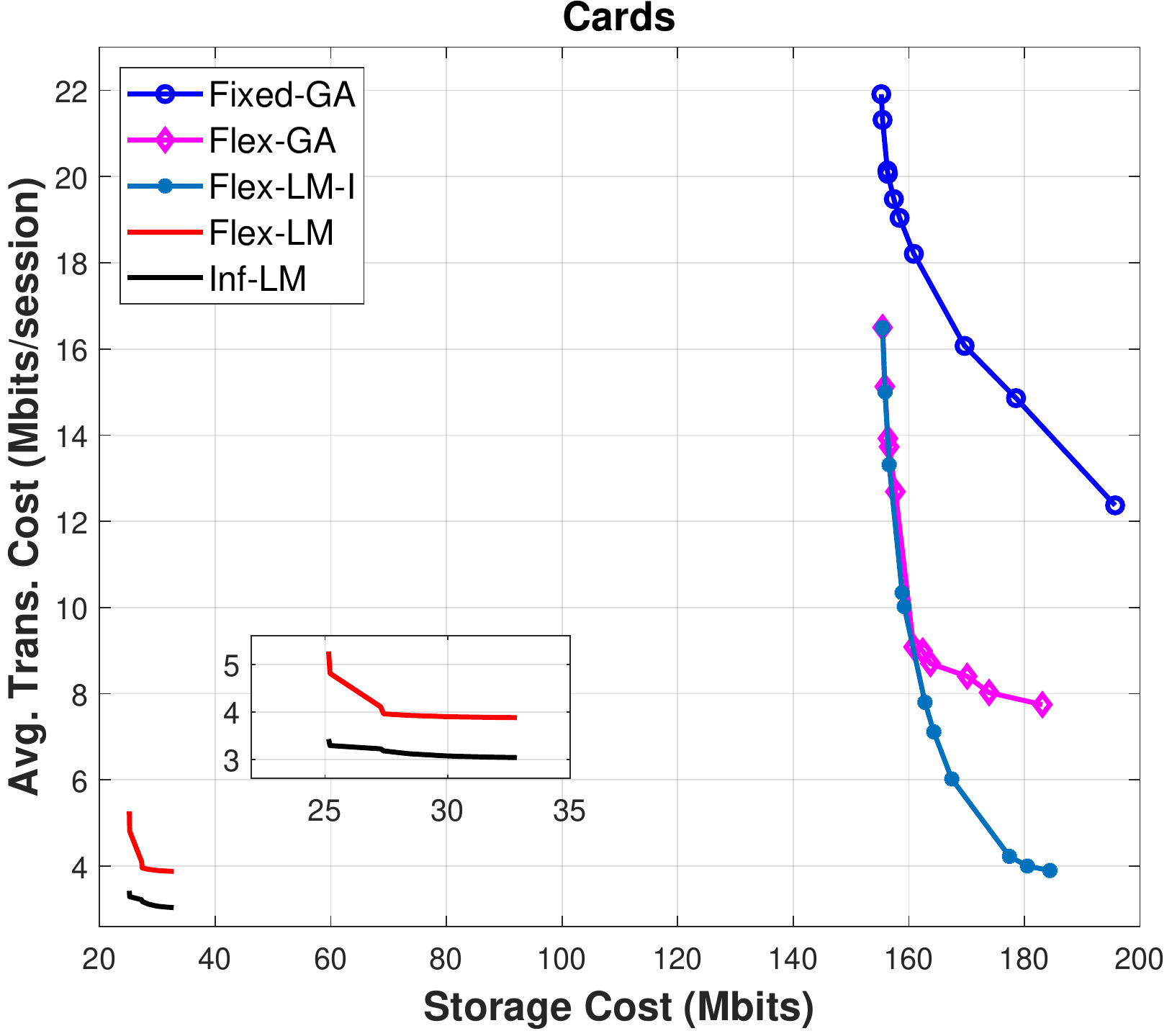} &
		\includegraphics[width=0.240\linewidth]{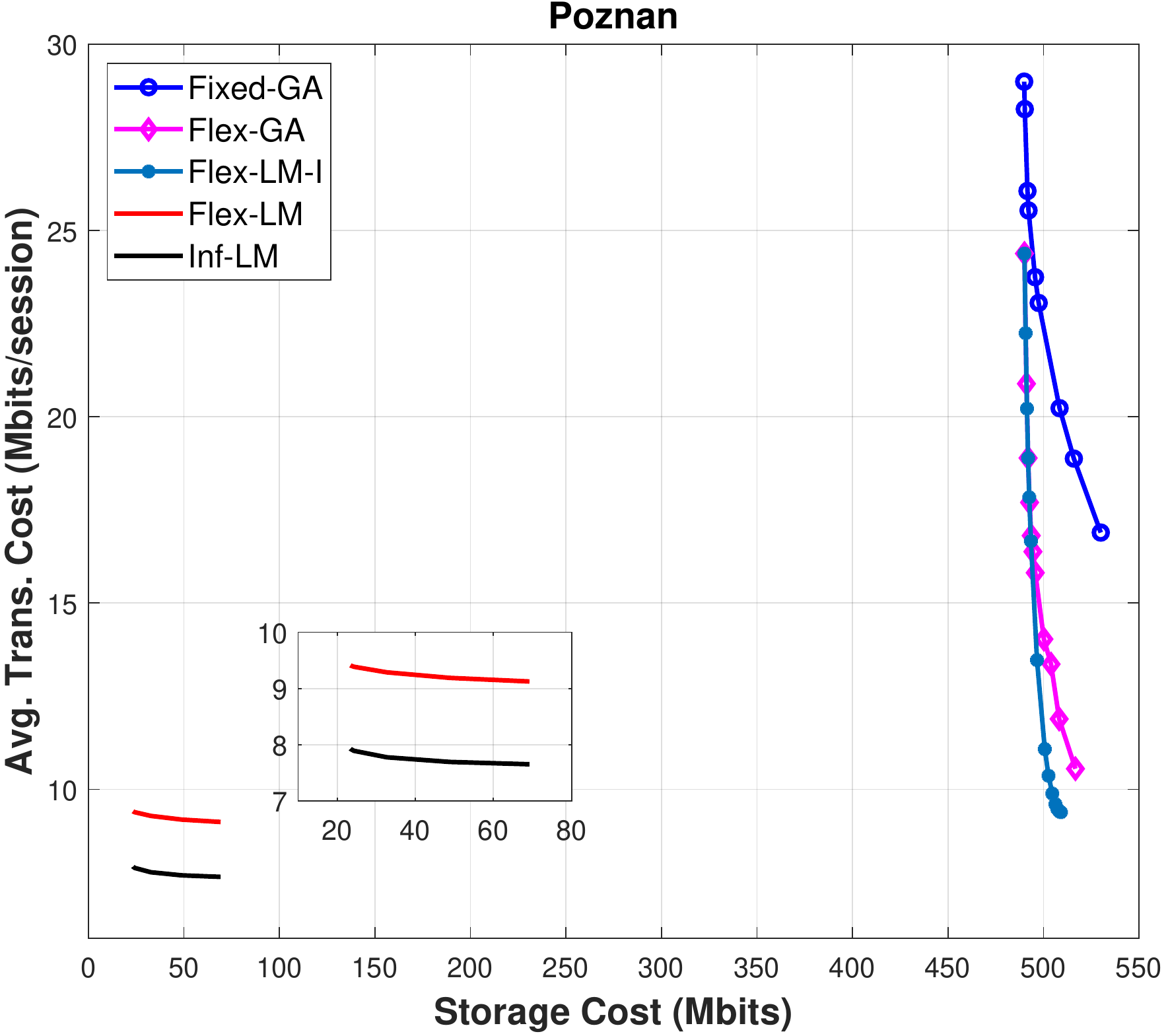} &
		\includegraphics[width=0.239\linewidth]{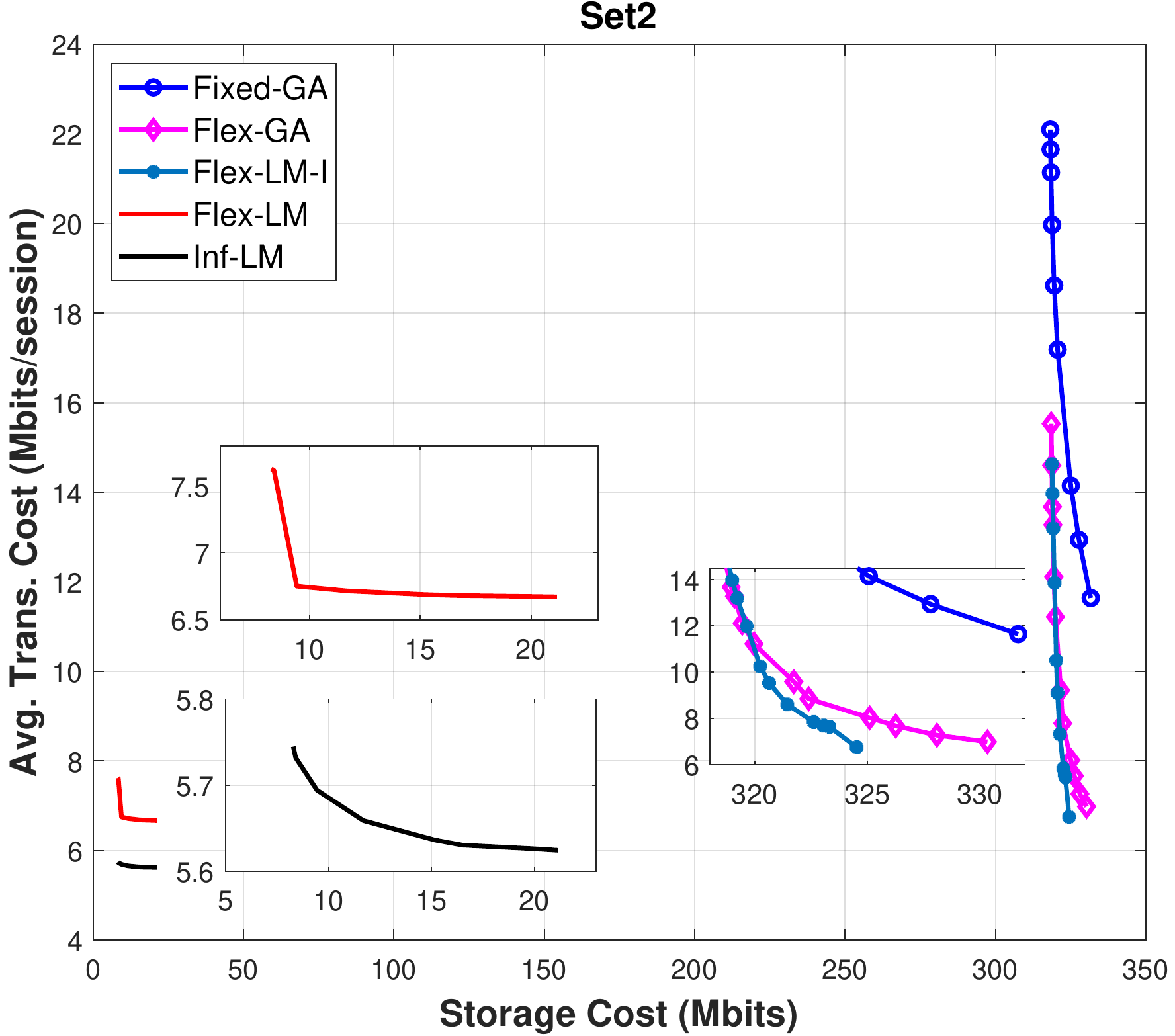} \\
		\footnotesize{(a) LF, \textsl{Bracelet}, $17 \times 17$} & \footnotesize{(b) LF, \textsl{Cards}, $17 \times 17$} & \footnotesize{(c) LF, \textsl{Poznan}, $31 \times 31$} & \footnotesize{(d) LF, \textsl{Set2}, $33 \times 11$} \\

		\includegraphics[width=0.242\linewidth]{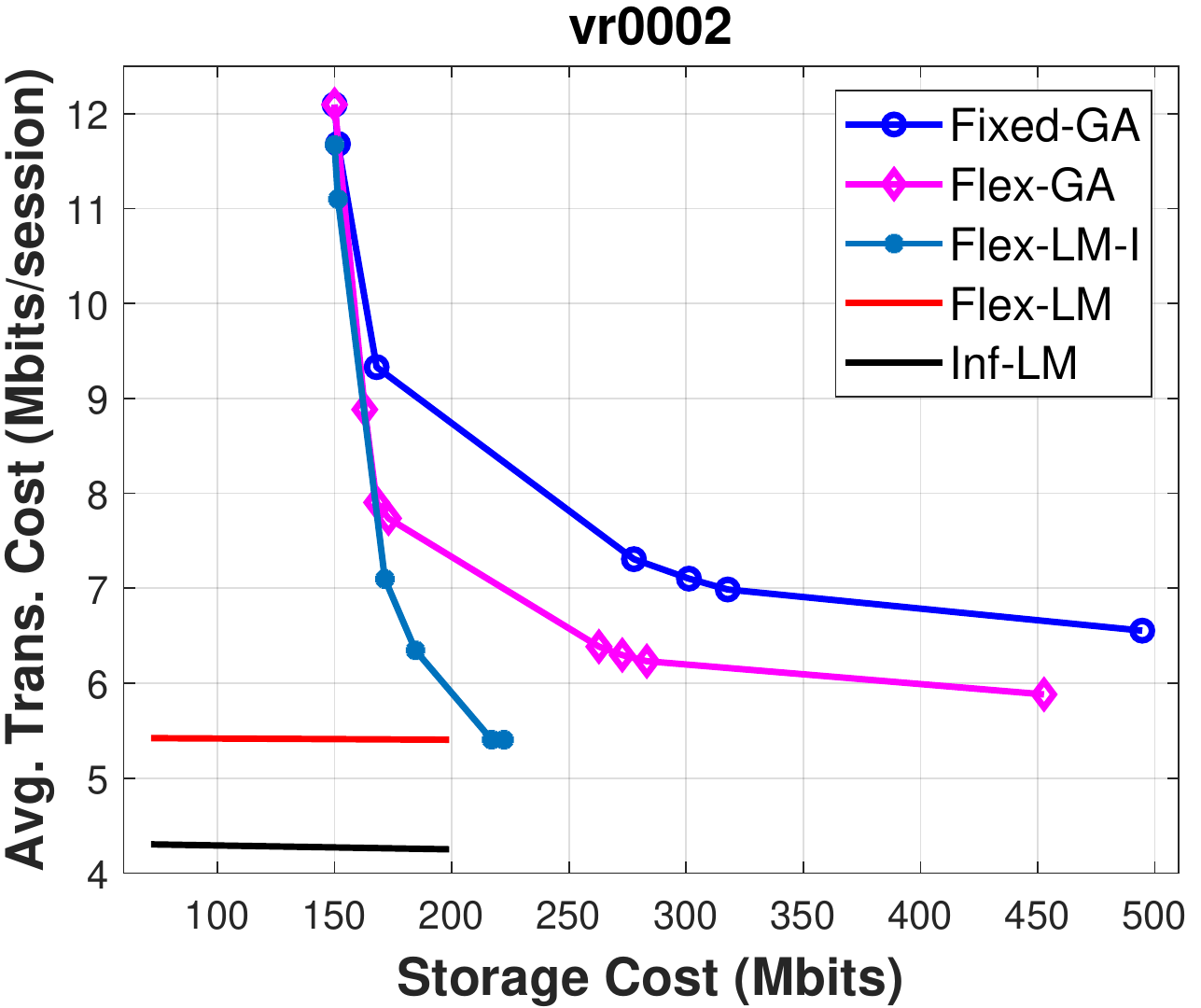} &
		\includegraphics[width=0.242\linewidth]{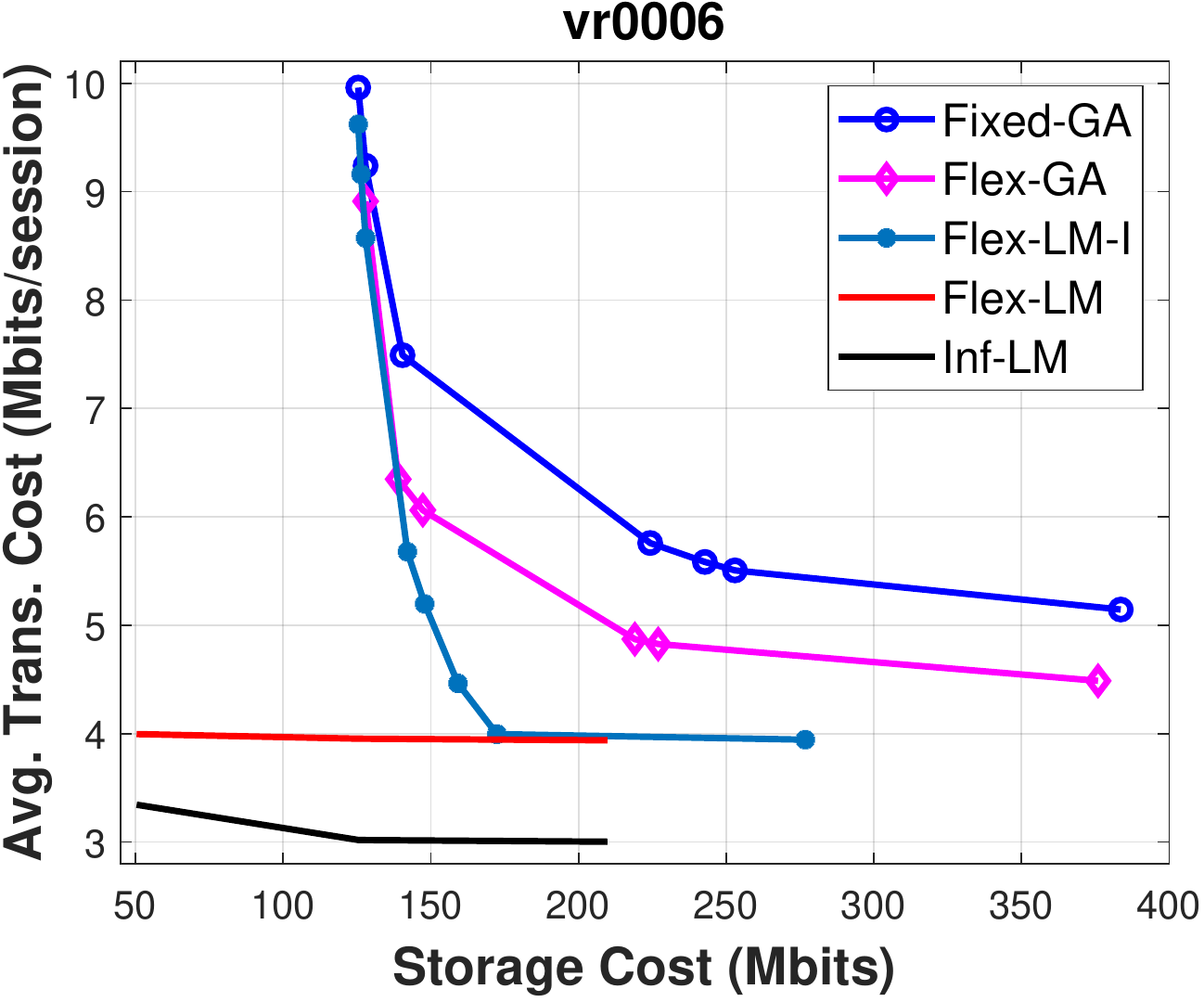} &
		\includegraphics[width=0.246\linewidth]{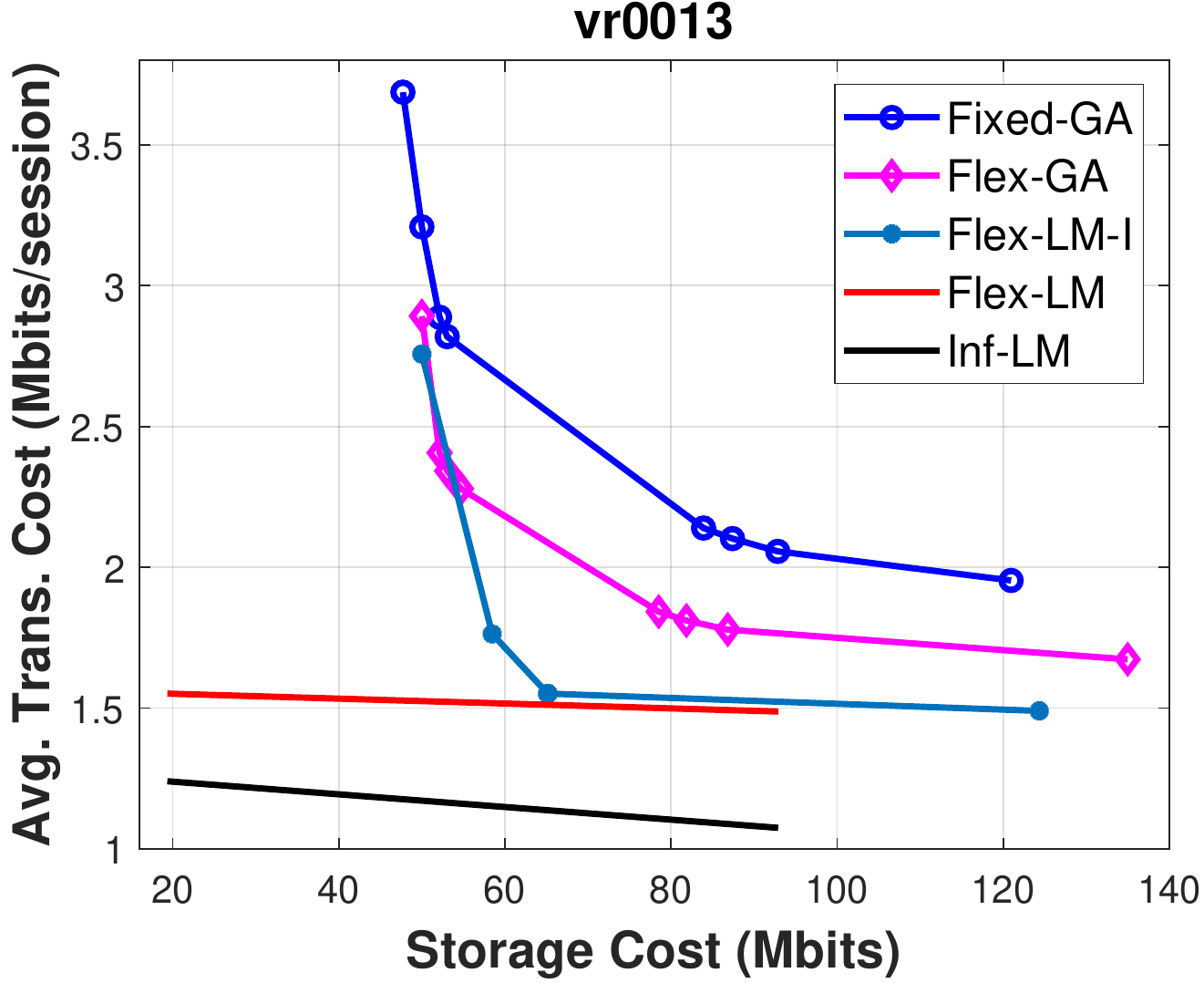} &
		\includegraphics[width=0.242\linewidth]{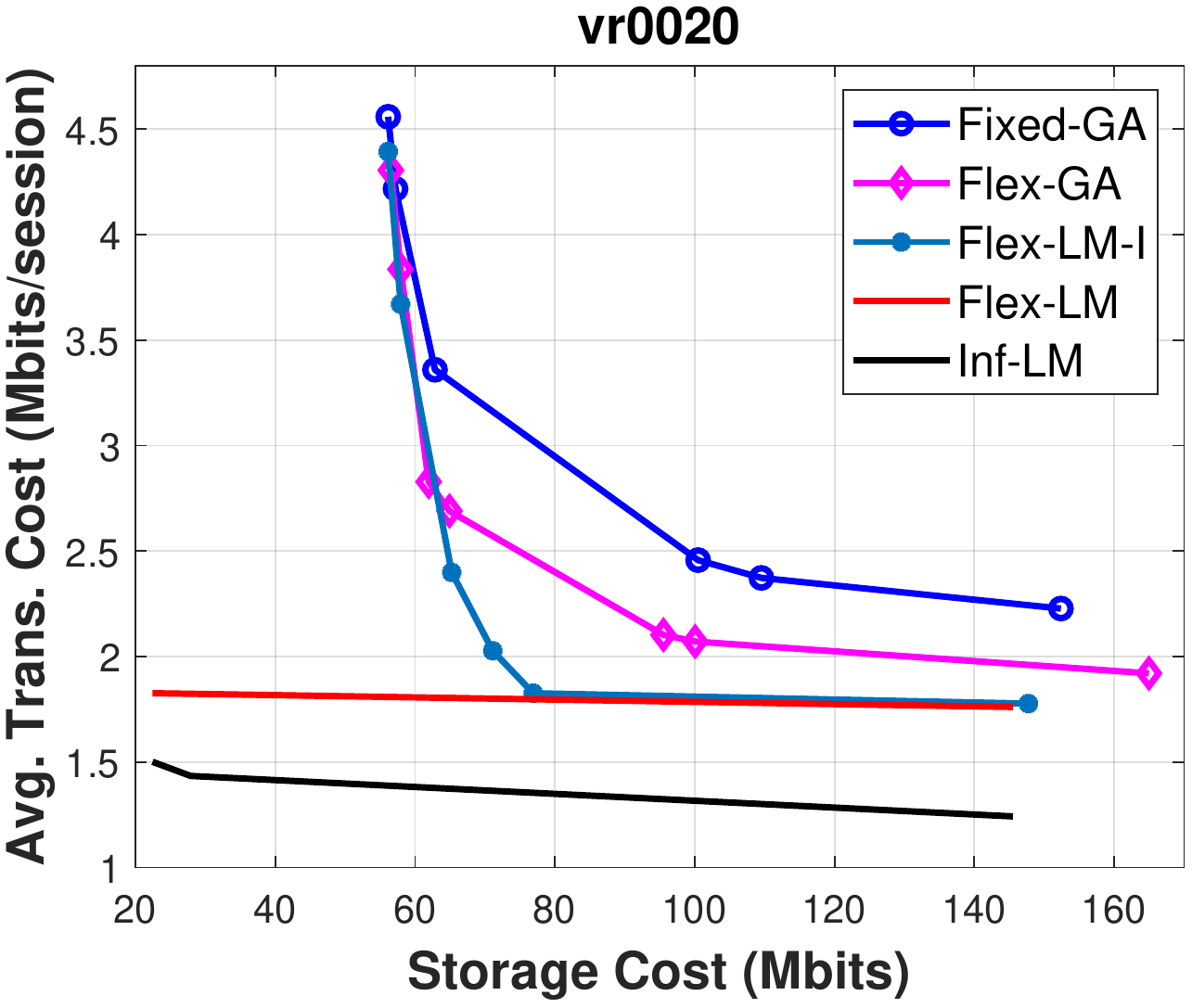} \\
		\footnotesize{(e) $360^\circ$, \textsl{vr0002}, 30 viewports} & \footnotesize{(f) $360^\circ$, \textsl{vr0006}, 30 viewports} & \footnotesize{(g) $360^\circ$, \textsl{vr0013}, 30 viewports} & \footnotesize{(h) $360^\circ$, \textsl{vr0020}, 30 viewports} \\
		\end{tabular}
	\end{center}
	\vspace{-0.15in}
	\caption{The storage cost (Mbits) \textit{vs.} the average transmission cost (Mbits/session) for LF data and $360^\circ$ images, respectively. The proposed method \texttt{Flex-LM} results in much smaller both storage and transmission cost than \texttt{Flex-GA} and \texttt{Fixed-GA}. More results are provided in the supplementary file.}
\label{fig:RD}
\end{figure*}

We first show our landmark insertion results.
Considering four nearest neighboring anchor views as an MDU, Fig.\;\ref{fig:lmkr} illustrated the MDU partitions and the landmark insertion results for LF data \textsl{Poznan} with $\lambda = 4.5$ (a) and $\lambda = 8.0$ (b), respectively.
The proposed TSVQ method divided the $30 \times 30$ MDUs into two partitions when $\lambda = 4.5$ and three partitions when $\lambda = 8.0$. 
As $\lambda$ became larger, according to \eqref{eq:object}, a smaller storage cost could reduce the total cost.
The number of landmarks increased and each landmark covered a smaller MDU neighborhood, such that P-MDUs used to transition to/from a landmark MDU could be smaller.
According to our proposed user behavior model, viewpoints around the center of a 2D grid had a larger probability to be observed.
Hence, the landmark MDUs were close to the starting viewpoint.

%\vspace{-0.08in}
\subsection{Experiment Results} 
Fig.\;\ref{fig:RD} showed the storage cost \textit{vs.} the expected transmission cost curves of the aforementioned 5 methods for LF and \360 images, respectively.
We enlarged the results of \texttt{Flex-LM} and \texttt{Inf-LM} for better visual experience.
We analyse the performance in the following.
For more results, please refer to the supplementary file.

\subsubsection{Overall Performance Analysis}
Compared with the other two MDU structure generating methods \texttt{Flex-GA} and \texttt{Fixed-GA}, the proposed \texttt{Flex-LM} achieved the best performance.
Even with a lower storage cost, \texttt{Flex-LM} could still result in a lower average transmission cost.
This result showed the efficiency and importance of our landmark based MDU structure optimization framework. 

We explain the results as follows.
Recall that \texttt{Flex-LM} initialized an MDU structure with I-MDUs of landmarks (a few) and $\mO(N)$ P-MDUs from landmarks to their neighborhood, where \texttt{Flex-GA} and \texttt{Fixed-GA} contained $\mO(N)$ I-MDUs.
Given the size of an I-MDU was much larger than a P-MDU,
the storage cost of \texttt{Flex-LM} was lower.
With the initialized structure and when a landmark was stored at the buffer, for any MDU-switch during navigation, the 1-hop transmission was very likely to be performed for \texttt{Flex-LM}. 
Even when MDU-switch crossed partition boundaries, the 2-hop transmission could enable switching from the current landmark to a new landmark. 
In contrast, the greedy algorithm did not generate landmarks in the MDU structure by making an immediately beneficial decision (considering one MDU at a time).
This was because a landmark MDU was not useful in reducing the objective only until a sufficient number of P-MDUs to neighboring MDUs were added. 
Since the 1-hop transmission cost was the smallest among the three possible transmission strategies, \texttt{Flex-LM} resulted in smaller expected transmission cost.
Note that the curves for \texttt{Flex-LM} were relatively ``flat" and ``short".
The reason was when decreasing $\lambda$, P-MDUs connecting two very close MDUs might be added into the structure to reduce the objective \eqref{eq:object}.
However, the expected transmission cost reduction was relatively small, since the initialized structure was already very efficient.
Moreover, when $\lambda$ became very small, greedily adding P-MDUs into the structure would no longer reduce the total cost.
Hence, iterative addition of P-MDUs terminated early.

%\vspace{-0.08in}
\renewcommand{\tabcolsep}{.1pt}
\begin{figure*}[htb]
	\begin{center}
		\begin{tabular}{cccc}
		
		\includegraphics[width=0.242\linewidth]{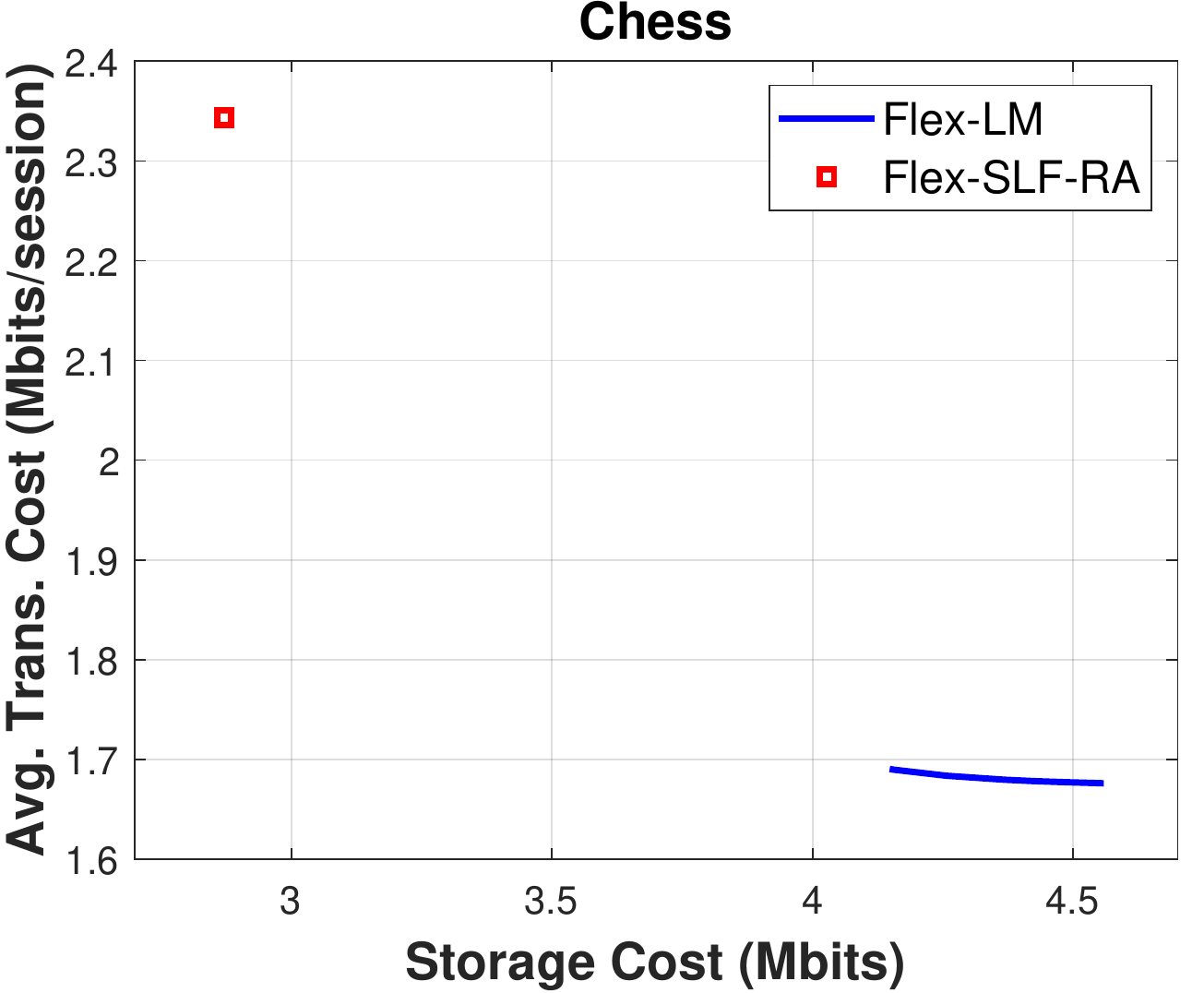} &
		\includegraphics[width=0.242\linewidth]{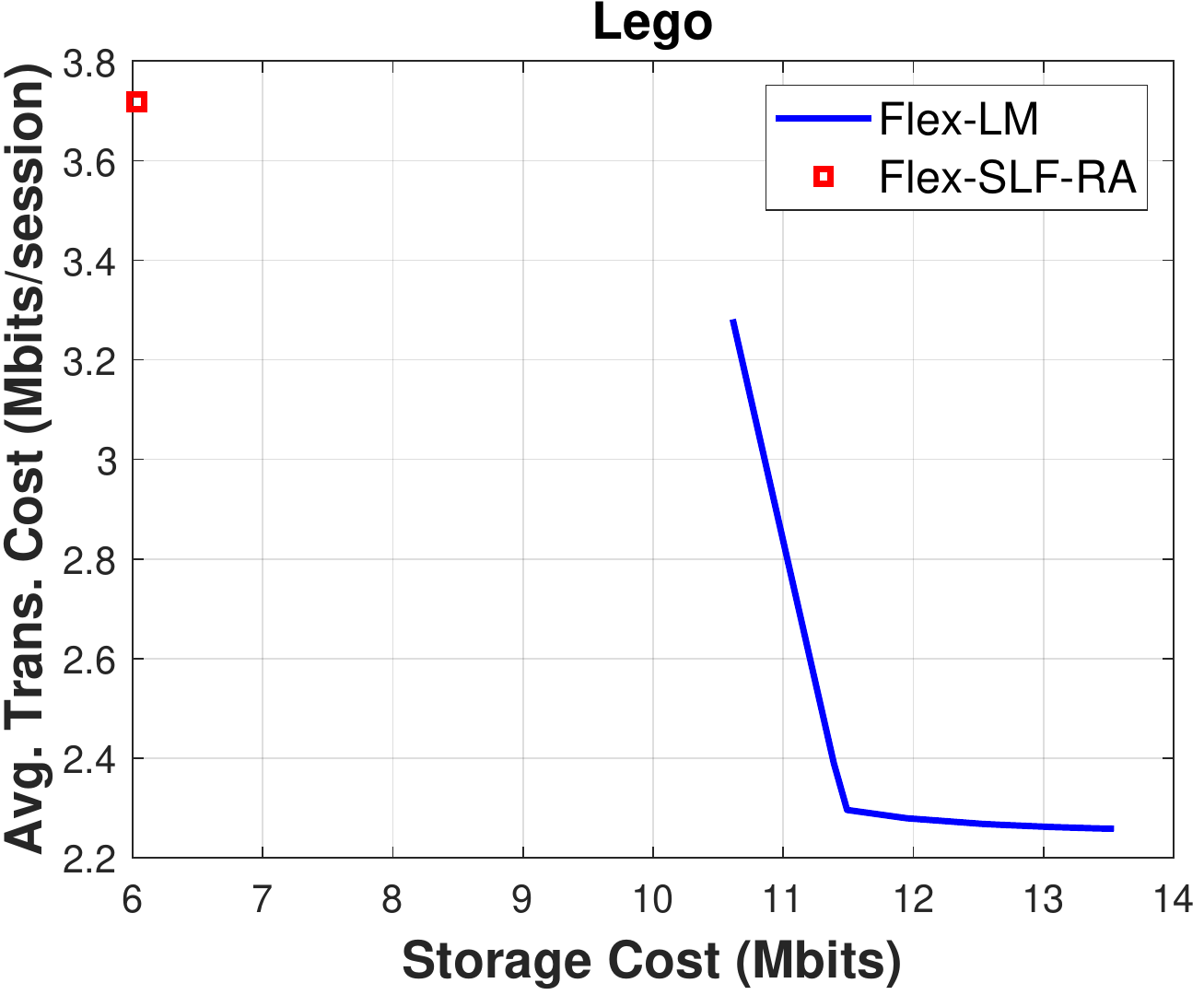} &
		\includegraphics[width=0.242\linewidth]{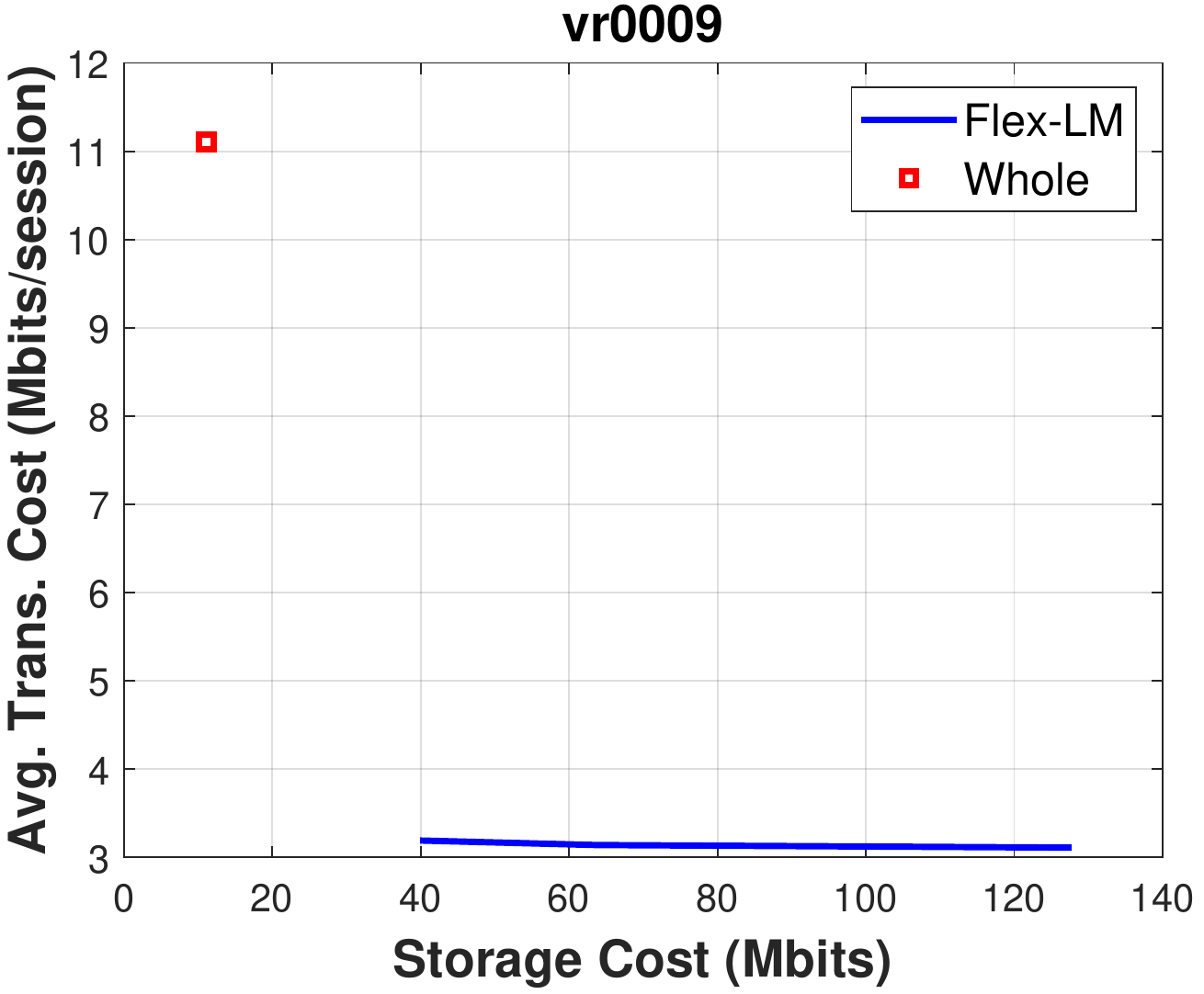} &
		\includegraphics[width=0.242\linewidth]{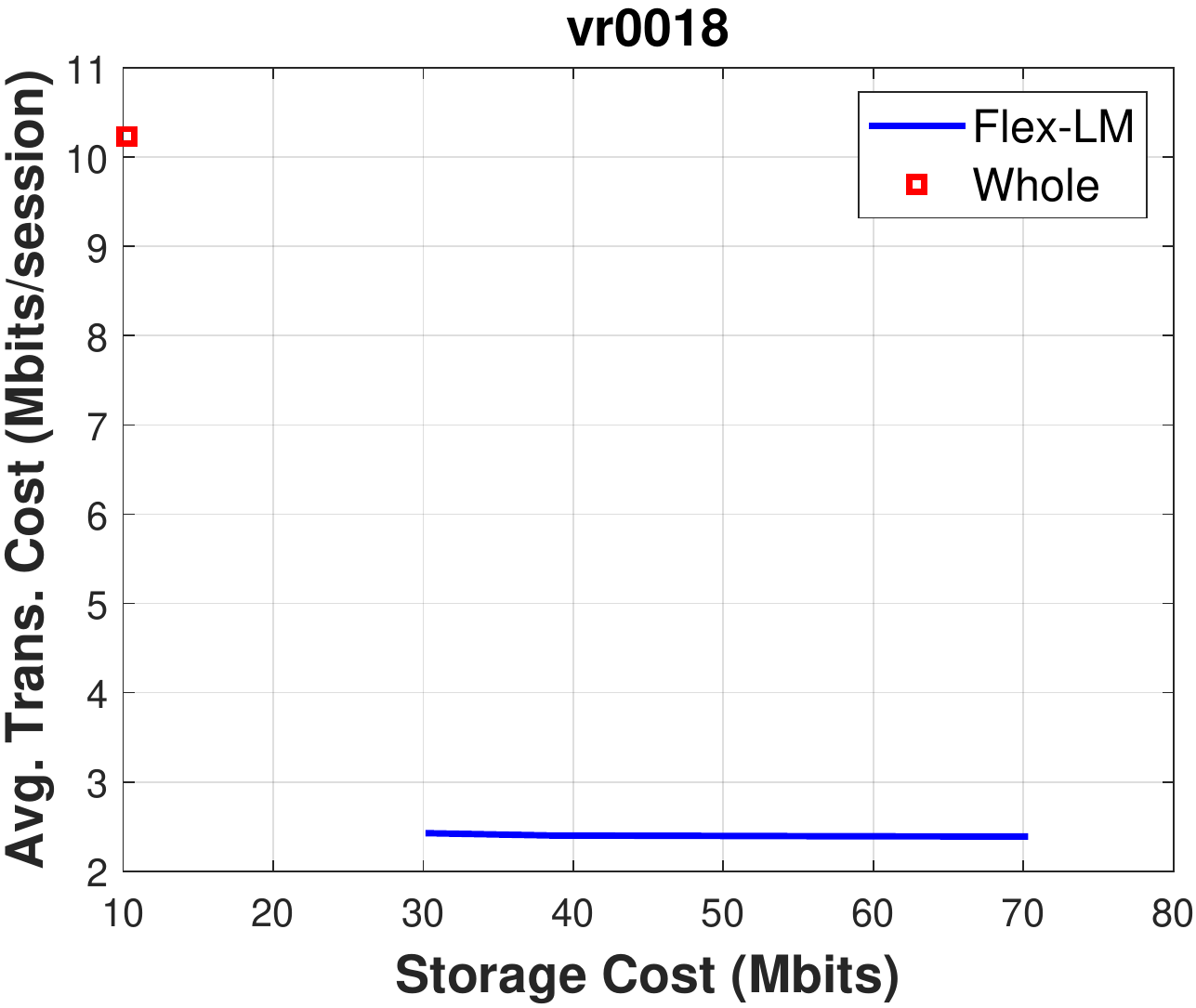}  \\
		\footnotesize{(a) LF, \textsl{Chess}} & \footnotesize{(b) LF, \textsl{Lego}}  & \footnotesize{(c) \360, \textsl{vr0009}} & \footnotesize{(d) \360, \textsl{vr0018}}\\
		\end{tabular}
	\end{center}
	\vspace{-0.15in}
	\caption{Comparing \texttt{Flex-LM} with other streaming strategies. \texttt{Flex-SLF-RA} \cite{monteiro2021light} is an LF image coding scheme that facilitates random access. \texttt{Whole} stands for encoding the \360 ERP image as a whole. \texttt{Flex-LM} can reduce transmission cost by exploiting extra storage space.}
\label{fig:LF_comp}
\end{figure*} 

\subsubsection{With or Without Landmarks}

To eliminate the influence of different I-MDU initialization, we compare \texttt{Flex-GA} with \texttt{Flex-LM-I}.
\texttt{Flex-LM-I} initialized the structure with I-MDUs for all the MDUs and P-MDUs from landmarks to their neighborhood.
Due to the redundancy during initialization, beyond greedily adding P-MDUs to the initialized structure, we also considered greedily subtracting P-MDUs.
We iteratively removed the most ``costly" P-MDU in the initialized structure one at a time, until the objective cost could not be further reduced.
We see from Fig.\;\ref{fig:RD} that the storage cost of \texttt{Flex-GA} and \texttt{Flex-LM-I} were in a similar range.
Given the same storage cost, \texttt{Flex-LM-I} resulted in a lower expected transmission cost.
The superiority of \texttt{Flex-LM-I} became more noticeable as storage cost increased. 
This was due to when the server could store more P-MDUs, P-MDUs connecting landmarks to their neighborhood MDUs were more likely to be pre-encoded and stored (with less subtraction). 
These P-MDUs facilitated the 1-hop transmission and thus reduced the transmission cost.
This comparison further shows the effectiveness of landmarking.

\subsubsection{Users' Decoder Buffer Size}
Compared with a flexible one-MDU and infinity buffer size, the MDU structures generated by a fixed one-MDU buffer, \textit{i.e.} \texttt{Fixed-GA} method, performed the worst. 
To achieve the same expected transmission cost, \texttt{Fixed-GA} required the largest storage cost.
This was reasonable since, with a flexible one-MDU buffer, one could select a beneficial MDU storing at buffer to facilitate future MDU-switches.
Given the same storage cost, \texttt{Inf-LM} resulted in the lowest transmission cost.
With infinity buffer size, any MDU that was previously traversed by a user did not require re-transmission and could be used as a reference to facilitated transmission.

\subsubsection{Comparing with Other Streaming Strategies}
To further show that our landmarking based MDU structure offers good tradeoffs between the expected transmission cost and storage size, we compared \texttt{Flex-LM} with some other streaming strategies for LF and \360 images, respectively.
Results are shown in Fig.\;\ref{fig:LF_comp}.

For LF images, the recent work \cite{monteiro2021light} proposed an LF image coding scheme that facilitates random access. 
It optimized the LF scan order and the selection of reference frames to trade off compression efficiency and viewpoint random access.
We used parameters provided in \cite{monteiro2021light} that offered the maximum random access capability to encode four LF image sets for comparison.
Specifically, we divided the $17 \times 17$ LF images into 9 regions.
The LF image in the middle of each region was intra-coded using HEVC, where the remaining images were inter-coded with the proposed scalability mask and a maximum of two dependency layers.
Given the compressed bitstream, we computed the storage cost and the expected transmission cost considering a flexible one-MDU reference buffer (labeled as \texttt{Flex-SLF-RA}).
As shown in Fig.\;\ref{fig:LF_comp}(a)(b), due to representation redundancy in our proposed structures, the storage costs of \texttt{Flex-LM} were larger than \texttt{Flex-SLF-RA}.
However, the expected transmission cost of \texttt{Flex-LM} could become less than half of \texttt{Flex-SLF-RA}. 

For \360 images, directly transmitting the entire image to users is a straightforward choice.
We encoded the entire \360 ERP images using the same VVC setup as encoding viewports.
The coding bitrates were then set as the storage and transmission cost, labeled as \texttt{Whole} in Fig.\;\ref{fig:LF_comp}(c)(d).
Similarly, the storage costs of \texttt{Flex-LM} were larger due to redundant representation.
But the expected transmission costs were more than 4 times lower than transmitting an entire \360 image.
When users only want an overview of \360 images, the proposed \texttt{Flex-LM} would be preferred than transmitting the entire images.

Results in Fig.\;\ref{fig:LF_comp} showed that \texttt{Flex-LM} was capable of exploiting extra storage space to create more redundant representations in order to lower the transmission rate.
Given that storage cost is generally cheaper than transmission cost, our proposed \texttt{Flex-LM} would be an important solution when bandwidth budget is the bottleneck.

%\vspace{-0.08in}
\renewcommand{\tabcolsep}{.1pt}
\begin{figure}[htb]
	\begin{center}
		\begin{tabular}{cc}
		
		\includegraphics[width=\swtwo]{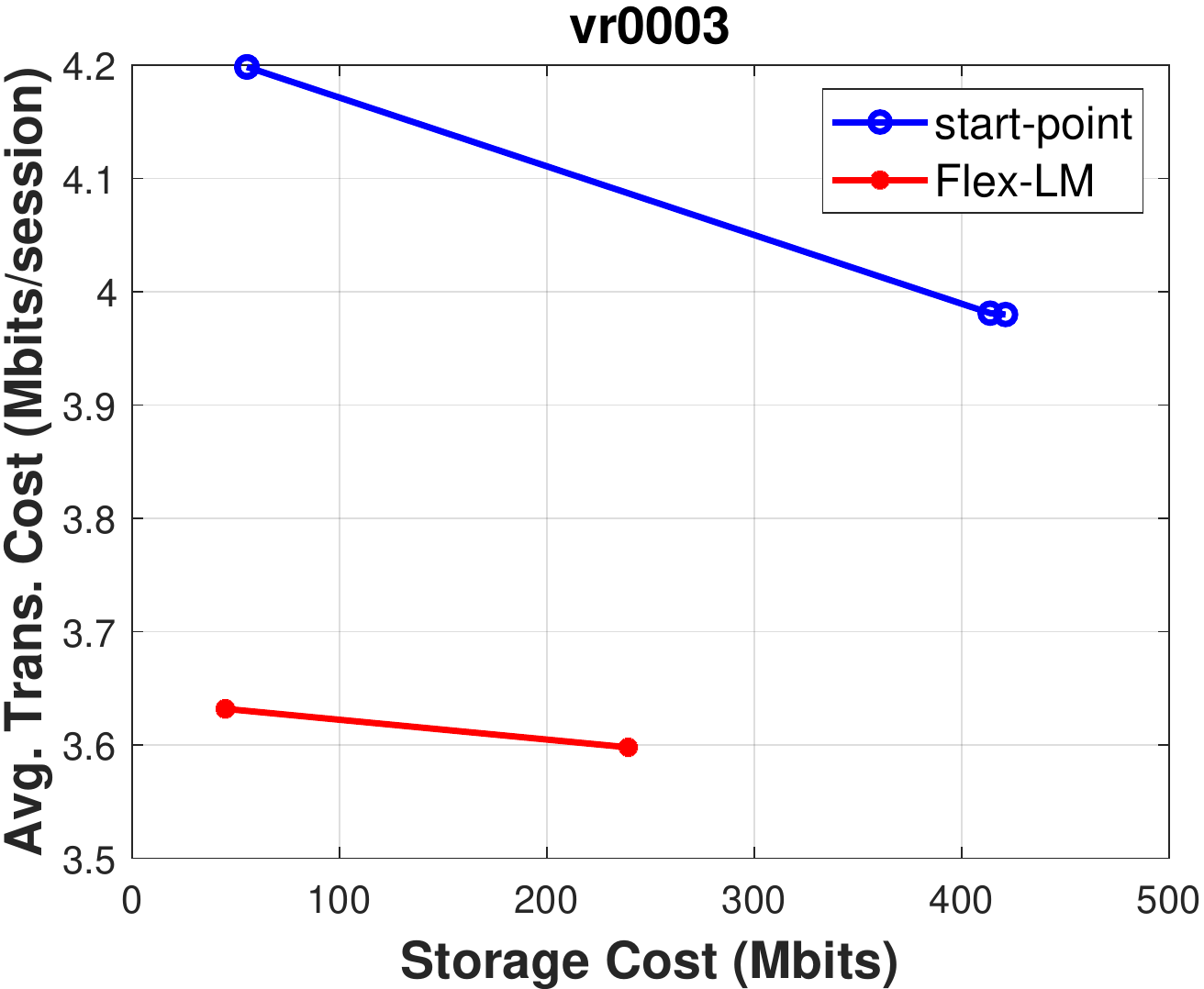} &
		\includegraphics[width=\swtwo]{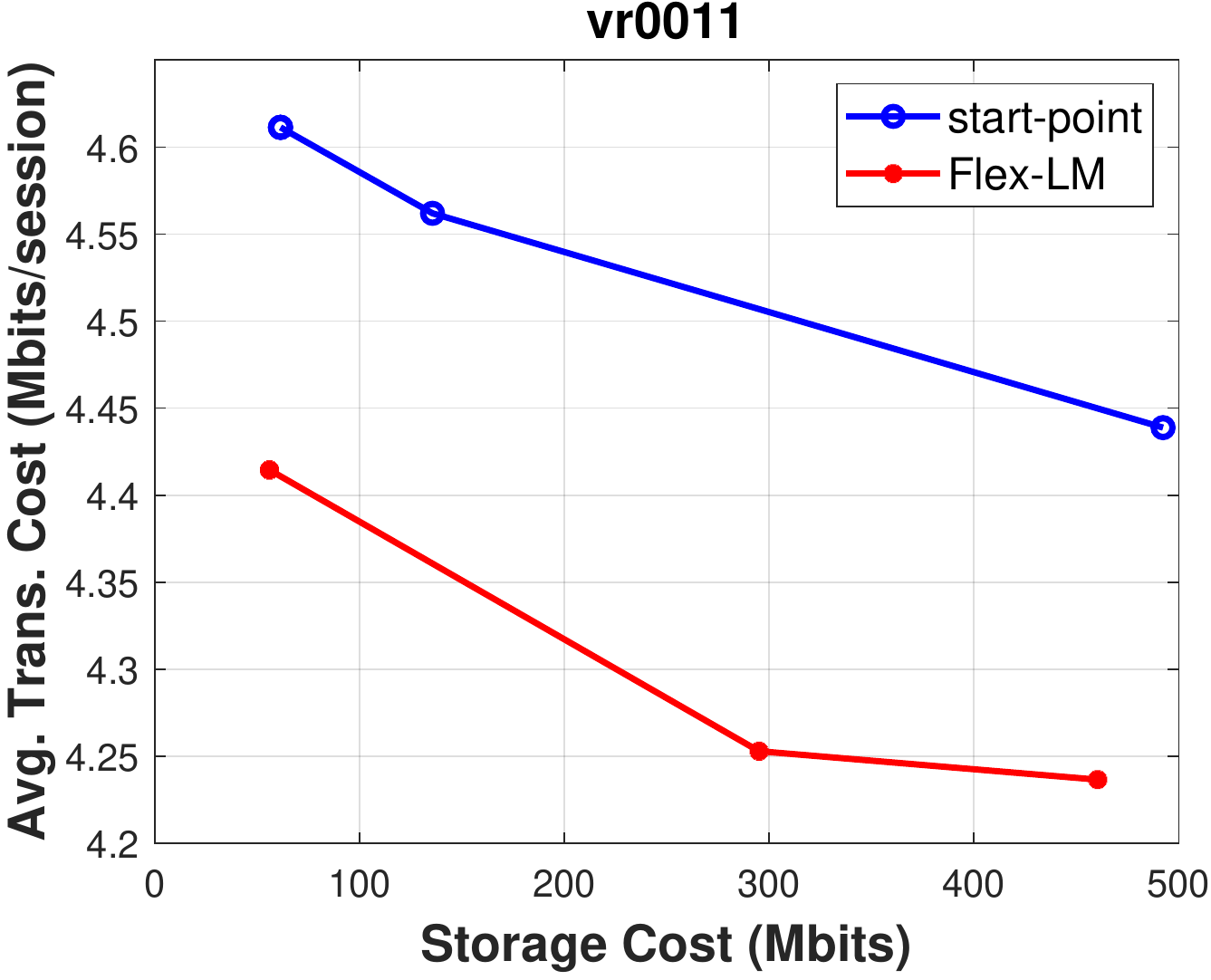}  \\
		\footnotesize{(a) $360^\circ$, \textsl{vr0003}} & \footnotesize{(b) $360^\circ$, \textsl{vr0011}} \\
		
		\end{tabular}
	\end{center}
	\vspace{-0.1in}
	\caption{The storage-transmission cost curves for $360^\circ$ images. \texttt{start-point} means adopting the starting viewport as the landmark directly, where \texttt{Flex-LM} means using our proposed landmark insertion algorithm to find the optimal landmark.}
\label{fig:comp}
\end{figure}

\subsubsection{Effectiveness of Landmark Insertion Algorithm}
Due to the small number of viewports in navigational \360 image streaming, there was always only one landmark in the generated viewport structure.
To show the effectiveness of our proposed optimal landmark insertion algorithm, we compared the storage-transmission cost curve by directly using the starting viewport as the landmark. 
The proposed MDU structure design algorithm was used in both these two cases.
Results were shown in Fig.\;\ref{fig:comp}, where \texttt{start-point} meant adopting the starting viewport as the landmark directly, and \texttt{Flex-LM} meant using our proposed algorithm to find the optimal landmark.
It showed that given the same storage cost, with our optimal landmark insertion algorithm, the expected transmission cost could be reduced.
These results proved the effectiveness of our optimal landmark insertion algorithm.

\vspace{\emptsp}
\section{Conclusion}
\label{sec:conclude}
%\vspace{-0.05in}
To enable bandwidth-friendly streaming of HD media, navigational streaming is desirable.
Towards efficient encoding, the problem is to simultaneously enable differential encoding among MDUs while achieving random access.
To address this problem, we propose a landmark-based MDU optimization framework with redundant representation: each MDU can be pre-encoded into intra-coded I-MDU as well as one or more inter-coded P-MDUs.
Specifically, we divide the media into non-overlapping partitions, where MDUs in each partition are inter-coded into P-MDUs using the chosen landmark as the predictor.
This means that any MDU can switch to another in the same neighborhood by transmitting just one P-MDU, assuming the landmark resides in the reference buffer. 
We first use tree-structured vector quantizer (TSVQ) to identify partitions and corresponding landmarks.
We then greedily add P-MDUs using a branch-and-bound algorithm to reduce the computation complexity.
Using light field images and $360^\circ$ images as examples, we conducted extensive experiments to show significant transmission cost reduction over structures without using landmarks. 
Our proposed algorithm can improve the coding performance and transmission efficiency of many real-world HD navigational streaming scenarios. 

\vspace{0.1in}
\noindent \textbf{Acknowledgement}. \;
We thank Zhiyou Ma from Kandao Technology for help projecting \360 spheres into viewports.



%\input{append}
\vspace{-0.1in}
\bibliographystyle{IEEEtran}
\bibliography{ref}

%\begin{IEEEbiography}[{\includegraphics[bb=0 0 750 1000, width=1in,height=1.25in,clip,keepaspectratio]{figures/yuan_yuan.jpg}}]{Yuan Yuan} (S'12, M'18) 
%received the B.Eng. degree from University of Science and Technology of China, Hefei, China, in 2011, and the Ph.D. degree in electrical and computer engineering from the University of Alberta, Canada, in 2017. 
%She also conducted her research in National Institute of Informatics in Tokyo, Japan (2013--2016). 
%Her research interests include 3D multimedia, image/video processing and computer vision.
%\end{IEEEbiography}

\begin{IEEEbiography}[{\includegraphics[width=1in,height=1.25in]{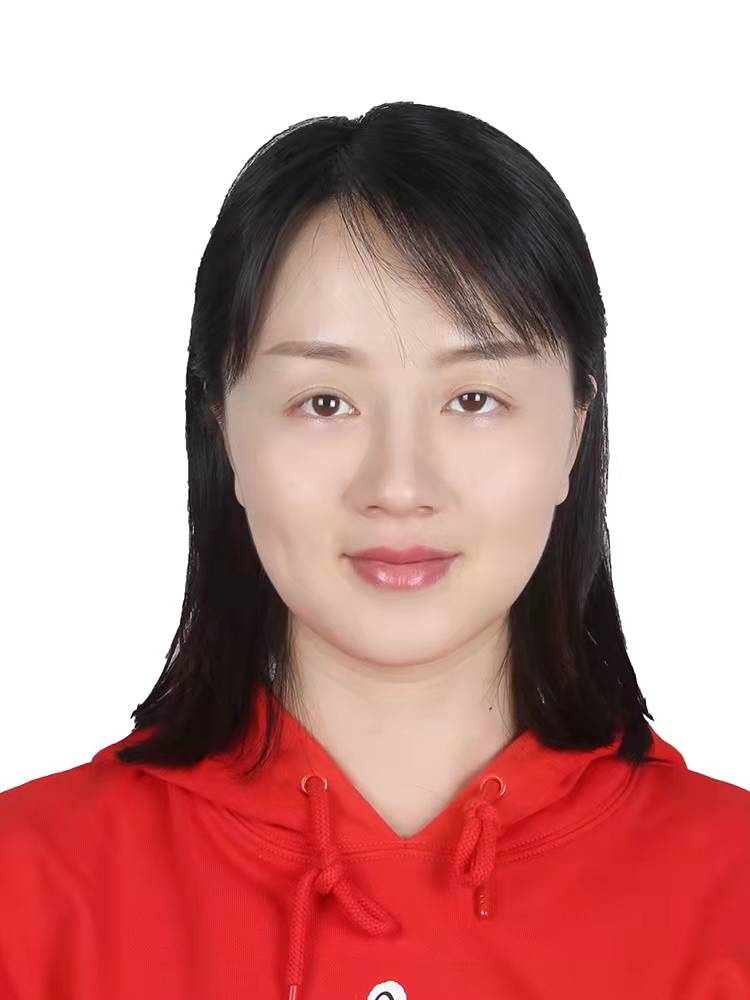}}]{Yuan Yuan} 
received the B.Eng. degree from the University of Science and Technology of China, Hefei, China, in 2011, and the Ph.D. degree in electrical and computer engineering from the University of Alberta, Canada, in 2017. 
She also conducted her research at the National Institute of Informatics in Tokyo, Japan (2013--2016).
She was an associate researcher in Shenzhen University, Shenzhen, China, from 2018 till 2021.
She is now an associate professor in Guangdong Polytechnic Normal University, Guangdong, China. 
Her research interests include 3D multimedia, image/video processing and computer vision.
\end{IEEEbiography}

\begin{IEEEbiography}[{\includegraphics[width=1in,height=1.25in]{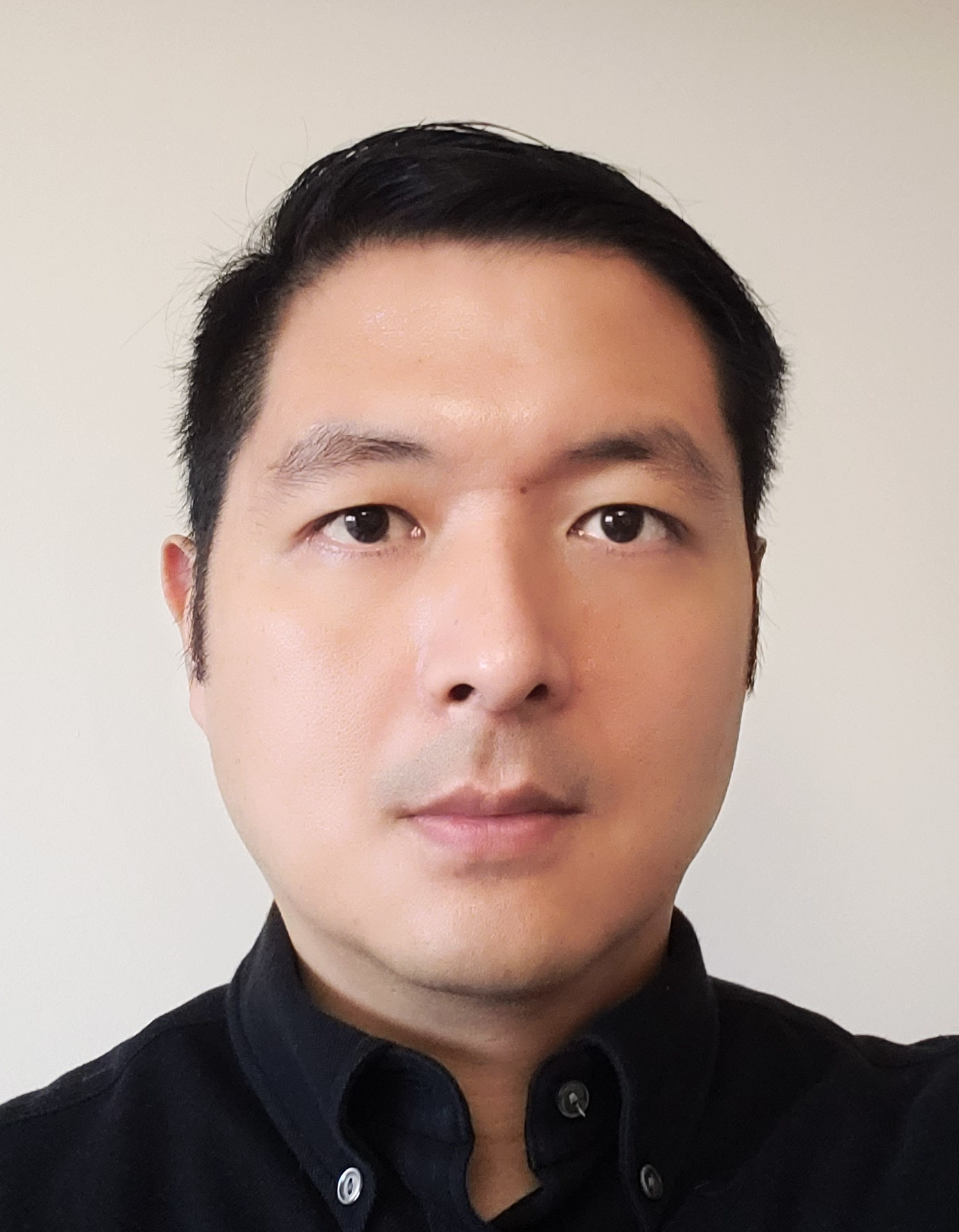}}]{Gene Cheung} (M'00---SM'07---F'21)
received the B.S. degree in electrical engineering from Cornell University in 1995, and the M.S. and Ph.D. degrees in electrical engineering and computer science from the University of California, Berkeley, in 1998 and 2000, respectively. 

He was a senior researcher in Hewlett-Packard Laboratories Japan, Tokyo, from 2000 till 2009. 
He was an assistant then associate professor in National Institute of Informatics (NII) in Tokyo, Japan, from 2009 till 2018. 
He is now an associate professor in York University, Toronto, Canada.

His research interests include 3D imaging and graph signal processing. 
He has served as an associate editor for multiple journals, including IEEE Transactions on Multimedia (2007--2011), IEEE Transactions on Circuits and Systems for Video Technology (2016--2017) and IEEE Transactions on Image Processing (2015--2019).
He currently serves as a senior associate editor for IEEE Signal Processing Letters (2021--present).
He served as a member of the Multimedia Signal Processing Technical Committee (MMSP-TC) in IEEE Signal Processing Society (2012--2014), and a member of the Image, Video, and Multidimensional Signal Processing Technical Committee (IVMSP-TC) (2015--2017, 2018--2020). 
He is a co-author of several paper awards and nominations, including the best student paper finalist in ICASSP 2021, best student paper award in ICIP 2013, ICIP 2017 and IVMSP 2016, best paper runner-up award in ICME 2012, and IEEE Signal Processing Society (SPS) Japan best paper award 2016.
He is a recipient of the Canadian NSERC Discovery Accelerator Supplement (DAS) 2019.
He is a fellow of IEEE.
\end{IEEEbiography}

\begin{IEEEbiography}[{\includegraphics[bb=0 0 216 250, width=1in,height=1.25in,clip,keepaspectratio]{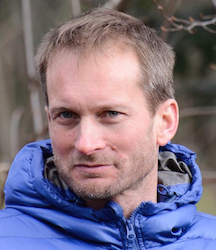}}]{Pascal Frossard} (Fellow, IEEE) 
has been a Faculty with \'{E}cole Polytechnique F\'{e}d\'{e}rale de Lausanne, since 2003, where he heads the Signal Processing Laboratory.
His research interests include network data analysis, image representation and understanding, and machine learning.
Between 2001 and 2003, he was a member of the Research Staff with IBM T. J. Watson Research Center, Yorktown Heights, NY, USA. 
Dr. Frossard was the recipient of the Swiss NSF Professorship Award in 2003, IBM Faculty Award in 2005, IBM Exploratory Stream Analytics Innovation Award in 2008, Google Faculty Award in 2017, IEEE Transactions on Multimedia Best Paper Award in 2011, and IEEE Signal Processing Magazine Best Paper Award in 2016. He is a fellow of ELLIS. 
\end{IEEEbiography}

\begin{IEEEbiography}[{\includegraphics[bb=0 0 853 1280, width=1in,height=1.25in,clip,keepaspectratio]{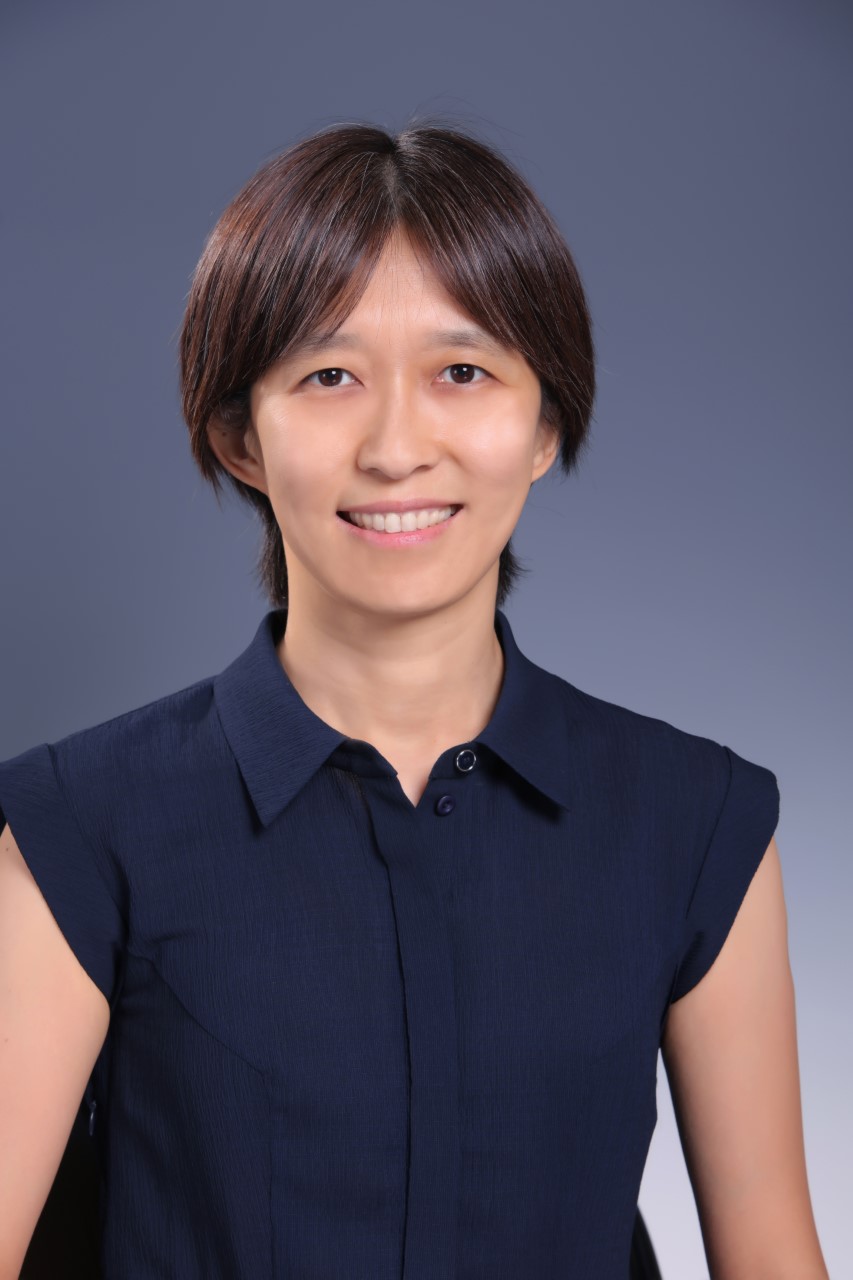}}]{H. Vicky Zhao} received the B.S. and M.S. degree from Tsinghua University, China, in 1997 and 1999, respectively, and the Ph. D degree from University of Maryland, College Park, in 2004, all in electrical engineering. She was a Research Associate with the Department of Electrical and Computer Engineering and the Institute for Systems Research, University of Maryland, College Park from Jan. 2005 to July 2006. From Aug. 2006 to Feb. 2016, she was with the Department of Electrical and Computer Engineering, University of Alberta, Edmonton, Canada, as an Assistant Professor (2006-2012) and then an Associate Professor (2012-2016). Since May 2016, she has been an Associate Professor with the Department of Automation, Tsinghua University, Beijing, China.

Dr. Zhao’s research interests include media-sharing social networks, information security and forensics, digital communications and signal processing. 
Dr. Zhao received the IEEE Signal Processing Society (SPS) 2008 Young Author Best Paper. 
She was a co-author of “Multimedia Fingerprinting Forensics for Traitor Tracing” (Hindawi, 2005), “Behavior Dynamics in Media-Sharing Social Networks” (Cambridge University Press, 2011), and “Behavior and Evolutionary Dynamics in Crowd Networks” (Springer, 2020). 
She was a member of IEEE Signal Processing Society Information Forensics and Security Technical Committee and Multimedia Signal Processing Technical Committee. 
She is the Senior Area Editor, Area Editor and the Associate Editor for IEEE Signal Processing Letters, IEEE Signal Processing Magazine, IEEE Information Forensics and Security, and IEEE Open Journal of Signal Processing. 
Dr. Zhao also actively participates in organizing international conferences as General Co-Chair, Technical Program Co-Chair, Finance Co-Chair, etc.
\end{IEEEbiography}

\begin{IEEEbiography}[{\includegraphics[width=1in,height=1.25in]{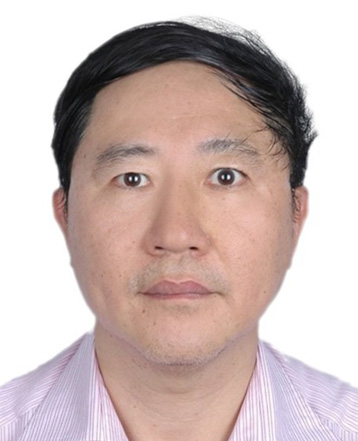}}]{Jiwu Huang} (Fellow, IEEE) 
received the B.S. degree from Xidian University, Xi’an, China, in 1982, the M.S. degree from Tsinghua University, Beijing, China, in 1987, and the Ph.D. degree from the Institute of Automation, Chinese Academy of Sciences, Beijing, in 1998. He is currently a Professor at the College of Electronics and Information Engineering, Shenzhen University, Shenzhen, China. His current research interests include multimedia forensics and security. He has coauthored more than 300 articles. He served as an Associate Editor for a few international journals, including IEEE TRANSACTIONS ON INFORMATION FORENSICS AND SECURITY, and the TPC chair for some international conferences.
\end{IEEEbiography}

\newpage
\textbf{SUPPLEMENTARY}
\IEEEpeerreviewmaketitle
\subsection{\textbf{The Starting MDU-switch Transition Probability $p_{s,j}$} }

In section \RNum{7}-B-2, when computing the MDU-switch probability for interactive light field streaming (ILFS), we only present the computation of the one-step memory MDU-switch probability $p_{k,i,j}$.
To compute the starting MDU-switch transition probability $p_{s,j}$ (viewpoints switching from starting MDU $s$ to MDU $j$), we define $f_{o,w}$ as the probability of switching from initial viewpoint $o$ to another viewpoint $w$,
\begin{eqnarray}
f_{o,w} = \beta \exp \left\{ - \frac{\norm{\mathbf{w} - \mathbf{o}}^2}{2\sigma^2} \right\}
\label{eq:fuv}
\end{eqnarray} 
where $\beta$ is a normalization parameter.
$\mathbf{o}$ and $\mathbf{w}$ are the 2D coordinates of view $o$ and $w$, respectively. 
The closer $w$ to $o$, the larger the switching probability.
The transition probability $p_{s,j}$ is then defined as:
\begin{align}
p_{s,j} =\int_{\mathbf{o} \in A_s} \int_{\mathbf{w} \in A_j} f_{o,w} \;\; d\mathbf{o}\; d\mathbf{w}.
\end{align}

\subsection{\textbf{\360 Images used in the Experiments}}
\begin{figure*}[htb]
	\begin{center}
		\begin{tabular}{cccc}
		
		\includegraphics[width=\swfour]{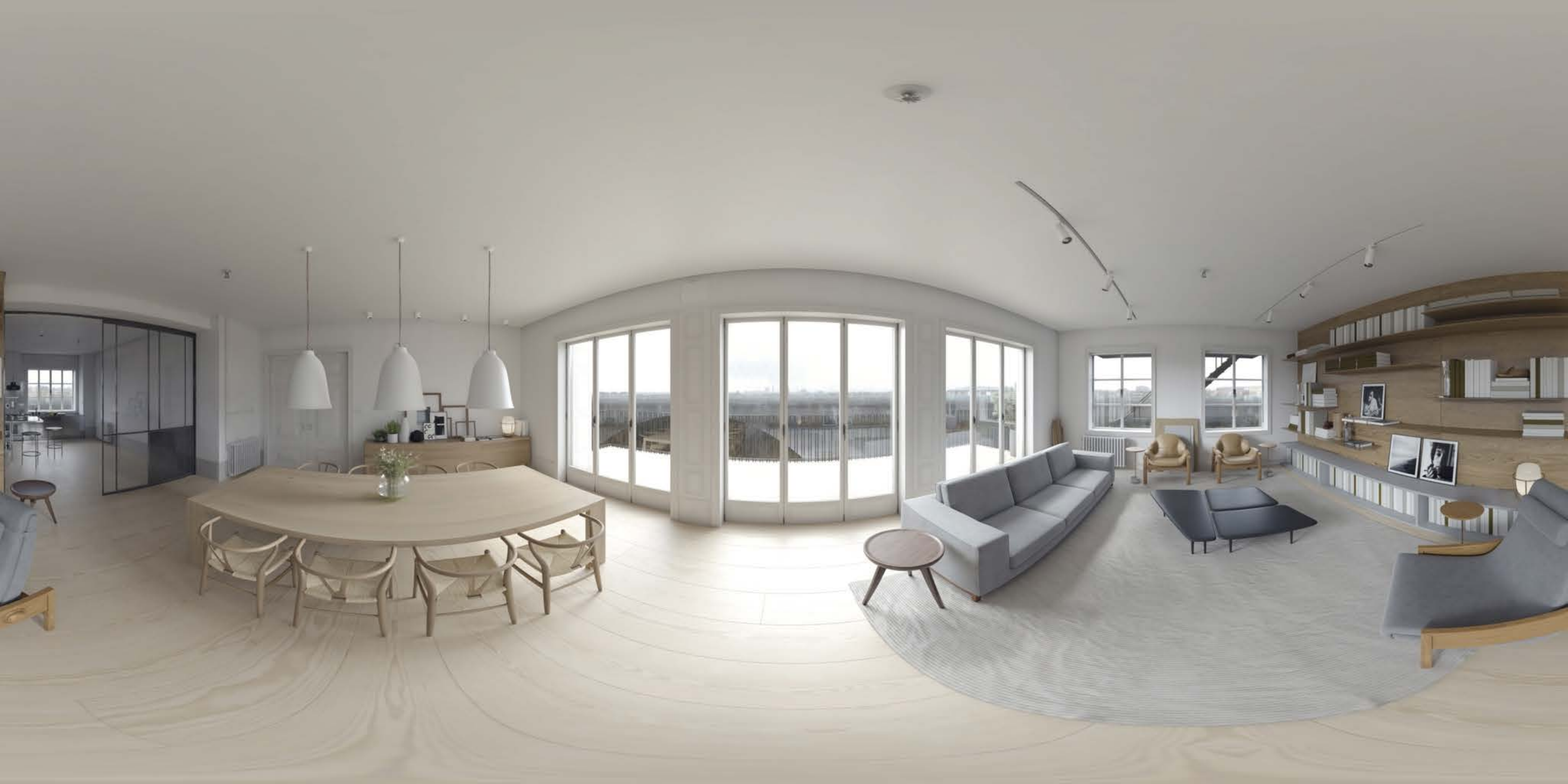} &
		\includegraphics[width=\swfour]{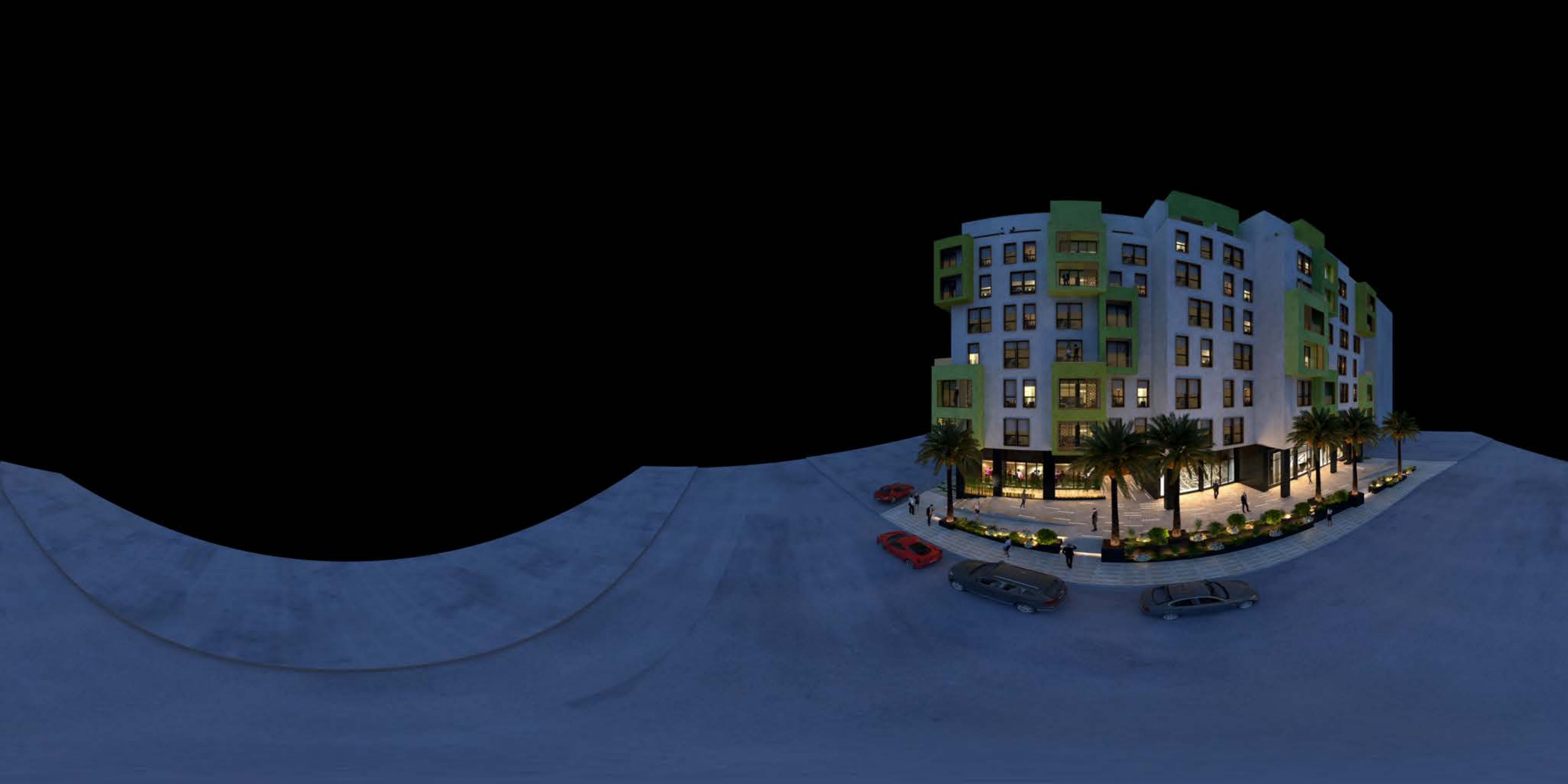} & 
		\includegraphics[width=\swfour]{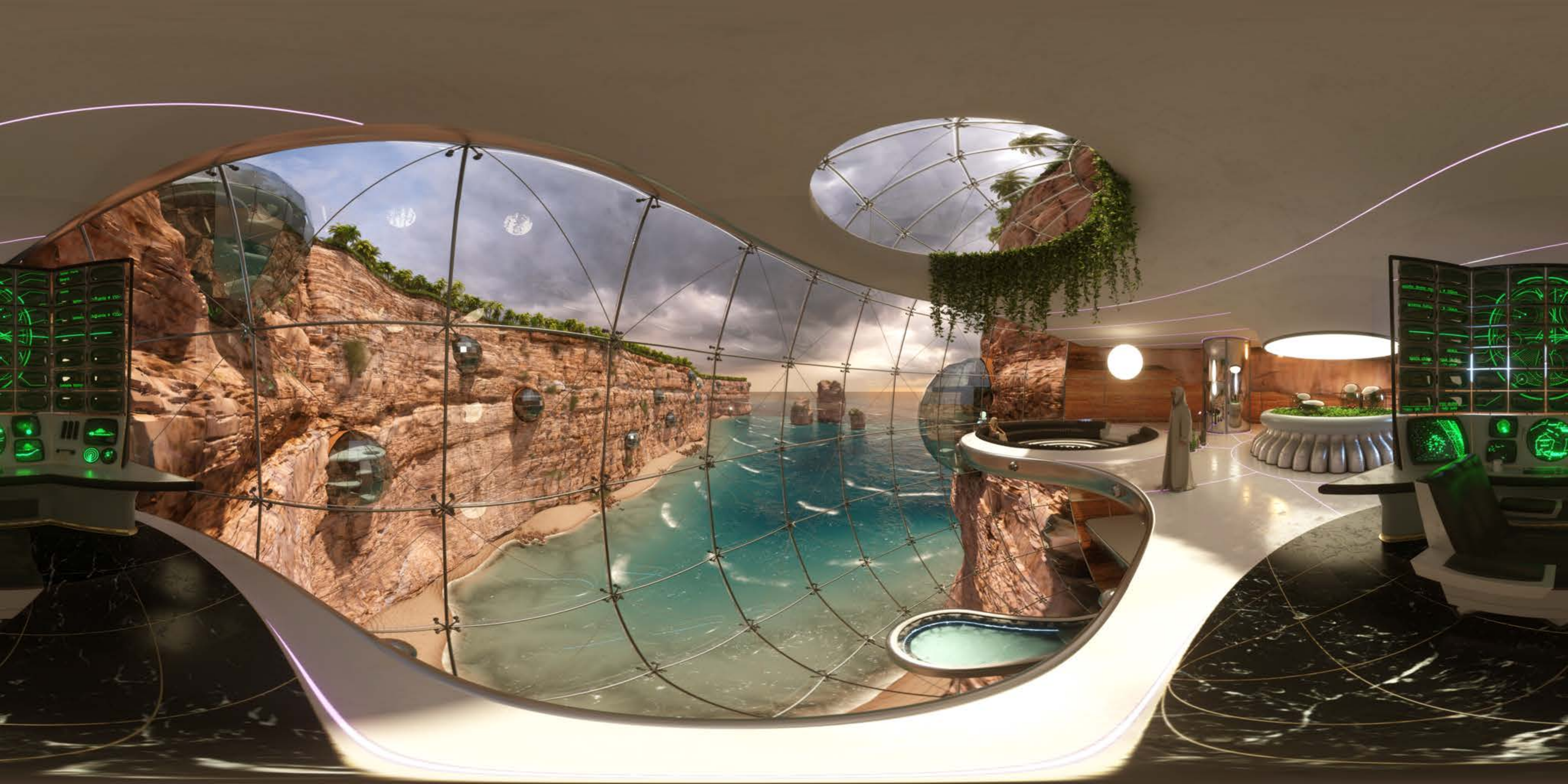} &
		\includegraphics[width=\swfour]{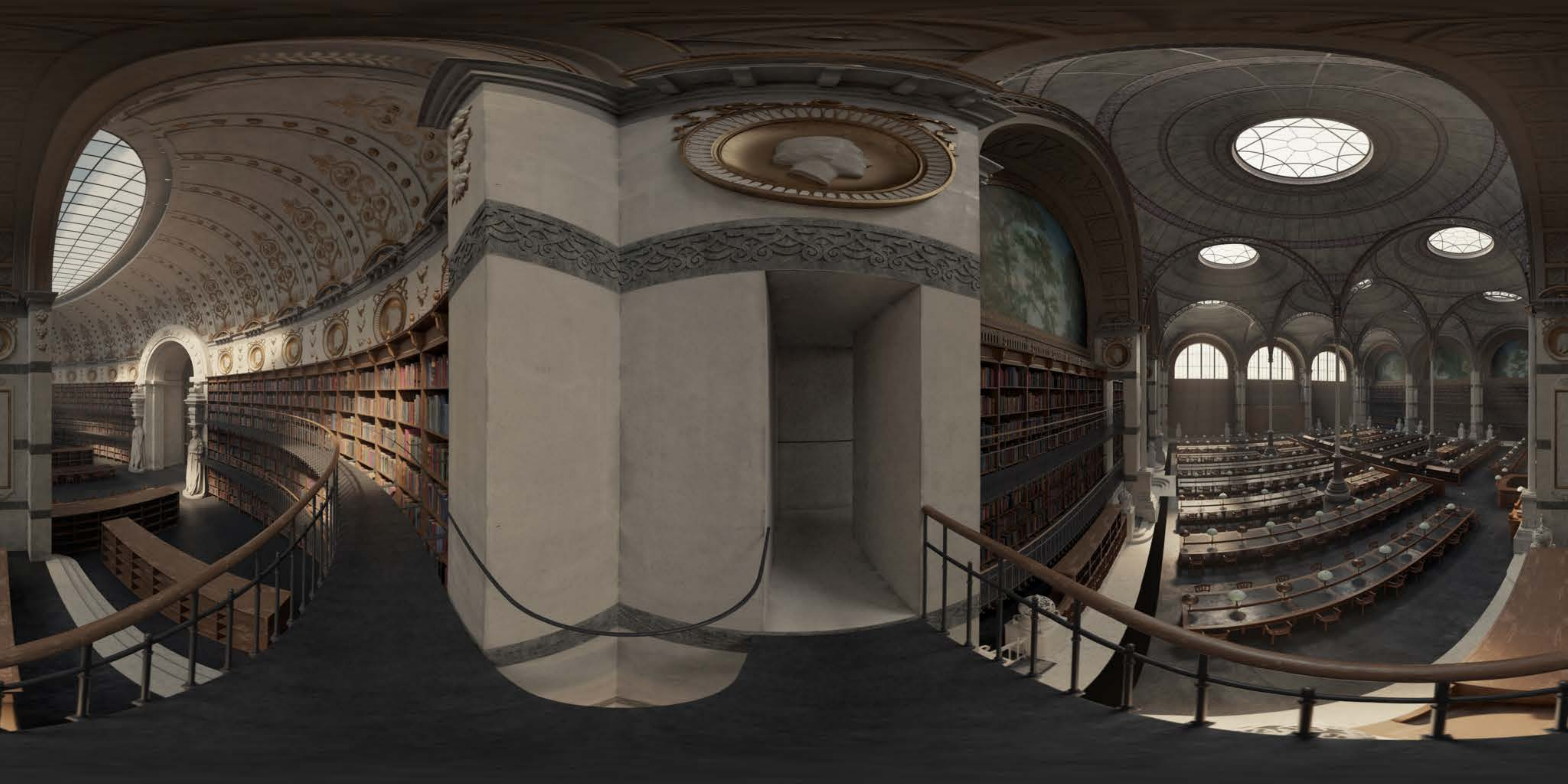} \\

     \footnotesize{\textsl{vr0000}} & \footnotesize{\textsl{vr0001}} & \footnotesize{\textsl{vr0002}} & \footnotesize{\textsl{vr0003}} \\		
		
		\includegraphics[width=\swfour]{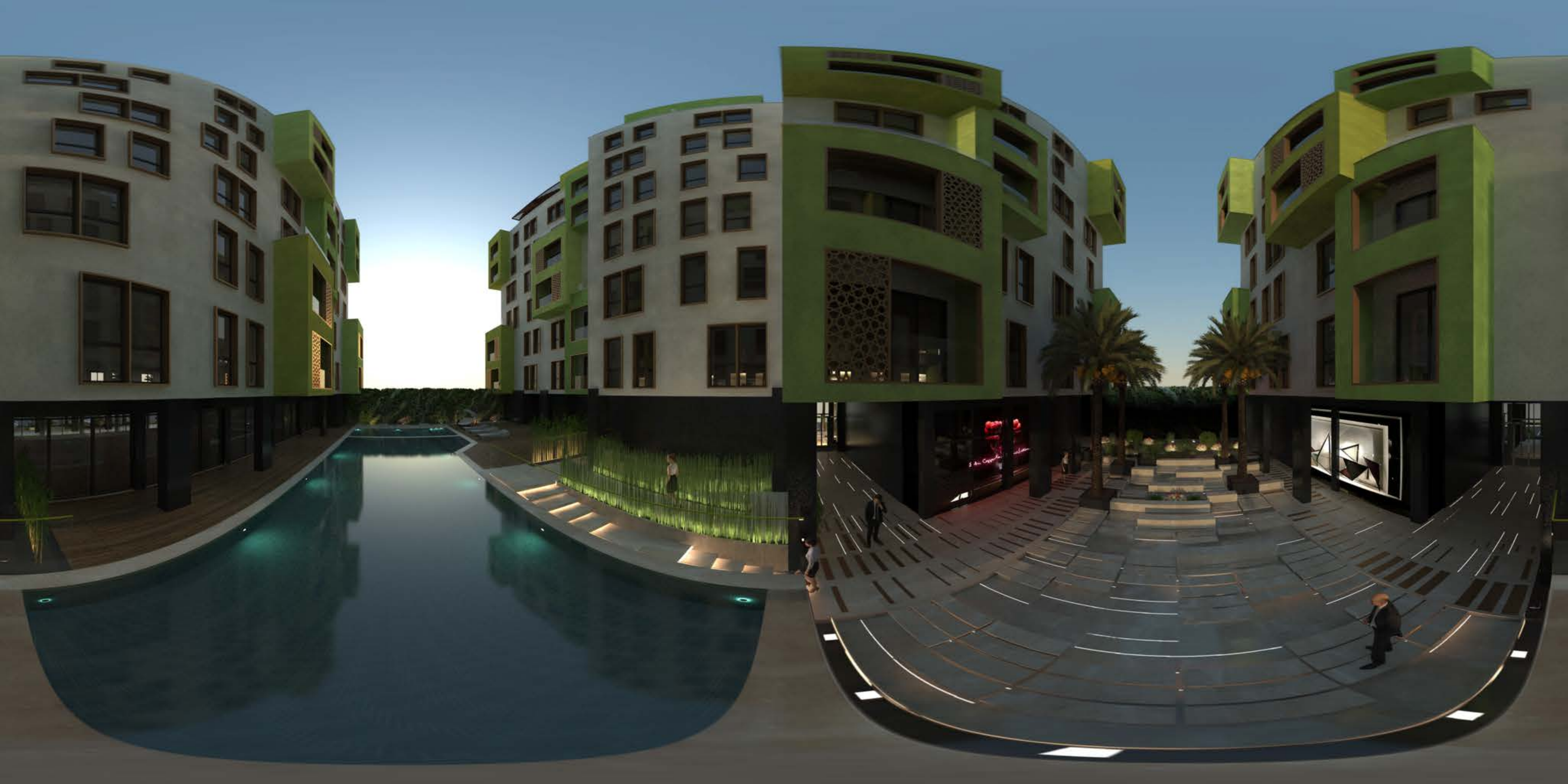} &		
		\includegraphics[width=\swfour]{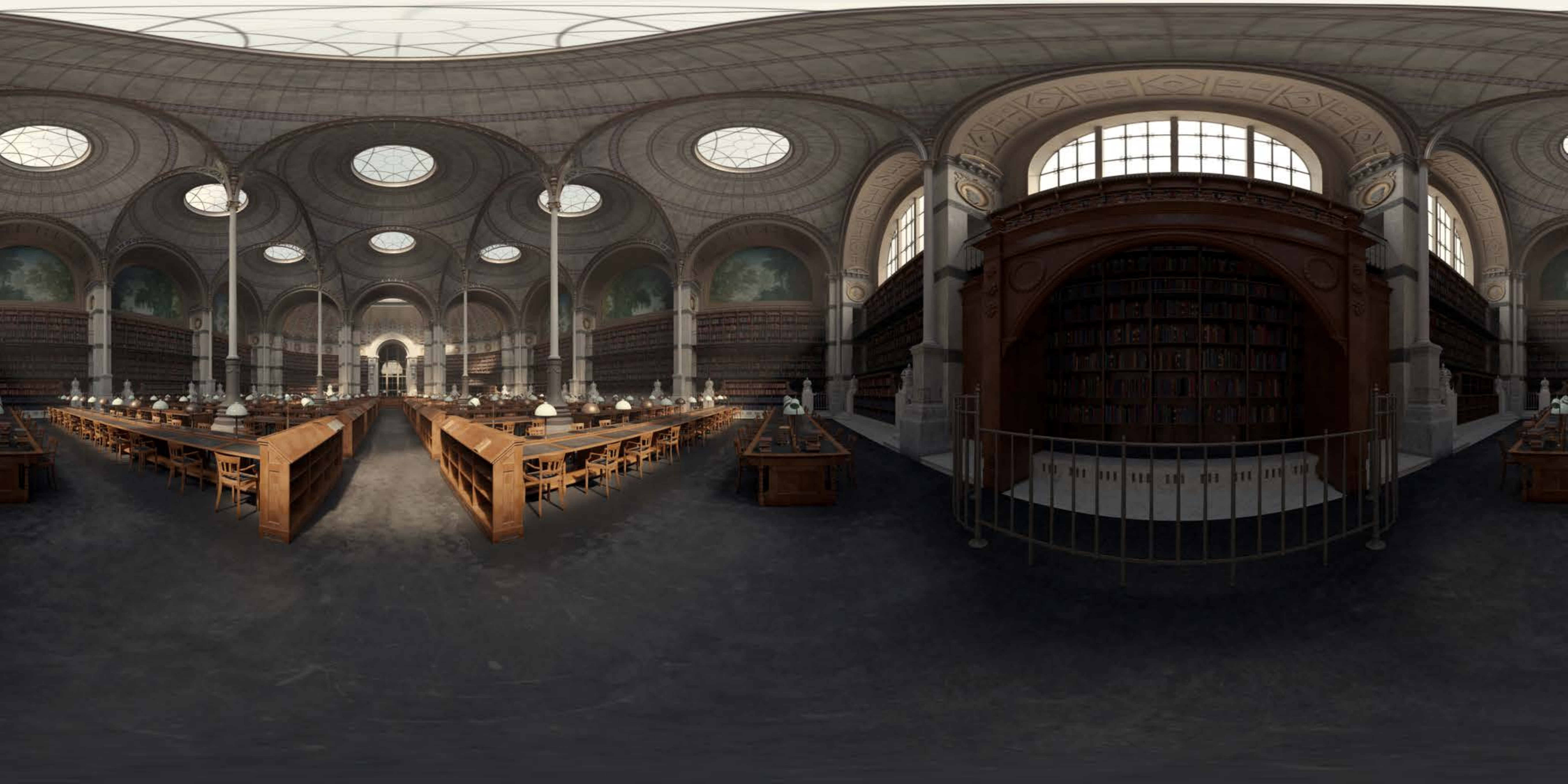} &
		\includegraphics[width=\swfour]{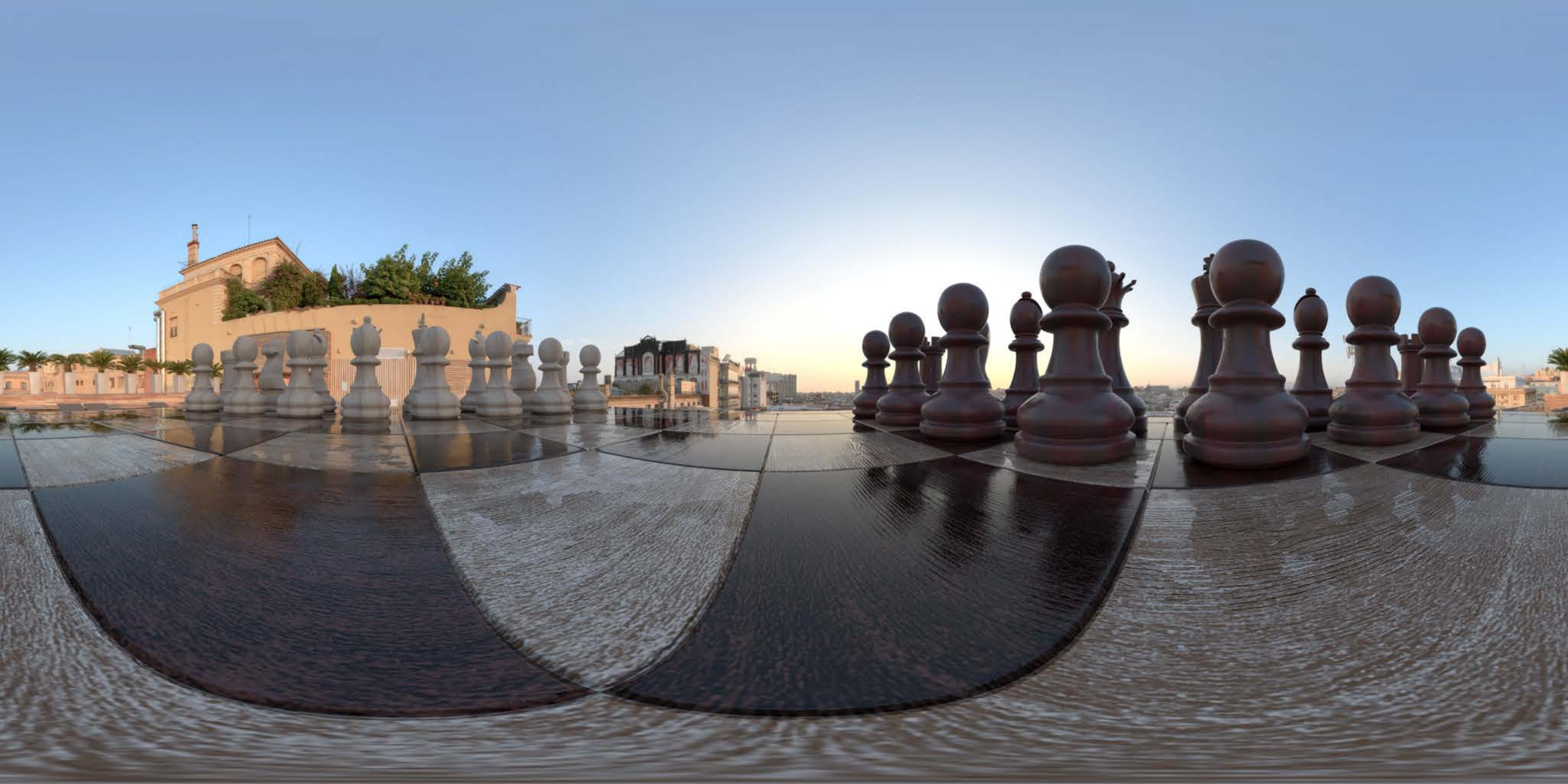} & 
		\includegraphics[width=\swfour]{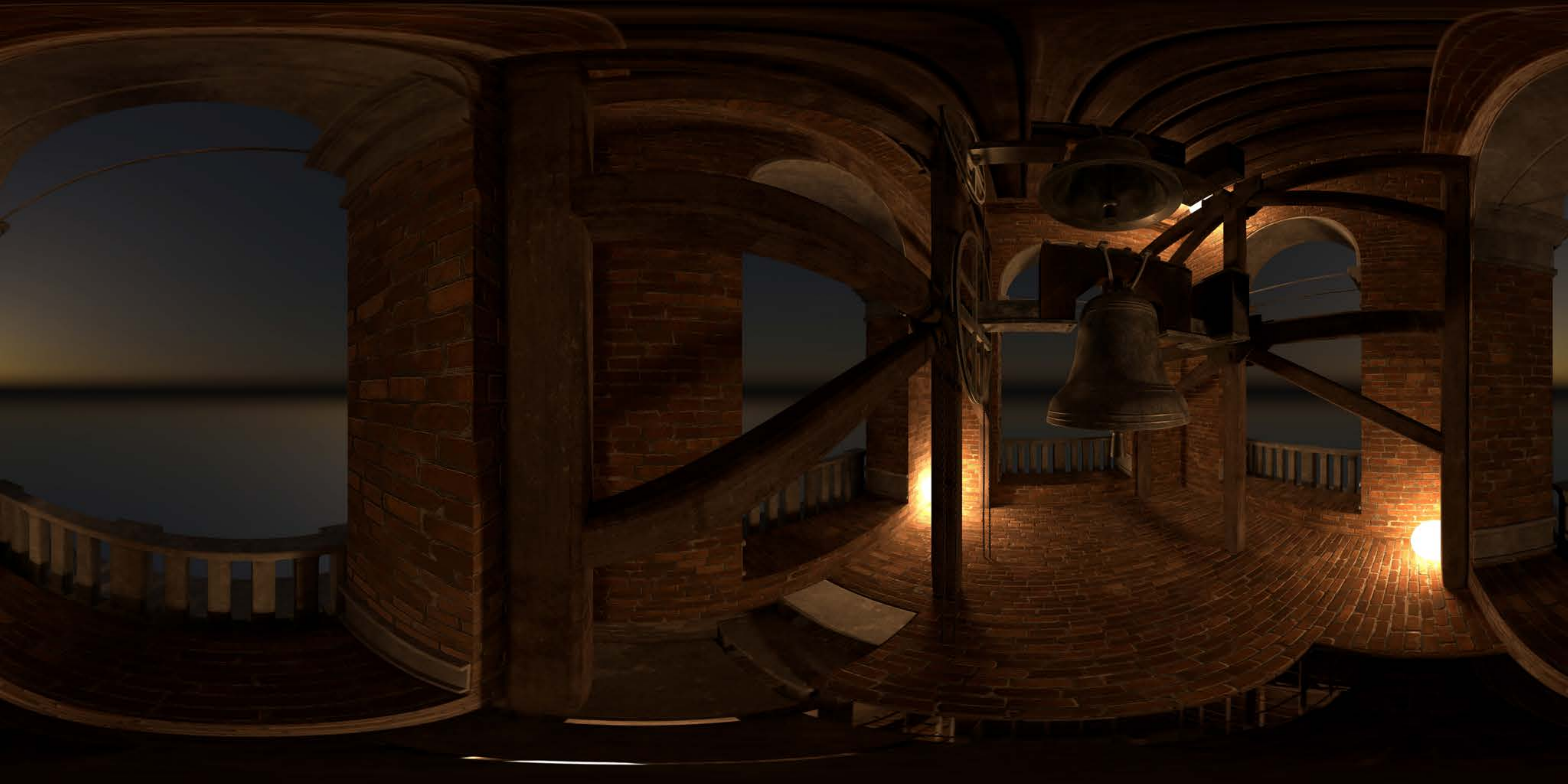} \\
		
		\footnotesize{\textsl{vr0004}} & \footnotesize{\textsl{vr0005}} & \footnotesize{\textsl{vr0006}} & \footnotesize{\textsl{vr0007}} \\
			
		\includegraphics[width=\swfour]{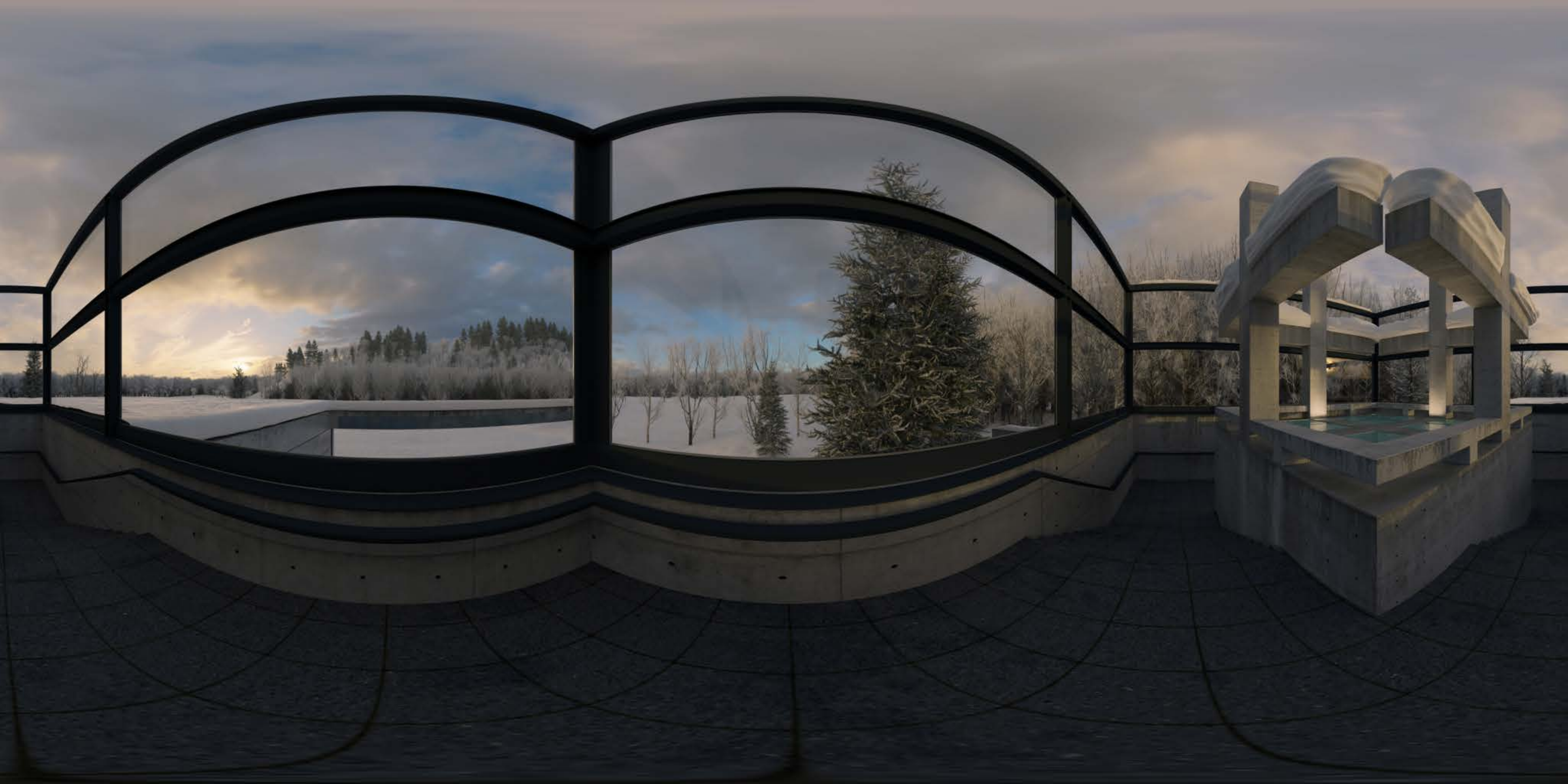} &
		\includegraphics[width=\swfour]{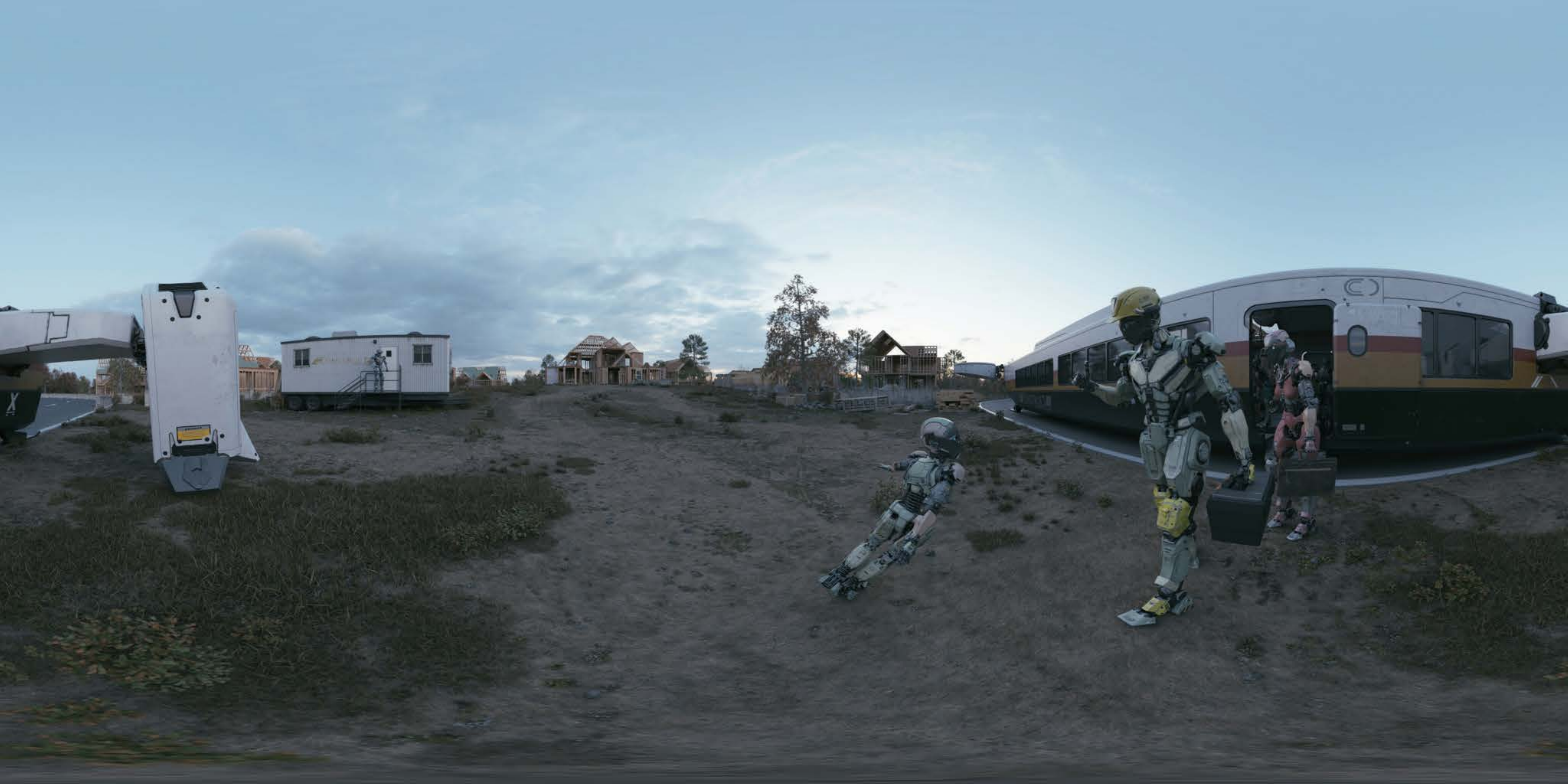} & 
		\includegraphics[width=\swfour]{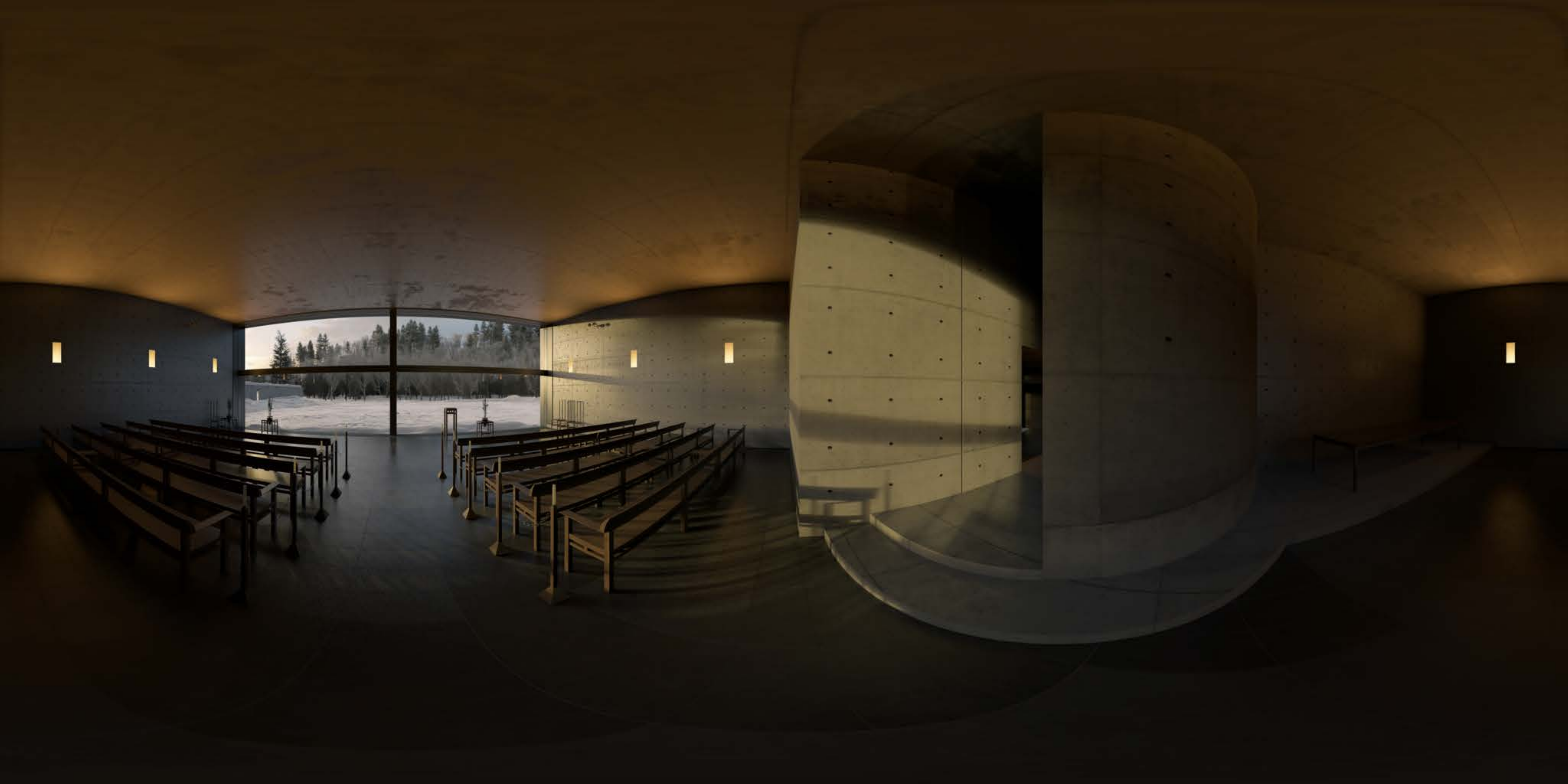} &
		\includegraphics[width=\swfour]{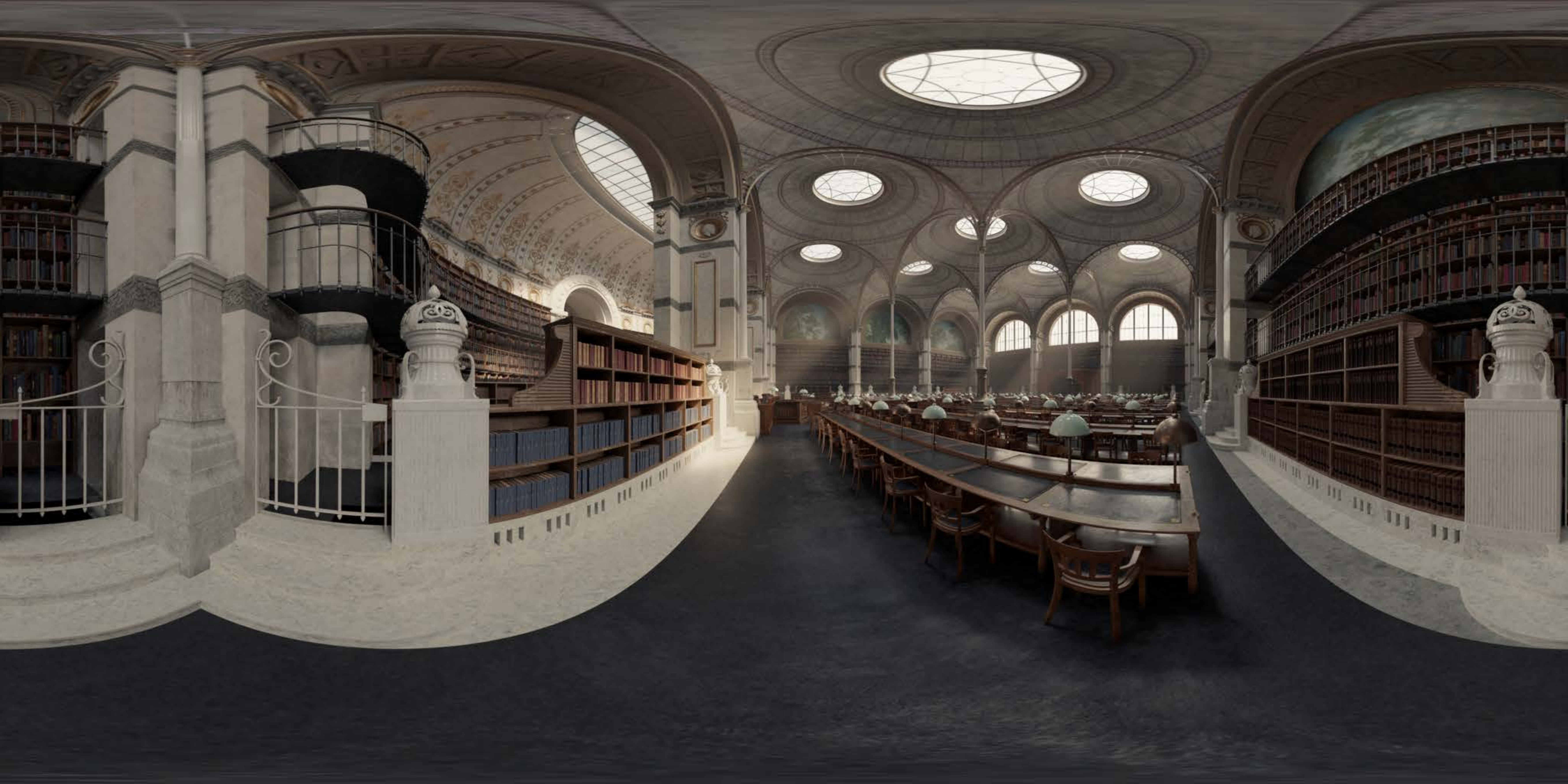} \\
		
		\footnotesize{\textsl{vr0008}} & \footnotesize{\textsl{vr0009}} & \footnotesize{\textsl{vr0010}} & \footnotesize{\textsl{vr0011}} \\	
		
		\includegraphics[width=\swfour]{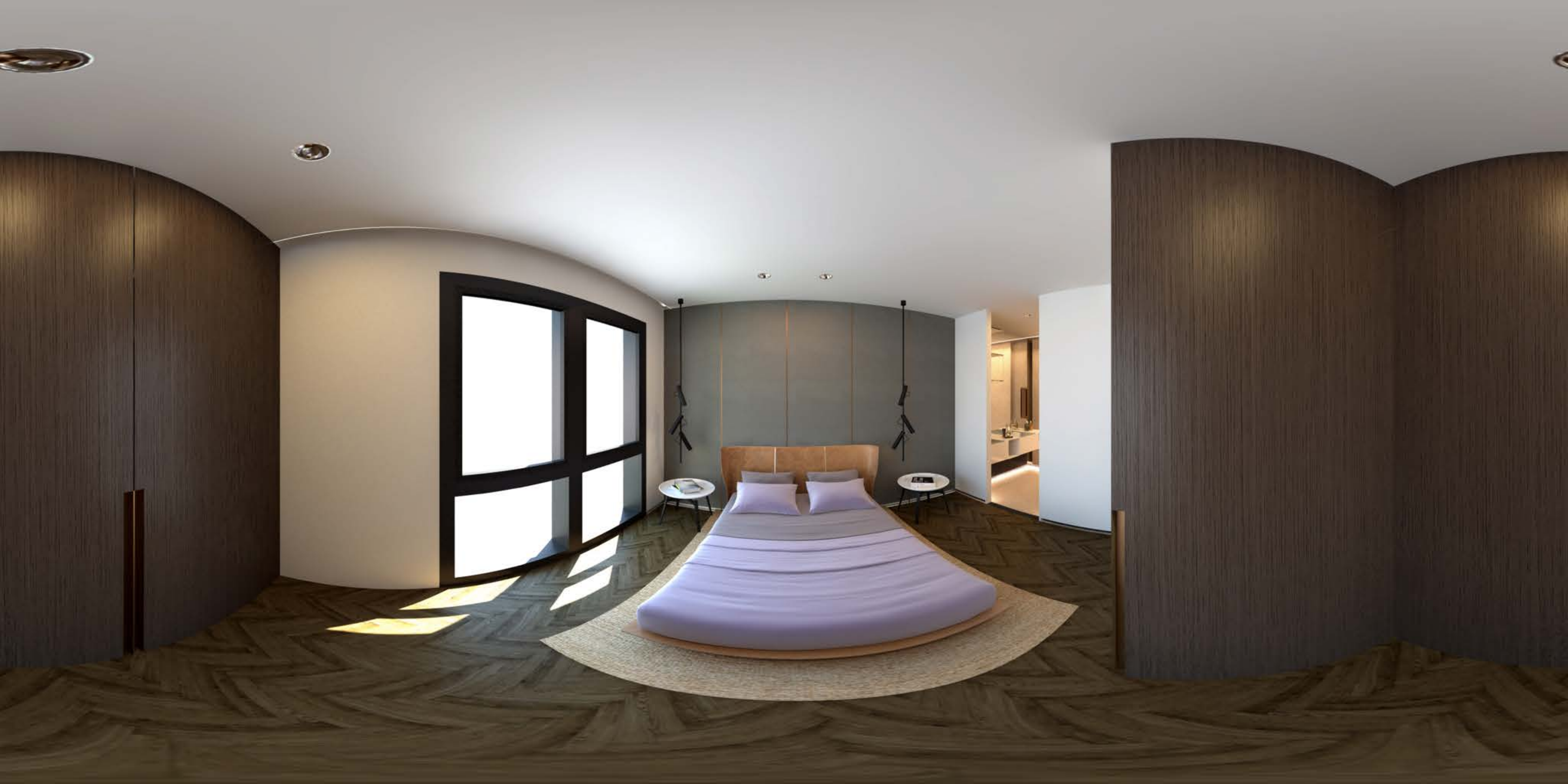} &
		\includegraphics[width=\swfour]{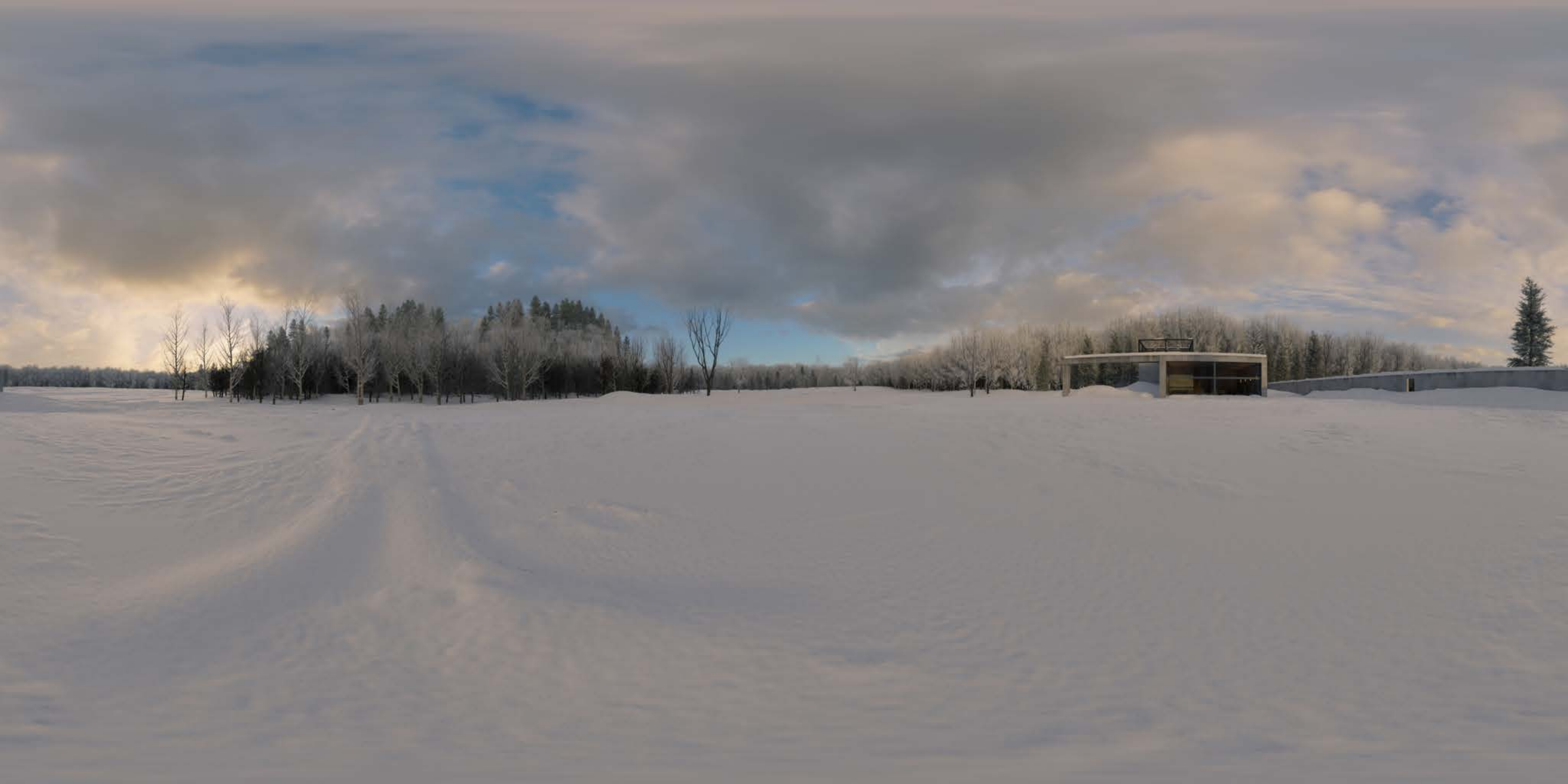} & 
		\includegraphics[width=\swfour]{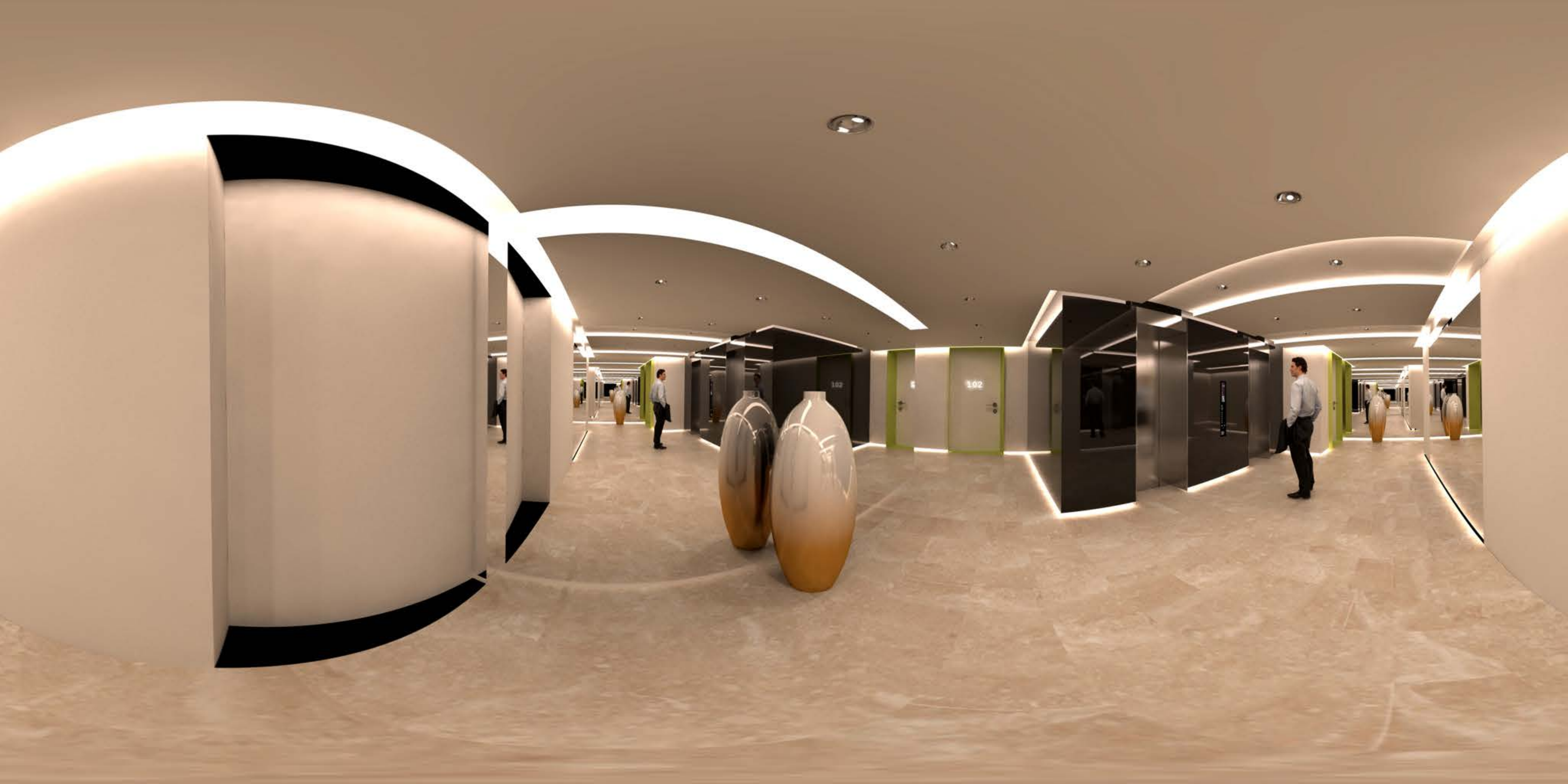} &
		\includegraphics[width=\swfour]{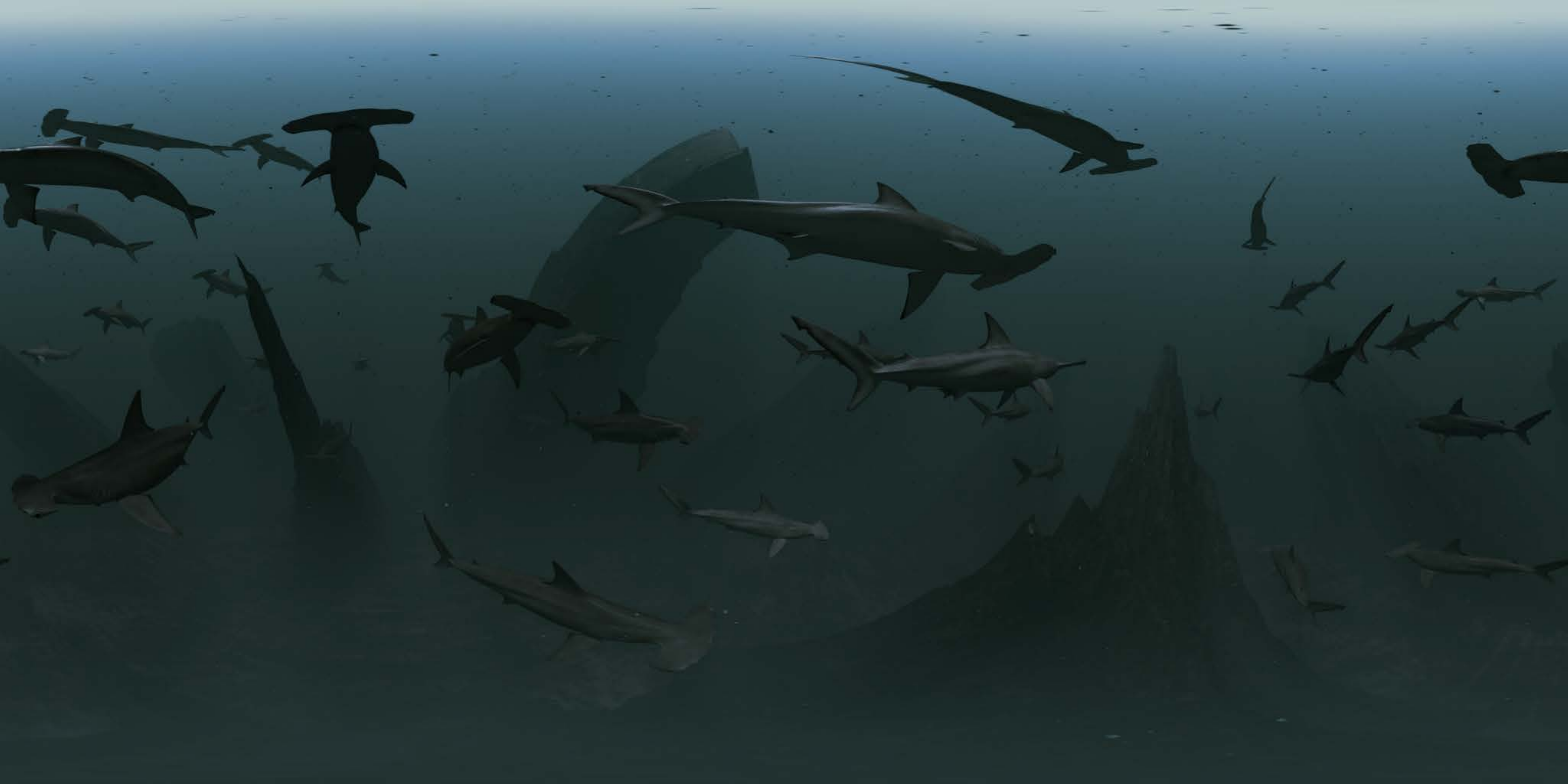} \\
		
		\footnotesize{\textsl{vr0012}} & \footnotesize{\textsl{vr0013}} & \footnotesize{\textsl{vr0014}} & \footnotesize{\textsl{vr0015}} \\
		
		\includegraphics[width=\swfour]{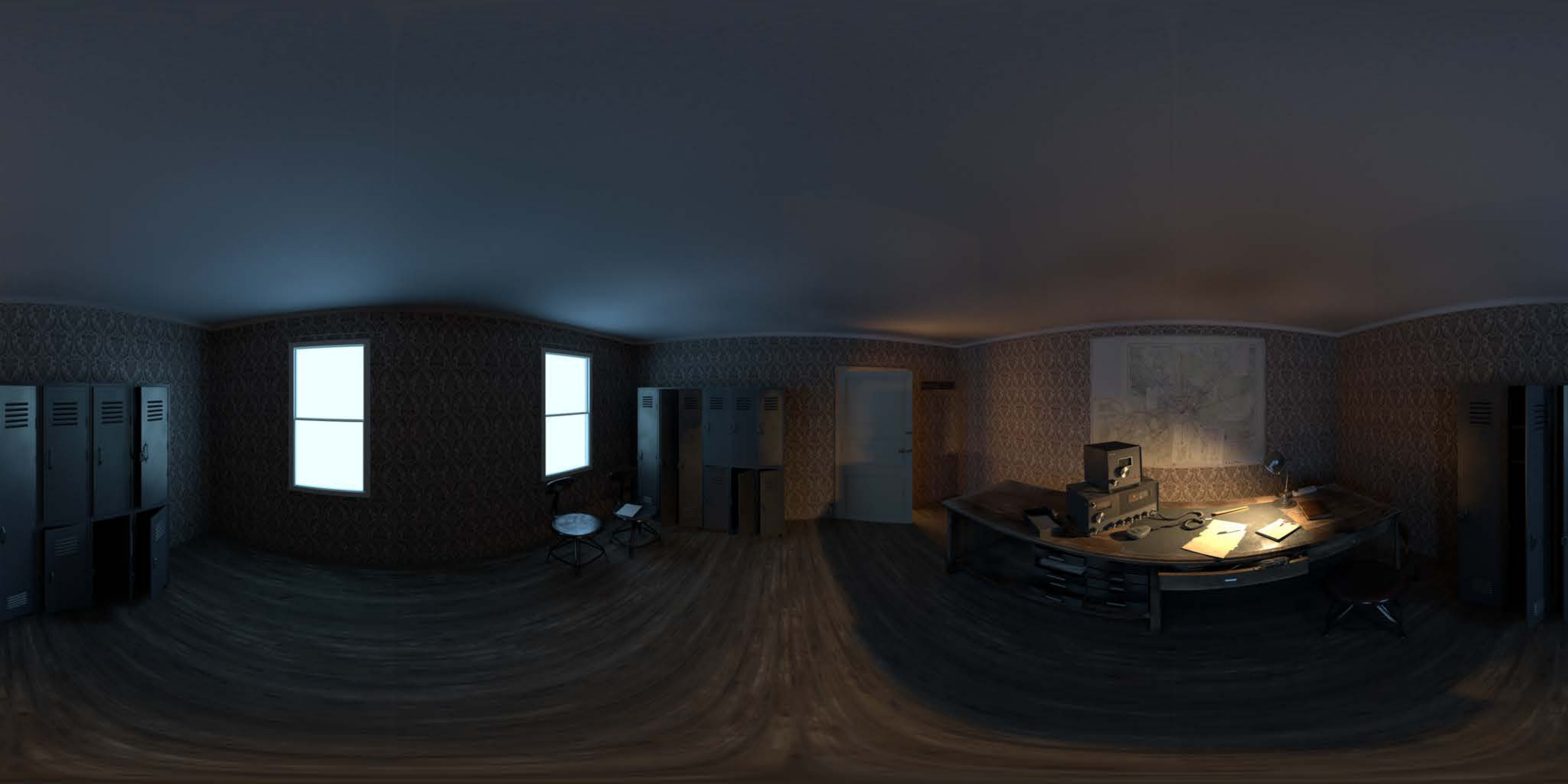} &
		\includegraphics[width=\swfour]{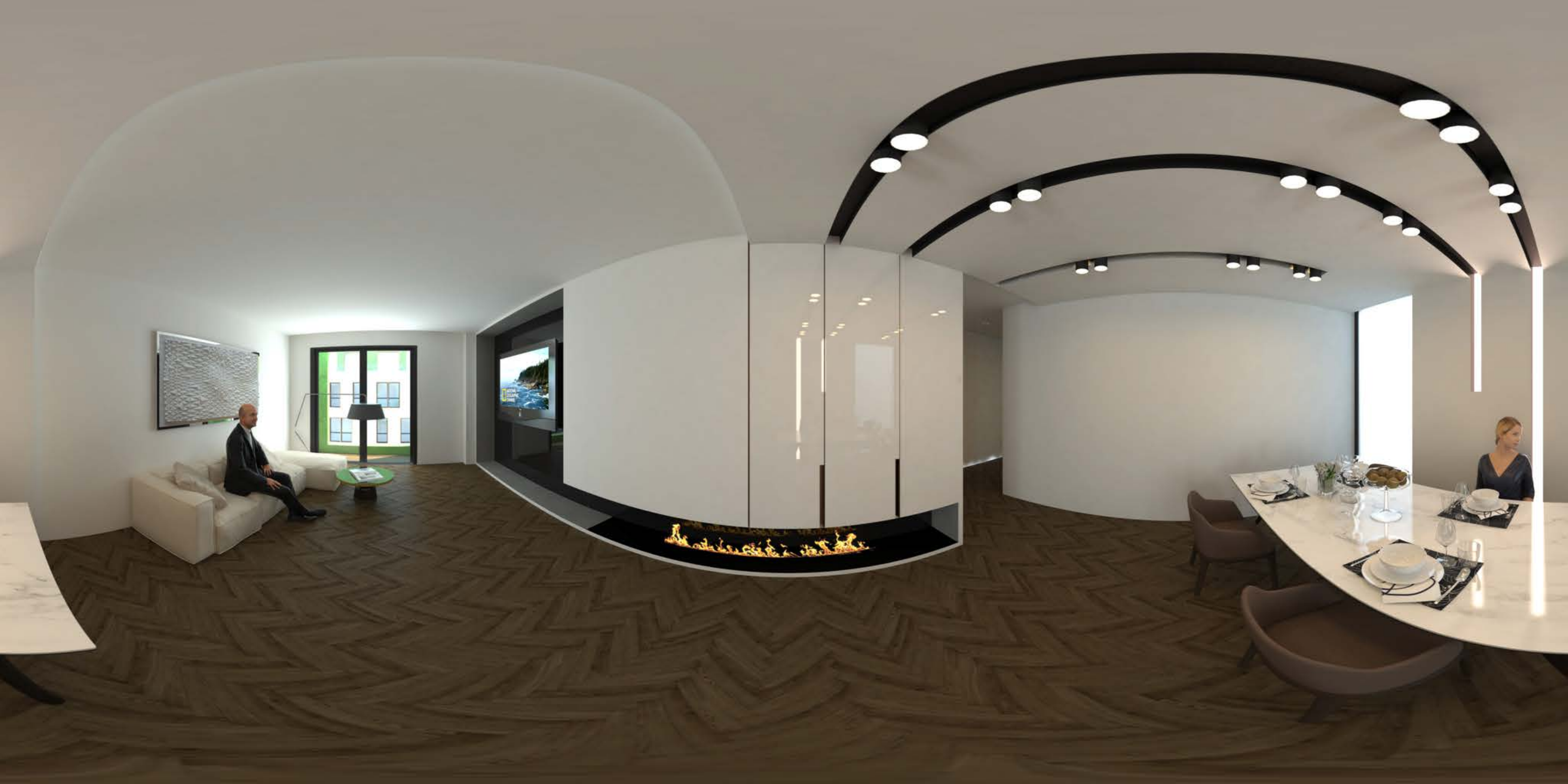} & 
		\includegraphics[width=\swfour]{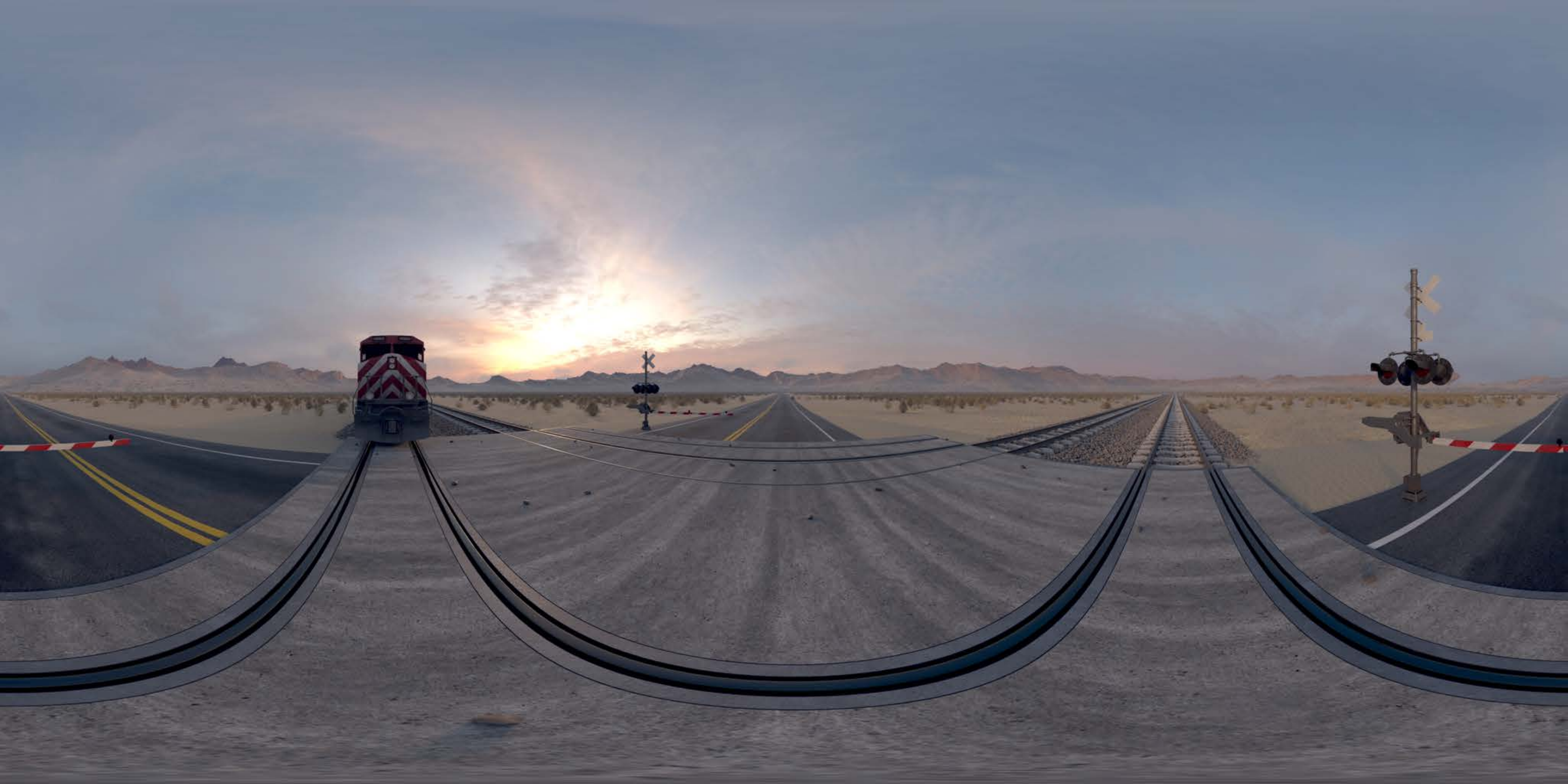} &
		\includegraphics[width=\swfour]{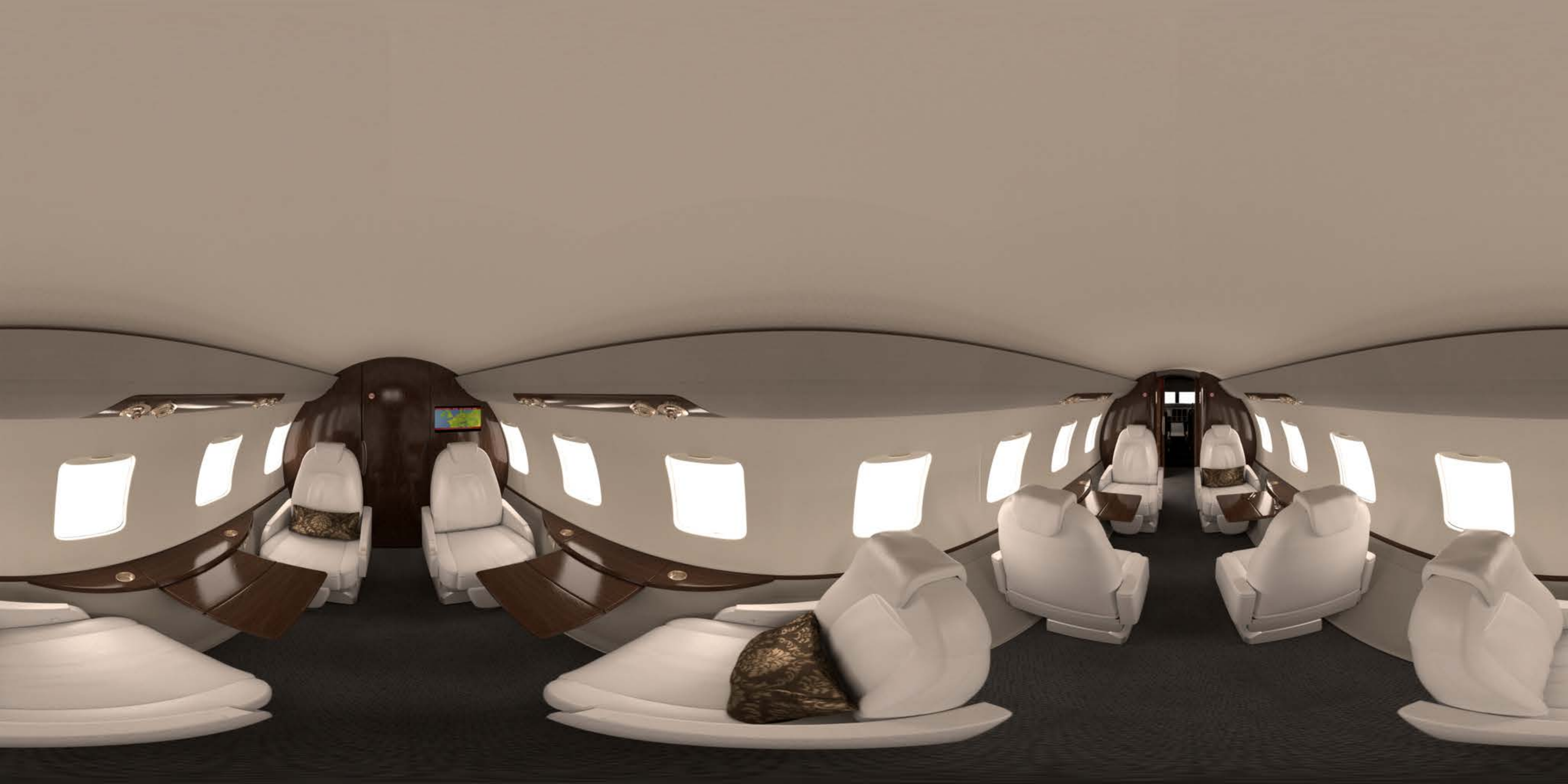} \\
		
		\footnotesize{\textsl{vr0016}} & \footnotesize{\textsl{vr0017}} & \footnotesize{\textsl{vr0018}} & \footnotesize{\textsl{vr0019}} \\
				
		\includegraphics[width=\swfour]{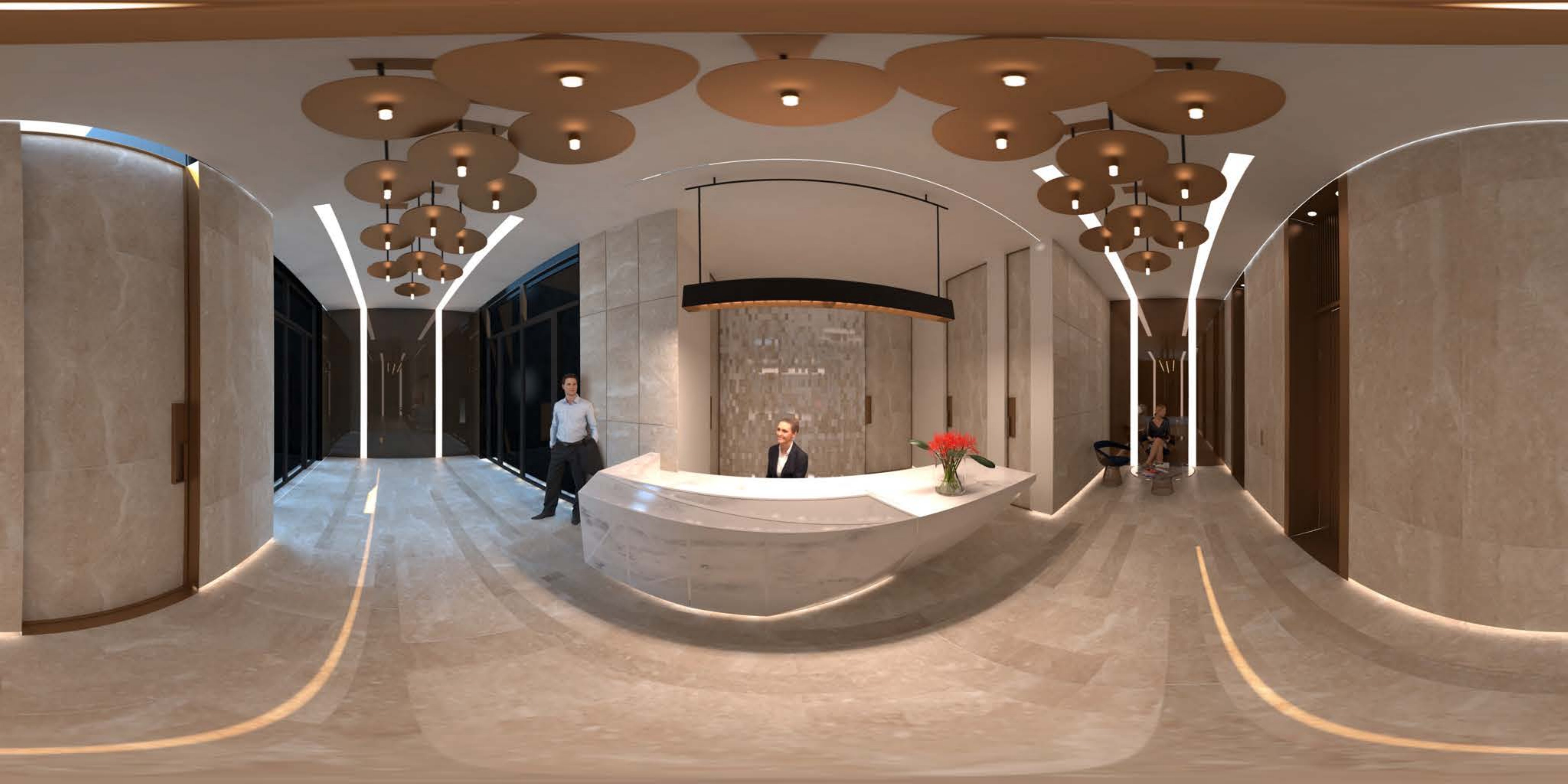} &
		\includegraphics[width=\swfour]{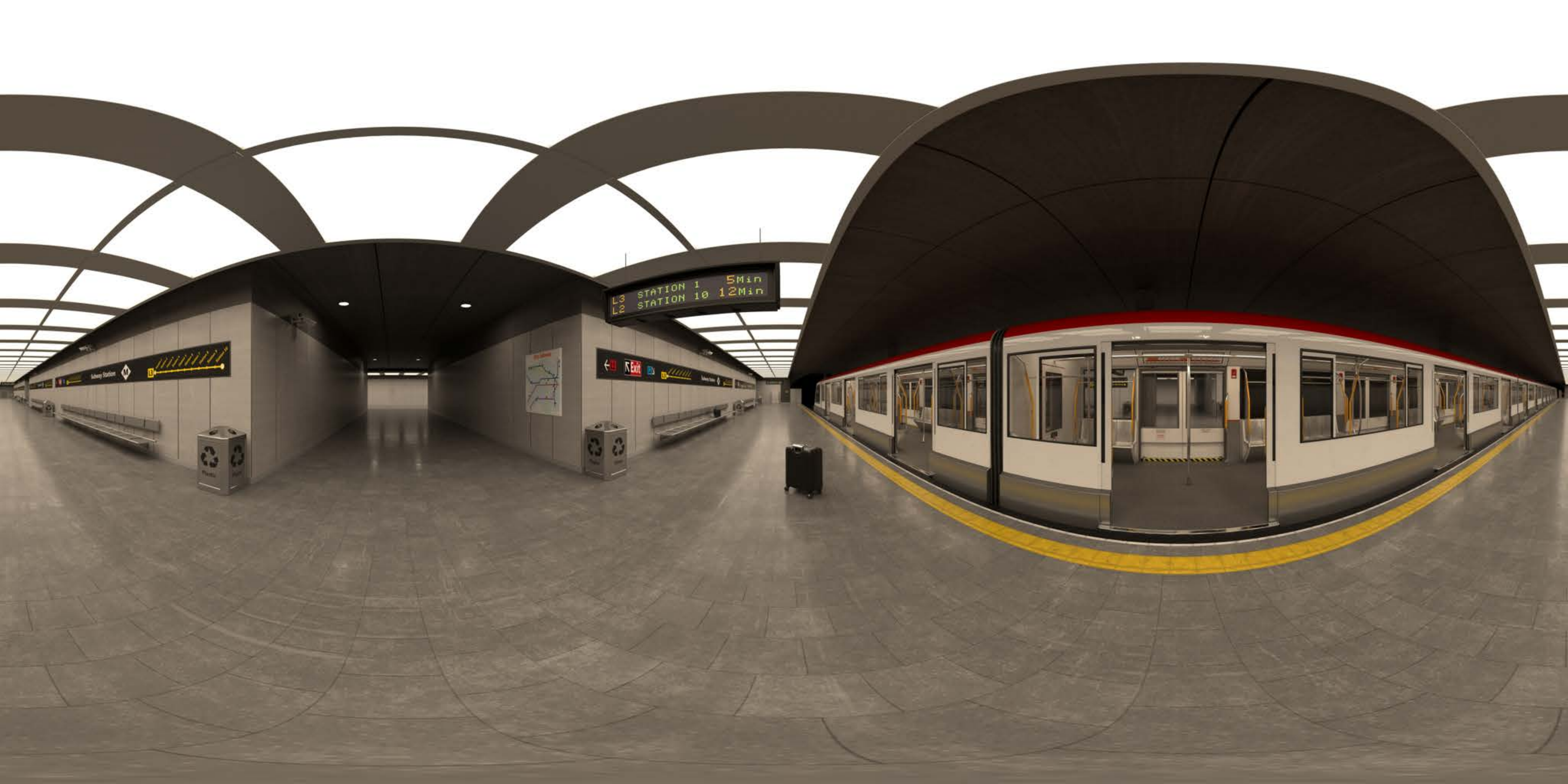} & 
		&
		 \\	
		 \footnotesize{\textsl{vr0020}} & \footnotesize{\textsl{vr0021}} &  \\
		\end{tabular}
	\end{center}
	\vspace{-0.1in}
	\center
	\caption{The 22 \360 images contained in the dataset by \cite{sitzmann2018saliency}, where we randomly selected 12 of them for experiments.}
\label{fig:saliency}
\end{figure*}

When running experiments for navigational \360 viewport streaming, we use \360 images from a dataset provided by \cite{sitzmann2018saliency}.
The dataset records roughly 2000 head and gaze trajectories from 169 users in 22 static stereoscopic \360 images, where we use the scanpath data to parameterize our navigation model.
In Fig.\;\ref{fig:saliency}, we show the 22 \360 images provided by  \cite{sitzmann2018saliency}.
Among the 22 images, we randomly selected 12 \360 images for experiments, which are \textsl{vr0000}, \textsl{vr0002}, \textsl{vr0003}, \textsl{vr0004}, \textsl{vr0006}, \textsl{vr0009}, \textsl{vr0011}, \textsl{vr0013}, \textsl{vr0015}, \textsl{vr0017}, \textsl{vr0018}, and \textsl{vr0020}.
As we can see, these images contain indoor and outdoor scenes, different light conditions, and with low or high saliency entropy, and so on.
We are not able to try other datasets since we require scanpath data to build our navigation model.

\renewcommand{\tabcolsep}{.1pt}

\subsection{\textbf{A Summary of the Streaming Strategies for Comparison}}
We run experiments on 6 light field (LF) image sets and 12 \360 images in total. 
Beyond the results provided in the paper, here we further show the rest results.

To show the superiority of the proposed landmarking insertion method, we compare it with several other streaming strategies.
We briefly describe the experimental setup for each strategy in Table \ref{tab:strategies}.
Among them, \texttt{Flex-LM} is our proposed method.
Comparing with \texttt{Flex-LM}, \texttt{Flex-LM-I} is with different initialized structure, where all the I-MDUs are pre-encoded and stored.
The only difference between \texttt{Flex-GA} and \texttt{Flex-LM-I} is the use or the lack of landmarks.
We show the results in the following.
\begin{table}
\begin{center}
\caption{The streaming strategies used in the experiments for comparison.}
 \label{tab:strategies}
\renewcommand\tabcolsep{2.5pt}
\renewcommand\arraystretch{1.3}
\begin{tabular}{|m{0.27\linewidth}|m{0.70\linewidth}|}
\hline
 \makecell[c]{Strategies} & \makecell[c]{Setup} \\ \hline
 \texttt{Flex-LM} & The proposed optimal landmark insertion method, with a flexible reference buffer. \\ \hline
 \texttt{Flex-GA} & Inserting P-MDUs using a greedy algorithm~\cite{motz16icip}, with a flexible reference buffer. \\ \hline
 \texttt{Fixed-GA} & The same greedy algorithm as \texttt{Flex-GA} for P-MDUs insertion, but with a fixed reference buffer. \\ \hline
 \texttt{Inf-LM} & Given the MDU structure generated by \texttt{Flex-LM}, considering users with an infinity reference buffer. \\ \hline
 \texttt{Flex-LM-I} & Considering all I-MDUs are pre-encoded and stored at the server, using the optimal landmark insertion method, and with a flexible reference buffer. \\ \hline
 \texttt{Flex-SLF-RA} & Encoding LF images using methods proposed in~\cite{monteiro2021light}, with a flexible reference buffer for streaming. \\ \hline
 \texttt{Whole} & Encoding and transmitting the entire \360 images directly. \\ \hline 
\end{tabular}
\end{center}
\end{table}

\subsection{\textbf{Experiments on Light Field Images}}
\begin{figure*}[htb]
	\begin{center}
		\begin{tabular}{ccc}
		
		\includegraphics[width=0.32\linewidth]{figures/bracelet_16x16.pdf} &
		\includegraphics[width=0.325\linewidth]{figures/cards_16x16.pdf} &
		\includegraphics[width=0.32\linewidth]{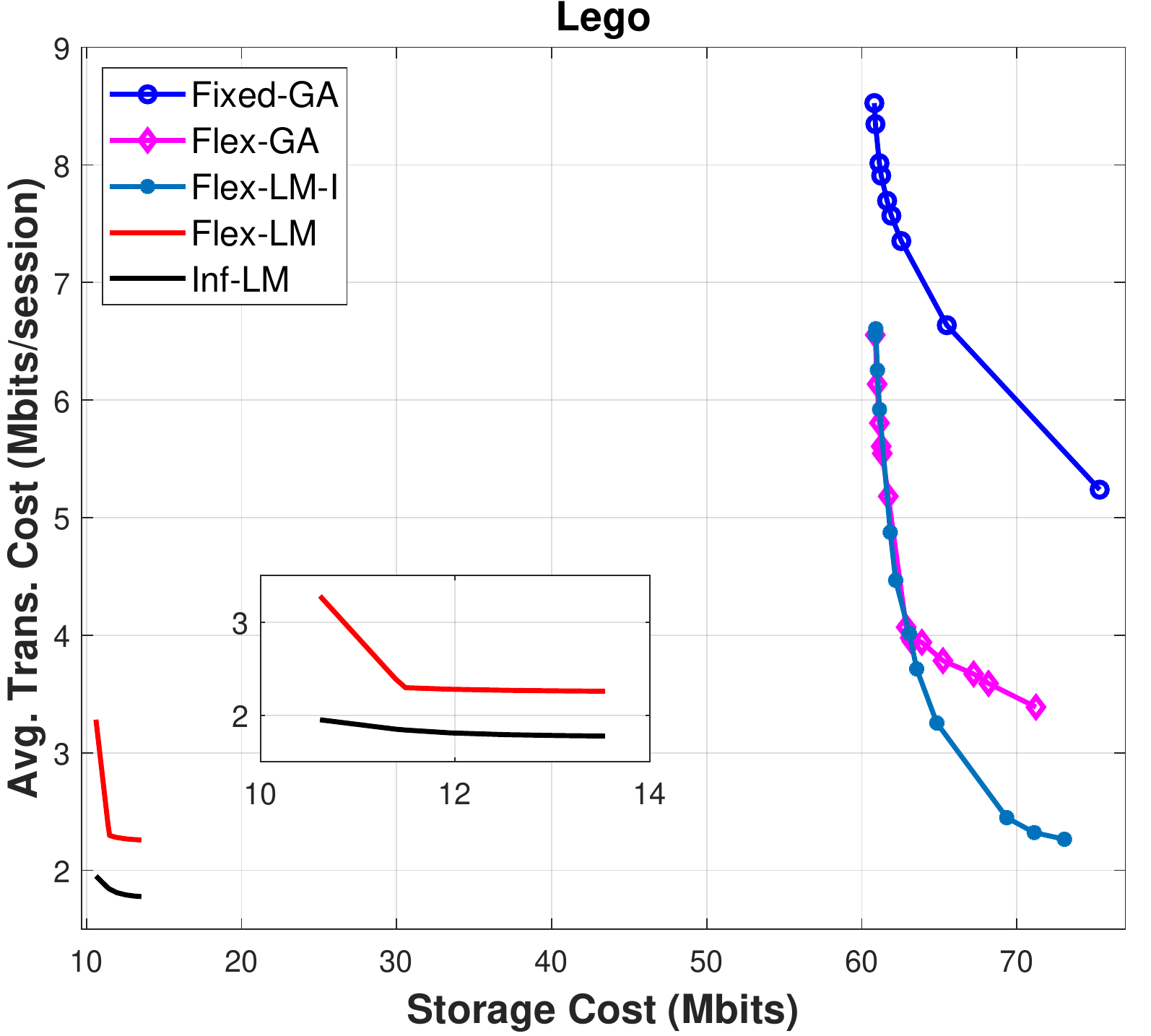} \\
		
		\footnotesize{(a) LF, \textsl{Bracelet}, $17 \times 17$} & \footnotesize{(b) LF, \textsl{Cards}, $17 \times 17$} & \footnotesize{(c) LF, \textsl{Lego}, $17 \times 17$} \\
		
\\		
		
		\includegraphics[width=0.32\linewidth]{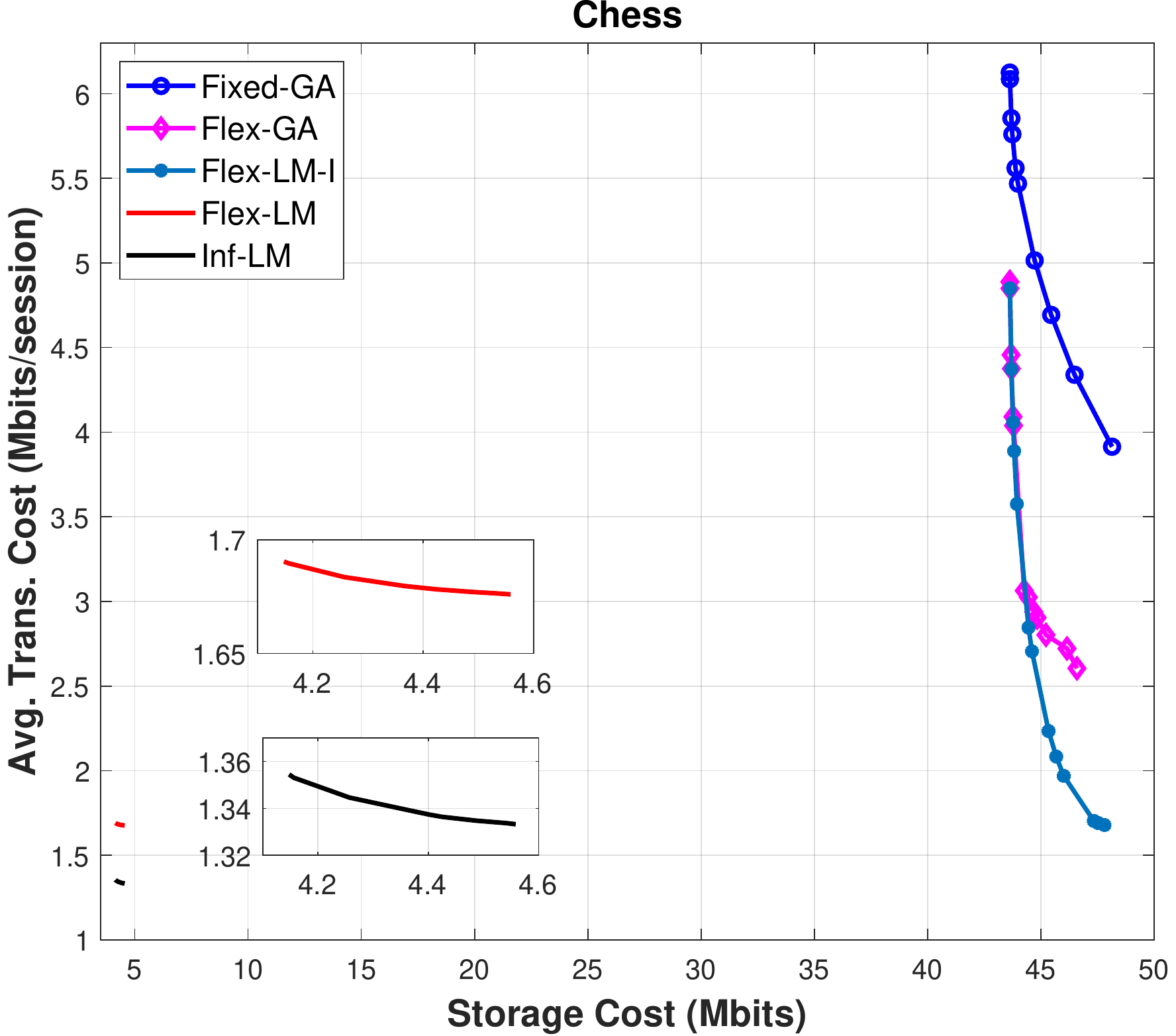} &
		\includegraphics[width=0.32\linewidth]{figures/Poznan_30x30.pdf} &
		\includegraphics[width=0.32\linewidth]{figures/Set2_32x10.pdf} \\
		\footnotesize{(d) LF, \textsl{Chess}, $17 \times 17$}  & \footnotesize{(e) LF, \textsl{Poznan}, $31 \times 31$} & \footnotesize{(f) LF, \textsl{Set2}, $33 \times 11$} \\		
	
		\end{tabular}
	\end{center}
%	\vspace{-0.15in}
	\caption{The storage cost (Mbits) \textit{vs.} the average transmission cost (Mbits/session) for the 6 LF image sets. The proposed method \texttt{Flex-LM} results in much smaller both storage and transmission cost than \texttt{Flex-GA} and \texttt{Fixed-GA}.}
\label{fig:RD2}
\end{figure*}

For LF images, we first show the storage cost \textit{vs}. the expected transmission cost curves of \texttt{Flex-LM}, \texttt{Flex-GA},
\texttt{Fixed-GA}, \texttt{FLex-LM-I}, and \texttt{Inf-LM} (corresponding to Figure 11 in the main paper).
Results are shown in Fig.\;\ref{fig:RD2}.
\renewcommand{\tabcolsep}{.6pt}
\renewcommand\arraystretch{1.2}

\begin{figure*}[htb]
	\begin{center}
		\begin{tabular}{cc}
		
		\includegraphics[width=0.4\linewidth]{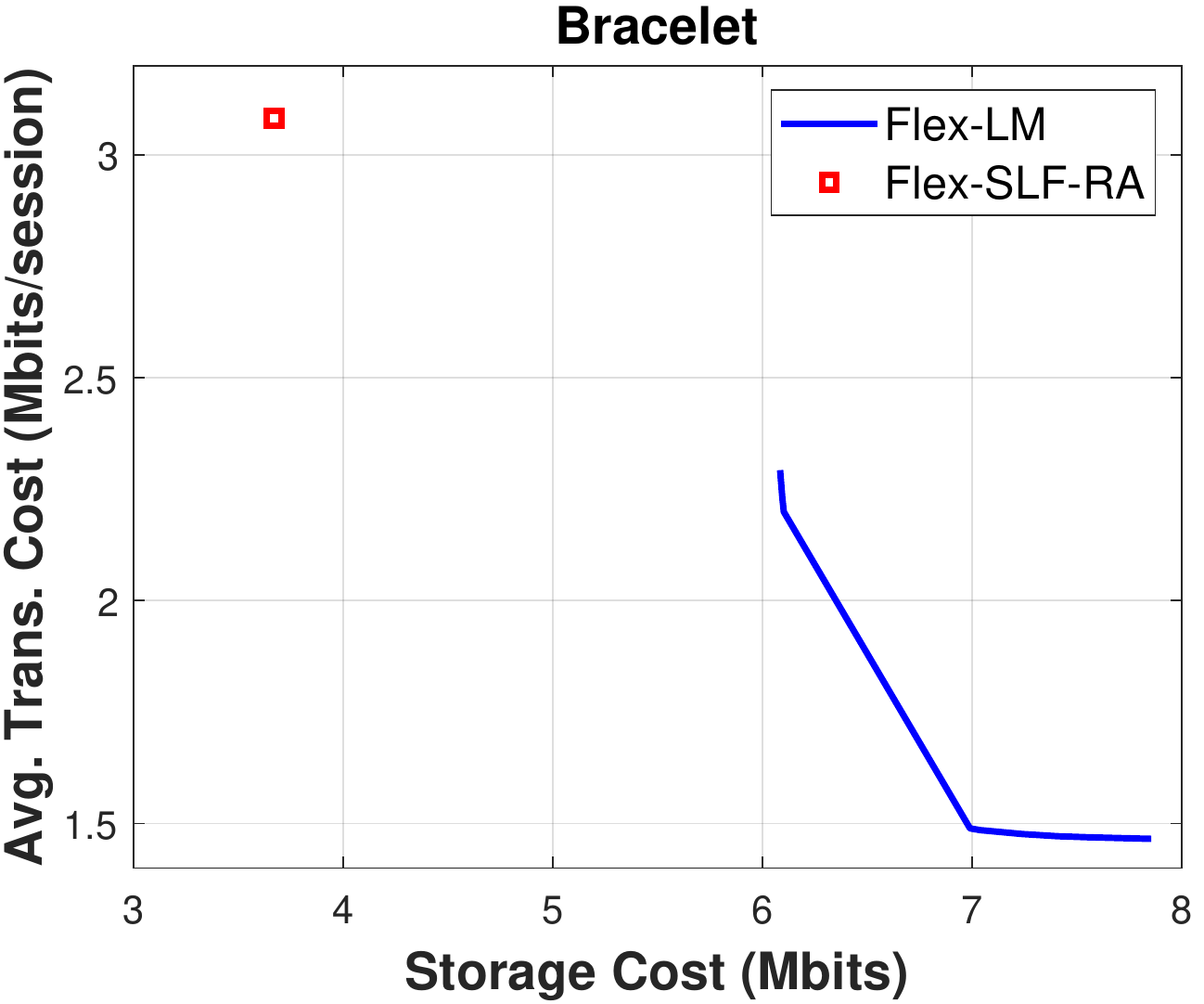} &
		\includegraphics[width=0.4\linewidth]{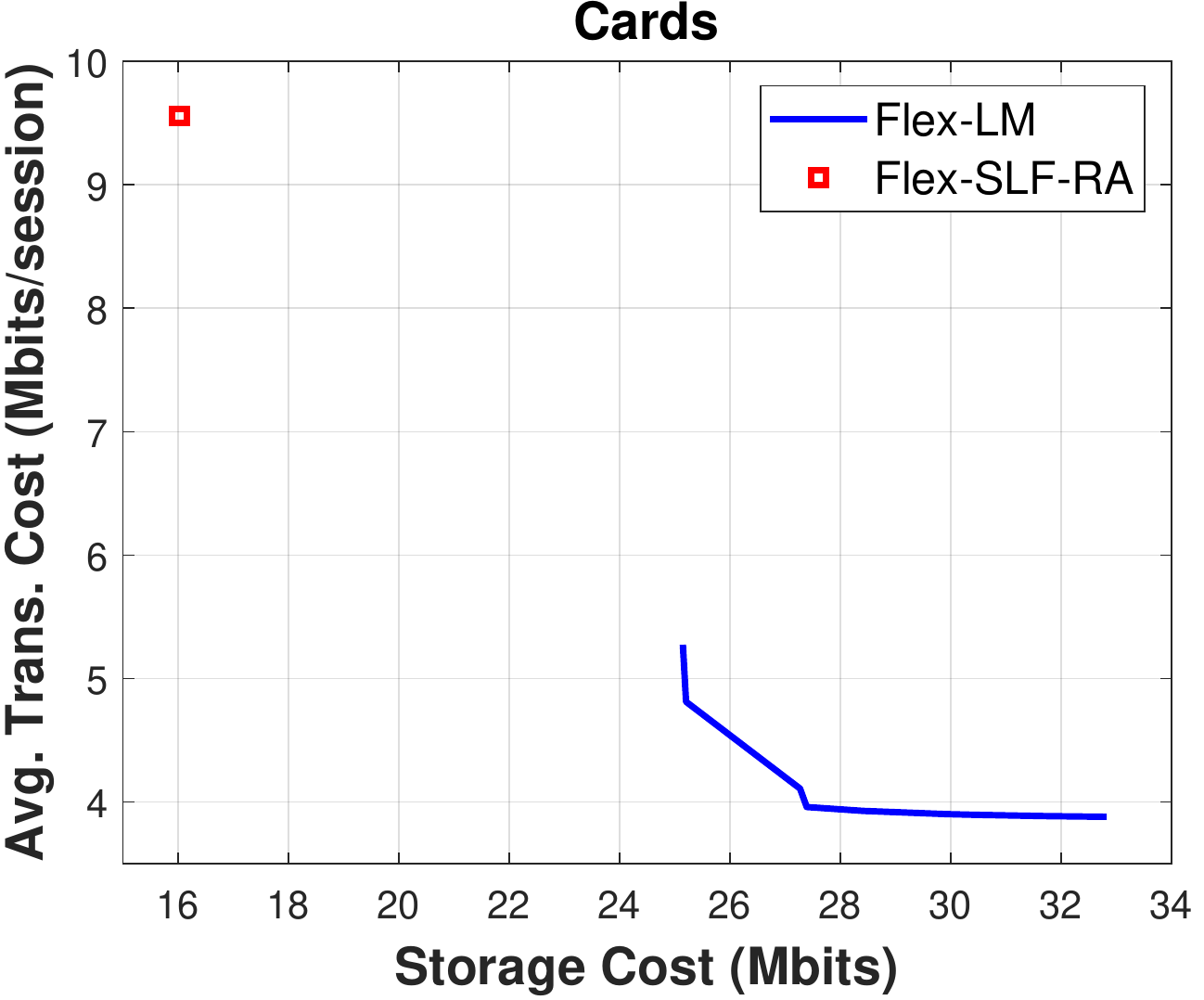} \\
		
		\footnotesize{(a) LF, \textsl{Bracelet}} & \footnotesize{(b) LF, \textsl{Cards}}  \\ 
		
		\\
		
		\includegraphics[width=0.4\linewidth]{figures/lego_compare.pdf} &
		\includegraphics[width=0.4\linewidth]{figures/chess_compare.pdf}  \\
		\footnotesize{(c) LF, \textsl{Lego}} & \footnotesize{(d) LF, \textsl{Chess}}  \\
		\end{tabular}
	\end{center}
	\vspace{-0.15in}
	\caption{Comparing \texttt{Flex-LM} with \texttt{Flex-SLF-RA} \cite{monteiro2021light} for interactive LF image streaming. \texttt{Flex-LM} can reduce transmission cost by exploiting extra storage space.}
\label{fig:LF_comp2}
\end{figure*}

We then compare \texttt{Flex-LM} with a recently proposed LF image coding scheme that facilitates random access~\cite{monteiro2021light}.
It is labelled as \texttt{Flex-SLF-RA} (corresponding to Figure 12 in the main paper).
The storage-transmission cost curves are shown in Fig.\;\ref{fig:LF_comp2}.
The result for \texttt{Flex-SLF-RA} is a point, since we keep the quality of MDUs the same with \texttt{Flex-LM}.
The expected transmission cost is computed according to its coding scheme.
Due to representation redundancy in our proposed structures, the storage costs of \texttt{Flex-LM} are larger than \texttt{Flex-SLF-RA}.
However, the expected transmission cost (proportional to bits needed to access a requested MDU) of \texttt{Flex-LM} can become less than half of \texttt{Flex-SLF-RA}.

\renewcommand{\tabcolsep}{.1pt}
\renewcommand\arraystretch{1.2}

\subsection{\textbf{Experiments on \360 Images}}
\begin{figure*}[htb]
	\begin{center}
		\begin{tabular}{ccc}
		
		\includegraphics[width=0.315\linewidth]{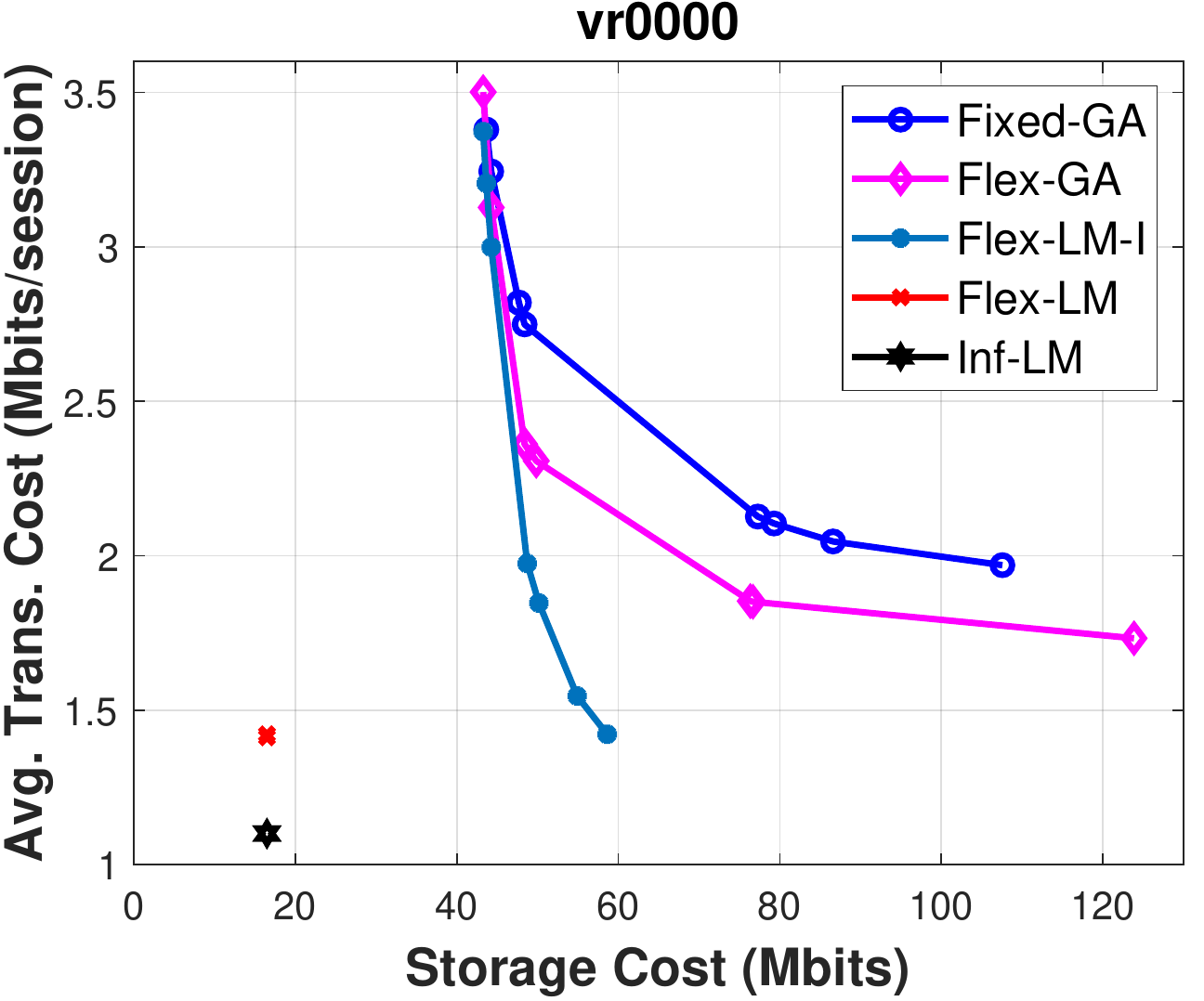} &
		\includegraphics[width=0.315\linewidth]{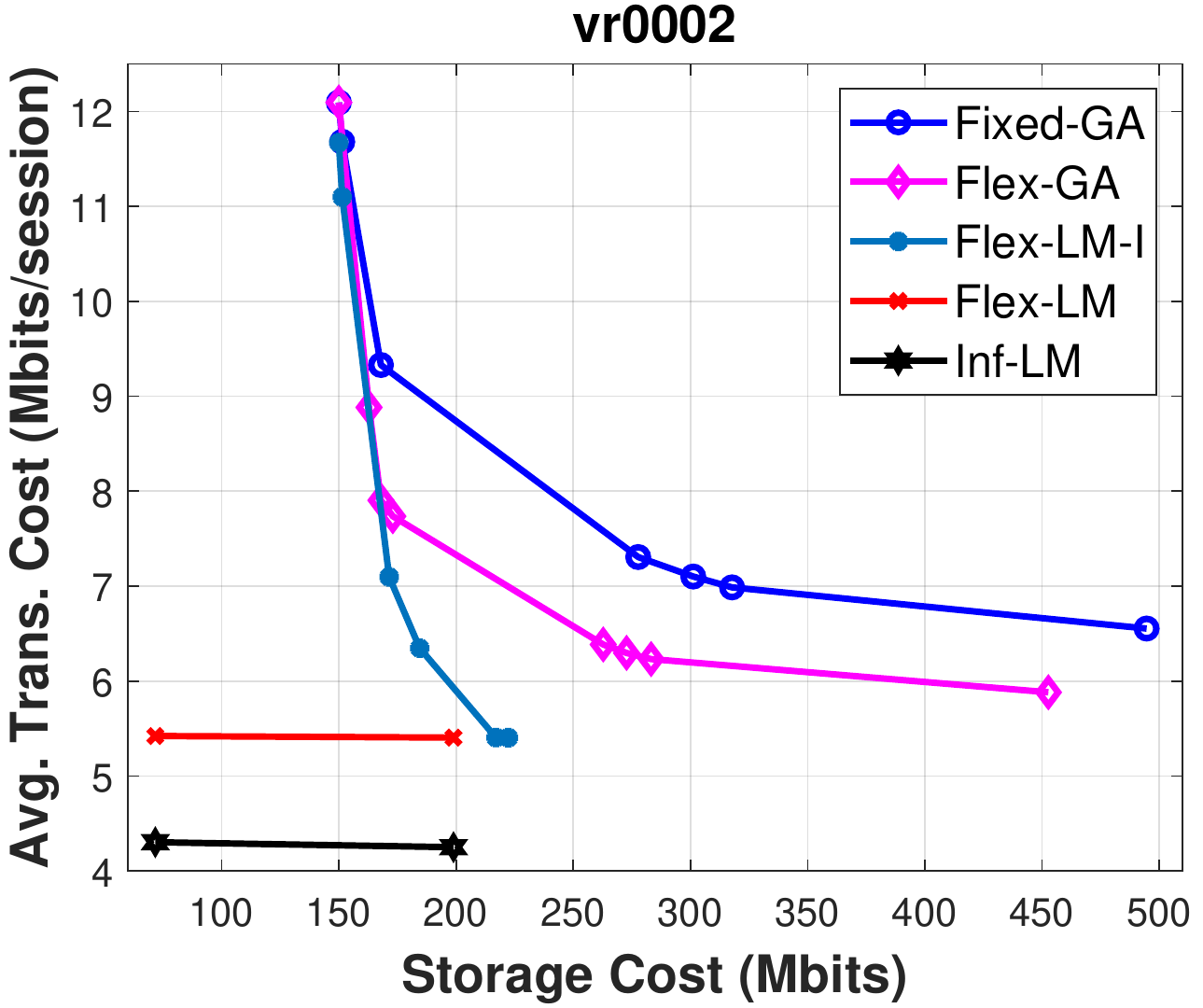} &
		\includegraphics[width=0.315\linewidth]{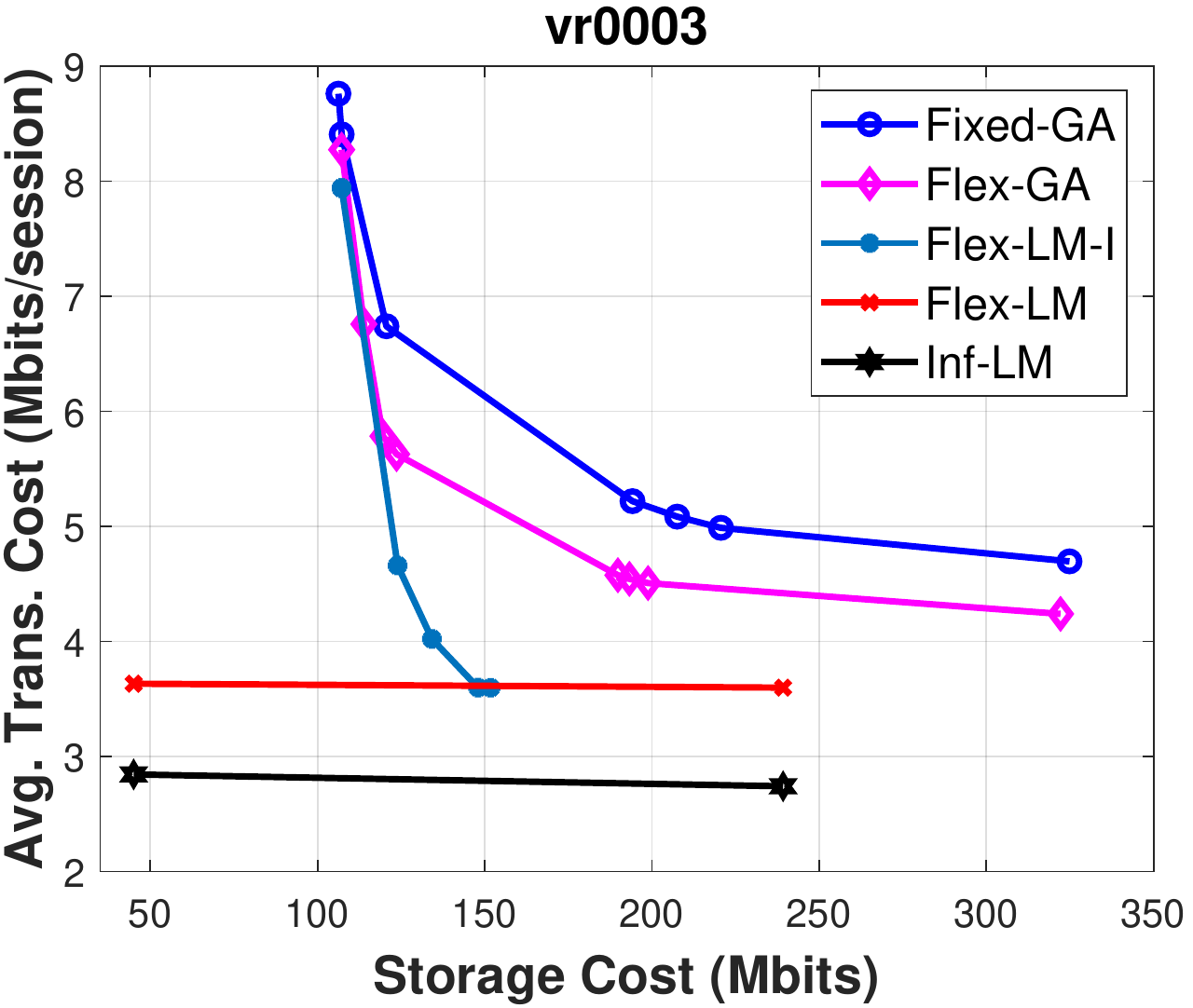} \\
		
		\footnotesize{(a) $360^\circ$, \textsl{vr0000}} &
		\footnotesize{(b) $360^\circ$, \textsl{vr0002}} &  
		\footnotesize{(c) $360^\circ$, \textsl{vr0003}} \\
		\\
		
		\includegraphics[width=0.317\linewidth]{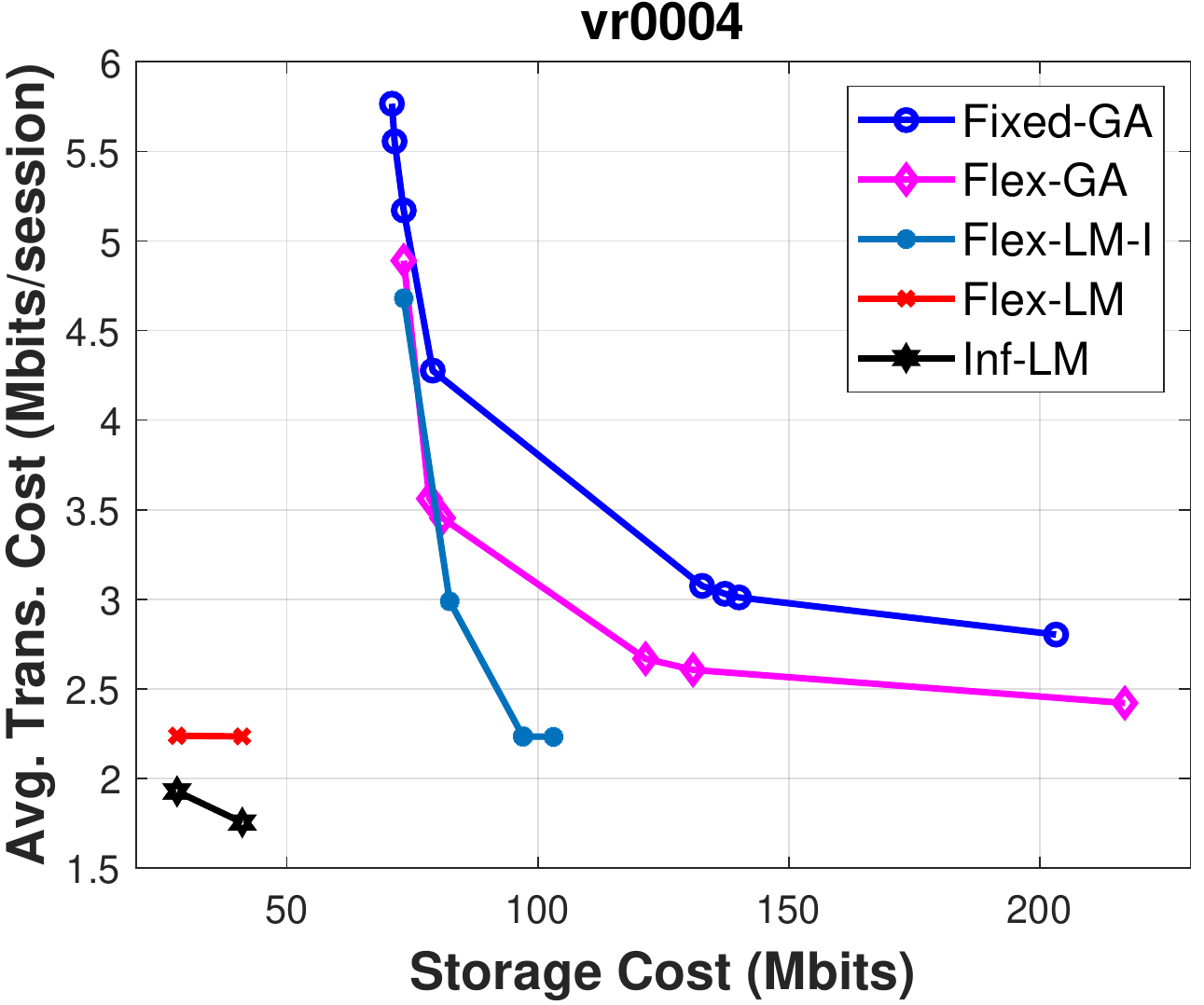} &
		\includegraphics[width=0.319\linewidth]{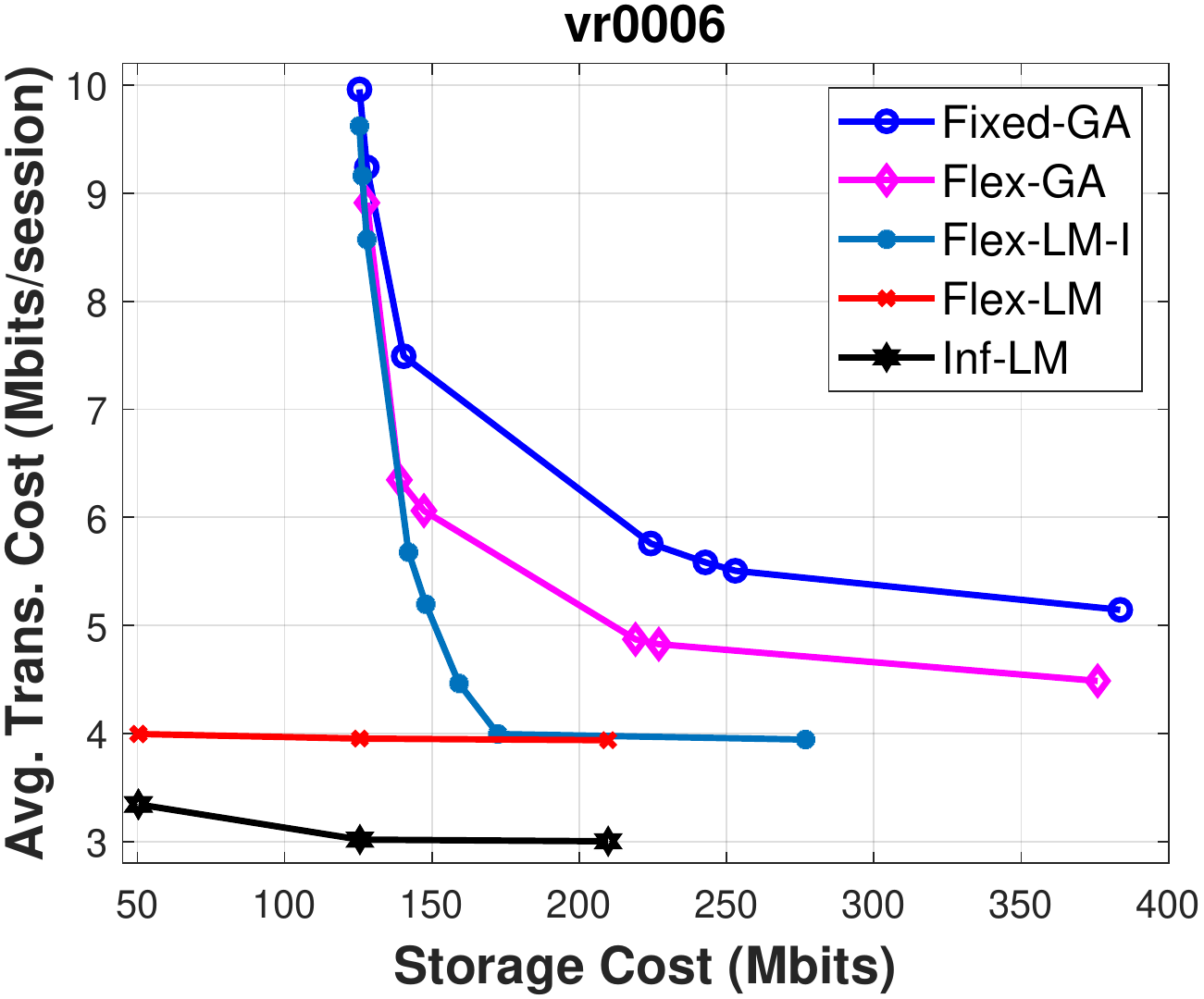} &
		\includegraphics[width=0.311\linewidth]{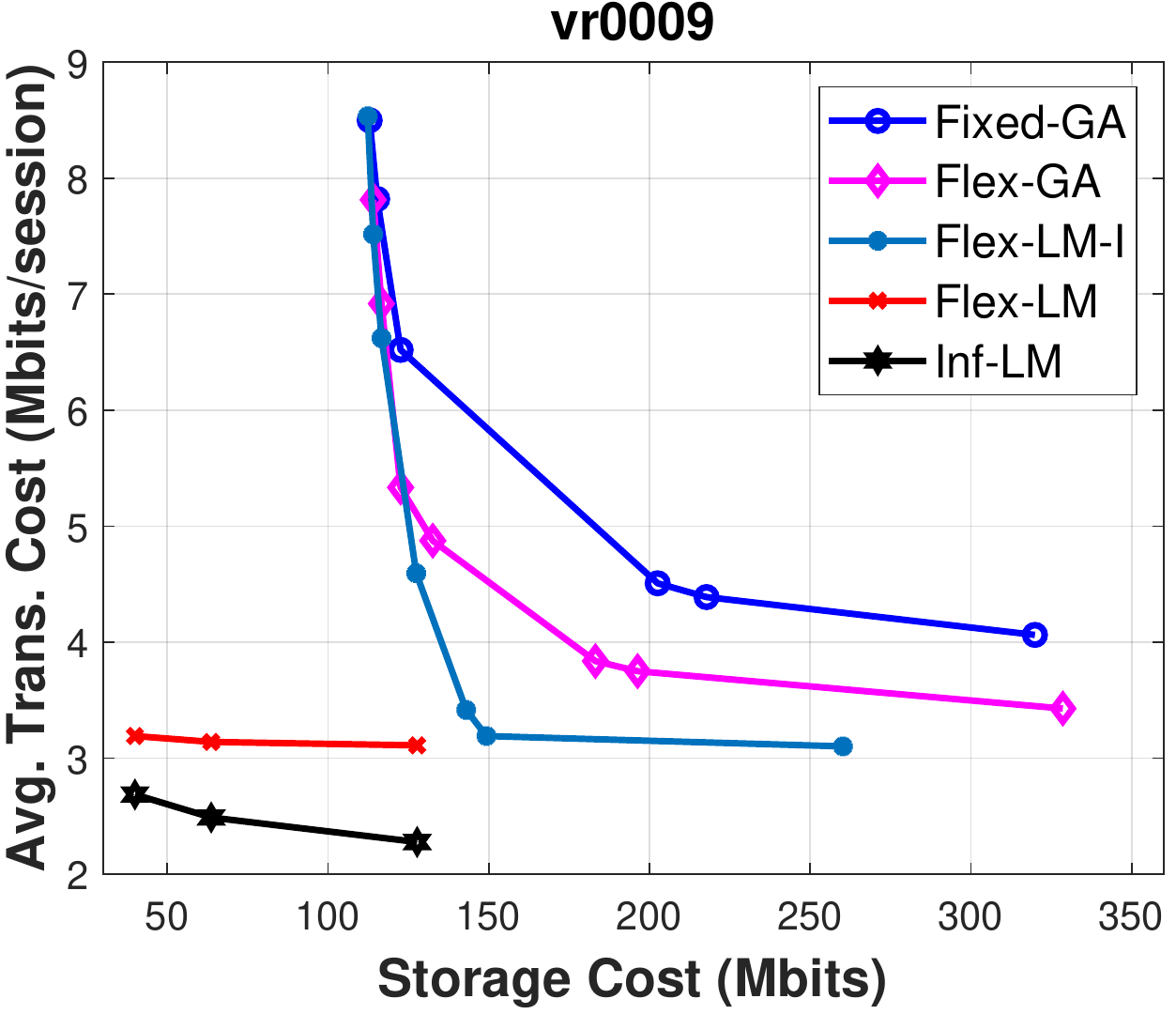} \\
		\footnotesize{(d) $360^\circ$, \textsl{vr0004}} &
		\footnotesize{(e) $360^\circ$, \textsl{vr0006}} &  
		\footnotesize{(f) $360^\circ$, \textsl{vr0009}} \\
		\\
		
		\includegraphics[width=0.31\linewidth]{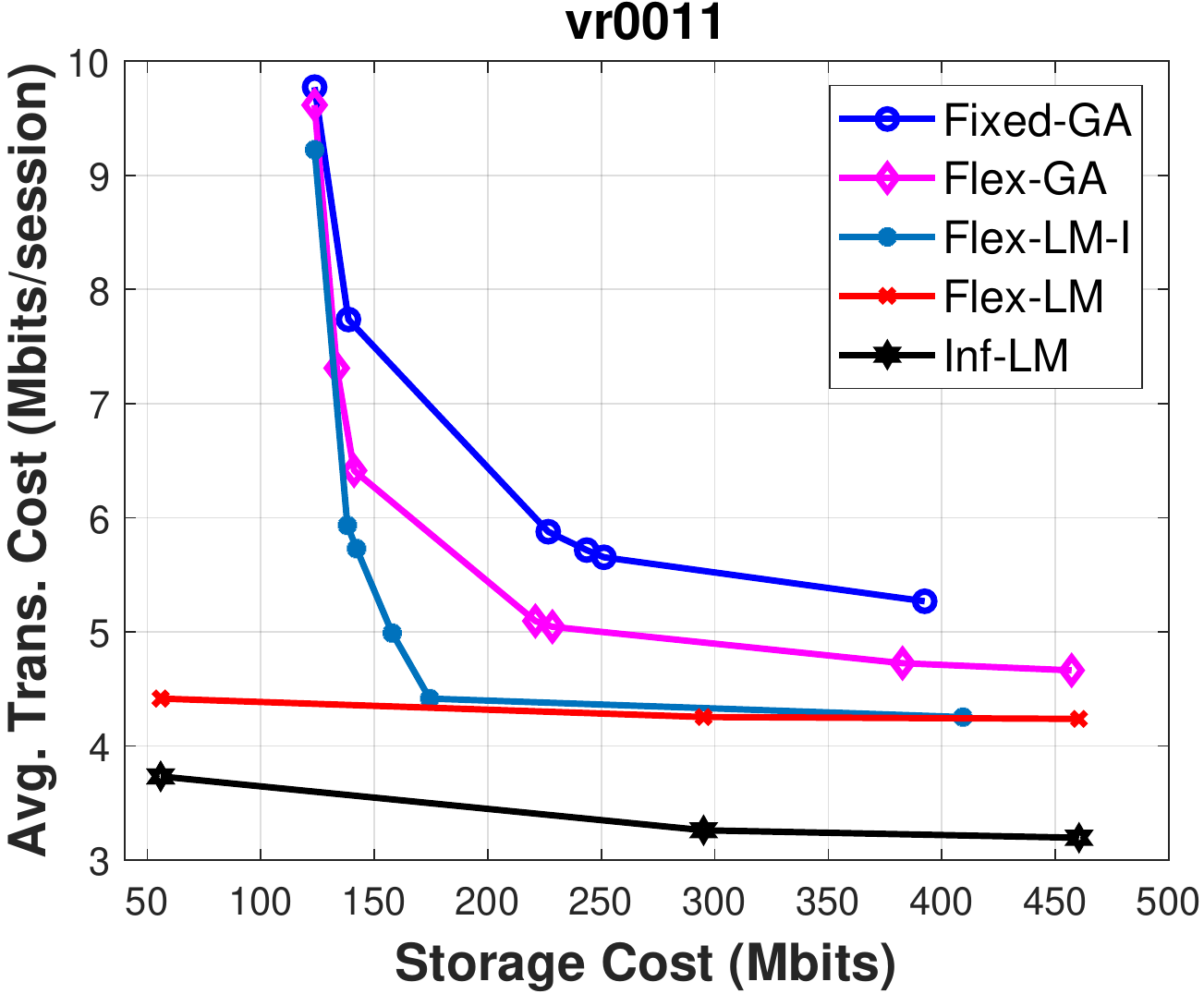} &
		\includegraphics[width=0.31\linewidth]{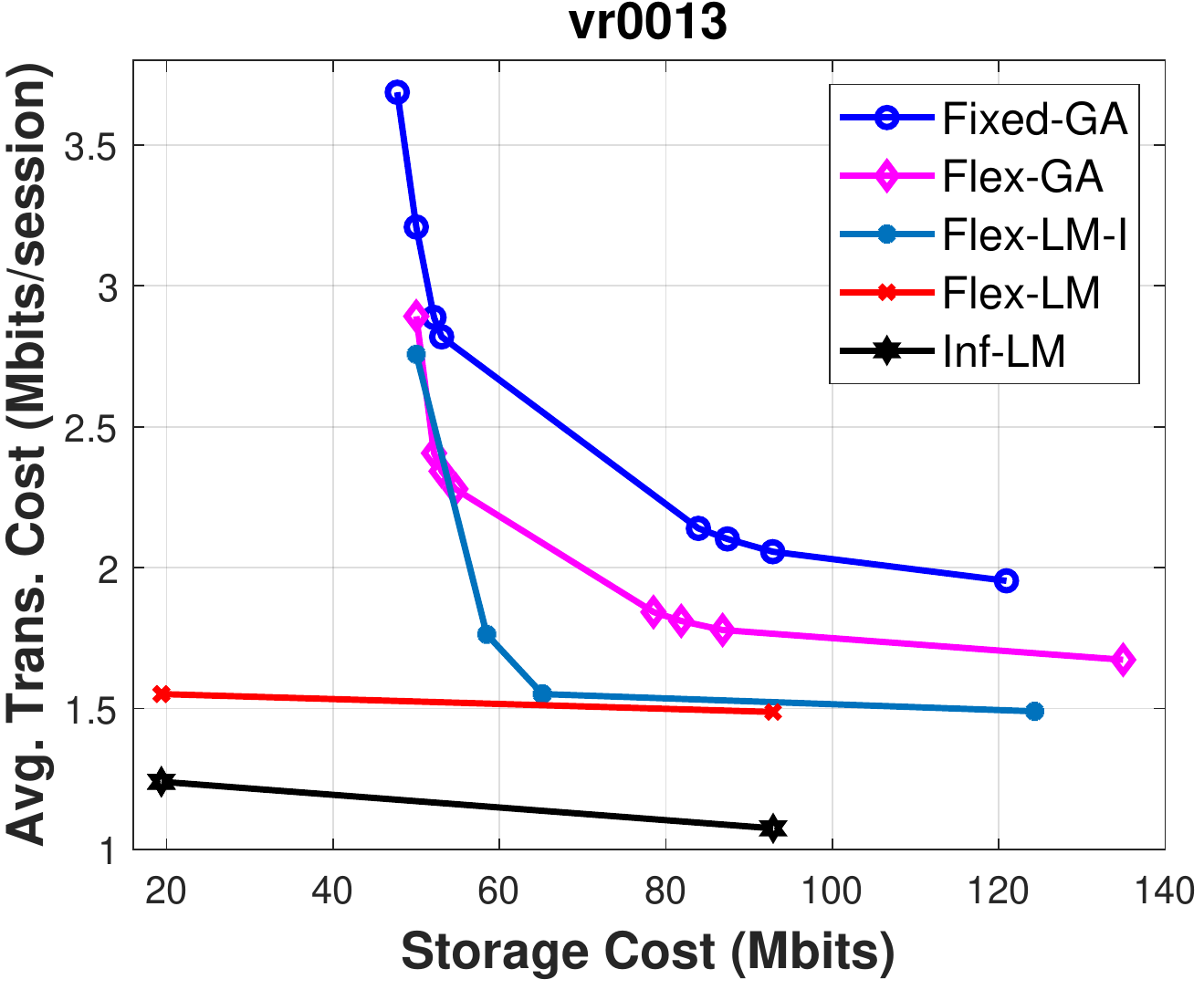} &
		\includegraphics[width=0.31\linewidth]{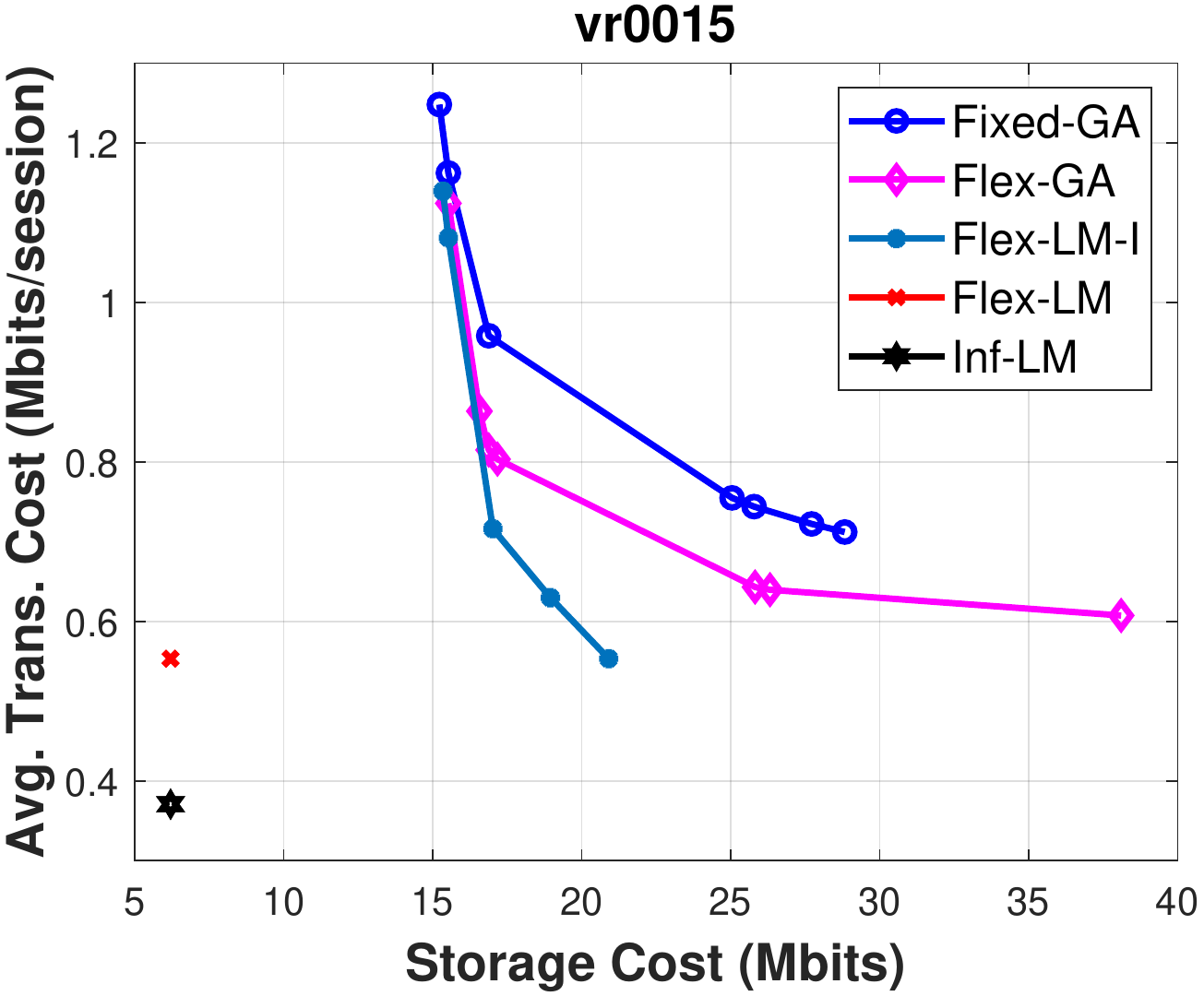} \\
		\footnotesize{(g) $360^\circ$, \textsl{vr0011}} &
		\footnotesize{(h) $360^\circ$, \textsl{vr0013}} &  
		\footnotesize{(i) $360^\circ$, \textsl{vr0015}} \\
		\\
		
		\includegraphics[width=0.315\linewidth]{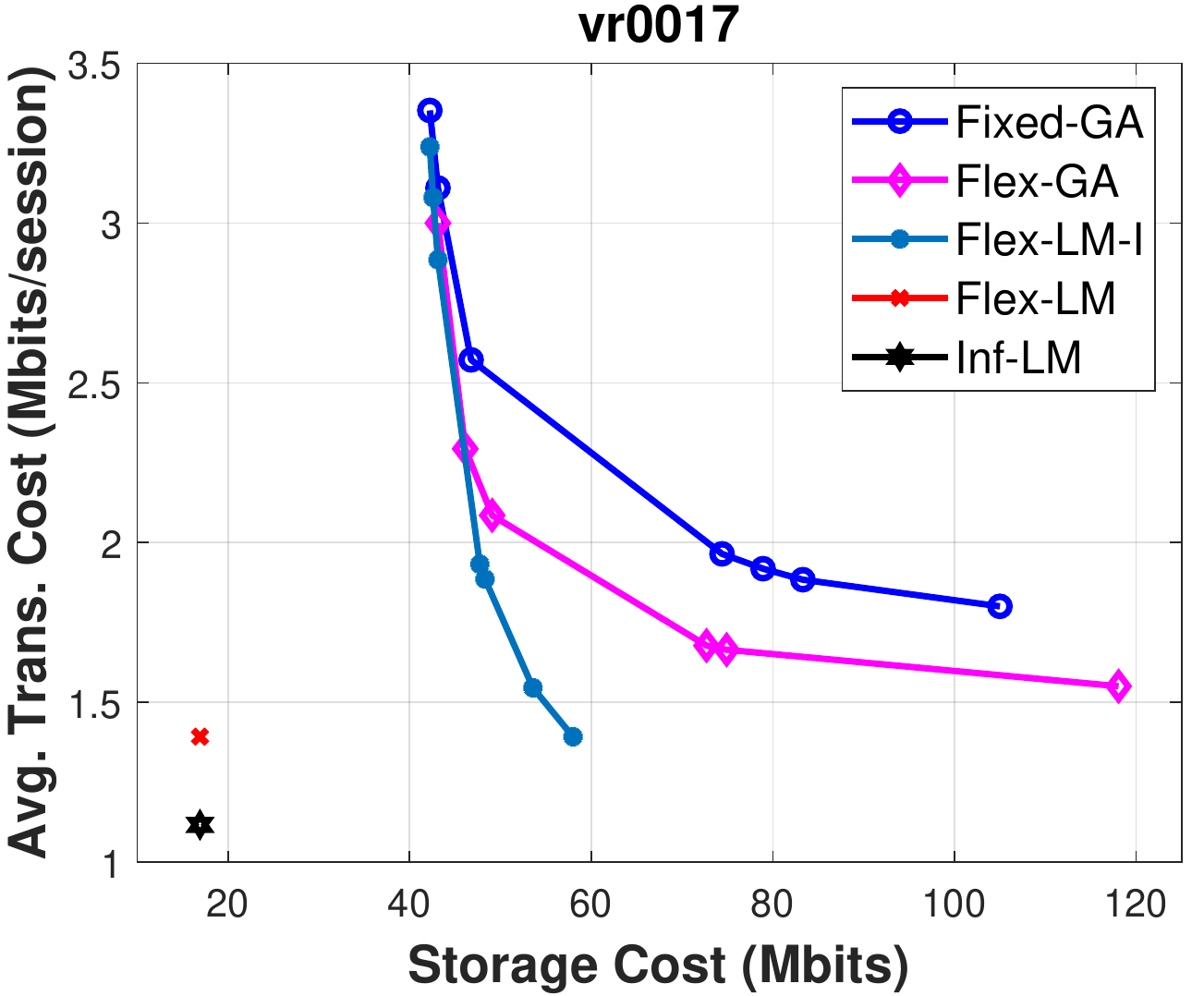} &
		\includegraphics[width=0.311\linewidth]{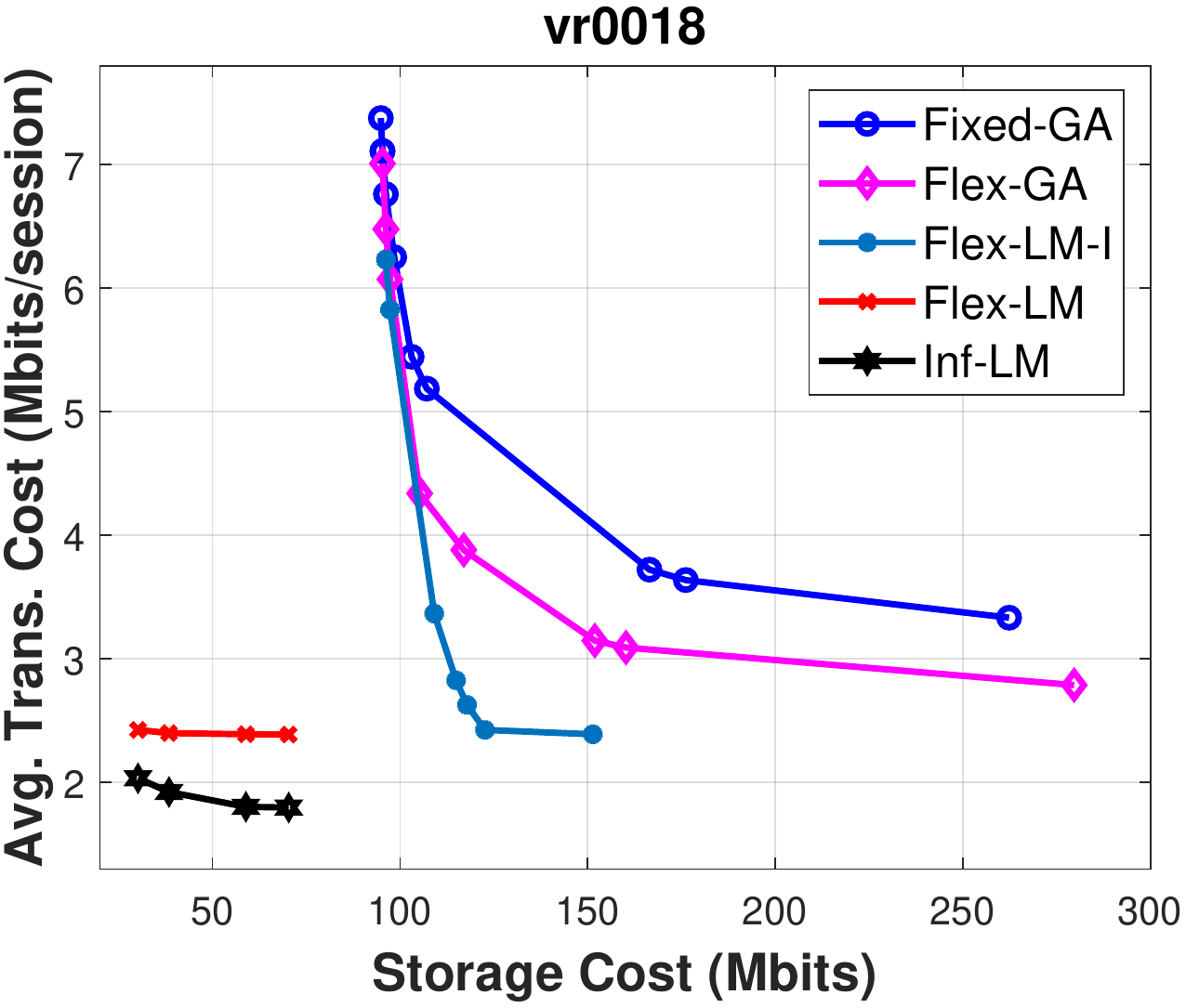} &
		\includegraphics[width=0.315\linewidth]{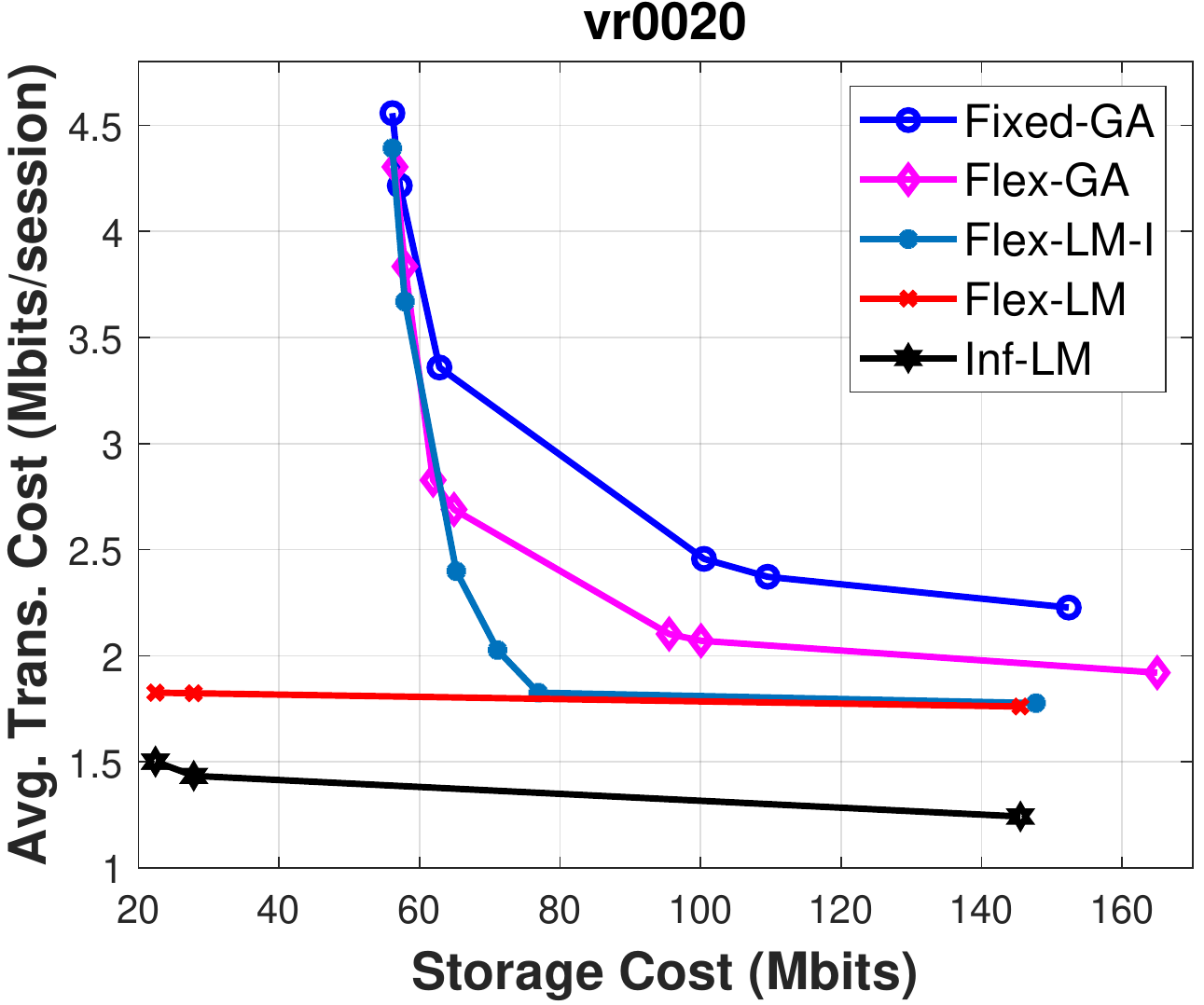} \\
		\footnotesize{(j) $360^\circ$, \textsl{vr0017}} &
		\footnotesize{(k) $360^\circ$, \textsl{vr0018}} &  
		\footnotesize{(l) $360^\circ$, \textsl{vr0020}} \\	
	
		\end{tabular}
	\end{center}
	\caption{The storage cost (Mbits) \textit{vs.} the average transmission cost (Mbits/session) for \360 images. The proposed method \texttt{Flex-LM} results in much smaller both storage and transmission cost than \texttt{Flex-GA} and \texttt{Fixed-GA}.}
\label{fig:RD_360}
\end{figure*}

We run experiments on \360 images similar to the experiments on LF images.
First, we show the storage cost \textit{vs}. the expected transmission cost curves (corresponding to Figure 11 in the main paper).
Results are shown in Fig.\;\ref{fig:RD_360}.
The names under the plots are correlated to the 22 \360 images shown in Fig.\;\ref{fig:saliency}.
The plots of the \texttt{Flex-LM} and \texttt{Inf-LM} for the images \textsl{vr0000}, \textsl{vr0015} and \textsl{vr0017} are dots instead of curves.
This is due to no more P-MDUs being added into the initialized structure no matter how small the $\lambda$ is.

\renewcommand{\tabcolsep}{.6pt}
\renewcommand\arraystretch{1}

\begin{figure*}[htb]
	\begin{center}
		\begin{tabular}{ccc}
		
		\includegraphics[width=0.32\linewidth]{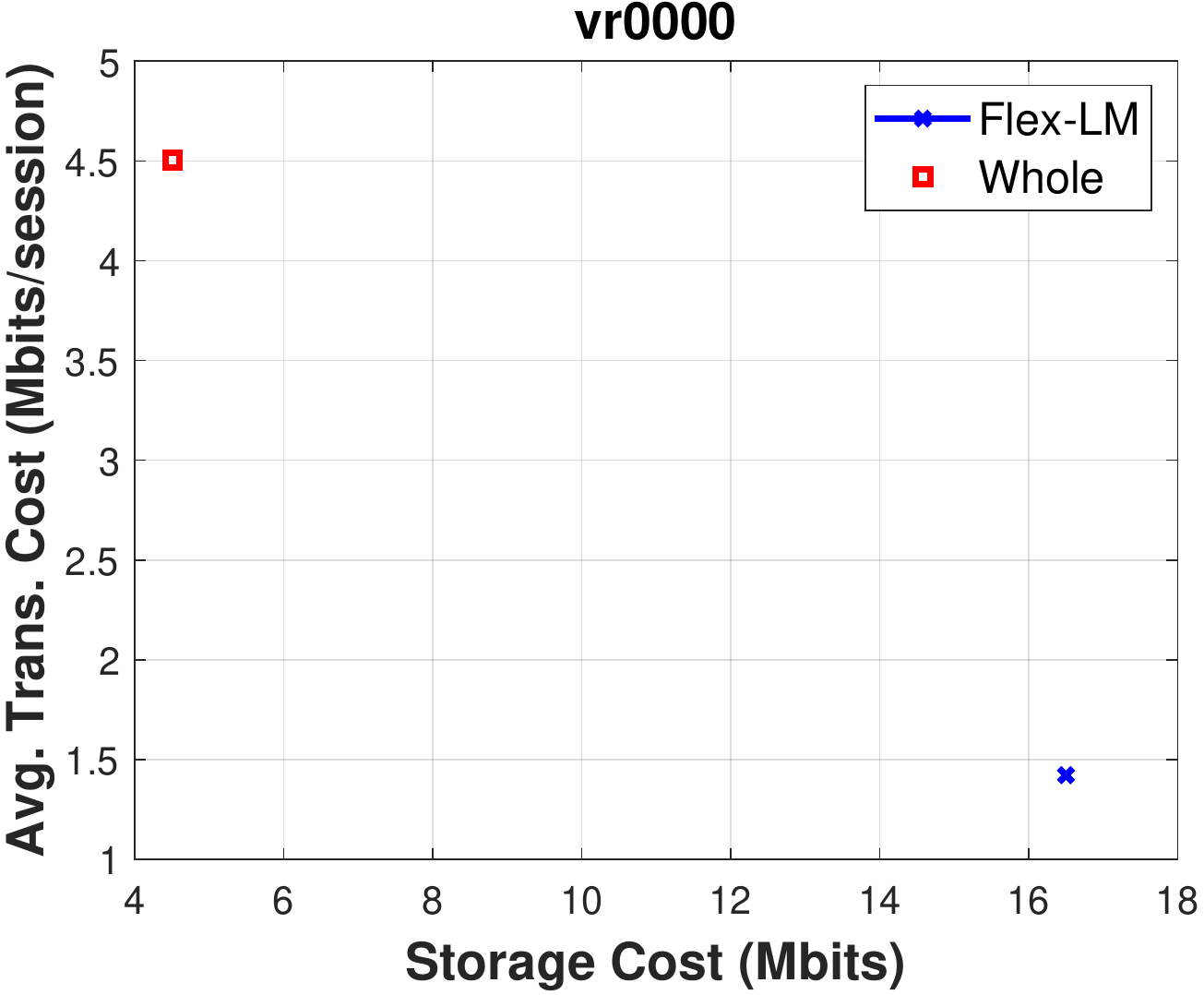} &
		\includegraphics[width=0.32\linewidth]{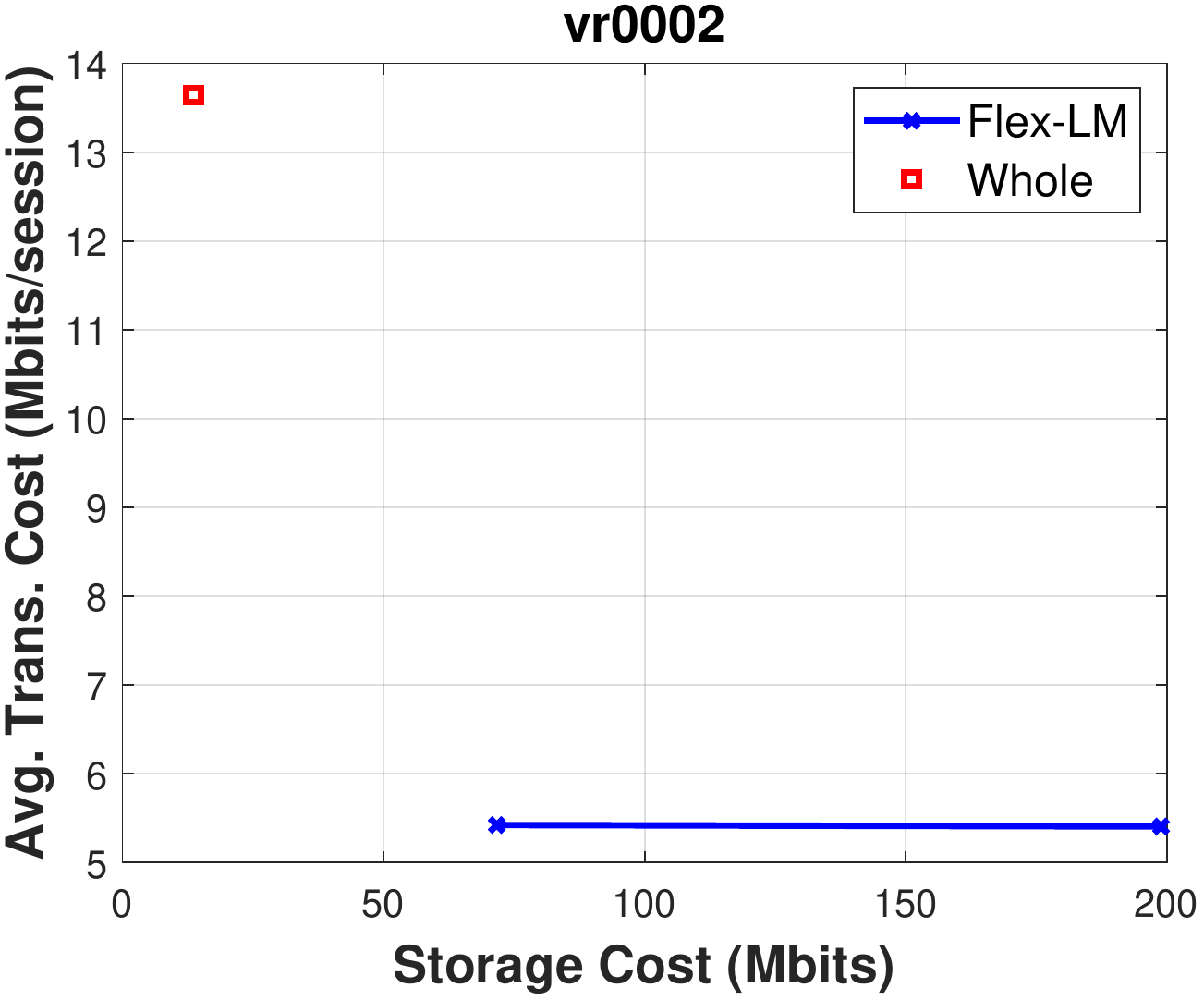} &
		\includegraphics[width=0.32\linewidth]{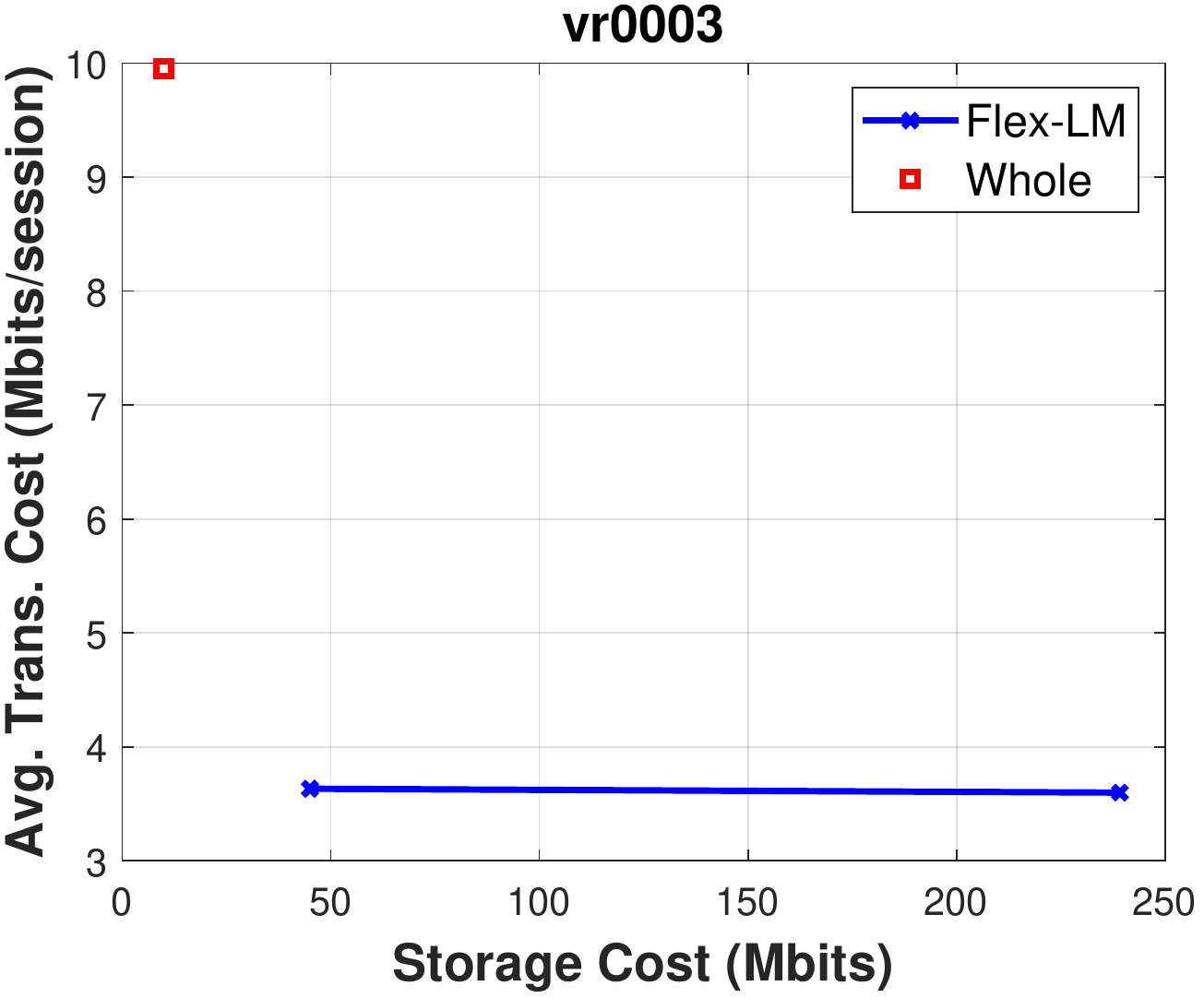} \\
		
		\footnotesize{(a) $360^\circ$, \textsl{vr0000}} &
		\footnotesize{(b) $360^\circ$, \textsl{vr0002}} &  
		\footnotesize{(c) $360^\circ$, \textsl{vr0003}} \\
		\\
		
		\includegraphics[width=0.312\linewidth]{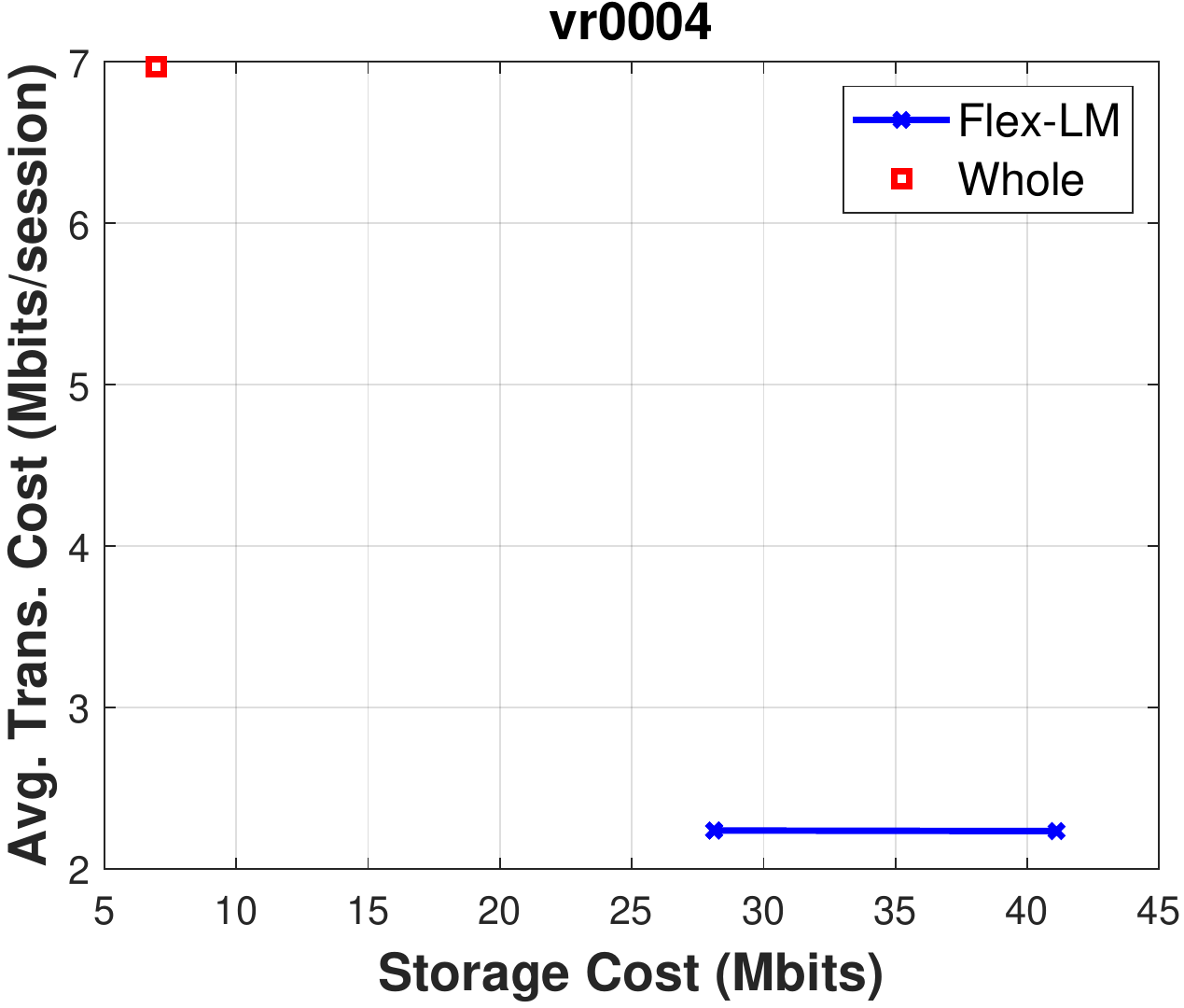} &
		\includegraphics[width=0.32\linewidth]{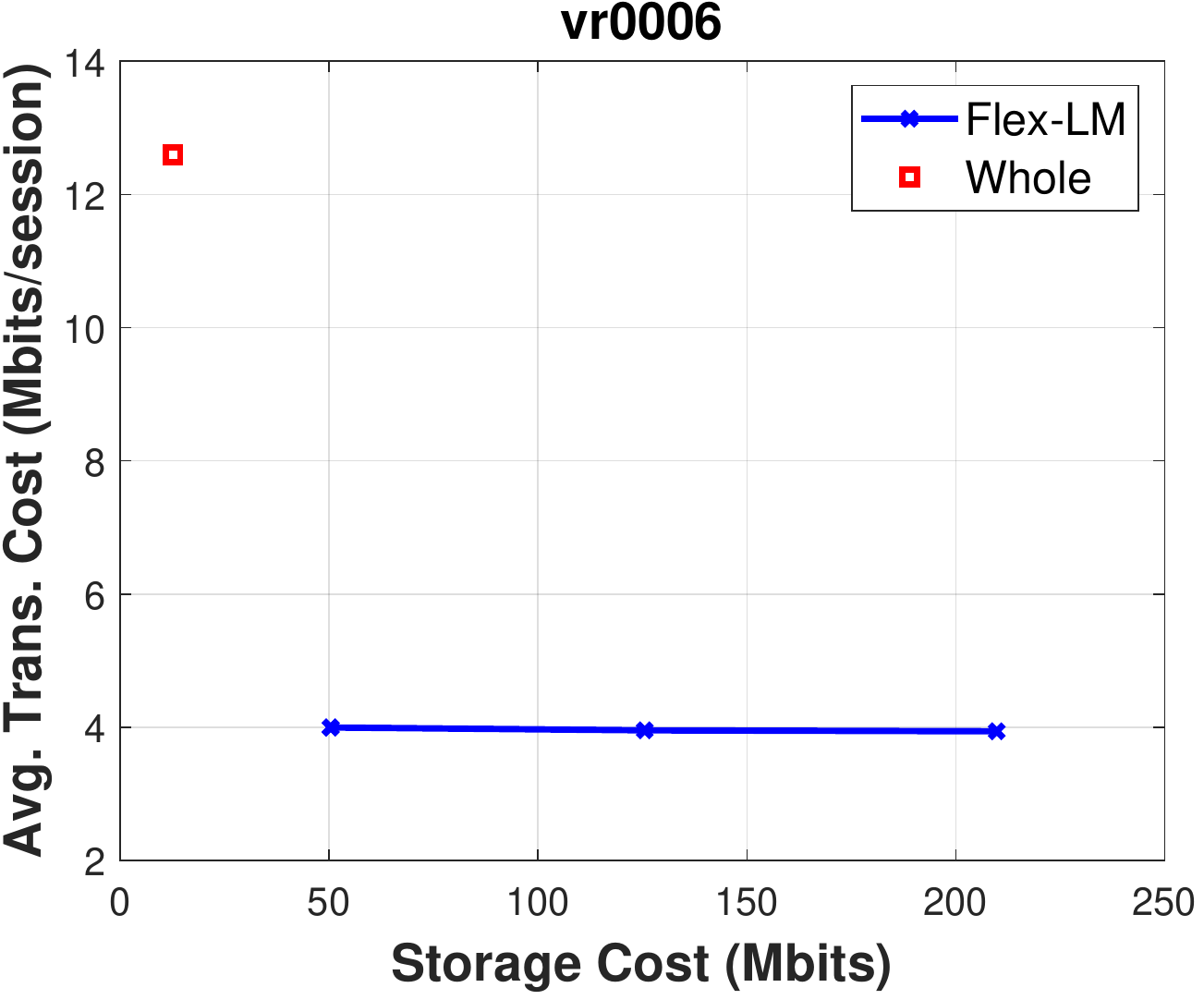} &
		\includegraphics[width=0.32\linewidth]{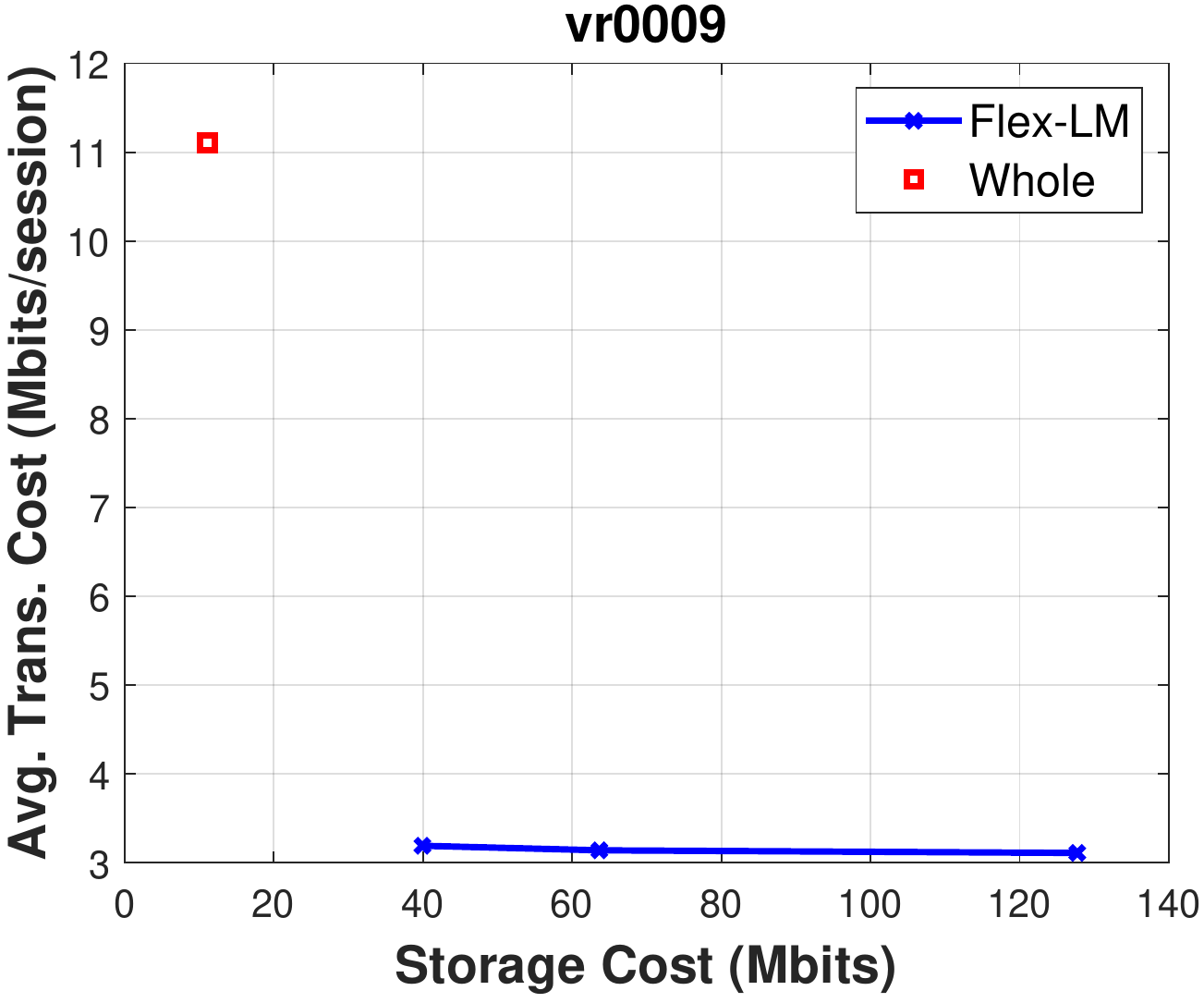} \\
		\footnotesize{(d) $360^\circ$, \textsl{vr0004}} &
		\footnotesize{(e) $360^\circ$, \textsl{vr0006}} &  
		\footnotesize{(f) $360^\circ$, \textsl{vr0009}} \\
		\\
		
		\includegraphics[width=0.32\linewidth]{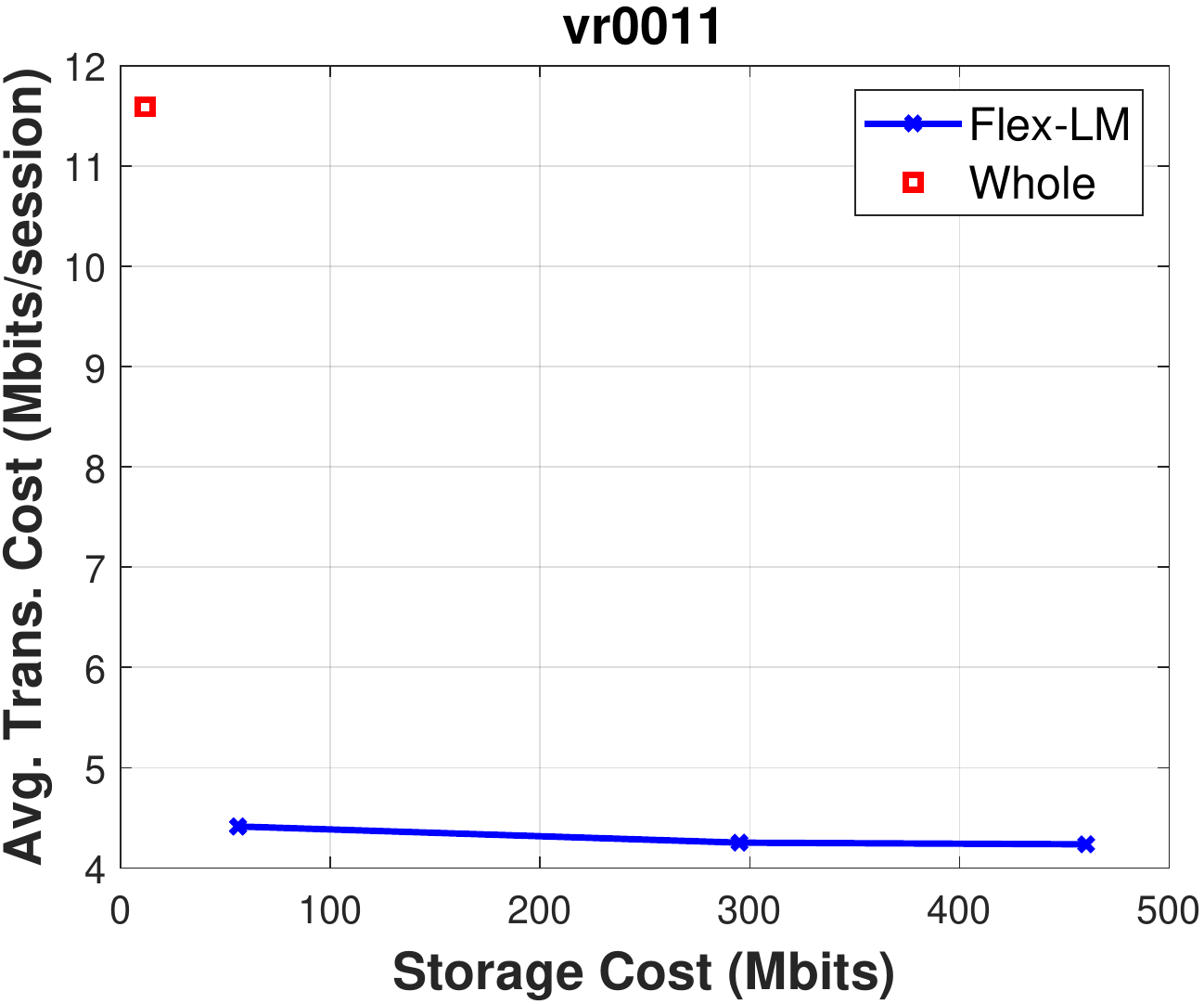} &
		\includegraphics[width=0.32\linewidth]{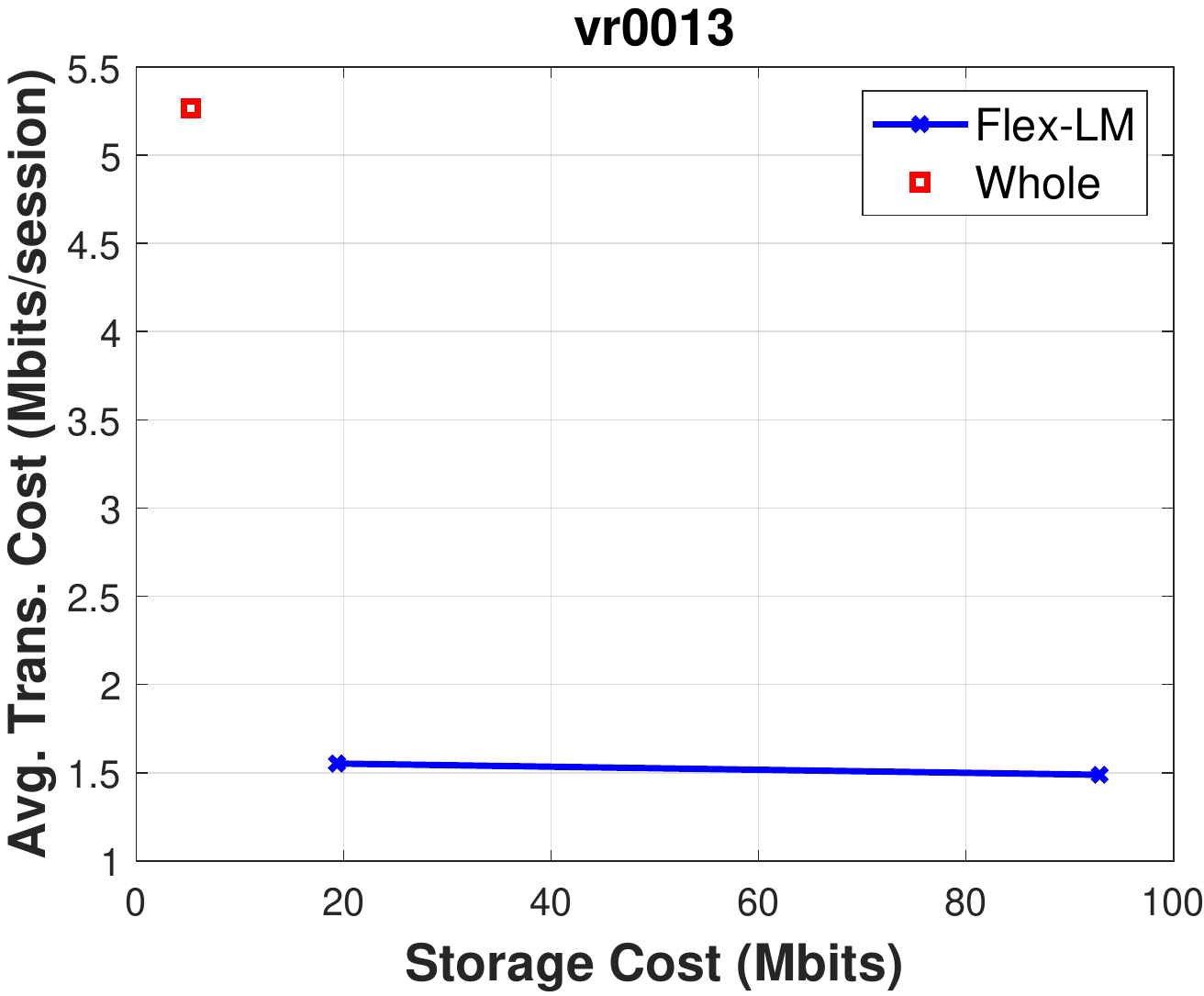} &
		\includegraphics[width=0.315\linewidth]{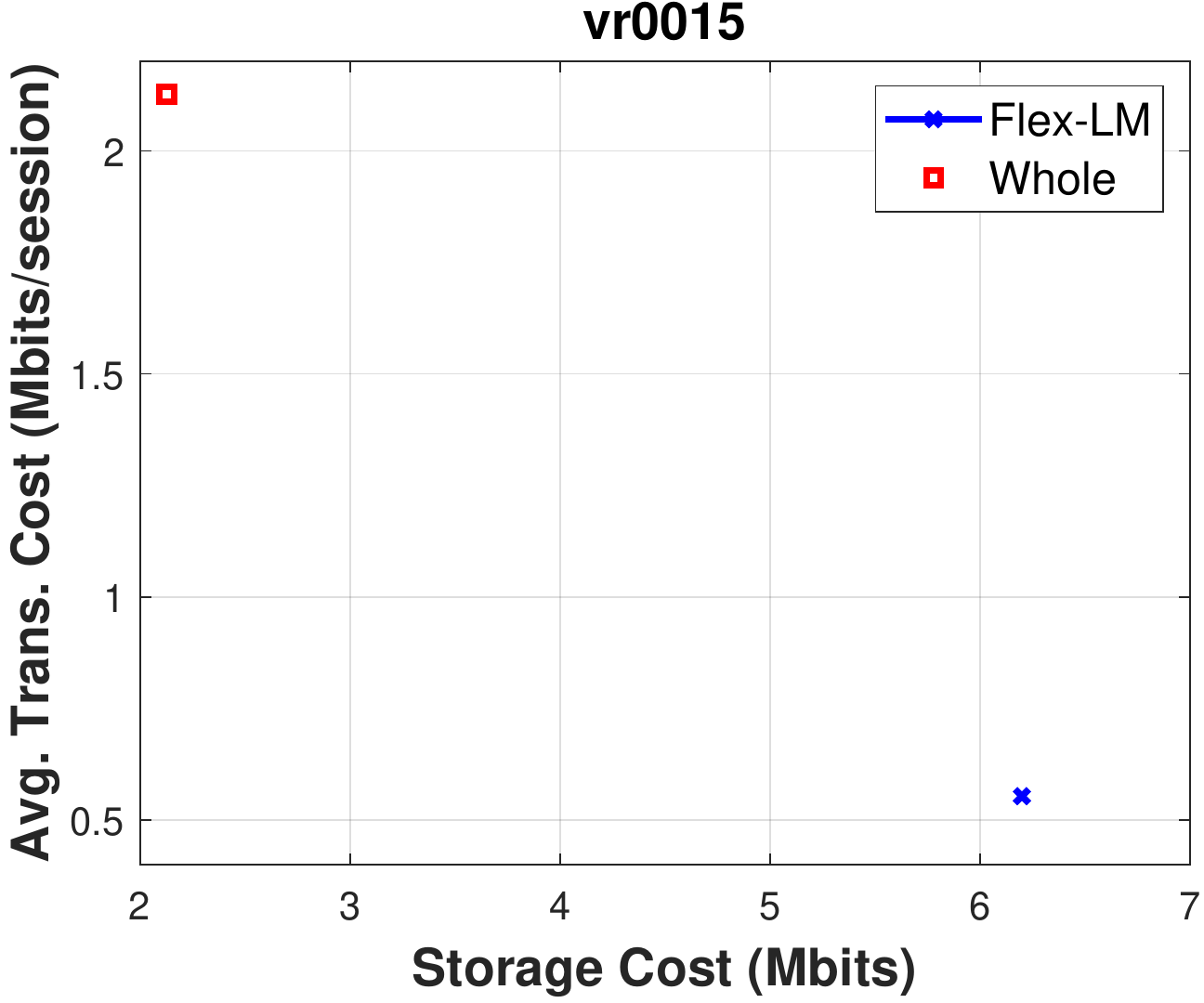} \\
		\footnotesize{(g) $360^\circ$, \textsl{vr0011}} &
		\footnotesize{(h) $360^\circ$, \textsl{vr0013}} &  
		\footnotesize{(i) $360^\circ$, \textsl{vr0015}} \\
		\\
		
		\includegraphics[width=0.32\linewidth]{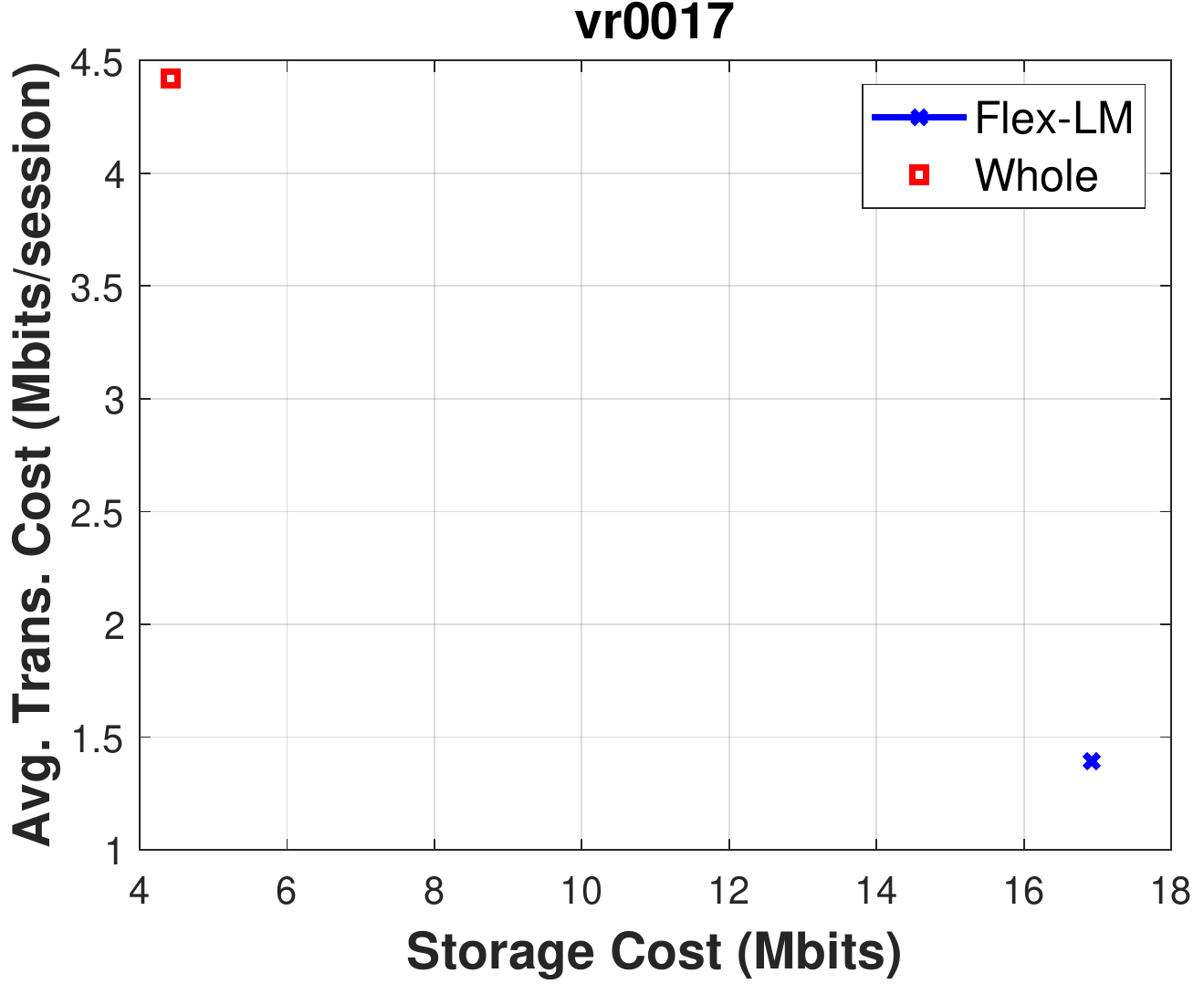} &
		\includegraphics[width=0.316\linewidth]{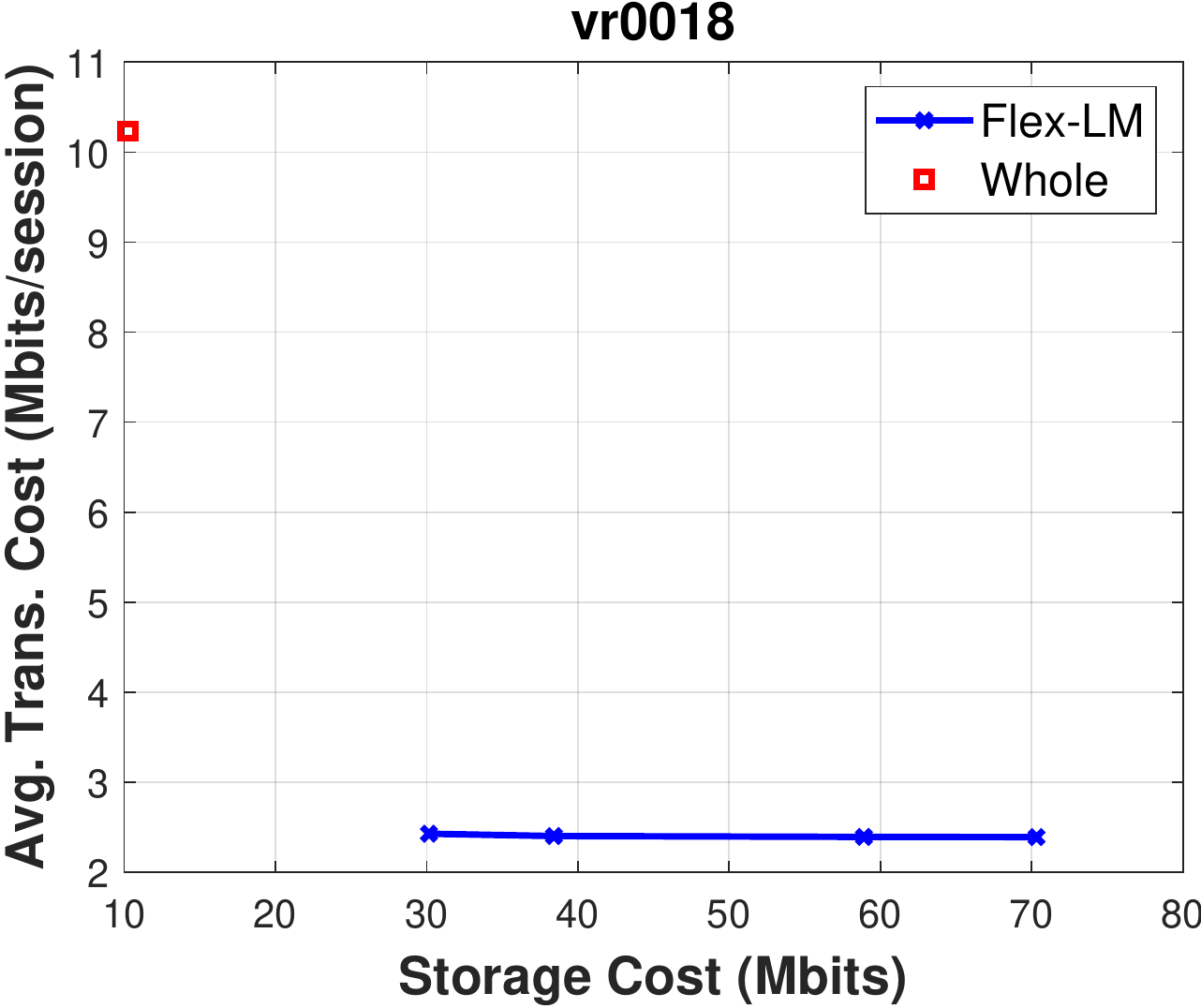} &
		\includegraphics[width=0.32\linewidth]{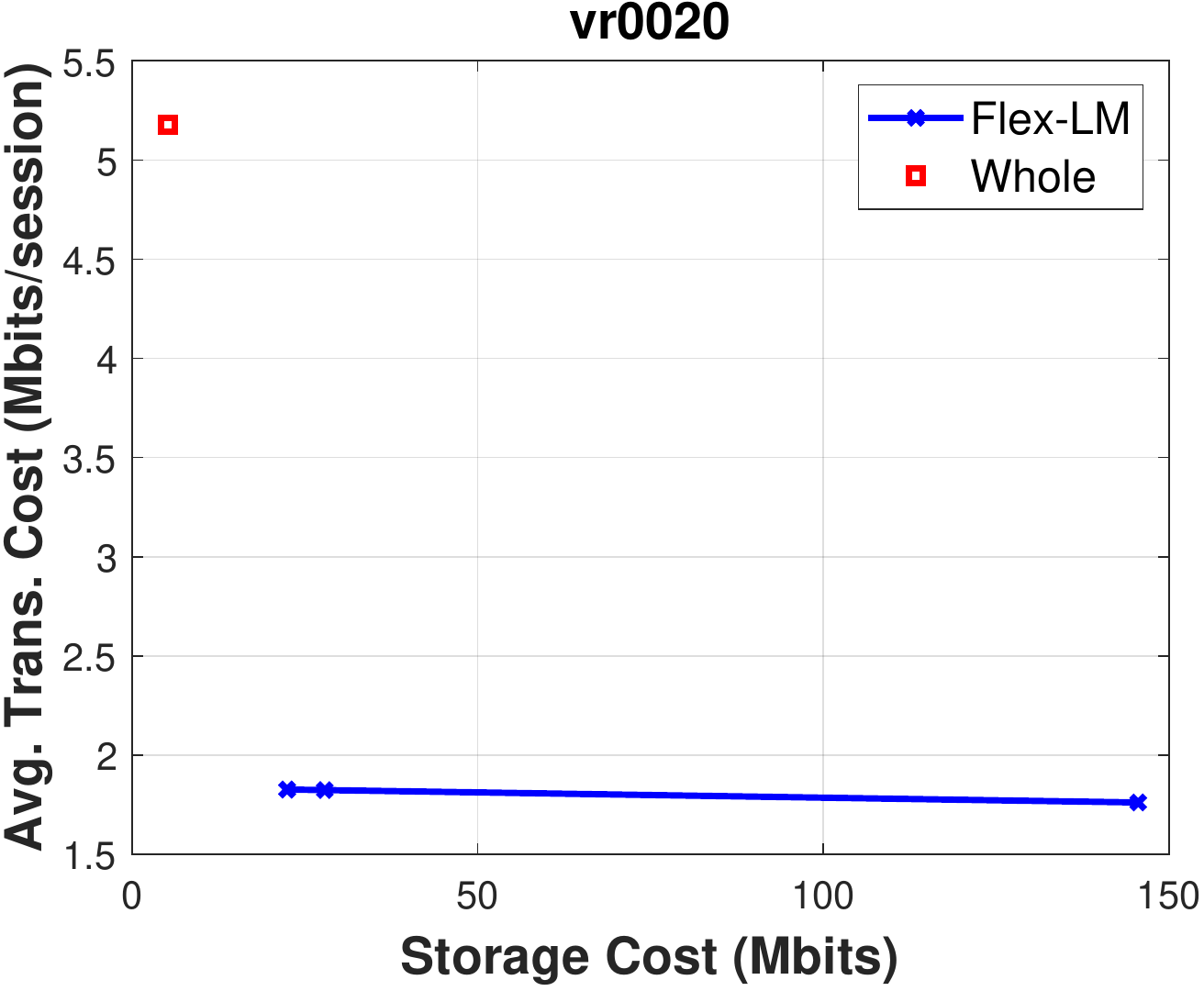} \\
		\footnotesize{(j) $360^\circ$, \textsl{vr0017}} &
		\footnotesize{(k) $360^\circ$, \textsl{vr0018}} &  
		\footnotesize{(l) $360^\circ$, \textsl{vr0020}} \\	
	
		\end{tabular}
	\end{center}
	\caption{Comparing \texttt{Flex-LM} with encoding and transmitting the entire \360 images directly (labelled as \texttt{Whole}). \texttt{Flex-LM} can reduce transmission cost by exploiting extra storage space.}
\label{fig:compare_360}
\end{figure*}

\begin{figure*}[htb]
	\begin{center}
		\begin{tabular}{ccc}
		
		\includegraphics[width=0.32\linewidth]{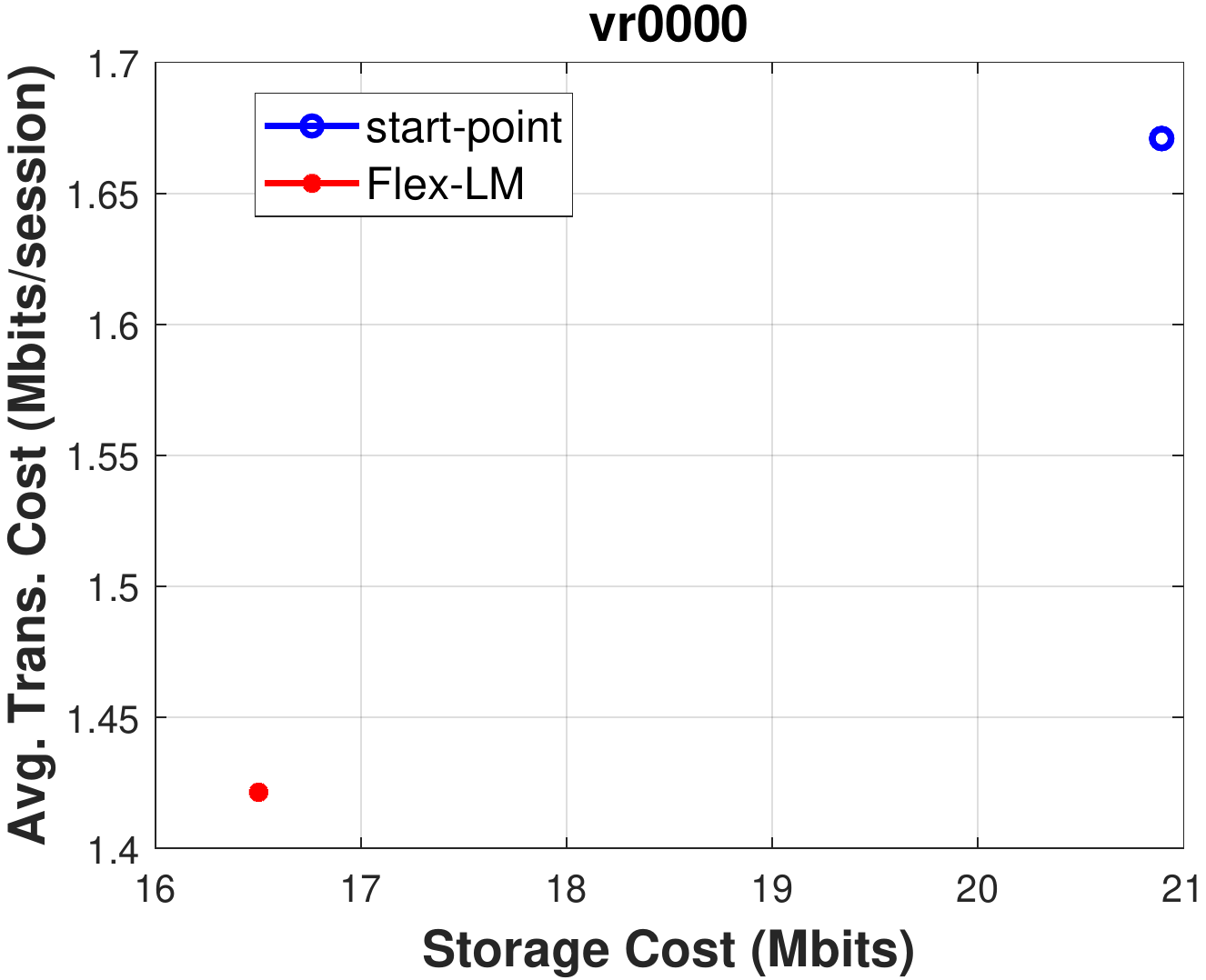} &
		\includegraphics[width=0.32\linewidth]{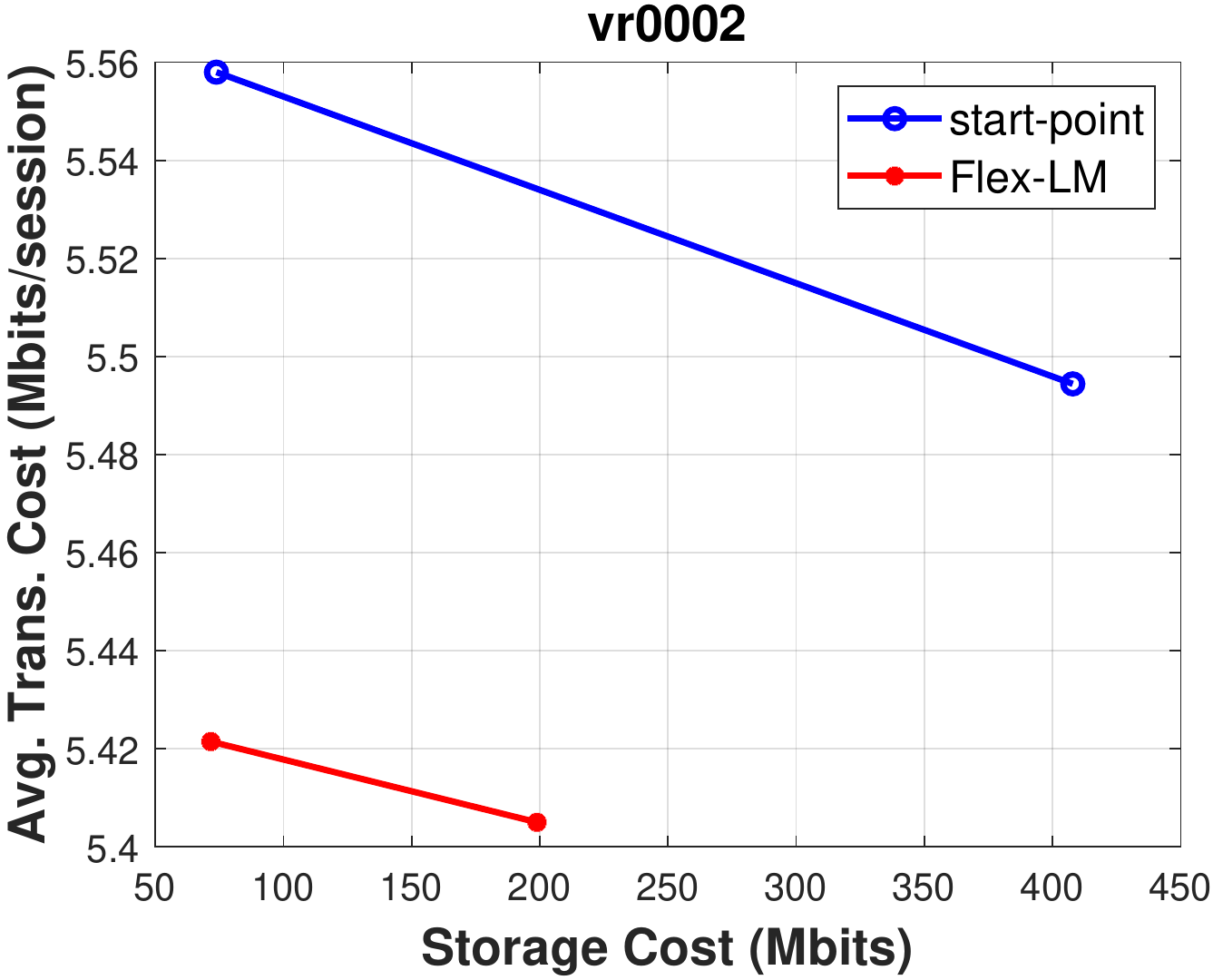} &
		\includegraphics[width=0.32\linewidth]{figures/vr0003_indS.pdf} \\
		
		\footnotesize{(a) $360^\circ$, \textsl{vr0000}} &
		\footnotesize{(b) $360^\circ$, \textsl{vr0002}} &  
		\footnotesize{(c) $360^\circ$, \textsl{vr0003}} \\
		\\
		
		\includegraphics[width=0.315\linewidth]{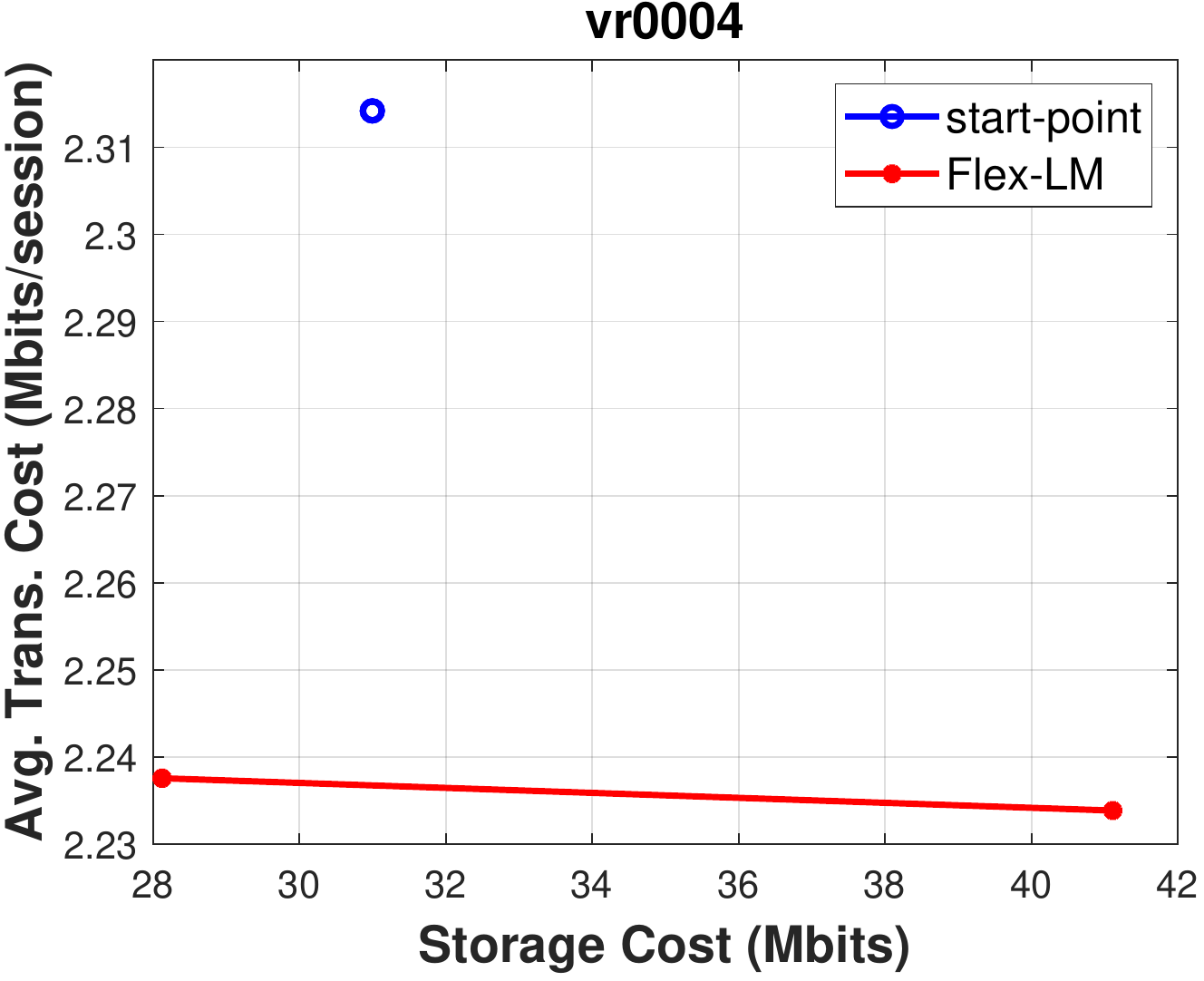} &
		\includegraphics[width=0.316\linewidth]{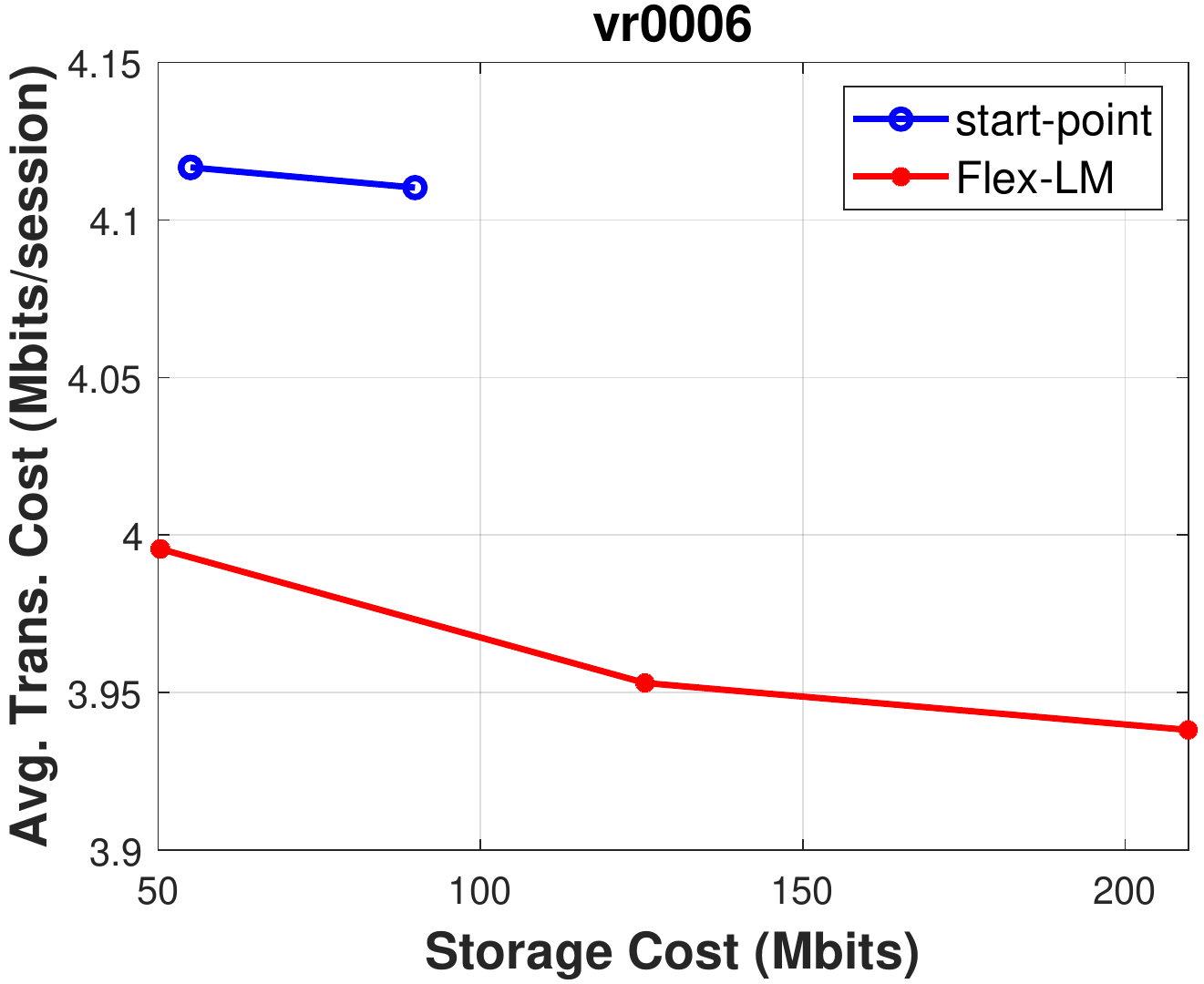} &
		\includegraphics[width=0.32\linewidth]{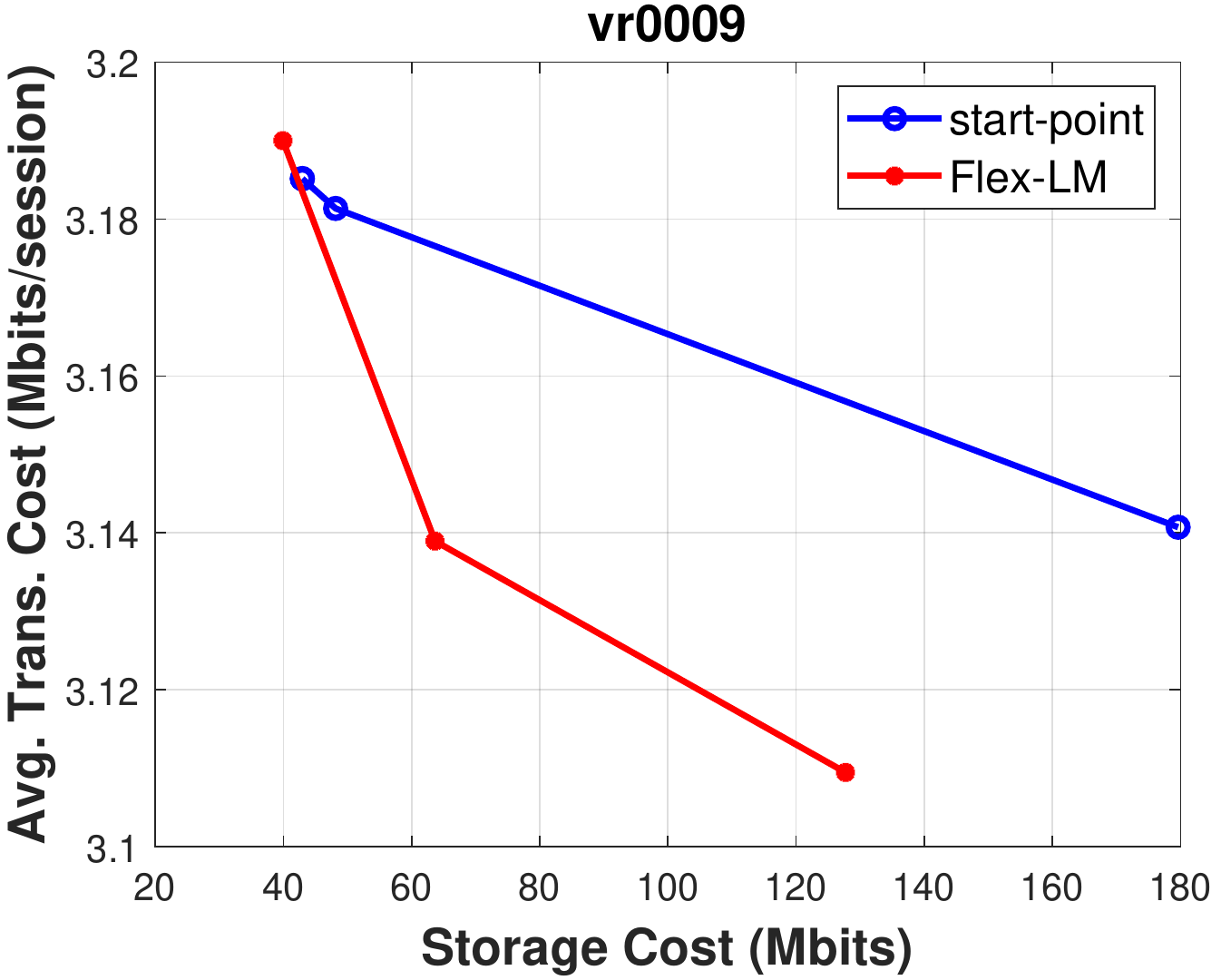} \\
		\footnotesize{(d) $360^\circ$, \textsl{vr0004}} &
		\footnotesize{(e) $360^\circ$, \textsl{vr0006}} &  
		\footnotesize{(f) $360^\circ$, \textsl{vr0009}} \\
		\\
		
		\includegraphics[width=0.32\linewidth]{figures/vr0011_indS.pdf} &
		\includegraphics[width=0.32\linewidth]{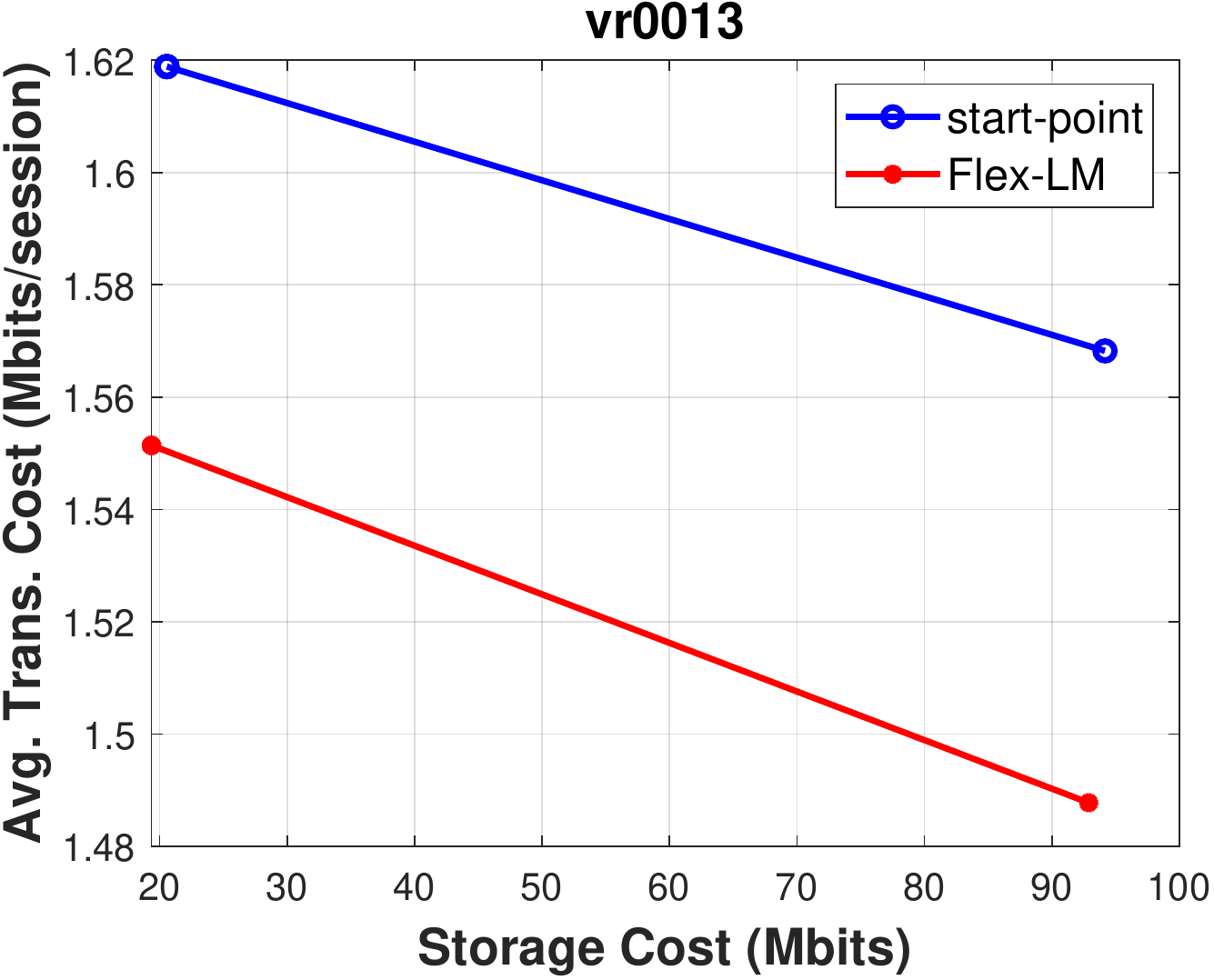} &
		\includegraphics[width=0.32\linewidth]{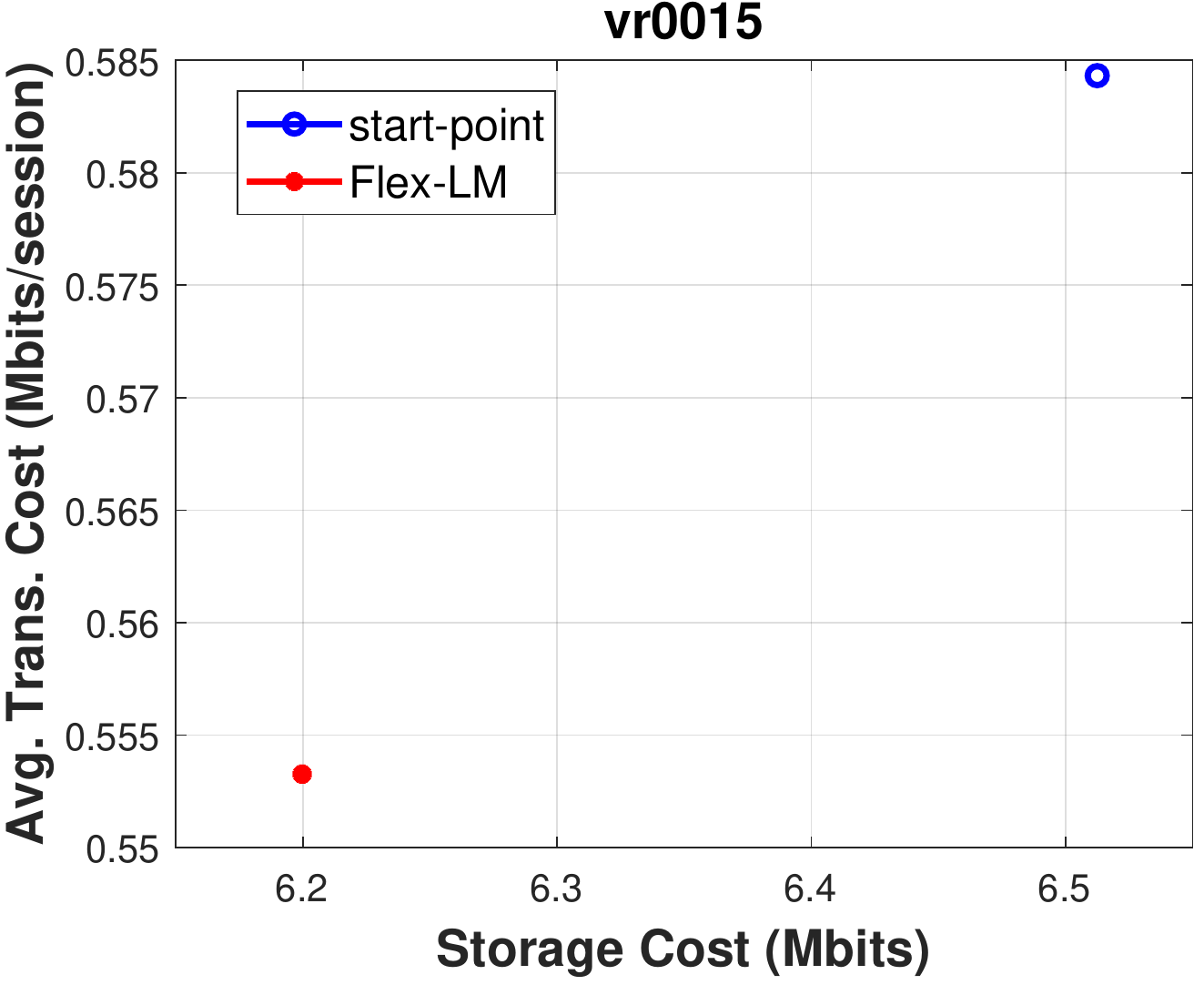} \\
		\footnotesize{(g) $360^\circ$, \textsl{vr0011}} &
		\footnotesize{(h) $360^\circ$, \textsl{vr0013}} &  
		\footnotesize{(i) $360^\circ$, \textsl{vr0015}} \\
		\\
		
		\includegraphics[width=0.32\linewidth]{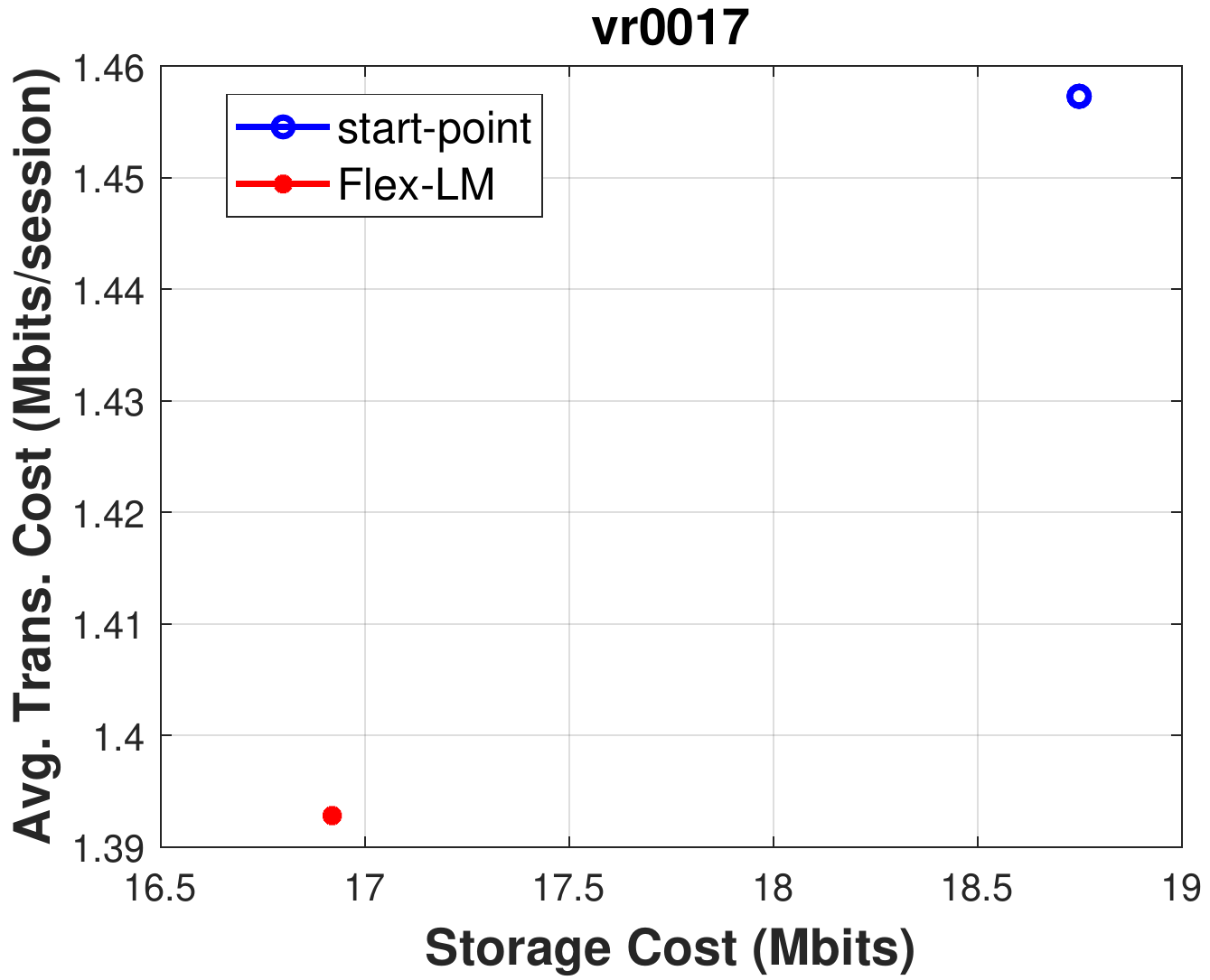} &
		\includegraphics[width=0.32\linewidth]{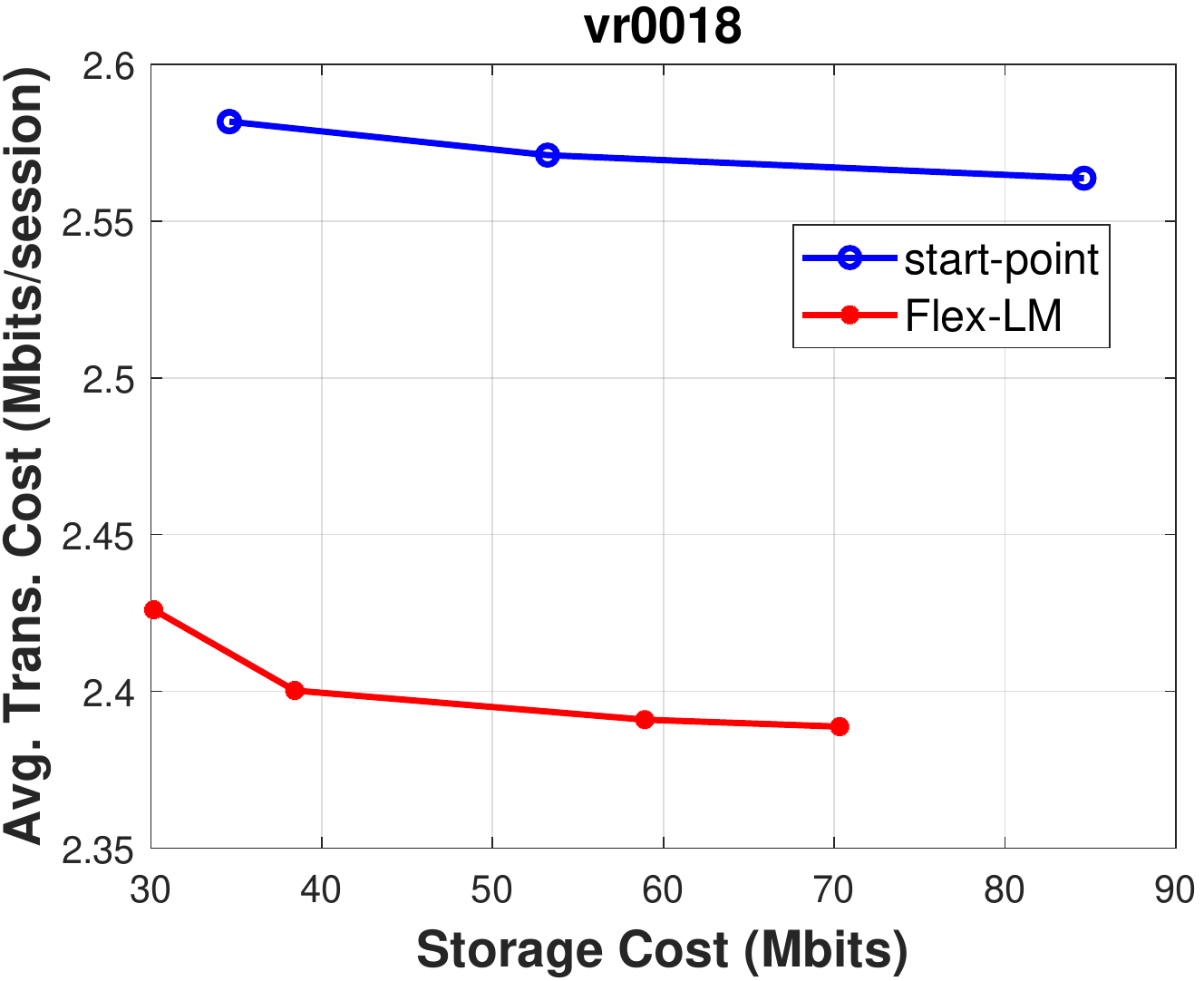} &
		\includegraphics[width=0.32\linewidth]{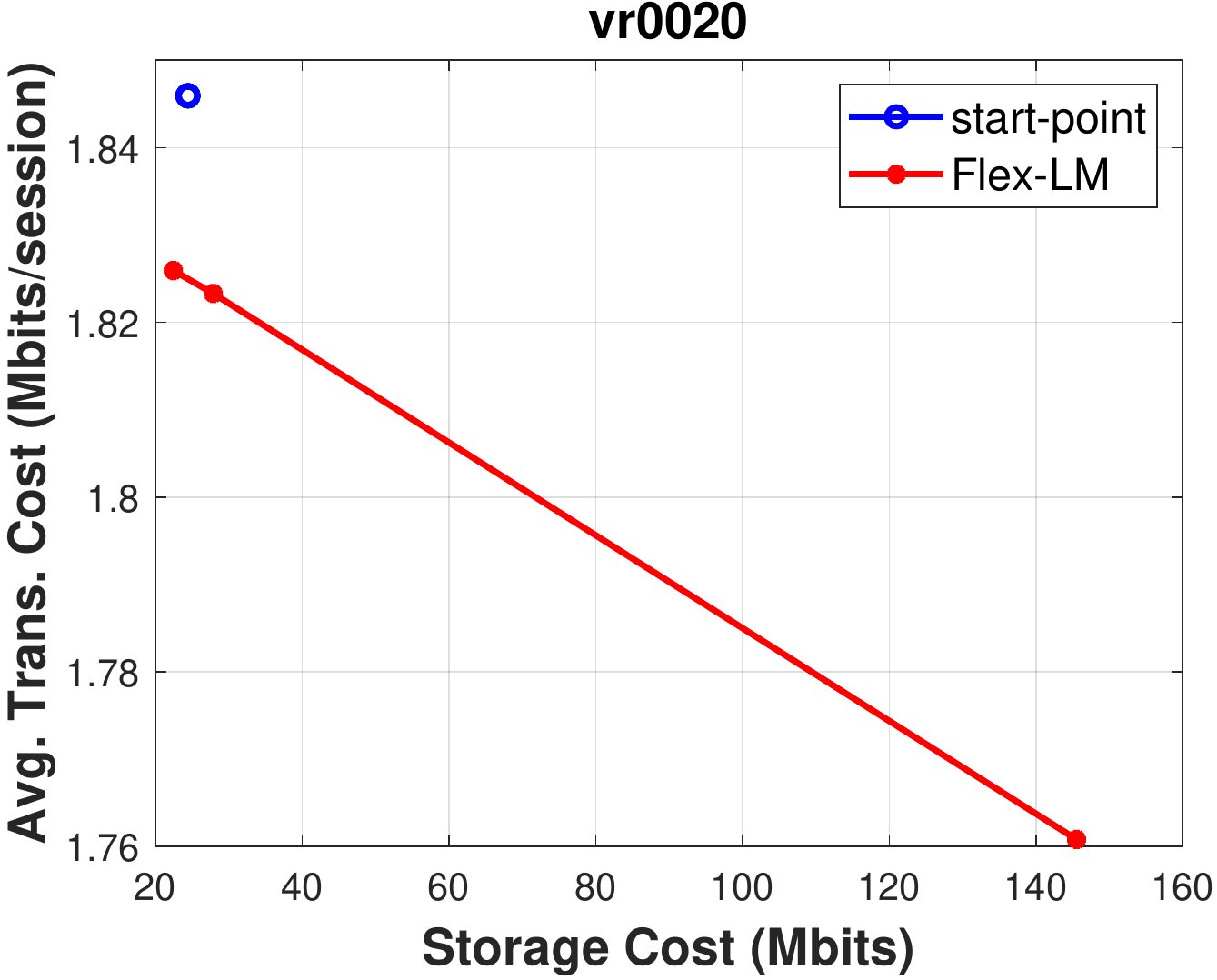} \\
		\footnotesize{(j) $360^\circ$, \textsl{vr0017}} &
		\footnotesize{(k) $360^\circ$, \textsl{vr0018}} &  
		\footnotesize{(l) $360^\circ$, \textsl{vr0020}} \\	
	
		\end{tabular}
	\end{center}
	\caption{The storage-transmission cost curves for $360^\circ$ images. \texttt{start point} means adopting the starting viewport as the landmark directly, where \texttt{Flex-LM} means using our proposed landmark insertion algorithm to find the optimal landmark. Given the same storage cost, with our optimal landmark insertion algorithm, the expected  transmission can be reduced.}
\label{fig:indS_360}
\end{figure*}

We next compare our proposed landmarking based MDU structure with a straightforward \360 image streaming strategy: encoding and transmitting the entire \360 images directly (corresponding to Figure 12 in the main paper).
Results are shown in Fig.\;\ref{fig:compare_360}.
Since the decrease of transmission cost of \texttt{Flex-LM} is much smaller than the value of \texttt{Whole}, the curves of \texttt{Flex-LM} look quite flat in the plots.
But one may see the trends that the transmission cost is reduced by increasing the storage cost from Fig.\;\ref{fig:RD_360} and Fig.\;\ref{fig:indS_360}.

\renewcommand{\tabcolsep}{.6pt}
\renewcommand\arraystretch{1}

In addition, we also show the effectiveness of the landmark insertion algorithm.
In the proposed \texttt{Flex-LM}, we optimally insert landmarks.
For comparison, we directly use the starting viewport as a landmark for \360 images.
We compare their storage-transmission cost curves.
Results are shown in Fig.\;\ref{fig:indS_360} (corresponding to Figure 13 in the main paper).
Interestingly, we find that for all the 12 testing \360 images, the optimal landmarks are not the starting viewports.
The curves for \texttt{start-point} are also quite short.
It is even a point in some cases (\textit{e.g., \textsl{vr0015}}).
This is also due to that the initialized MDU structure (with I-MDUs of landmarks and P-MDUs connecting landmarks to their neighbourhoods) is already very efficient.
Greedily adding P-MDUs will soon stop since the total cost would no longer be reduced. 
Given the same storage cost, \texttt{Flex-LM} with optimal inserted landmarks results in a smaller expected transmission cost.
This figure also clearly shows that the expected transmission cost decreases as the storage cost increases for \texttt{Flex-LM}.

\renewcommand{\tabcolsep}{.6pt}
\renewcommand\arraystretch{1}

\end{document}